\documentclass[aps,prd,
twocolumn,
showpacs,superscriptaddress,preprintnumbers,groupedaddress,nofootinbib,longbibliography]{revtex4-1}  

\usepackage{supertabular}
\usepackage{placeins}
\usepackage{afterpage}

\usepackage{graphicx}
\usepackage{xcolor}
\usepackage{amsmath}
\usepackage{amsfonts}
\usepackage{amssymb}
\usepackage{amsthm}
\usepackage{fixmath}
\usepackage{braket}
\usepackage{slashed}
\usepackage{fontenc}
\usepackage{dsfont}
\usepackage{bm}
\usepackage{lineno} 

\allowdisplaybreaks
\renewcommand{\bm}{\mathbold}
\newcommand{\nn}{\nonumber}

\newcommand{\abs}[1]{|#1|}
\newcommand{\as}{\alpha_{s}}
\newcommand{\aqq}{\alpha_{Q\bar Q}}
\newcommand{\Fqq}{F_{Q\bar Q}}

\newcommand{\Fs}{F_S}
\newcommand{\fs}{f_S}
\newcommand{\Vs}{V_S}
\newcommand{\Fq}{F_Q}
\newcommand{\fq}{f_Q}
\newcommand{\qbq}{Q\bar Q}
\newcommand{\md}{m_D}
\newcommand{\tc}{T_c}
\newcommand{\nc}{N_c}
\newcommand{\cf}{C_F}
\newcommand{\cdf}{\chi^2/{\rm d.o.f.}}

\newcommand{\ileq}[1]{\begin{equation}#1\end{equation}}
\newcommand{\ilal}[1]{\vskip-3ex\begin{align}#1\end{align}}
\newcommand{\bmat}{\left(\begin{array}}
\newcommand{\emat}{\end{array}\right)}

\definecolor{red}      {rgb}{0.8,0.0,0.0}
\definecolor{green}    {rgb}{0.0,0.6,0.0}
\definecolor{darkblue} {rgb}{0.0,0.1,0.7}
\definecolor{brown}    {rgb}{0.6,0.1,0.0}
\definecolor{gray}     {rgb}{0.6,0.6,0.6}
\definecolor{darkgreen}{rgb}{0.0, 0.545098, 0.0}
\definecolor{orange}   {RGB}{238,80,25}
\definecolor{purple}   {rgb}{0.5,0.0,0.5}
\definecolor{babypink} {rgb}{0.64, 0.44, 0.44}
\definecolor{spartandarkgreen} {RGB}{24, 69, 59}

\begin{document}

\title{Color screening in (2+1)-flavor QCD}

\author{A. Bazavov$^a$, N. Brambilla$^{b,c}$, P. Petreczky$^d$, A. Vairo$^b$, J. H. Weber$^{a,b,e}$ \\ (TUMQCD Collaboration)}
\affiliation{
$^a$ Department of Computational Mathematics, Science and Engineering and Department of Physics and Astronomy, Michigan State University, East Lansing, MI 48824, USA\\
$^b$ Physik Department, Technische Universit\"{a}t M\"{u}nchen, D-85748 Garching, Germany\\
$^c$ Institute of Advanced Studies, Technische Universit\"{a}t M\"{u}nchen, D-85748 Garching, Germany\\
$^d$ Physics Department, Brookhaven National Laboratory, Upton, New York 11973, USA\\
$^e$ Exzellenzcluster Universe, Technische Universit\"{a}t M\"{u}nchen, D-85748 Garching, Germany
}
\date{\today}

\begin{abstract}
We study correlation functions of spatially separated static quark-antiquark 
pairs in (2+1)-flavor QCD in order to investigate onset and nature of color 
screening at high temperatures.
We perform lattice calculations in a wide temperature range, 
$140 \le T \le 5814\,{\rm MeV}$, 
using the highly improved staggered quark action and several lattice 
spacings to control discretization effects. 
By comparing at high temperatures our lattice results to weak-coupling 
calculations 
as well as to the zero temperature result for the energy of a static 
quark-antiquark pair, we observe that color screening sets in at $rT \approx 0.3$. 
Furthermore, we also observe that in the range $0.3 \lesssim r T \lesssim 0.6$ 
weak-coupling calculations in the framework of suitable effective field 
theories provide an adequate picture of color screening.
\end{abstract}

\preprint{TUM-EFT 81/16}
\maketitle

\section{Introduction}
\label{sec:intro}

As the temperature of strongly interacting matter increases to a 
pseudocritical temperature $\tc$, a transition to a state with 
different properties than the vacuum at zero temperature occurs. 
Deconfinement of gluons and quarks, restoration of chiral symmetry and 
screening of color charges are key properties of this thermal medium 
(\mbox{cf.} recent reviews, \mbox{e.g.}~\cite{Bazavov:2015rfa, 
Petreczky:2012rq, Ding:2015ona}). 

For pure Yang-Mills theory (SU($\nc$) gauge theory), the Polyakov loop 
and the correlator of two Polyakov loops are the order parameters of the 
deconfinement transition, in which the $Z(\nc)$~center 
symmetry~\cite{McLerran:1981pb,Kuti:1980gh} of the pure Yang-Mills vacuum is broken. 
Since the center symmetry is broken by dynamical quarks already at zero 
temperature, neither the Polyakov loop nor the Polyakov loop correlator 
are order parameters of full QCD~\cite{Bazavov:2009zn, Bazavov:2011nk}.
Nevertheless, the Polyakov loop correlator is a sensitive probe of deconfinement and color screening. 
In particular, it provides insight into the intricate screening properties of the thermal medium. 
The properly renormalized correlator of two Polyakov loops \(C_P\) is 
related to the free energy of a static quark-antiquark ($\qbq$) pair 
\(\Fqq\) at separation $r$ and temperature \(T\)~\cite{McLerran:1981pb,Kaczmarek:2002mc},

\ileq{
C_P(r,T)=\exp{\left[-\frac{\Fqq(r,T)}{T}\right]}.
\label{eq:defFqq}
}

We choose to renormalize $C_P$ in such a way that $\Fqq$ for 
infinite separation matches the sum of the free energies of a decoupled 
static quark and antiquark using the renormalization of 
\mbox{Ref.}~\cite{Bazavov:2016uvm}. 
This is the same renormalization scheme that is usually used for matching the 
static energy at zero temperature for different lattice spacings. 

Polyakov loop correlators have been calculated on the lattice
for a long time. Calculations have been performed in SU(2) gauge
theory~\cite{Irback:1991eh,LaCock:1991hh,Datta:1999yu,Fiore:2003yw, 
Digal:2003jc,Bazavov:2008rw} and SU(3) gauge theory \cite{Karkkainen:1992jh, 
Kaczmarek:1999mm,Petreczky:2001pd,Kaczmarek:2002mc,Akerlund:2013cga}
as well as in QCD with two quark flavors \cite{Detar:1998qa, 
Karsch:2000kv,Kaczmarek:2005ui} and three quark flavors~\cite{Petreczky:2004pz}. 
More recent calculations  of the Polyakov loop correlator and of the color 
singlet correlator have been performed with (2+1) dynamical flavors of sea 
quarks using different types of staggered quark formulations at or close to 
the physical point~\cite{Petreczky:2010yn,Bazavov:2012fk,Borsanyi:2015yka}. 
In Ref. \cite{Borsanyi:2015yka} results in the continuum limit have been presented. 
Very recently the Polyakov loop correlators have been calculated at nonzero 
baryon density \cite{Andreoli:2017zie}.

At sufficiently high temperatures and short distances, the Polyakov loop 
correlators can be calculated using the weak-coupling expansion, \mbox{i.e.} 
the expansion in the gauge coupling $g=\sqrt{4\pi\as}$.
In this regime, besides the scale $1/r$ and the Coulomb potential $\as/r$, 
there are three relevant thermodynamical scales that need to be considered: 
$\pi T$, $g T$ (electric mass) and $g^2 T$ (also called the magnetic mass).
The leading order weak-coupling result for the Polyakov loop correlator has 
been known for a long time \cite{Gross:1980br,McLerran:1981pb} and implies a 
non-Coulombic behavior of the static $Q \bar Q$ free energy due to the 
cancellation of the singlet and octet contributions~\cite{Gross:1980br, 
McLerran:1981pb}. 
The next-to-leading-order calculation of the Polyakov loop correlator was 
performed by Nadkarni in the large distance regime, 
$r \sim 1/(gT)$~\cite{Nadkarni:1986cz}, whereas the corresponding calculation 
in the short distance regime, $r T \ll 1$, 
has been performed relatively recently~\cite{Brambilla:2010xn}.
In~\cite{Berwein:2017thy} one can find the state of the art expression for the 
Polyakov loop correlator at short distances, which is valid up to order 
\(g^7\) and \((rT)^3\) in the multipole expansion. 
In this regime it turns out to be useful to organize the calculation by means of 
the effective field theory called potential nonrelativistic QCD 
(pNRQCD)~\cite{Pineda:1997bj, Brambilla:1999xf, Brambilla:2004jw}.
For distances $r \gtrsim 1/(gT)$ the behavior of the Polyakov loop correlator
can be best understood in the framework of the dimensionally reduced effective 
field theory called electrostatic QCD (EQCD) \cite{Braaten:1995jr} (see 
references therein for related works). 
At very large distances, $r>1/(g^2 T)$, the perturbative expansion breaks down 
due to the well-known Linde problem~\cite{Linde:1980ts}.
In \mbox{Refs.}~\cite{Rebhan:1994mx,Braaten:1994pk}, using EQCD, it was shown 
indeed that the very large distance behavior of the Polyakov loop correlator 
is affected by chromomagnetic screening and its correct asymptotic behavior 
is given by the lowest glueball mass at high 
temperatures~\cite{Braaten:1994qx, Arnold:1995bh}. 
The screening mass that governs the large distance behavior of the Polyakov 
loop correlator has been calculated in EQCD using lattice techniques and 
also investigated with weak-coupling methods in~\cite{Laine:2009dh}.

Most of the lattice QCD studies have focused on the large distance behavior of the 
Polyakov loop correlators and the extraction of the corresponding screening masses. 
The aim of this paper is to study the Polyakov loop correlator from small to 
large distances and see in what temperature and distance regime contact to 
weak-coupling calculations can be made. 
Such an analysis will also help to clarify at which distances color screening 
effects set in. 
As it was pointed out in the past, for understanding the temperature behavior
of the Polyakov loop correlator it is useful to consider the singlet 
correlator~\cite{Kaczmarek:2002mc,Brambilla:2010xn}. 
Therefore in this paper we will also present a detailed study of the singlet 
correlator and its comparison to weak-coupling 
calculations~\cite{Burnier:2009bk,Berwein:2017thy}.

The question of the onset of color screening and applicability of 
weak-coupling expressions to the Polyakov loop and singlet correlators is not only of academic interest. 
There is an ongoing theoretical effort to understand quarkonium production 
in heavy-ion collisions (see \mbox{e.g. Refs.} \cite{Mocsy:2013syh,Blaizot:2017ypk}), which requires the knowledge of in-medium quarkonium properties. 
Full QCD lattice determinations of in-medium quarkonium properties turn out to be difficult  
(see, \mbox{e.g., Refs.} \cite{Wetzorke:2001dk,Datta:2003ww}),
which leads to the use of effective field theories~\cite{Burnier:2007qm,Brambilla:2008cx,Petreczky:2010tk,Burnier:2015tda,Burnier:2016kqm,Brambilla:2016wgg,Brambilla:2017zei}. 
Those, on the other hand require one to know which scale hierarchy is relevant for the physically interesting case.
The analysis presented in this paper aims at shedding some light on this issue.

The rest of the paper is organized as follows. 
In \mbox{Sec.}~\ref{sec:setup}, we cover technicalities of the lattice 
simulations.  
In \mbox{Sec.}~\ref{sec:results}, we present our numerical results, the 
extraction of the continuum limit, and discuss some general features of
the studied correlators, which do not require reference to explicit 
weak-coupling expressions. 
In the following two Secs.~\ref{sec:singlet} and~\ref{sec:plc}, we 
conduct a systematic and quantitative comparison of the lattice 
results with the predictions from weak-coupling calculations. 
We finally use this knowledge for a discussion of the regime of asymptotic 
screening in \mbox{Sec.}~\ref{sec:screening}. 
The last section of the paper contains our conclusions. 
Many technical details of the calculations are discussed in the Appendixes.
Some preliminary results from the present study have been reported in
conference proceedings \cite{Bazavov:2016qod,Petreczky:2016siu,Weber:2017dmi}.

\section{Lattice QCD Setup and renormalization}
\label{sec:setup}

We perform calculations of the Polyakov loop correlator as well as of the 
singlet correlator of static $\qbq$ in (2+1)-flavor QCD at nonzero temperature
on $N_{\sigma}^3 \times N_{\tau}$ lattices with $N_\tau=4,~6,~8,~10,~12$, and 
$16$ and aspect ratios of $N_{\sigma}/N_{\tau}=4$ and $6$ using the highly 
improved staggered quark (HISQ) action~\cite{Follana:2006rc}. 
The calculations have been performed at the physical value of the strange 
quark mass $m_s$ and the light quark masses $m_l=m_s/20$. 
The latter corresponds to a pion mass of $m_\pi \sim160\,{\rm MeV}$ in the continuum limit. 
We performed calculations in a wide beta range $5.9\le\beta=10/g_0^2\le9.67$ 
with $g_0$ being the lattice bare gauge coupling.
The lattice spacing $a$ has been fixed by the $r_1$ scale and we use the
parametrization of $r_1/a$ given in Ref. \cite{Bazavov:2014pvz}.
Using this parametrization we find that the above $\beta$ range corresponds
to a temperature range of $116 \le T \le 5814\,{\rm MeV}$.
The gauge configurations used in this study have been generated by 
the HotQCD~Collaboration~\cite{Bazavov:2011nk,Bazavov:2014pvz}. 
We also used the gauge configurations generated by the study of quark number 
susceptibilities at high temperatures~\cite{Ding:2015fca, Bazavov:2013uja}. 
Additional gauge configurations have been generated specifically for this study. 
Some of these gauge configurations have been used in the study of the 
renormalized Polyakov loop with the HISQ action~\cite{Bazavov:2013yv,Bazavov:2016uvm}. 
A detailed account of the used gauge configurations, including the new gauge 
configurations by our (TUMQCD) collaboration is presented in Appendix~\ref{app:A}.
Furthermore, in the high temperature region $T>350$\,MeV additional gauge 
configurations have been generated with $N_{\tau}=4,~6,~8,~10,~12$, and $16$, 
$N_{\sigma}/N_{\tau}=4$ for $\beta=7.03,~7.825,~8.0,~8.2$, and $8.4$ and light quark mass $m_l=m_s/5$.  
This light quark mass corresponds to a pion mass of $m_\pi \approx 320\,{\rm MeV}$ in the continuum limit. 
The aim of these calculations was to extend the continuum results on the correlators to higher temperatures.
For temperatures $T>400$ MeV, quark mass effects are expected to be negligible. 
We checked this explicitly in Appendix \ref{app:A}.
Finally, to study finite size effects for the large distance behavior of the correlators, 
we generated additional gauge configurations on $24^3 \times 4$ lattices (see Appendix \ref{app:A}). 

We have generated the new gauge ensembles using the rational hybrid 
Monte-Carlo (RHMC) algorithm and the MILC Collaboration~code. 
Details on the HISQ action implementation in the MILC~code can be 
found  in \mbox{Ref.}~\cite{Bazavov:2010ru}. 

In this paper, we study the correlation functions of the Polyakov loop.
The (bare) Polyakov loop is defined in lattice QCD as the normalized trace of a 
temporal Wilson line $ W(N_\tau,\bm x) $ wrapping around the time direction once,

\ilal{
 P(N_\tau,\bm x)  &= 
 \frac{1}{3}\, \mathrm{Tr}\, W(N_\tau,\bm x),
 \label{eq:defP}\\ 
 W(N_\tau,\bm x) &= 
 \prod_{\tau/a=1}^{N_\tau} U_0(\tau,\bm x),
 \label{eq:defW}
}

where $U_0(x=(\tau,\bm x))$ are the temporal link variables. 
We denote the expectation value of the bare Polyakov loop averaged 
over the spatial lattice volume as 

\ileq{
L^{\rm bare}(\beta,N_\tau)=\braket{P(N_\tau,\bm x)}.
\label{eq:defLbare}
} 

The bare Polyakov loop correlator is defined as the ensemble average (spatial 
average over $ \bm x $ included) 

\ileq{
  C_P^{\rm bare}(\beta,N_\tau,r) = 
  \Braket{ P(N_\tau,\bm x)
   P^\dagger(N_\tau,\bm x+\mathbf{r}) }.
  \label{eq:defCPbare}
}

The superscript ``bare'' refers to the fact that the quantity is not renormalized.
In addition we study the singlet correlator in the Coulomb 
gauge \cite{Kaczmarek:2002mc,Digal:2003jc}, defined as

\ileq{
  C_S^{\rm bare}(\beta,N_\tau,r) = 
  \frac13\, \Braket{ \mathrm{Tr}\, \left[ W(N_\tau,\bm x) 
  W^\dagger(N_\tau,\bm x+\mathbf{r}) \right] }, 
  \label{eq:defCSbare}
}

as well as the cyclic Wilson loops \cite{Jahn:2004qr,Bazavov:2008rw,Berwein:2012mw}

\begin{widetext}
\begin{equation*}
  W_S^{\rm bare}(\beta,N_\tau,r) = 
  \frac13\, \Braket{ \mathrm{Tr}\, \left[ W(N_\tau,\bm x) 
  S(N_\tau,\bm x;\mathbf{r}) 
  W^\dagger(N_\tau,\bm x+\mathbf{r})  
  S^\dagger(0,\bm x;\mathbf{r}) \right] }  
  \label{eq:defWSbare},
\end{equation*}
\end{widetext}

where \(S(\tau/a,\bm x;\mathbf{r})\) is a spatial Wilson line between the 
points \((\tau,\bm x)\) and \((\tau,\bm x+\bm r)\). 
These quantities can be considered as meson correlation functions composed of 
a static quark and antiquark separated by a distance $r$ \cite{Jahn:2004qr, 
Bazavov:2008rw} and evaluated at Euclidean time $\tau=1/T$~\cite{Jahn:2004qr,Bazavov:2008rw}.
The subscript bare refers again to the fact that these are unrenormalized quantities.
At very large distances $C_P^{\rm bare}(\beta,N_\tau,r)$, 
$C_S^{\rm bare}(\beta,N_\tau,r)$ and $W_S^{\rm bare}(\beta,N_\tau,r)$ 
approach $[L^{\rm bare}(\beta,N_\tau)]^2$ in the infinite volume limit~\cite{Kaczmarek:2002mc, 
Bazavov:2008rw,Bazavov:2013zha}.\footnote{Here we take into account that in QCD $L$ is real.} 
These bare quantities contain ultraviolet (UV) divergences that are regularized on the lattice.
The continuum limit requires looking for combinations of bare quantities that are free of UV divergences.

The renormalization of $C_P^{\rm bare}(\beta,N_\tau,r)$ is simple: the only  
divergences here are the ones associated with the two Polyakov loops. 
Therefore, the normalized Polyakov loop correlator 

\ileq{
  C_P^{\rm sub}(\beta,N_\tau,r) = 
  \frac{C_P^{\rm bare}(\beta,N_\tau,r)}{[L^{\rm bare}(\beta,N_\tau)]^2},
  \label{eq:defCPsub}
}

is free of UV divergences and the continuum limit can be taken. 
In particular, this implies that \mbox{Eq.}~\eqref{eq:defCPsub} can be written 
directly in terms of renormalized quantities, namely, 

\ileq{
 C_P^{\rm sub}(\beta,N_\tau,r)=\frac{C_P(\beta,N_\tau,r)}{[L(\beta,N_\tau)]^2},
\label{eq:renCPsub}
} 

where \(C_P\) is the renormalized Polyakov loop correlator and \(L\) is the renormalized Polyakov loop. 
The normalized Polyakov loop correlator \(C_P^{\rm sub}\) is simply the 
connected part of the Polyakov loop correlator and contains the complete 
information about the in-medium modification of the \(\qbq\) interaction. 
The disconnected part of the renormalized correlator, \mbox{i.e.} \(L^2\), 
contains information about infinitely separated static quarks or antiquarks 
that interact exclusively with the medium, but not with each other. 

A similar observation holds for the singlet correlator and we can use the 
normalized correlation function also in this case,

\ileq{
  C_S^{\rm sub}(\beta,N_\tau,r) =
  \frac{C_S^{\rm bare}(\beta,N_\tau,r)}{[L^{\rm bare}(\beta,N_\tau)]^2}
  = \frac{C_S(\beta,N_\tau,r)}{[L(\beta,N_\tau)]^2}.
  \label{eq:defCSsub}
}

The above expression, which is defined in the Coulomb gauge, is free of 
divergences and has a well-defined continuum limit. 
Such a statement does not hold for general gauges: 
in covariant gauges, the normalized correlator will be UV divergent~\cite{Burnier:2009bk}.
Another reason for using the Coulomb gauge is that weak-coupling calculations are available 
in this case~\cite{Brambilla:2010xn,Burnier:2009bk,Berwein:2017thy}. 

Since the normalized Polyakov loop and singlet correlators are ultraviolet 
finite, we can define the subtracted free energy of a static $Q\bar Q$ pair 
and the so-called (subtracted) singlet free energy by taking the logarithm of these correlators,

\ilal{
\Fqq^{\rm sub}(r,T,a) &=-T \ln C_P^{\rm sub}(\beta,N_\tau,r),
\label{eq:defFqqsub}\\
\Fs^{\rm sub}(r,T,a) &=-T \ln C_S^{\rm sub}(\beta,N_\tau,r)
\label{eq:Fssub}.
}

Here we have traded $\beta$ and $N_{\tau}$ for the physical temperature 
$T=1/(N_{\tau} a)$ and the lattice spacing $a$. 
Up to a finite constant, \mbox{Eqs.}~\eqref{eq:defFqq} and \eqref{eq:defFqqsub} define the same free energy.
From \mbox{Eq.}~\eqref{eq:renCPsub}, it follows that this constant is given by \(-2T\ln L\).
The expectation value of the renormalized Polyakov loop \(L\) is 
related to the free energy of an isolated static quark \(\Fq\) at temperature \(T\),

\ileq{
L(T)=\exp{\left[-\frac{\Fq(T)}{T}\right]}.
\label{eq:defFq}
}

The constant \(-2T\ln L\) is therefore the free energy of two 
static quarks or antiquarks that are at infinite separation and, thus, can 
only interact with the medium.
The regularized UV divergence of \(L^{\rm bare}\) is exactly the square root 
of the regularized UV divergence of \(C_P^{\rm bare}\) or \(C_S^{\rm bare}\).
Hence, by adding twice the free energy of an isolated static quark $\Fq$, 
determined in \mbox{Ref.}~\cite{Bazavov:2016uvm}, we get the free energy of 
a static $\qbq$ pair as well as the singlet free energy,

\ilal{
\Fqq&=\Fqq^{\rm sub} +2\Fq, 
\label{eq:normFqq}\\
\Fs&=\Fs^{\rm sub} +2\Fq.
\label{eq:normFs}
} 

\mbox{Equation}~\eqref{eq:normFqq} trivially follows from \mbox{Eqs.}~\eqref{eq:defFqq},~\eqref{eq:renCPsub},~\eqref{eq:defFqqsub}, and~\eqref{eq:defFq}. 
\mbox{Equation}~\eqref{eq:normFs} is the analog for the singlet free energy.
Thus, the problem of renormalizing the Polyakov loop and singlet 
correlators is reduced to the problem of renormalizing $L^{\rm bare}$.
For the latter, we use the procedure described in 
\mbox{Ref.}~\cite{Bazavov:2016uvm} and update it with the most recent lattice results. 
The updated renormalization of $L^{\rm bare}$ is presented in Appendix~\ref{app:A}.
The above procedure also turns out to be very convenient when performing
the continuum extrapolations discussed in the next section.

Another way to study in-medium modifications of the interaction
of static quark and antiquark is to consider cyclic Wilson 
loops \cite{Jahn:2004qr,Bazavov:2008rw,Bazavov:2013zha,Berwein:2013xza}.
The renormalization of the cyclic Wilson loops, however, is more complicated. 
In this case, in addition to the self-energy divergences present in the 
Polyakov loop correlators, there are also self-energy divergences associated 
with the spatial Wilson lines \(S(N_\tau,\bm x;\mathbf{r})\) and \(S^\dag(0,\bm x;\mathbf{r})\), as well as intersection divergences~\cite{Berwein:2012mw}.
On the lattice there are also cusp divergences for off-axis separations. 
Rectangular Wilson loops evaluated for $\tau< 1/T$ have cusp divergences 
at the four corners of the loop. 
These cusp divergences are absent for cyclic Wilson loops (\mbox{i.e.} for 
$\tau=1/T$) in the continuum, because the corresponding contour is smooth. 
However, on the lattice, one can also consider cyclic Wilson loops for 
noninteger multiples of the lattice spacing by using Wilson lines that are 
not aligned with any of the spatial lattice axes.
Such off-axis spatial separations, \mbox{e.g.}, $r/a=\sqrt{2}$, involve cusps. 
In other words, for off-axis separation, the cyclic Wilson loops have 
additional cusp divergences, while for on-axis separation, such divergences are absent.
To simplify the analysis, we only consider on-axis separation.
Constructing the spatial part of the cyclic Wilson loops from the smeared 
gauge links can reduce the size of the UV divergences. 
Therefore, we calculate the cyclic Wilson loops with spatial lines obtained from smeared gauge links. 
We apply several steps of hypercubic (HYP) smearing \cite{Hasenfratz:2001hp} 
and use the same parameters for HYP smearing as in \mbox{Ref.}~\cite{Hasenfratz:2001hp}.

\section{Lattice results}
\label{sec:results}

In this section, we discuss the dependence of the Polyakov loop correlator, 
defined in Eq. \eqref{eq:defCPbare} and the singlet correlator, given in 
Eq. \eqref{eq:defCSbare}, on the separation of the static 
quark-antiquark pair at different temperatures.
The lattice results for the Polyakov loop correlator and the singlet
correlator are functions of three variables:
the distance $r$, the lattice spacing $a$, and the temporal extent $N_\tau$. 
The temperature $T$ is related to $N_\tau$ and $a$ through $a N_\tau=1/T$; 
by trading $a$ for the temperature $T$, we parametrize cutoff effects by 
the $N_\tau$ dependence of the correlators at fixed temperature. 
In total, we analyze a full set of more than 10000 data 
(expectation values) from more than 150 gauge ensembles. 
The temperature range covers the confined phase, the crossover region, and 
a strongly as well as weakly coupled quark-gluon plasma phase. 
The separation range covers asymptotic freedom at short and color screening 
at large distances.

\subsection{Outline of the analysis}\label{sec:analysis}

The correlator data for different $N_\tau$ are available only for 
heterogeneous sets of temperatures and for heterogeneous sets of separations 
at each temperature. 
Therefore, the comparison of correlators obtained for different $N_\tau$ 
and eventually the continuum extrapolation requires either a 
three-dimensional, fully global fit or, in one way or another, an interpolation 
in the separation $r$ (or in $rT$) and in the temperature $T$ using 
smoothing functions in order to obtain an intermediate, homogeneously spaced 
grid of mock data. 
We perform separate fits in the separation and the temperature sequentially 
(local fits) or simultaneous fits in the separation and the temperature 
(global fits). 
Smoothness in both separation and temperature is a physical constraint in the 
absence of sharp phase transitions, but inevitably leads to a shrinkage of 
the errors in the process.
Hence, we have to expect that errors of the mock data may be underestimated. 

Since no analytic results are available that cover the whole range spanned by 
the available lattice data, any interpolation introduces some degree of bias. 
In order to minimize this bias, we identify classes of apparently sensible 
model functions. 
For small separations, such model functions are constrained by asymptotic 
freedom and expectations based on the leading-order weak-coupling results. 
For larger separations, such model functions are constrained by the dominance 
of massive one-particle exchange processes, \mbox{i.e.} exponential damping 
due to color screening. 
We do not enforce exponential damping for lower temperatures, arguing that 
confinement may still be unabated or that we may fail to resolve the damping 
due to a lack of data with sufficient accuracy at large enough separations.
For even larger separations, subtracted free energies must approach zero due 
to cluster decomposition, unless statistical fluctuations (for smaller 
ensembles) or artifacts of the finite volume lead to an incomplete 
cancellation and, thus, to a constant value. 
We account for all of these effects in the interpolations discussed in  
Appendix~\ref{app:C} and use a data-driven procedure to identify the most 
adequate (\mbox{i.e.} most simple) interpolating functions. 
We note that in the case of \(\Fqq^{\rm sub}\) we actually find that the 
lattice data successfully override our inherent bias at low temperatures. 
Namely, we identify Coulombic interactions despite our explicit bias towards 
non-Coulombic interactions (see the discussion in 
\mbox{Sec.}~\ref{sec:effective coupling}).

As ensembles at different temperature \(T\) (or at different 
\(N_\tau\)) are statistically independent, correlated fluctuations that 
affect the size of error estimates are relevant only for an interpolation 
in the separation.
Since we are interested in the behavior of the correlators in a wide
range of distances $r$, down to the shortest ones, it is important
to take care of the lattice artifacts present at short distances
before performing the interpolation in $r$.
To do this, we utilize the results of the lattice calculation of the
static \(\qbq\) energy at zero temperature \(\Vs\) (published in 
\mbox{Refs.}~\cite{Bazavov:2014pvz,Bazavov:2017dsy} and the procedure 
described in \mbox{Ref.}~\cite{Bazavov:2014soa}).
We give a detailed account of the underlying procedure in Appendix~\ref{app:B}. 
When discussing the \(T=0\) static energy throughout this section, we always 
use these nonperturbatively improved lattice results.  
This improvement procedure, however, introduces systematical uncertainties 
at short distances that we have to estimate heuristically.  
Such systematical uncertainties are manifestly uncorrelated and orders of 
magnitude larger than the corresponding statistical errors. 
That is why the total uncertainties of the data underlying the interpolations 
(including short distances) are always dominated by uncorrelated errors. 
Hence, uncorrelated fits aiming for \(\cdf \simeq 1\) are an appropriate 
procedure for analyzing these data, where \(\rm d.o.f.\) represents degrees 
of freedom.

We generalize the improvement procedure to correlators with \(T>0\). 
Here, in particular, for ensembles with smaller \(N_\tau\), we have to be 
aware that medium effects may become important in the short distance range, 
where we apply the nonperturbative improvement. 
For this reason, we have to expect that we may very well underestimate the 
associated uncertainties and account for this by enlarging the errors if we 
cannot interpolate smoothly between data at short distances with 
\(\cdf \simeq 1\). 
After correcting for short distance lattice artifacts we perform 
interpolations of $r\Fs^{\rm sub}$ and $r^2T\Fqq^{\rm sub}$ in 
$rT$ and $T$ using splines or polynomial fits with exponentially decaying 
prefactor of the form \(\exp{[-M rT]}\), where \(M\) is a fit parameter. 
We perform uncorrelated fits on the individual jackknife bins and 
determine the errors of the fit from the respective variance of the fit model. 
These interpolations are only weakly constrained for the largest separations. 

In a secondary interpolation at fixed \(rT\), we use in the absence 
of justified model assumptions, smoothing splines to interpolate the mock data 
in \(T\) using uncorrelated fits, since ensembles at different \(T\) are 
statistically independent. 
Correlators at lower temperatures usually have consistently larger 
uncertainties than correlators at higher temperatures. 
For this reason, we have to separately fit in overlapping low and high 
temperature ranges in order to be sensitive to the lower temperatures. 
As we typically have about 30 different temperatures for each \(N_\tau\), we 
find that smoothness under variation of the temperature can provide tighter 
constraints for the \(rT\) dependence than the actual correlations in each 
ensemble at a fixed temperature.
This justifies the previous use of uncorrelated fits. 
In order to directly incorporate the smoothness under variation of \(T\) in 
the analysis of the \(rT\) dependence, we also perform global fits. 
Namely, we fit the \(rT\) and \(T\) dependence simultaneously in separate but 
overlapping low and high temperature intervals for each \(N_\tau\). 
In these global fits, we model the temperature dependence heuristically with 
(inverse) monomials in \(T\) and with a \(T\)-dependent \textit{Ansatz} for the 
screening mass. 
The global fits yield much smaller uncertainties at larger distances than 
the local fits, but are less effectively constrained by the data at short 
distances.
Namely, the few data at short distances with large systematic errors have 
only minor weights in the global fits. 
Eventually, we estimate systematic uncertainties from the differences between 
local and global fits, since their strengths and deficiencies are 
complementary. 

Using the mock data for these quantities at the intermediate values 
of $rT$ and $T$ obtained from interpolations, we perform fits of the form 
$a+b/N_{\tau}^2+c/N_{\tau}^4$ to obtain the continuum limit. 
This form is appropriate for the HISQ action, which has cutoff 
effects only at even powers of the lattice spacing. 
We encounter cases where the expected scaling behavior is observed on a set 
of coarser lattices, \mbox{i.e.} smaller \(N_\tau\), but apparently not 
realized for finer lattices. 
We understand that this may be caused by underestimating the errors of the 
mock data and rescale the errors in order to obtain a more realistic 
(\mbox{i.e.} larger) uncertainty of the continuum result in these cases. 
As the lattice data cover vastly different physical regimes in the 
\((rT,T)\) plane, it is unsurprising that the cutoff effects vary 
significantly over the whole data set. 
In the absence of prior knowledge\footnote{This is the reason why we did 
not attempt a fully global fit.} about the variation of cutoff effects, we 
conservatively perform a separate fit for each point in the \((rT,T)\) plane.
The coefficients \(b\) and \(c\) in these fits sometimes were set to zero
and the range in $N_{\tau}$ was restricted to estimate the systematic errors 
of the extrapolations. 
These systematic errors can be accidentally large for individual points. 
However, only subsets of the \(N_\tau\) range permit access to the various  
regions in the \((rT,T)\) plane. 
As a consequence, we have to combine different continuum extrapolations in 
various regions and, thus, obtain minor discontinuities in the continuum 
result. 
The details of this analysis are discussed in Appendix \ref{app:C}.
The procedure works as long as the statistical errors on $\Fs^{\rm sub}$ 
and $\Fqq^{\rm sub}$ are sufficiently small. 
At large distances, $\Fs^{\rm sub}$ and $\Fqq^{\rm sub}$ approach zero and the 
relative statistical errors become very large. 
For these distances our procedure does not work and we cannot provide a 
continuum result for the free energy and the singlet free energy. 
This limits the range of $rT$ where continuum results can be provided. 
Finally, to obtain the free energy of a static $\qbq$ pair and the singlet 
free energy, we add the continuum limit of $2 \Fq$ to the continuum results 
of $\Fs^{\rm sub}$ and~$\Fqq^{\rm sub}$.
We note that for most of the range the overall error of \(\Fs\) and 
\(\Fqq\) is dominated by the uncertainty of their asymptotic value $2 \Fq$.

\subsection{Continuum results}

\begin{figure*}
\hspace{-3cm}
\includegraphics[width=10.4cm]{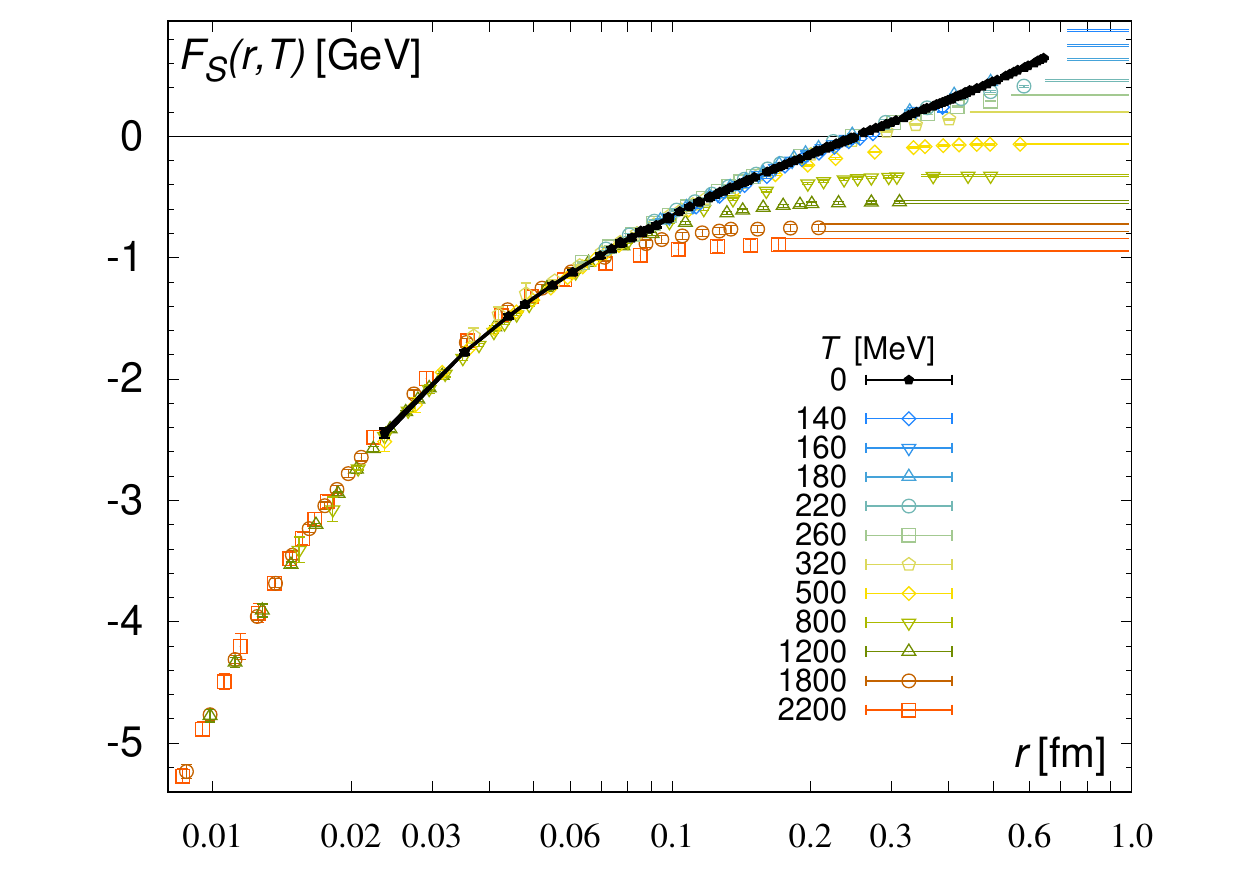}
\hspace{-1.4cm}
\includegraphics[width=10.4cm]{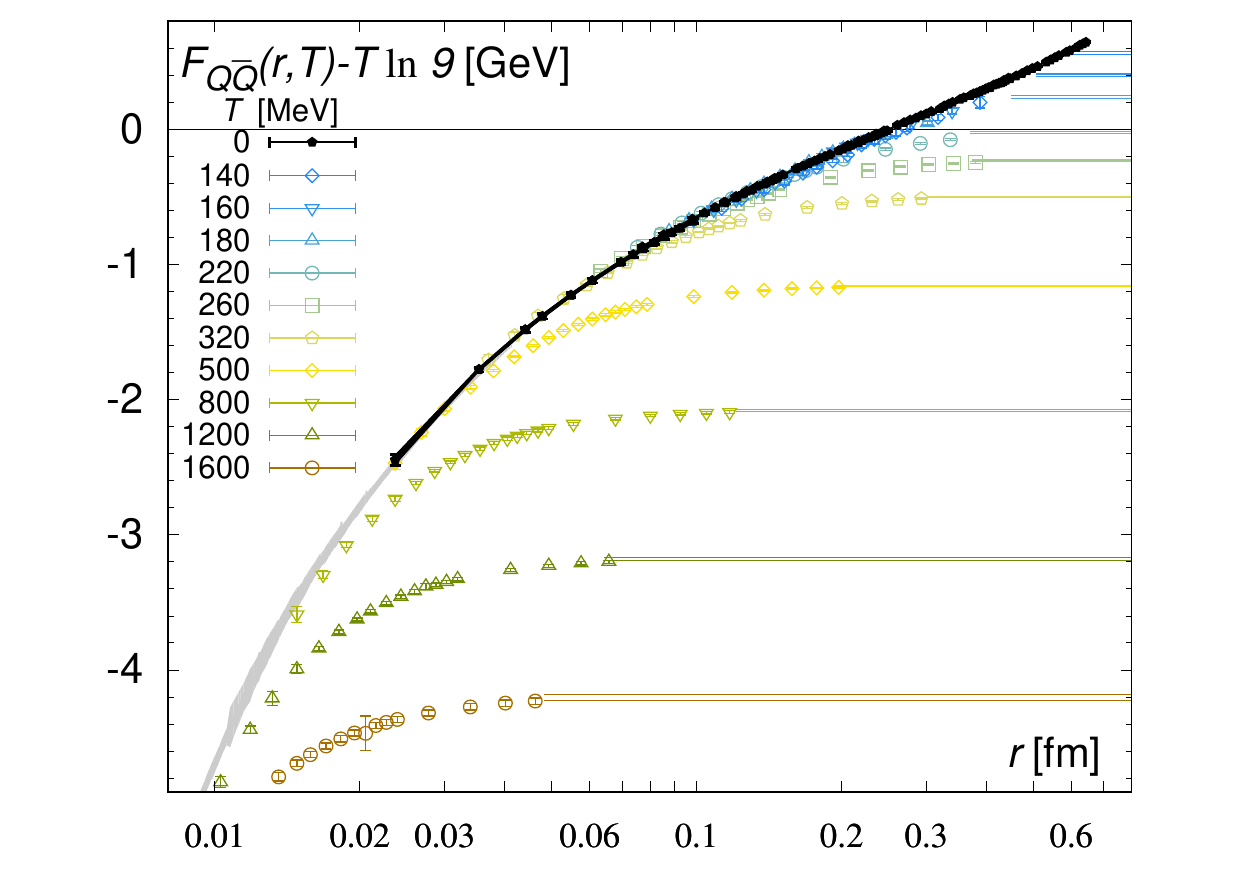}
\hspace{-3cm}
\caption{\label{fig:main}
The continuum limit of the singlet free energy $\Fs$ (left) and of the free 
energy $\Fqq$ (right). 
On the right and left, the black band and symbols show the $T=0$ static 
$\qbq$ energy \(\Vs\); 
(right) the light gray band shows \(\Fs\) at high temperatures and 
very short distances, 
where finite temperature effects are smaller than the statistical errors. 
The horizontal bands correspond to the $r \rightarrow \infty$ limit of $\Fs$ 
and $\Fqq$, i.e., $2 \Fq$.
The subtraction of \(T\ln 9\) in the right panel is due to the normalization convention used for \(\Fqq\) as discussed in the text.
}
\end{figure*}

Our continuum results for $\Fs$ are shown in \mbox{Fig.}~\ref{fig:main} 
(left) and compared to the zero temperature static $\qbq$ energy \(\Vs\), 
shown as black pentagons connected by a thin black band. 
These represent the nonperturbatively improved static energy for 
five different lattice spacings in overlapping intervals. 
The details of the composition of these data are discussed in Appendix~\ref{app:C}.
The numerical data for the zero temperature static energy are taken from 
Refs. \cite{Bazavov:2011nk,Bazavov:2014pvz,Bazavov:2017dsy}.
The asymptotic values, \mbox{i.e.}, $2 \Fq$, are shown as horizontal bands. 
As one can see from the figure, at high temperatures, $T>200$ MeV, continuum 
results are available up to sufficiently large distances to see the approach to $2 \Fq$.  
At short distances, the singlet free energy is temperature independent and 
agrees with the vacuum static $\qbq$ energy. 
At larger distances, we see the onset of medium effects. 
As the temperature is increased, the onset of the medium effects happens at shorter and shorter distances, 
but there is still an extended region in $r$ where $\Fs$ agrees with the zero temperature static energy. 
This is expected from weak-coupling calculations in pNRQCD~\cite{Berwein:2017thy}.

The continuum results for $\Fqq$ are shown in \mbox{Fig.} \ref{fig:main} (right). 
Again, only for $T>200$ MeV our continuum extrapolation extends to large
enough distances to make connection to the asymptotic value of $2 \Fq$.
To compare to the zero temperature static energy, we subtract the constant 
$T \ln N_c^2=T \ln 9$, which is due to the standard normalization convention 
of the Polyakov loop correlator defined by \mbox{Eqs.} \eqref{eq:defP} and 
\eqref{eq:defCPbare}, since usually the Polyakov loop correlator 
is defined without the $1/N_c^2=1/9$ normalization of the trace (see, e.g., \cite{Jahn:2004qr}).
For the \(T=0\) static energy, we use the same composite band as in the figure for \(\Fs\). 
We extend this band to even shorter distances using the continuum limit of 
\(\Fs\) at very short distances as an estimate of \(\Vs\).
We show this estimate with a light gray color to keep the distinction clear 
and discuss the details of this band in Appendix~\ref{app:C}. 
For $\Fqq$, we see a much larger temperature dependence already at relatively 
short distances and for temperatures $T>800$ MeV no clear agreement with the vacuum static $\qbq$ energy can be seen.
Qualitatively this behavior of the free energy and singlet free energy
of $Q\bar Q$ is very similar to the results of previous calculations 
performed at fixed lattice spacing \cite{Karsch:2000kv,Kaczmarek:2005ui, 
Petreczky:2004pz,Petreczky:2010yn,Bazavov:2012fk,Borsanyi:2015yka}.
The reason for this behavior of the $Q\bar Q$ free energy is related to
the presence of a color octet contribution and will be discussed in \mbox{Sec.}~\ref{sec:plc}.
Finally, we  mention that our continuum results for $F_{Q \bar Q}$ agree well
with the continuum extrapolated results presented in Ref. \cite{Borsanyi:2015yka} for
$T \le 400$ MeV. Our analysis, however, extends to higher temperatures.

\subsection{The effective coupling}\label{sec:effective coupling}

At $T=0$ it is customary to study the force between the static quark and 
antiquark defined in terms of the static \(\qbq\) energy $V_S(r)$ or, 
equivalently, the corresponding effective coupling 

\ileq{
\aqq(r) = \frac1\cf r^2 \frac{\partial V_S(r)}{\partial r},
\label{eq:aqq}
}

where \(C_F=4/3\) is the Casimir operator in the fundamental representation of SU(3). 
Following \mbox{Ref.}~\cite{Kaczmarek:2004gv} we can generalize this approach to the free energy and singlet free energy 
and obtain the corresponding effective couplings. 
The reasoning behind this generalization is that, as we see in \mbox{Fig.}~\ref{fig:main}, 
\(\Fs\) and \(\Fqq-T\ln 9 \) are similar to \(\Vs\) at sufficiently small distances. 
At such distances, a sensible definition of an effective coupling seems possible. 

\begin{figure}[ht]
\hspace{-1.8cm}
 \includegraphics[height=7.2cm,clip]{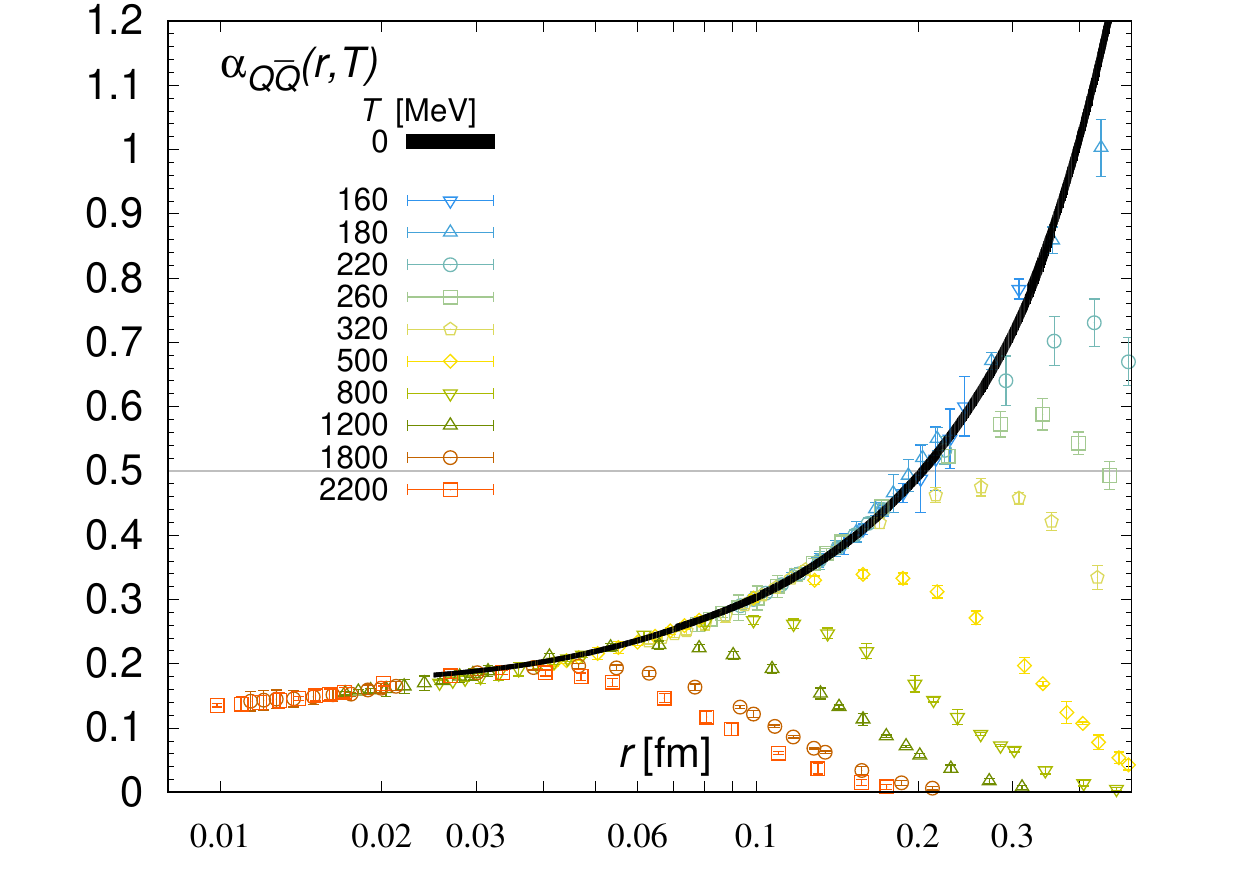}
\hspace{-1.8cm}
\caption{
The effective coupling obtained from the singlet free energy in the continuum limit and compared to the vacuum result (black band).
}
\label{fig:aqq}
\end{figure}

First, we discuss our numerical results for $\aqq(r)$ from the singlet free energy, which are shown in \mbox{Fig.}~\ref{fig:aqq}. 
The effective coupling defined in terms of $F_S$ is temperature independent
at short distances and agrees with the result obtained from a numerical derivative of the static energy $V_S$ 
(details are given in Appendix~\ref{app:C}).
At short distances, the running of the coupling is controlled by the scale $1/r$.  
The effective coupling reaches a maximum at some distance $r=r_{\rm max}(T)$ 
and then decreases, indicating the onset of color screening. 
Furthermore, it turns out that $r_{\rm max}$ approximately scales like $0.4/T$.
The distance $r_{\rm max}$ roughly separates the region that is dominated by vacuum physics from the one where screening sets in.\footnote{
Note that screening starts affecting the effective coupling somewhat below $r_{\rm max}$.
}  
This will be addressed more quantitatively in the next section.
It is interesting to see that the value of $\aqq(r_{\rm max},T)$ is quite large for $T<320$ MeV, 
implying that the physics is strongly coupled. 
At higher temperatures, $\aqq(r_{\rm max},T)<0.5$, indicated by a horizontal 
line in the figure, so that weak-coupling methods might be applicable. 

\begin{figure}[ht]
\hspace{-1.8cm}
 \includegraphics[height=7.2cm,clip]{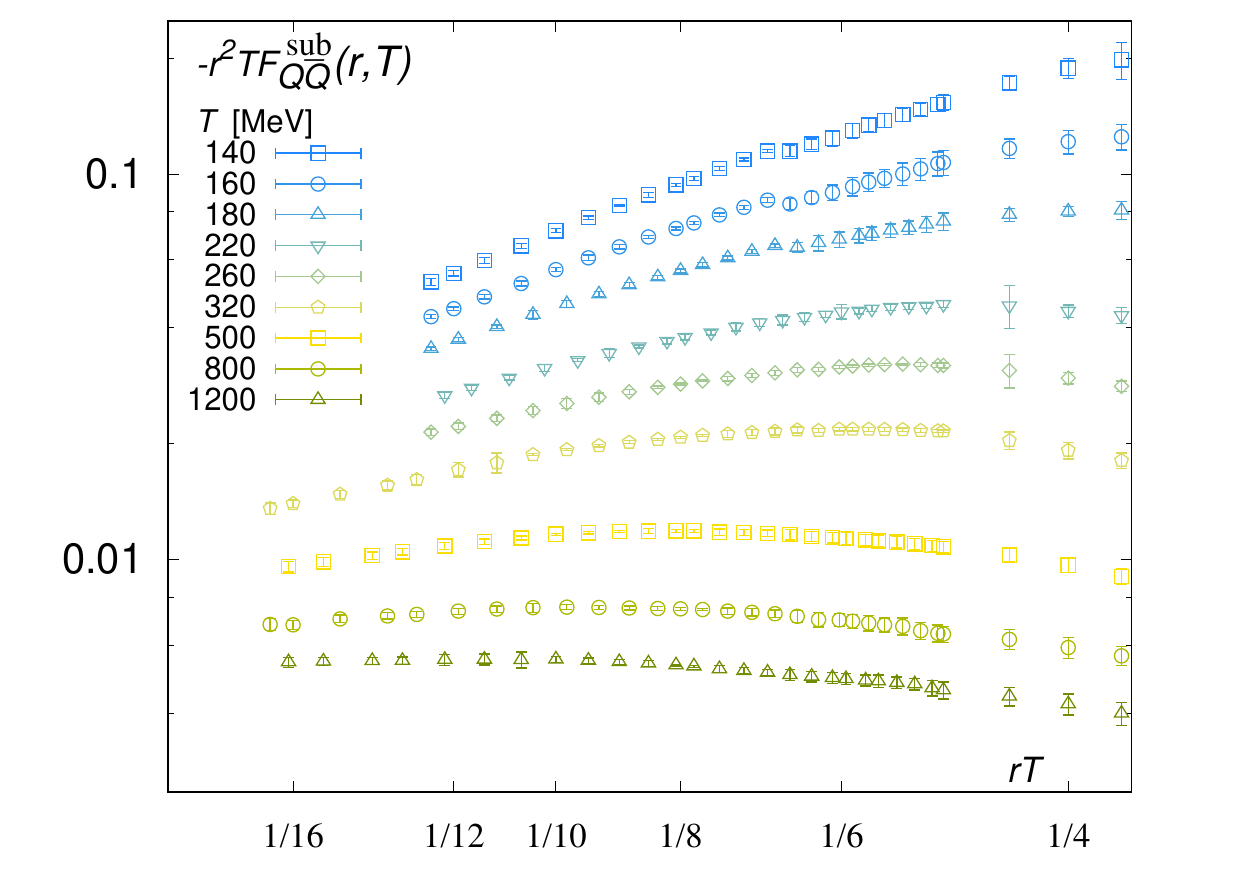}
\hspace{-1.8cm}
\caption{
The subtracted free energy multiplied by \(-r^2T\) at short distances 
in a log-log plot.  
}
\label{fig:fqqs short}
\end{figure}

In the case of the free energy $\Fqq$, we expect at short distances 
and for weak coupling two possible regimes. 
When $T \ll \as/r$, $\Fqq$ should exhibit a typical Coulombic behavior 
$1/r$, 
whereas for $T \gg \as/r$, due to a cancellation between color singlet 
and color octet contributions, 
the behavior of $\Fqq$ as a function of the distance should be like $1/r^2$.
Indeed, the subtracted free energy, defined in \mbox{Eq.}~\eqref{eq:defFqqsub}, 
multiplied by $-r^2T$, shown in \mbox{Fig.}~\ref{fig:fqqs short},\footnote{
The apparent small discontinuities and pronounced variations of errors are 
an artifact of the analysis (\mbox{cf. Sec.}~\ref{sec:analysis}).} 
seems to exhibit both regimes.
At short distances and for $T < 500$ MeV, $-r^2T \Fqq^{\rm sub}$ raises 
with the distance $r$, a behavior consistent with $\Fqq^{\rm sub} \sim 1/r$, 
while for $T > 500$ MeV, the available data for $-r^2T \Fqq^{\rm sub}$ appear 
less sensitive to $r$, a behavior consistent with $\Fqq^{\rm sub} \sim 1/r^2$.
Analytical results and a detailed quantitative analysis can be found in 
\mbox{Sec.}~\ref{sec:reconstruction}. 
Here our purpose is to justify the use of \mbox{Eq.}~\eqref{eq:aqq}, at short 
distances and for $T \lesssim 500$ MeV, 
to define an effective coupling also in the case of the free energy $\Fqq$.
Eventually, also this effective coupling can be compared with the 
coupling obtained from the $\qbq$ static energy at zero temperature.

\begin{figure}[ht]
\hspace{-1.8cm}
\includegraphics[height=7.2cm,clip]{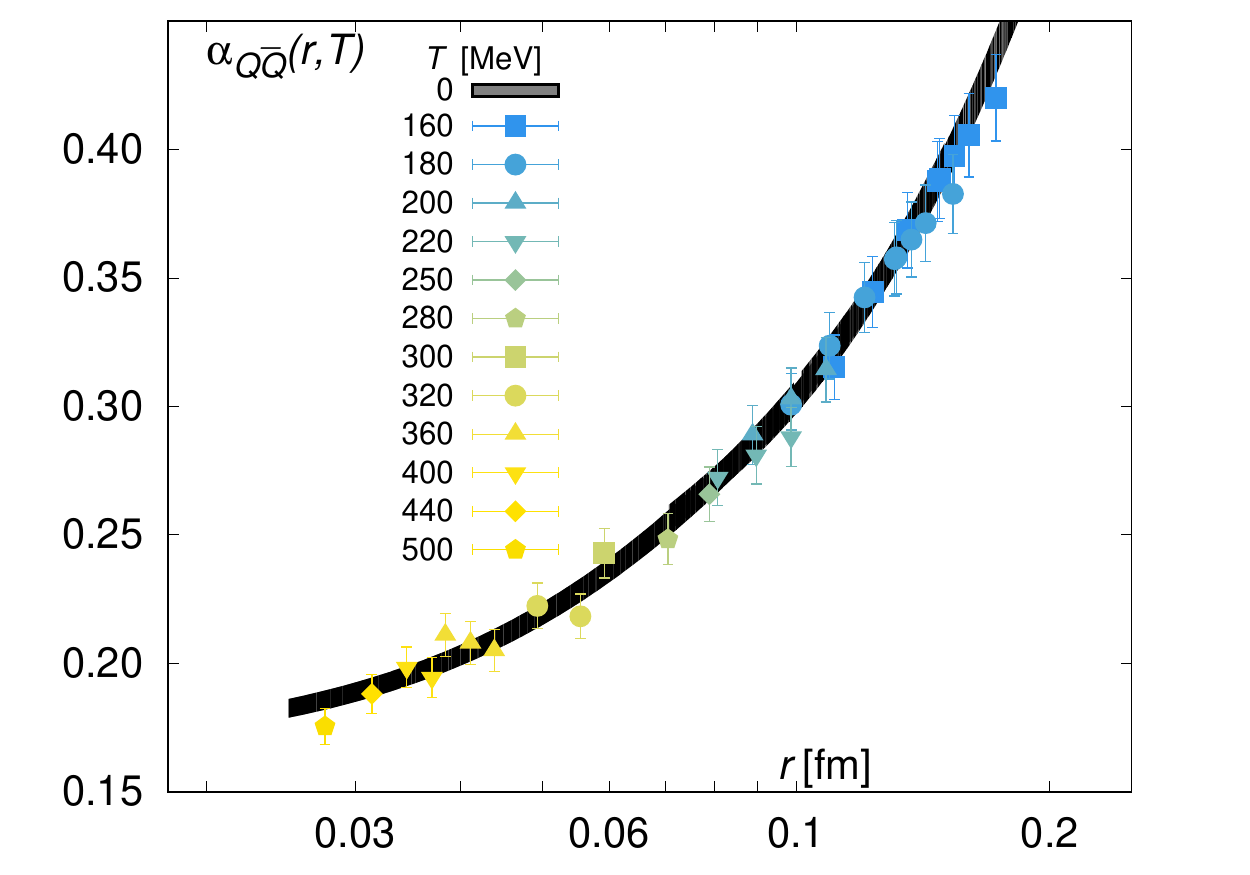}
\hspace{-1.8cm}
\caption{
The effective coupling obtained from the free energy of a $Q\bar{Q}$ pair 
compared with the effective coupling obtained from the $Q\bar{Q}$ static 
energy at zero temperature shown as a black band. 
}
\label{fig:aqqa}
\end{figure}

The statistical errors on our lattice results for \(\Fqq\) are much larger 
than in the $F_S$ case and therefore we calculate the corresponding effective 
coupling only for finite~$N_{\tau}$. 
In Fig. \ref{fig:aqqa}, we show results for $N_{\tau}=12$ and $16$. 
Following the above discussion, we look at temperatures $T \lesssim 500$ MeV 
and at distances short enough that $-r^2T \Fqq^{\rm sub}$ raises with the 
distance. 
Because of lack of accuracy of the derivative at the first data point for each 
temperature, we have to exclude derivatives for $rT$ smaller than the center 
of the interval between the first two data points for each temperature.
The points in Fig.~\ref{fig:aqqa} satisfy the above requirements.
The figure supports a description where, in the considered temperature range 
and for the shortest distances, the free energy of a $\qbq$ pair is dominated 
by the singlet contribution and agrees well with the $T=0$ static energy.

\subsection{Cyclic Wilson loops}

\begin{figure*}
\includegraphics[width=8.4cm]{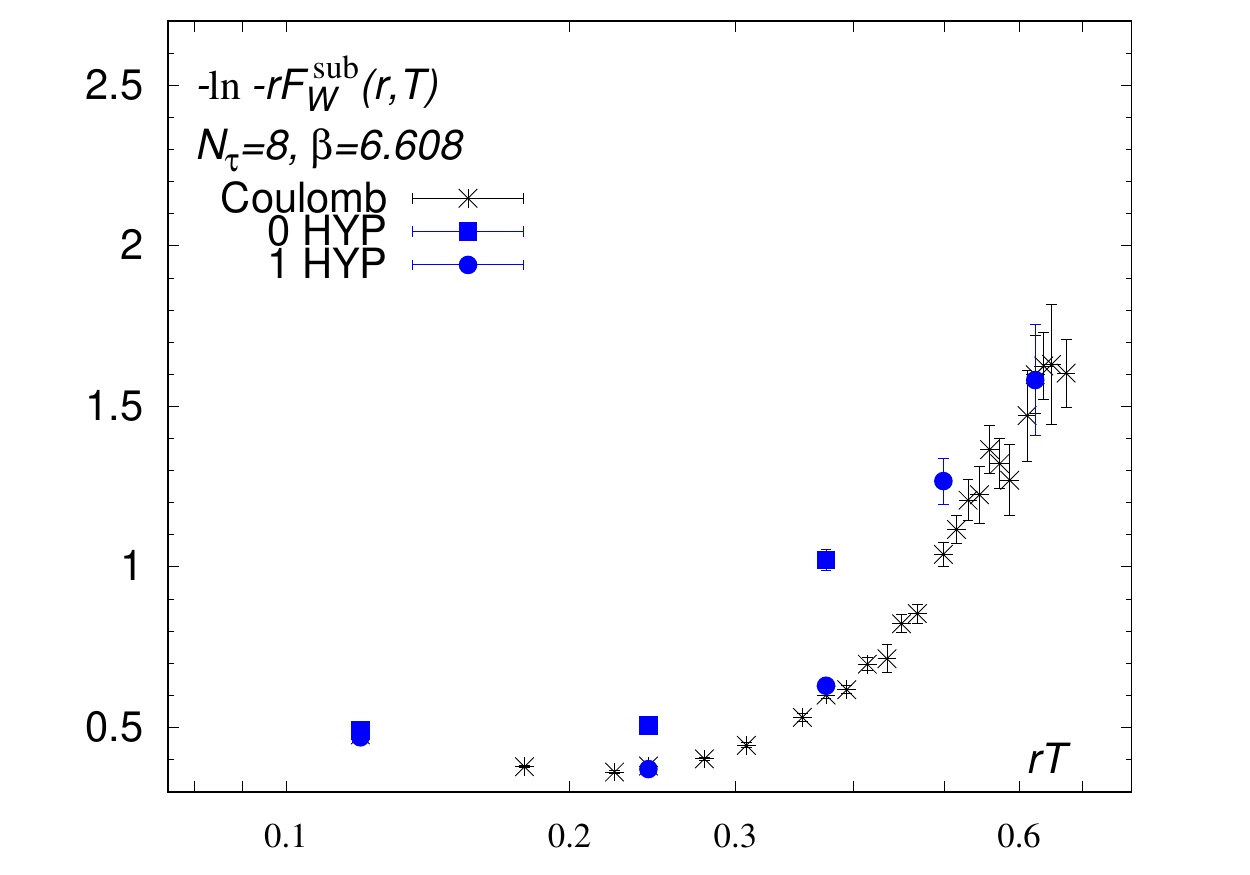}
\includegraphics[width=8.4cm]{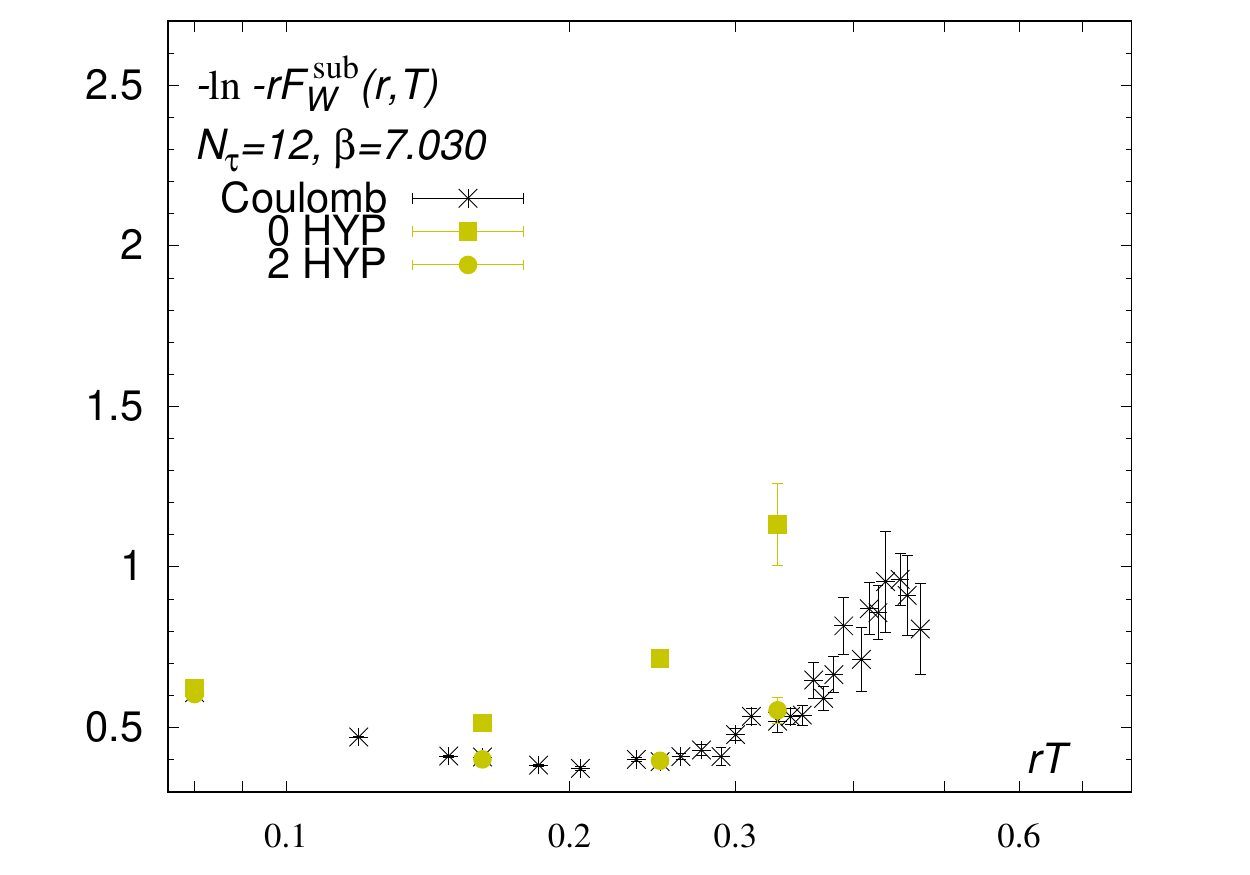}
\includegraphics[width=8.4cm]{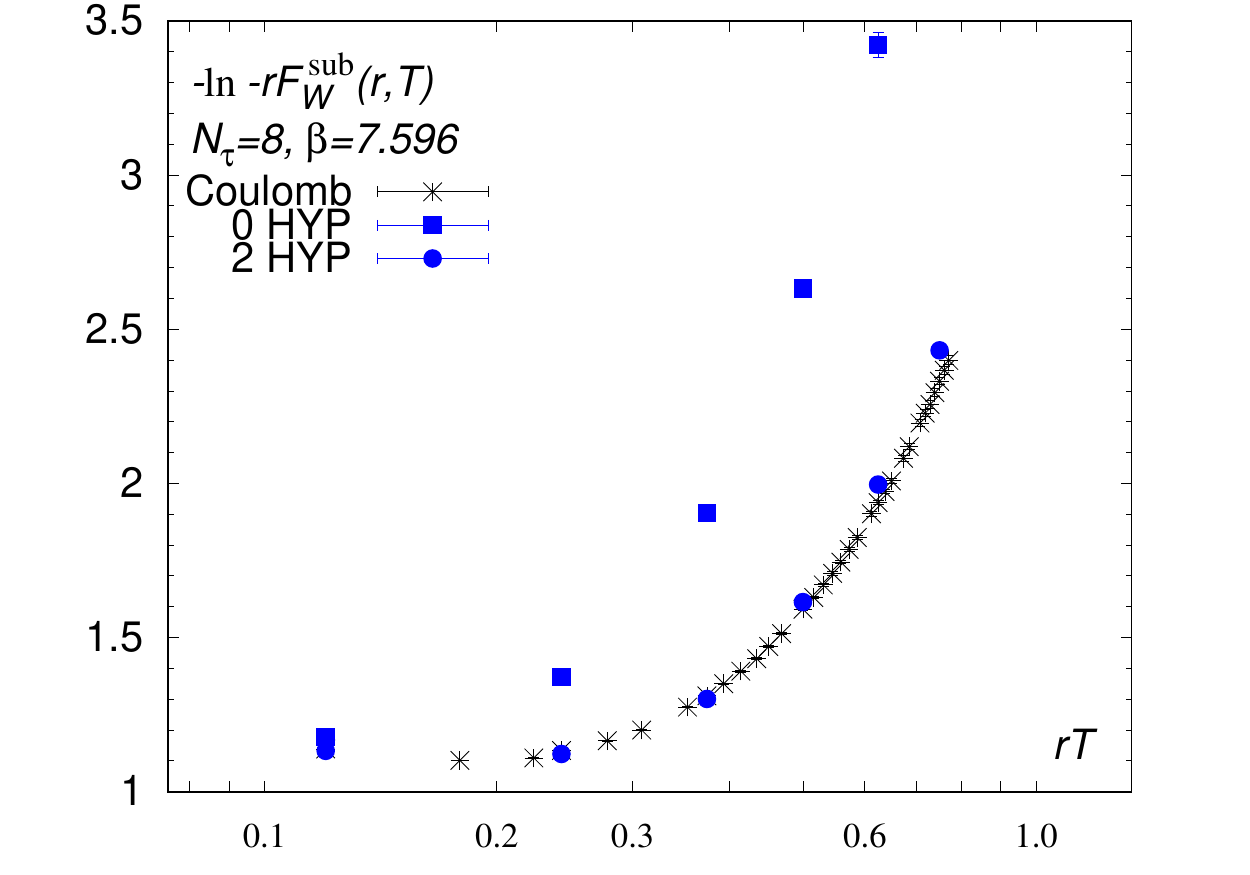}
\includegraphics[width=8.4cm]{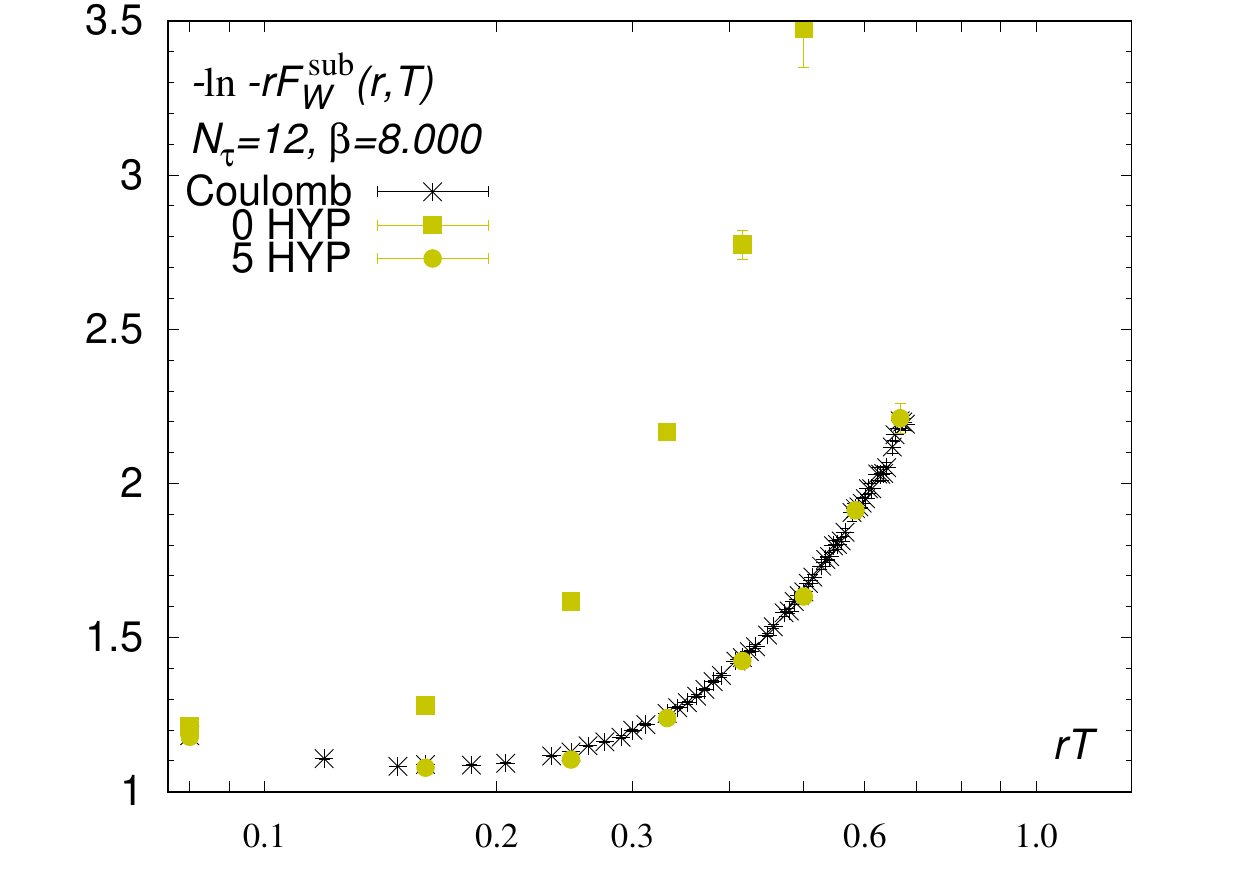}
\caption{
The logarithm of $-r F_W^{\rm sub}$ calculated from smeared and unsmeared Wilson loops. 
The black bursts show the result for the singlet correlator in the Coulomb gauge.
(Upper) Correspond to $T=200$ MeV. (Lower) Correspond to $T=480$ MeV. 
(Left) Show the $N_{\tau}=8$ results. (Right) Show the $N_{\tau}=12$ result.
}
\label{fig:Wilson_loops}
\end{figure*}

In this subsection, we present our numerical results for cyclic Wilson loops $W_S^{\rm bare}$. 
Cyclic Wilson loops being gauge invariant offer an attractive
possibility to study the medium modifications of $\qbq$ interactions.
From previous lattice results, we know that $W_S^{\rm bare}$ approaches 
$[L^{\rm bare}]^2$ at large distances \cite{Bazavov:2009zn,Bazavov:2013zha}. 
This is also expected based on weak-coupling calculations~\cite{Burnier:2009bk}. 
Like the Coulomb gauge singlet correlator, $W_S^{\rm bare}$ is also dominated 
by the singlet contribution at small distances~\cite{Berwein:2012mw,Berwein:2013xza}. 
Moreover, weak-coupling calculations show that $W_S^{\rm bare}$ is screened at 
large distances like the Coulomb gauge singlet correlator. 
However, the study of cyclic Wilson loops is complicated by the presence of UV divergences.
In the continuum theory, the renormalization of cyclic Wilson loops has been 
worked out for short distances, $rT\ll1$~\cite{Berwein:2012mw,Berwein:2013xza}.
However, on the lattice, we do not have enough data points at truly short distances to test these ideas. 
In part, this is due to the fact that off-axis separation cannot be used in 
the analysis (see the discussion in \mbox{Sec.}~\ref{sec:setup}). 
Therefore, we attempt to compare the unrenormalized smeared cyclic Wilson loops with the Coulomb gauge correlator. 
The smearing of the spatial links in $W_S^{\rm bare}$ is expected to reduce 
the size of UV divergences due to the self-energy, which are the largest divergences that affect the spatial lines. 
As we will see, for smeared cyclic Wilson loops, the UV divergences are 
sufficiently small at the considered values of $N_{\tau}$ to allow for a meaningful comparison.

We discuss our numerical results for cyclic Wilson loops in terms of \(F_W^{\rm sub}=-T \ln(W_S^{\rm bare}/[L^{\rm bare}]^2)\). 
The analogy with \mbox{Eq.}~\eqref{eq:Fssub} is obvious. 
We show the logarithm of $-r F_W^{\rm sub}$ in \mbox{Fig.}~\ref{fig:Wilson_loops}. 
There, we also show the usual Coulomb gauge singlet correlator in direct comparison. 
It is evident that $-r F_W^{\rm sub}$ from the unsmeared Wilson loop decays 
much more rapidly than the singlet correlator in the Coulomb gauge. 
This is due to the self-energy divergence of the spatial lines of the loop, 
which effectively increases the screening mass. 
As we increase the number of HYP smearing iterations, the size of the 
self-energy divergence in the spatial line is gradually reduced. 
After a few steps of HYP smearing, we obtain results for $-r F_W^{\rm sub}$ 
that agree with the results from the singlet correlator in the Coulomb gauge. 
The number of HYP smearing that is needed to sufficiently reduce the size of 
the self-energy divergence is larger for smaller lattice spacing.
For higher temperatures, and thus finer lattice spacings at fixed $N_\tau$, 
more smearing iterations are required in general, which can be seen by 
comparing the lower temperature ($T\approx 200\,{\rm MeV}$, upper 
panel in \mbox{Fig.}~\ref{fig:Wilson_loops}) with the higher temperature 
($T\approx 480\,{\rm MeV}$, lower panel). 
We see that larger $N_\tau$ (finer lattices) requires larger numbers 
of smearing iterations to achieve a comparable effect.
These findings are in qualitative agreement with the study performed
in the SU(2) gauge theory \cite{Bazavov:2008rw}. 

Though both correlators are affected by different technical difficulties, 
the accessible information regarding the $\qbq$ interaction and the 
properties of the quark-gluon plasma seem to be similar.

\section{Comparison of the singlet free energy to weak coupling}
\label{sec:singlet}

In this section, we compare the lattice results for the singlet free energy 
with the predictions of the weak-coupling calculations. 
These comparisons are the key for a deeper understanding of the mechanisms 
behind the onset of thermal effects with increasing separation of the 
$\qbq$ pair and provide a tool to distinguish between electric and magnetic screening on the lattice.

The numerical results discussed in the previous section suggest that 
medium effects for the singlet free energy are small for $r T<0.4$. 
Furthermore, weak-coupling calculations have been worked out up to 
order~$g^5$ for \(rT \ll 1\)~\cite{Berwein:2017thy} and to next-to-leading order (NLO) for $r \sim 1/m_D$ \cite{Burnier:2009bk}, 
which is accurate up to order $g^4$ \cite{Berwein:2017thy}.
Hence, we will compare to the weak-coupling results separately for short ($rT \ll 1$) and large distances ($r\md \sim 1$). 
This separation of different regimes is, strictly speaking, valid only for asymptotically 
small couplings $g\ll 1$. 
However, as we will see below, the scale separation and the corresponding 
effective field theory calculations can also help to understand 
the lattice results when $g \sim 1$.

\subsection{Thermal corrections for small $r$}
\label{sec:vminusf}

Here we compare the lattice results with the weak-coupling calculations for $r T \ll 1$. 
In this case the singlet free energy is given by the energy of a static $\qbq$ 
pair in the vacuum $\Vs$, plus thermal corrections, which, however, are expected to be small. 
For this reason, it makes sense to perform the comparison with the lattice results in terms of the difference $\Vs-\Fs$. 
The comparison of the lattice and weak-coupling results in terms of this difference has several advantages. 
First, one does not have to deal with the complications associated with the
perturbative calculation of \(\Vs\), among others, infrared logarithms (see 
\mbox{Refs.}~\cite{Brambilla:1999qa, Bazavov:2014soa}). 
Second, lattice artifacts largely cancel out in this difference, making the 
continuum extrapolations of this quantity simpler.
Finally, the ultraviolet additive renormalization does not enter in $\Vs-\Fs$.

We show our continuum results for $\Vs-\Fs$ in Fig. \ref{fig:vs-fs}.
The details of the continuum extrapolations are discussed in Appendix~\ref{app:C}. 
The difference $\Vs-\Fs$ in physical units is small at short distances
and roughly temperature independent. 
As the temperature increases, the distances where medium effects are significant become smaller. 
Since we expect that thermal corrections should be proportional to 
the temperature, we present our results in temperature units in
the right panel of \mbox{Fig.}~\ref{fig:vs-fs}. 
We see that for $T>300$ MeV the difference $\Vs-\Fs$ in temperature units 
approaches at short distances a constant.
As we will explain in the next paragraph, this behavior is expected based on 
the weak-coupling calculations.

\begin{figure*}
\includegraphics[width=8.7cm]{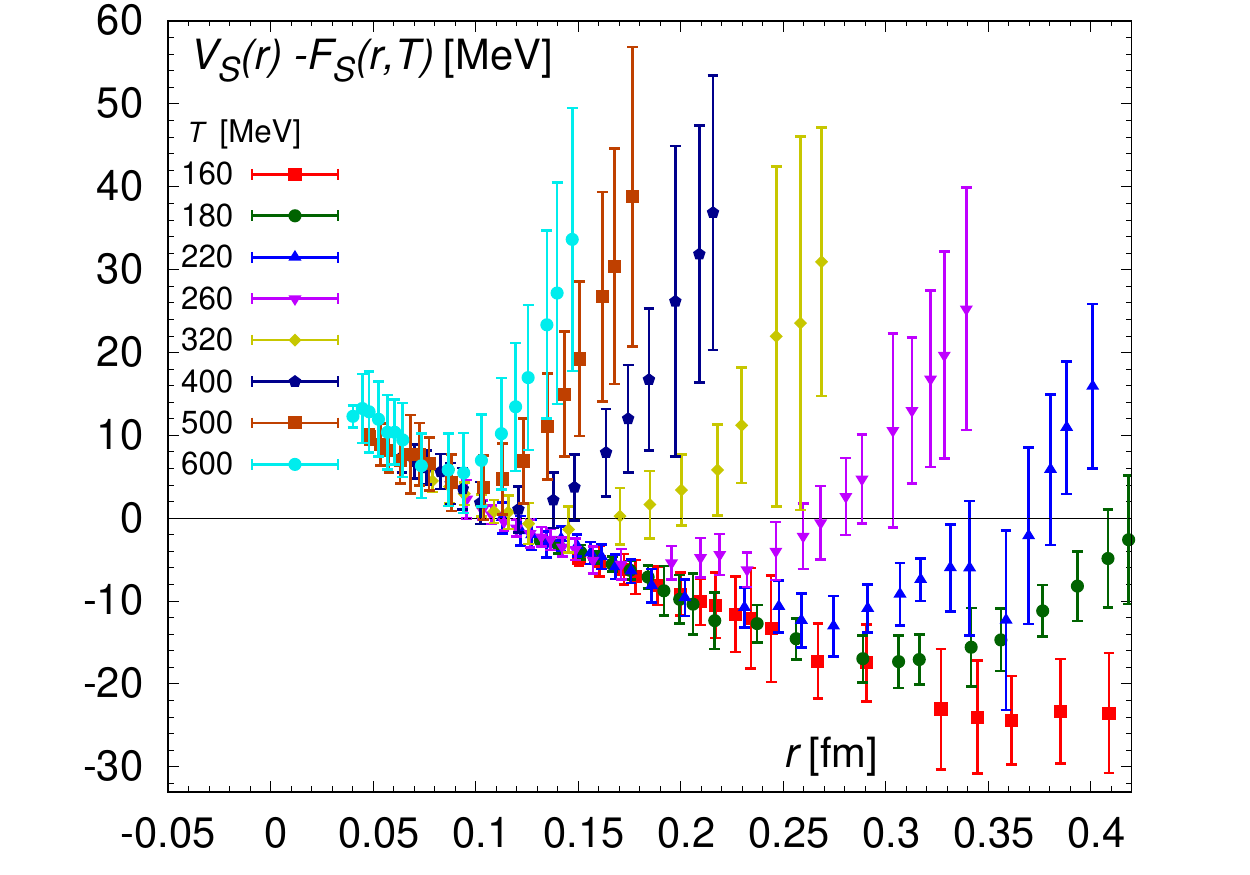}
\includegraphics[width=8.7cm]{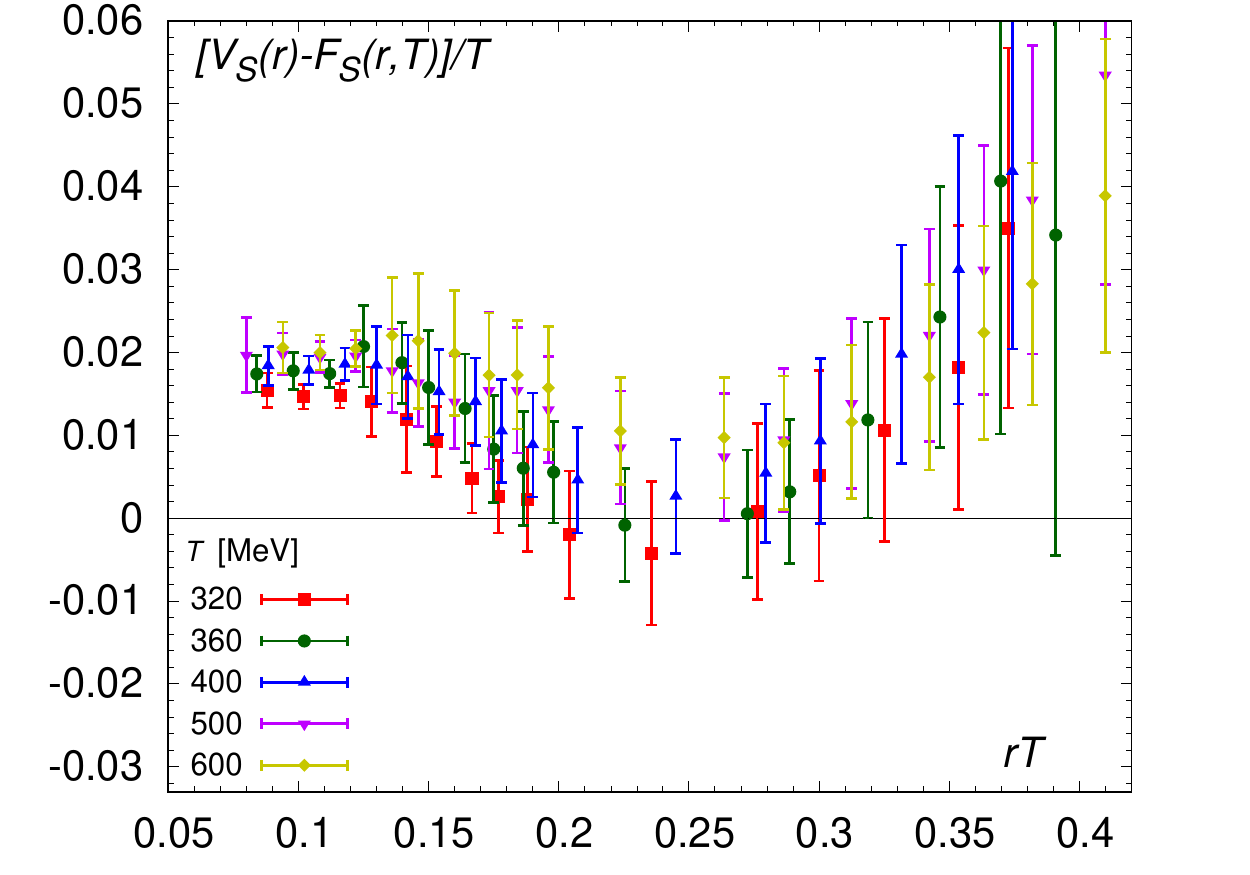}
\caption{
Continuum results for the difference $\Vs-\Fs$ in MeV (left) and in 
temperature units (right) as function of the distance $r$ between the 
static quark and antiquark.
}
\label{fig:vs-fs}
\end{figure*}

\begin{figure}
\includegraphics[width=8.5cm]{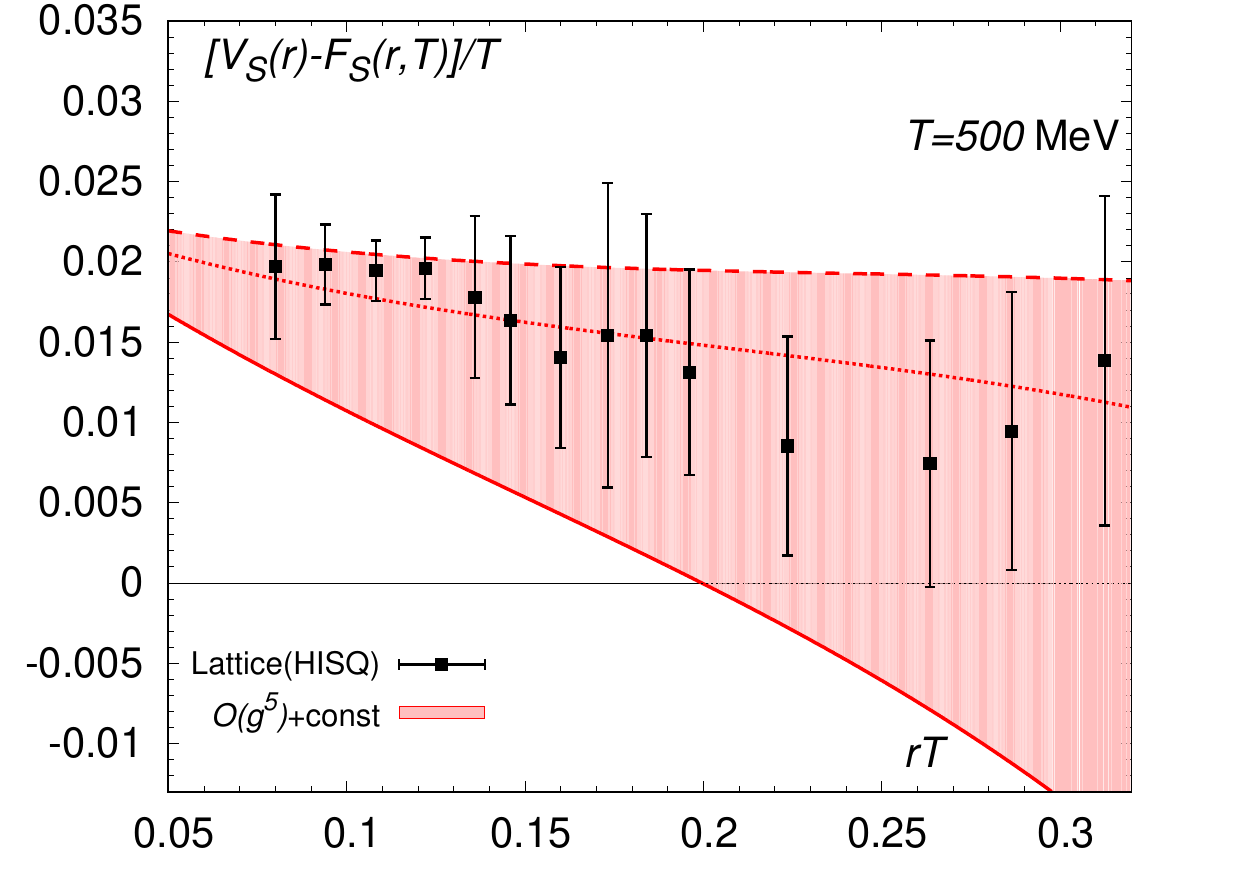}
\caption{Comparison of lattice and weak-coupling results for $\Vs-\Fs$. 
The weak-coupling results have been shifted by a small constant to match 
them to the lattice results at the shortest distance (see text). 
The dotted line corresponds to the renormalization scale $\mu=2\pi T$, 
while the band corresponds to its variation from $\pi T$ (solid) to 
$4\pi T$ (dashed).
}
\label{fig:vs-fs_pert}
\end{figure}
 
Now we discuss the comparison of $\Vs-\Fs$ with the weak-coupling calculations. 
From the point of view of weak-coupling calculations, short distances imply the scale hierarchy $1/r \gg T$.
In this regime, pNRQCD can be used.
Moreover, the above scale hierarchy allows two further possible scale hierarchies: $\as/r \gg T$ and $T \gg \as/r$. 
We discuss the second case, since here explicit calculations for $\Vs-\Fs$ 
are available up to unknown corrections of order $g^4(rT)^5$ and 
$g^6$~\cite{Berwein:2017thy}. 
The result reads

\ileq{
 \Vs-\Fs 
 =\left(\Delta F_g+\Delta F_f+\Delta \Fs\right) T
 +\mathcal O(g^4(rT)^5,g^6), 
\label{eq:vmf weak}
}

with contributions from the nonstatic gluons $\Delta F_g$, from the quarks 
(\(N_f\) massless quark flavors) $\Delta F_f$, and from the soft (Debye mass) scale $\Delta \Fs$,

\ilal{
 \Delta F_g
 &= \as^2 \nc\cf\left[-\frac\pi9\,(rT) +\frac43\zeta(3)\,(rT)^2-
 \frac{22\pi^3}{675}\,(rT)^3\right],
\label{eq:dFg}
 \\
 \Delta F_f
 &= \as^2 N_f\cf\left[\zeta(3)\,(rT)^2-
 \frac{7\pi^3}{270}\,(rT)^3\right],
\label{eq:dFf}
 \\
 \Delta \Fs
 &=-\as^{5/2} \frac\cf6\left[4\pi \frac{\nc+\frac{N_f}2}3\right]^{\frac32}\,(rT)^2\, .
\label{eq:dFs}
}

It is clear from the above expression that $\Vs-\Fs$ vanishes in the $r \rightarrow 0$ limit. 
This seems to be in apparent contradiction with the lattice results shown in \mbox{Fig.}~\ref{fig:vs-fs}. 
However, as remarked in \mbox{Ref.}~\cite{Berwein:2017thy} (see also the singlet normalization in~\cite{Brambilla:1999xf}), 
there is a possible order $g^6$ contribution, which is proportional to the temperature and independent of $r$. 
Such a contribution would naturally explain the trend seen in the lattice results in Fig. \ref{fig:vs-fs} (right).
Note that the $(rT)^2$ term in \mbox{Eq.}~\eqref{eq:dFs} has the opposite sign 
with respect to the $(rT)^2$ terms in \mbox{Eqs.}~\eqref{eq:dFg} and~\eqref{eq:dFf}, leading to a partial cancellation. 
This partial cancellation is in fact responsible for relatively
small medium effects in the considered distance region.
In \mbox{Fig.}~\ref{fig:vs-fs_pert}, we compare the lattice QCD results
for $T=500\,{\rm MeV}$ to the perturbative calculation reproduced in \mbox{Eq.}~\eqref{eq:vmf weak}.
We use $\Lambda_{\overline{\rm MS} }=320\,{\rm MeV}$ and the two-loop running coupling for this comparison. 
This choice of $\Lambda_{\overline{\rm MS} }$ and running coupling
gives $\as^{N_f=3}(\mu=1.5~{\rm GeV})=0.35$, which is very close to the 
lattice QCD determination of $\as^{N_f=3}$ based on the static 
energy~\cite{Bazavov:2014soa}, as well as to the lattice determination based 
on the charmonium correlators~\cite{Maezawa:2016vgv}.
For the renormalization scale, we choose $\mu=2 \pi T$ and add a small
constant to the above weak-coupling results, which mimics the possibly missing $g^6$ 
contribution such that the weak-coupling calculation agrees with the lattice at the shortest distance.
We vary the renormalization scale by a factor of 2 around the central
value to estimate the uncertainty of the perturbative result. 
This uncertainty is shown as a band in \mbox{Fig.}~\ref{fig:vs-fs_pert}.
Because of the partial cancellations between $\Delta F_g+\Delta F_f$ and $\Delta \Fs$, 
the net temperature shift of $\Vs-\Fs$ is relatively small for $r T \lesssim 0.3$ in agreement with the lattice results. 
Thus weak-coupling results obtained for $r T \ll 1$ enable us to explain the 
lattice findings on $\Fs$ for small $r$, \mbox{i.e.} for $r \lesssim 0.3/T$, in 
particular, their small temperature dependence. 
For larger distances, $\Delta \Fs$ becomes numerically the dominant correction, 
even though it is parametrically small (order $g^5$)
compared to $\Delta F_g$ and $\Delta F_f$, which are of order $g^4$.
This means that screening effects become important, and the hierarchy of
scales $1/r \gg T \gg m_D$ is not fulfilled anymore. 
We need to consider a weak-coupling expansion that does not assume such a hierarchy and resums screening effects right from the start.
This is discussed in the next subsection.

There is no perturbative calculation of $\Fs-\Vs$ for the case $\as/r \gg T$.
However, from the power counting, we expect thermal corrections to be small~\cite{Brambilla:2008cx}. 
Furthermore, we also expect an $r$-independent term proportional to $T$ appearing at order $g^6$, 
since such a term possibly arises from the matching of pNRQCD to nonrelativistic QCD (NRQCD)~\cite{Berwein:2017thy}. 
Therefore, several qualitative features of $\Fs-\Vs$ discussed above should also hold in this hierarchy of scales.
This is quite different from the case of $\Fqq$, where the two regimes $\as/r \ll T$ and  $\as/r \gg T$
differ at leading order in the functional dependence on $r$.

\subsection{Electric screening of the singlet free energy}
\label{sec:electric}

\begin{figure}
\vspace{0.2ex}
\includegraphics[width=8.8cm]{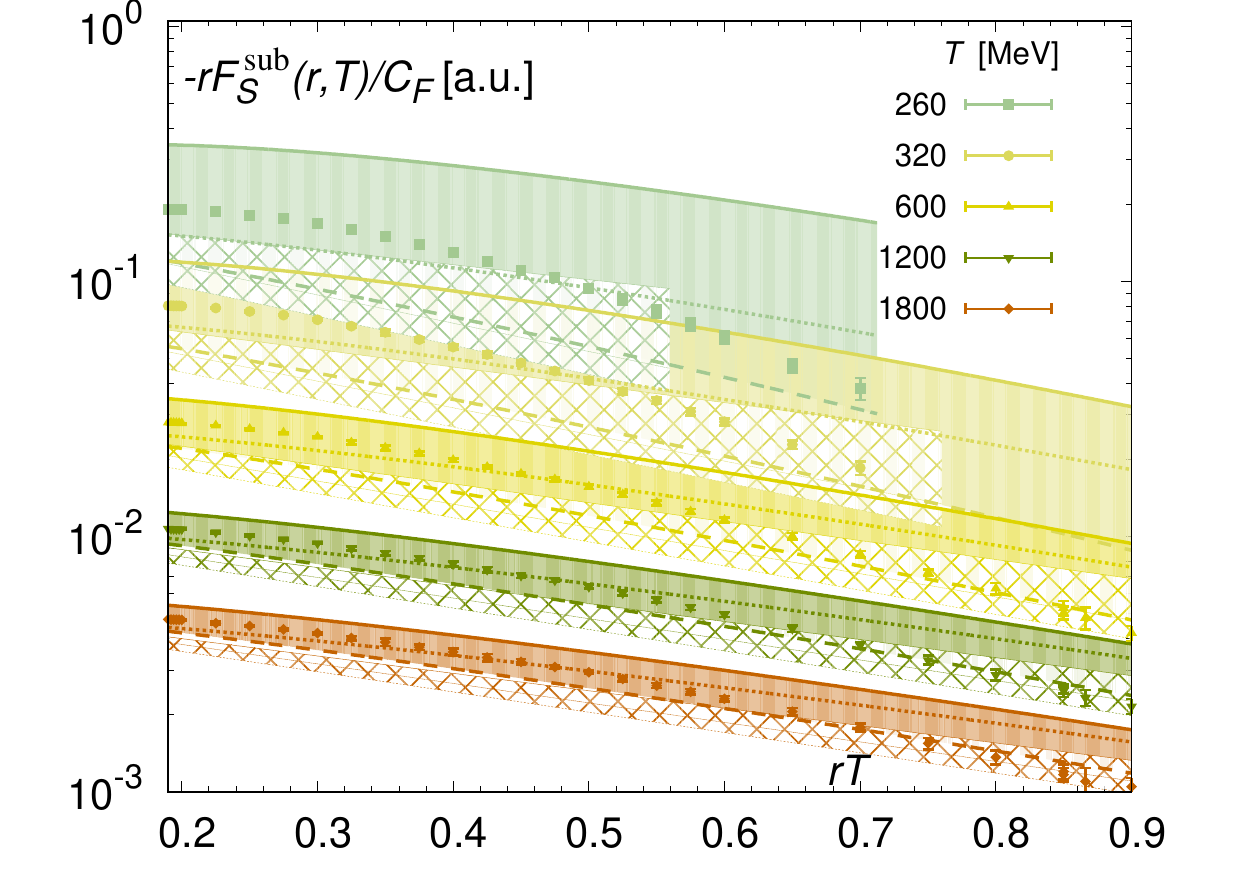}
\caption{Continuum results for $-r\Fs^{\rm sub}$ as function of $rT$ at different temperatures. 
For visibility, we shifted the results for different temperatures. 
The uncertainty bands represent the leading-order (LO, hashed) and NLO (solid) results; 
the renormalization scale is varied from $\mu=\pi T$ (solid) to $2\pi T$ (dotted) and $4\pi T$ (dashed). 
}
\label{fig:fs_sub}
\end{figure}

In this subsection, we compare the lattice results
for the singlet free energy with the weak-coupling prediction
in the regime $r \sim  1/\md$, \mbox{i.e.} in the electric screening regime.
As mentioned in the Introduction, in this case EQCD can be used
and the correlator can be calculated perturbatively. 
The physics in this regime is controlled to a large extent by the mass
parameter of EQCD, $\md$, which is often called the Debye mass.
At leading order

\ileq{
\md|_{\rm LO}(\mu)=g(\mu)T\sqrt{\frac{2N_c+N_f}6}.
\label{eq:mdlo}
}

The comparison of the lattice and the weak-coupling results is performed 
in terms of the subtracted free energy $\Fs^{\rm sub}$ defined in 
\mbox{Eq.}~\eqref{eq:Fssub}. 
The comparison in terms of \(\Fs^{\rm sub}\) is most convenient, since $\Fs$ 
is very close to $2\Fq$ in this distance regime, and the weak-coupling 
calculations of $2\Fq$ are difficult \cite{Berwein:2015ayt,Berwein:2015ayt}.
Our continuum results for $\Fs^{\rm sub}$ are shown in 
\mbox{Fig.}~\ref{fig:fs_sub}.
We clearly see that at $rT>0.25$ $\Fs^{\rm sub}$ decreases with 
increasing $r$. 
This fact is consistent with the observation made in the previous section 
that medium effects become large at $rT >0.3$. 
The subtracted singlet free energy in the Coulomb gauge has been calculated at 
NLO in \mbox{Ref.}~\cite{Burnier:2009bk} using EQCD.
We use \mbox{Eq.~(3.22)} from that reference and omit the two-gluon
exchange contribution proportional to $g^4/r^2$, which is of order $g^6$
in the regime $r\sim 1/\md$ \cite{Berwein:2017thy}. 
The result reads

\begin{widetext}
\ilal{
  \Fs^{\rm sub}|_{\rm NLO}(r,T) &=-\cf \frac{\as(\mu)}{r} e^{-r\md(\mu)} \Big(
  1+ \as(\mu)\nc rT\ \Big[2-\ln\big(2r\md(\mu)\big)-\gamma_E+
  e^{+2r\md(\mu)}E_1(2r\md(\mu)) \Big] \Big),
\label{eq:S1 nlo}
}
\end{widetext}

where \(E_1(z)\) is the exponential integral, \mbox{i.e.}

\ileq{
E_1(z) = \int\limits_{z}^\infty \frac{dt}{t} e^{-t}.
}

There are two corrections to the above result that are formally of higher
order but numerically important. The first one is the correction
due to the running coupling \cite{Burnier:2009bk,Berwein:2017thy}, 
which can be written in the form \cite{Berwein:2017thy}

\ilal{
 \delta \Fs^{\rm sub}(r,T)&=-\frac{\cf \as(\mu)}{r} e^{-r\md(\mu)}
 \nn\\&\hspace{12pt}
 \times\left(1-\frac{r\md(\mu)}{2}\right)\delta Z_1(\mu),
}

with

\ilal{
 \delta Z_1(\mu)&=\as(\mu)\Big[\frac{11 N_c}{3}+\frac{2}{3}
 \left(1-4\ln 2\right)
 \nonumber\\[2mm]
 &\hspace{43pt}+2 \beta_0 \left(\gamma_E+\ln\frac{\mu}{4 \pi T} \right) \Big],
 \nonumber\\[2mm]
 \beta_0&=\left(11 N_c/3-2 N_f/3\right).
\label{eq:Z1 nlo}
}

The other correction is the correction to $\md$.
The NLO result for the Debye mass, which has been calculated in \mbox{Ref.}~\cite{Braaten:1995jr}, reads

\ilal{\nonumber
& \!\!\!\! \md^2|_{\rm NLO}(\mu) = 
 \md^2|_{\rm LO}(\mu) \Big(
 1+\frac{\as(\mu)}{4\pi}\Big[
 2 \beta_0 \left(\gamma_E+\ln\frac{\mu}{4 \pi T} \right)
 \nonumber\\[2mm]&\quad
+ \frac{5\nc}{3} 
+ \frac{2N_f}{3} (1-4\ln 2)
 \Big]\Big)
 -\cf N_f \as^2(\mu) T^2 .
\label{eq:md nlo}
}

In \mbox{Fig.}~\ref{fig:fs_sub}, we compare the weak-coupling results at LO and NLO 
with the continuum extrapolated lattice results.
For the LO one, we use $\Fs^{\rm sub}|_{\rm LO}+\delta \Fs^{\rm sub}$ 
and for the NLO one, we use $\Fs^{\rm sub}|_{\rm NLO}+\delta \Fs^{\rm sub}$. 
For $\md$, we always use the above NLO result.
As before, we use $\Lambda_{\rm \overline{\rm MS}}=320\,{\rm MeV}$ and the two-loop running coupling. 
We vary the scale $\mu$ from $\pi T$ to $4\pi T$ to account for the 
uncertainty of the weak-coupling result. 
First of all, we see that the scale dependence is much reduced for higher temperatures.
For $T \gtrsim 600$ MeV, it is sufficiently small to allow for an accurate comparison with the lattice.
Moreover, the central value of the NLO result is always higher than the central value of the LO result.
Within the uncertainty band, due to the variation of the scale, we see good 
agreement between the NLO result and the lattice data up to at least $rT \approx 0.6$. 
For $rT\sim 0.3-0.6$, the shape of the lattice results and the
weak-coupling results is quite similar. 
At larger distances, however, the falloff of the lattice results is faster than the one of the NLO prediction. 
This is most likely due to the fact that, for these distances, the physics is 
affected by static chromomagnetic fields and, thus, it is nonperturbative. 
Based on previous calculations performed in the pure gauge theory,
we expect that the effective screening mass governing the large distance 
behavior of the singlet correlator becomes significantly larger than $\md$ 
due to chromomagnetic effects~\cite{Heller:1997nqa,Karsch:1998tx,Digal:2003jc,Kaczmarek:2004gv}. 
This would explain the change in the slope at large distances that we observe. 
The behavior of $\Fs$ in the large distance regime will be discussed in \mbox{Sec.}~\ref{sec:screening}. 
However, it is important to stress that chromoelectric screening effects 
are important already at distances $rT < 0.6$.

\section{Comparison of the Polyakov loop correlator to weak coupling}
\label{sec:plc}

In this section, we compare the lattice results for the Polyakov loop 
correlator \(C_P\) or, equivalently, the free energy of a static $\qbq$ pair 
\(\Fqq\), with the predictions of the weak-coupling calculations. 
Here and in the following, we use small letters for the indices to mark 
color singlet or color octet objects defined in the framework of weak-coupling 
calculations and capital letters for the indices to mark color singlet or 
color octet objects defined in the context of nonperturbative lattice calculations. 
As in the case of the singlet free energy, we will perform this comparison 
separately for the short distance regime and for the screening regime, where $r \sim 1/\md$.

\subsection{The Polyakov loop correlator at short distances}
\label{sec:reconstruction}

We start by assuming the case \(1/r \gg T \gg \md\). 
As it was discussed in \mbox{Sec.}~\ref{sec:results}, the free energy of a static $\qbq$ pair, 
\(\Fqq-T\ln 9\), shows large deviations from the zero temperature static energy already at short distances. 
One way to understand this is in the language of pNRQCD.
In the pNRQCD framework the Polyakov loop correlator at short distances 
can be written up to corrections of order \(\as^3(rT)^4\) as~\cite{Berwein:2017thy}

\ileq{
C_P(r,T) \equiv e^{-F_{Q\bar Q}/T}=\frac{1}{\nc^2} e^{-f_s/T}+\frac{\nc^2-1}{\nc^2} e^{-f_o/T},
\label{eq:pnrqcd fsfo}
}

where $f_s$ and $f_o$ are gauge-invariant singlet and octet free energies
defined in terms of pNRQCD correlators [see Eqs. (71) and (72) of \mbox{Ref.}~\cite{Berwein:2017thy}]. 
A good approximation of \eqref{eq:pnrqcd fsfo} is

\ileq{
e^{-F_{Q\bar Q}/T}=\frac{1}{\nc^2} e^{-V_s/T}+\frac{\nc^2-1}{\nc^2} L_A e^{-V_o/T},
\label{pnrqcd_decomp}
}

where \(V_s=-\cf\as/r+\ldots\) and \(V_o=-V_s/(\nc^2-1)+\ldots\) are, respectively, the 
quark-antiquark color singlet and octet potentials at $T=0$, and $L_A$ is the adjoint Polyakov loop expectation value.
\mbox{Equation}~\eqref{pnrqcd_decomp} is valid up to corrections of order \(\as^3\) 
[\mbox{cf. Eqs.}~(79) and (80) in \mbox{Ref.}~\cite{Brambilla:2010xn}].
From these equations, it is clear that the temperature dependence of the 
Polyakov loop correlator, and thus of the $\qbq$ free energy comes, through the color octet contribution, mostly from~$L_A$.

\begin{figure*}[ht]
\includegraphics[width=8.6cm]{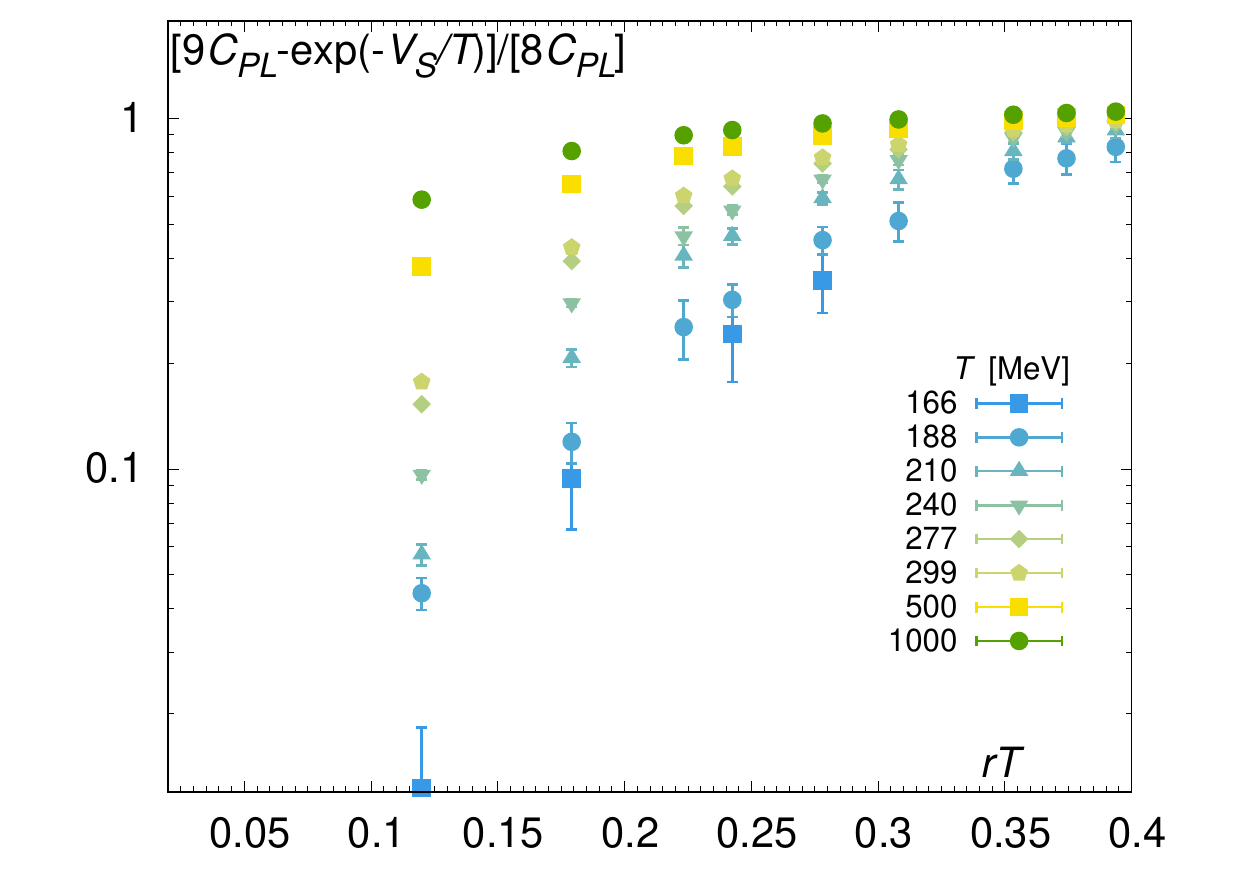}
\includegraphics[width=8.6cm]{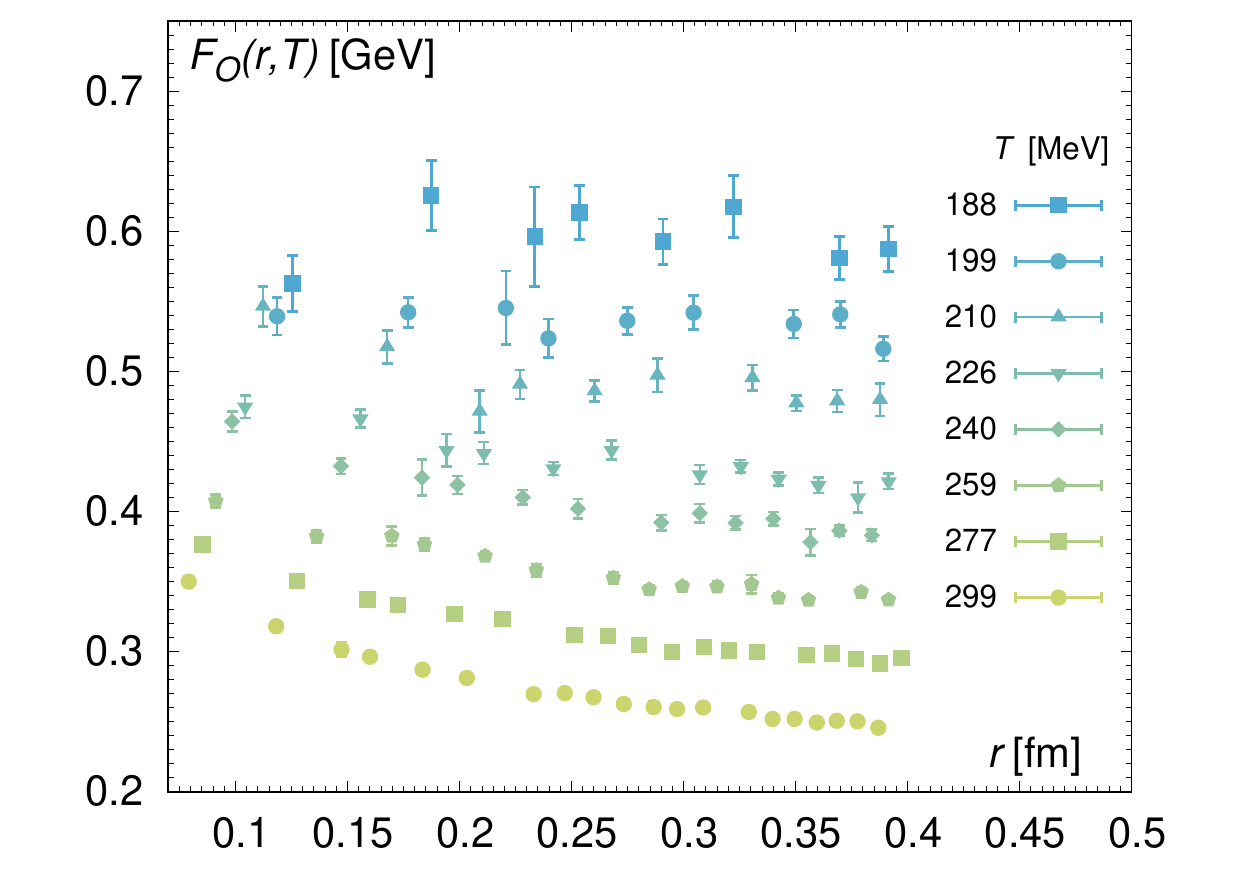}
\caption{The octet contribution to the Polyakov loop correlator (left) and 
the octet free energy (right) at different temperatures as function of the 
distance $r$ calculated on $N_{\tau}=8$ lattices; see text.}
\label{fig:oc}
\end{figure*}

To test to what extent the above formulas work in the temperature and distance 
ranges considered in this study, we define the octet contribution to the 
Polyakov loop correlator as \(C_o \equiv C_P-\exp(-V_s/T)/\nc^2\). 
In \mbox{Fig.}~\ref{fig:oc} (left panel) we show \(C_O\) (normalized by 
$8C_P/9$), which is a proxy for \(C_o\) directly defined on the lattice. 
Specifically, this means that we are using lattice data for \(C_P\) (on 
\(N_\tau=8\) lattices) and the zero temperature static energy \(\Vs\), 
calculated on the lattice in \mbox{Refs.}~\cite{Bazavov:2014pvz, 
Bazavov:2017dsy}, as a proxy for $V_s$.
In this way, we avoid the complications associated with the perturbative 
calculation of \(V_s\).
We see that \(C_O\) increases with increasing temperatures and increasing distances. 
This can be understood by considering that \(V_o\) is positive. 
Furthermore, at low temperatures, \(L_A\) is small, thus leading to additional
suppression of the octet contribution, while at high temperatures, $L_A$ is 
of order one~\cite{Petreczky:2015yta}.
According to \mbox{Eq.}~\eqref{eq:pnrqcd fsfo}, we can also parametrize \(C_o\) as 

\ileq{
C_o=\frac{N_c^2-1}{N_c^2} e^{-f_o/T}.
}

Thus, the proxy \(C_O\) defines a proxy \(F_O\) for the octet free energy \(f_o\). 
The results for \(F_O\) are also shown in \mbox{Fig.}~\ref{fig:oc} (right panel). 
We see that $F_O$ decreases with increasing temperatures. 
Moreover, for distances $r<0.1\,{\rm fm}$, we see hints for the presence of the 
repulsive tail \(V_o\).

To validate the pNRQCD framework on a quantitative level we reconstruct the 
subtracted free energy of a static \(\qbq\) pair using the 
pNRQCD decomposition formula~\eqref{pnrqcd_decomp} 
and lattice results for the zero temperature static \(\qbq\) energy \(\Vs\), 
taken from \mbox{Refs.}~\cite{Bazavov:2014pvz,Bazavov:2017dsy}, as a proxy 
for $V_s$, and for the Polyakov loop expectation value in the fundamental 
representation, \(L_F\equiv L\), taken from \mbox{Ref.}~\cite{Bazavov:2016uvm}.  
First, we write for the adjoint Polyakov loop, 

\ileq{
L_A=L_F^{C_A/C_F} \cdot (1-\delta_8)^{C_A/C_F},
}

where \(C_F=(\nc^2-1)/(2\nc)\) and $C_A=\nc$ are the Casimir operators of 
the fundamental and adjoint representations. 
The parameter $\delta_8$ is the measure of the breaking of Casimir scaling 
introduced in \mbox{Ref.}~\cite{Petreczky:2015yta}.
Casimir scaling implies $\delta_8=0$ and $L_A=L_F^{C_A/C_F}$. 
It was found that $\delta_8$ is small above $T_c$ and compatible with zero 
for $T>300$ MeV \cite{Petreczky:2015yta}.
Taking into account that $L_F=\exp[-\Fq/T]$ and $C_A/\cf=2 N_c^2/(N_c^2-1)$, 
we can then write for the subtracted energy 

\ilal{
&e^{-\Fqq^{\rm sub}/T}=
\frac{1}{\nc^2} \exp\left[-\frac{\Vs-2 \Fq}{T}\right]+\nonumber\\[2mm]
&
\frac{\nc^2-1}{\nc^2} \exp \left[ \frac{\Vs-2 \Fq}{(\nc^2-1)T} \right] \cdot (1-\delta_8)^{C_A/\cf} \cdot
e^{-\delta V_o/T},\nonumber\\[2mm]
&
\delta V_o=V_o+V_s/(\nc^2-1).
\label{eq:fqqrec}
}

In the above equation, we have separated out the factor $\delta V_o$ that breaks Casimir 
scaling for \(V_o\), which is the only purely perturbative input to \mbox{Eq.}~\eqref{eq:fqqrec}. 
This correction was calculated in \mbox{Ref.}~\cite{Kniehl:2004rk},

\ileq{
  \delta V_o = \frac{\nc}8\frac{\as^3}r\left(\frac{\pi^2}4-3\right) 
 +\mathcal{O}(\as^4).
\label{eq:dVo}
}

\begin{figure*}
\includegraphics[width=8.4cm]{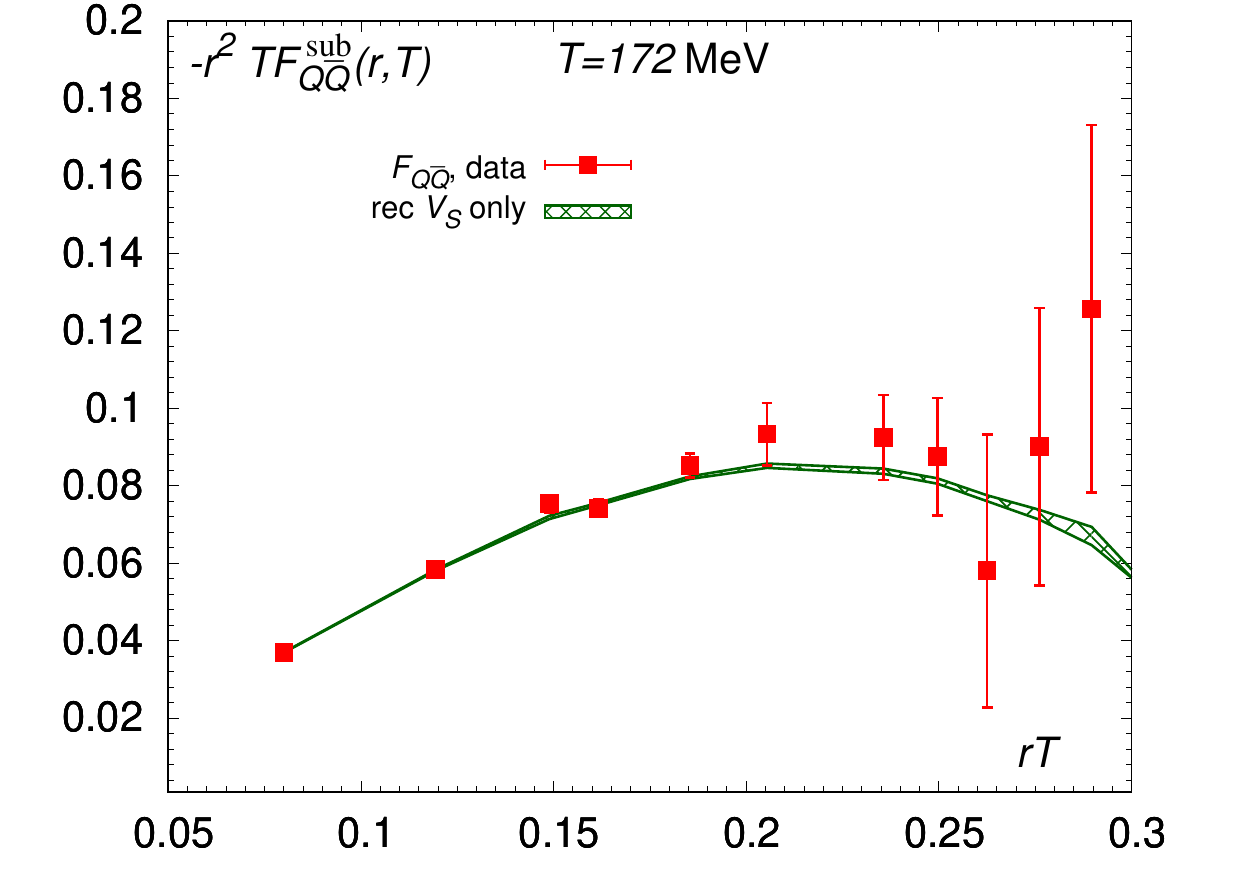}
\includegraphics[width=8.4cm]{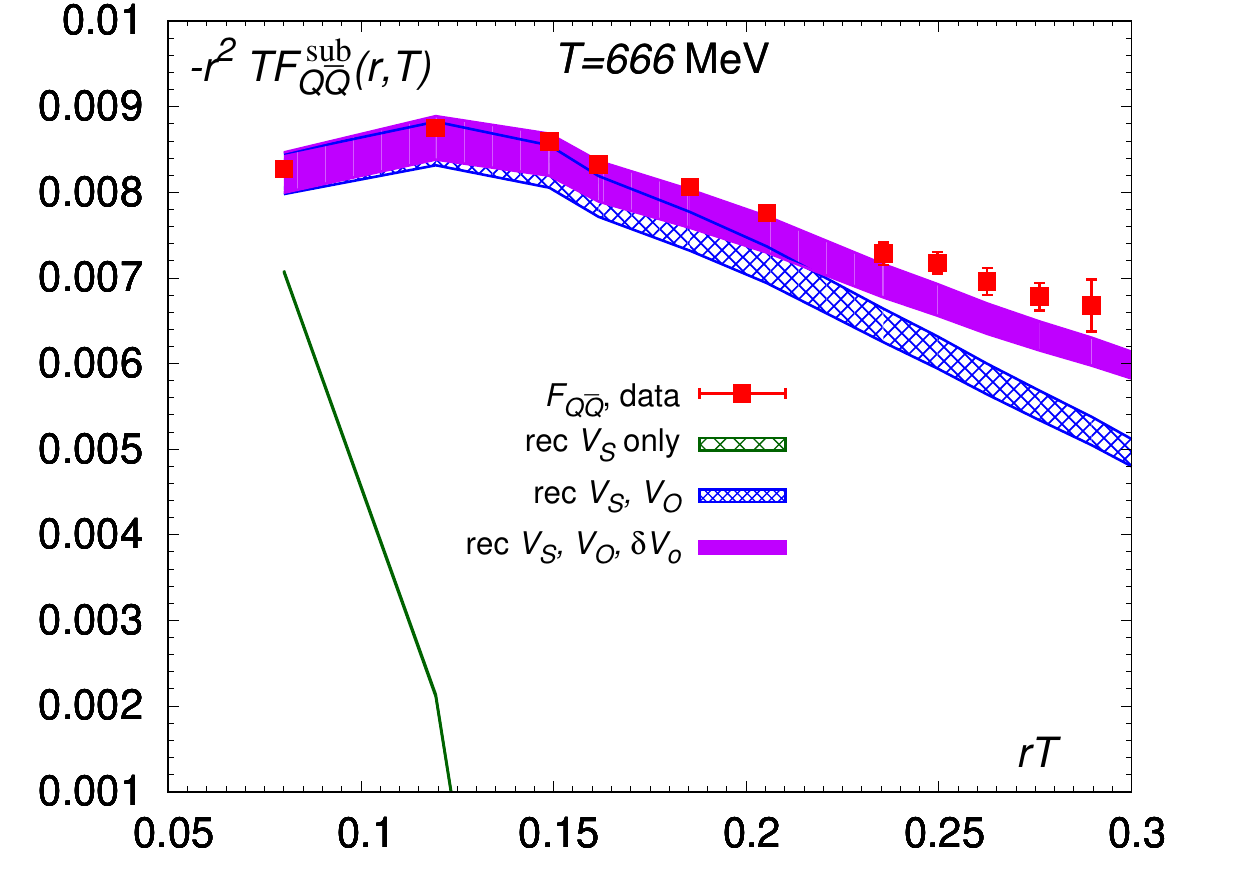}
\caption{The subtracted free energy multiplied by $-r^2 T$ calculated on $N_{\tau}=12$ lattices (squares)
and compared to the reconstruction based on the pNRQCD formula given by \mbox{Eq.} \eqref{eq:fqqrec} (bands).
(Left) The result for $T=172$ MeV 
(Right) The result for $T=666$ MeV
(the fully reconstructed result is in magenta, while the blue band ignores Casimir scaling violating 
contributions to the octet potential and the green one ignores the octet contribution).
}
\label{fig:fqqrec}
\end{figure*}

Now, using the lattice results for \(\Vs\) and \(\Fq\), we can calculate 
\(\Fqq^{\rm sub}\) according to \mbox{Eq.}~\eqref{eq:fqqrec} and compare this 
to the results of direct lattice calculations of \(\Fqq^{\rm sub}\). 
This comparison is shown in \mbox{Fig.}~\ref{fig:fqqrec} for $N_{\tau}=12$ at the two temperatures $T=172$ and $T=666$ MeV. 
The use of $N_{\tau}=12$ data for this comparison is justified because they are quite close to the continuum limit.
As one can see from the figure, the \mbox{pNRQCD} decomposition formula works very well up to distances $rT \approx 0.3$.
For $T=172$ MeV, the octet contribution is very small and in fact can be omitted.
For high temperatures, we could also use $\Fs$ as a proxy for $V_s$ and obtain very similar results. 
We note that for high temperatures not only the octet contribution but also 
the breaking of Casimir scaling for $V_o$ in \mbox{Eq.}~\eqref{eq:dVo} is numerically important.
If this breaking would be neglected, then the reconstruction of the Polyakov loop 
correlator in the right panel of \mbox{Fig.}~\ref{fig:fqqrec} would only work for $rT<0.2$.

We have seen that pNRQCD can explain the lattice results for the Polyakov loop correlator at short distances. 
Up to distances of about $0.3/T$, the behavior of the Polyakov loop correlator can be 
understood without invoking medium modification of the static quark-antiquark interactions [cf. Eq. \eqref{pnrqcd_decomp}].
This is in agreement with our findings for the singlet free energy 
$\Fs$, which also shows that the interaction between the static quark and 
antiquark is only mildly affected by the medium up to distances $rT \approx 0.3$. 

\begin{figure}[ht]
\includegraphics[width=8.6cm]{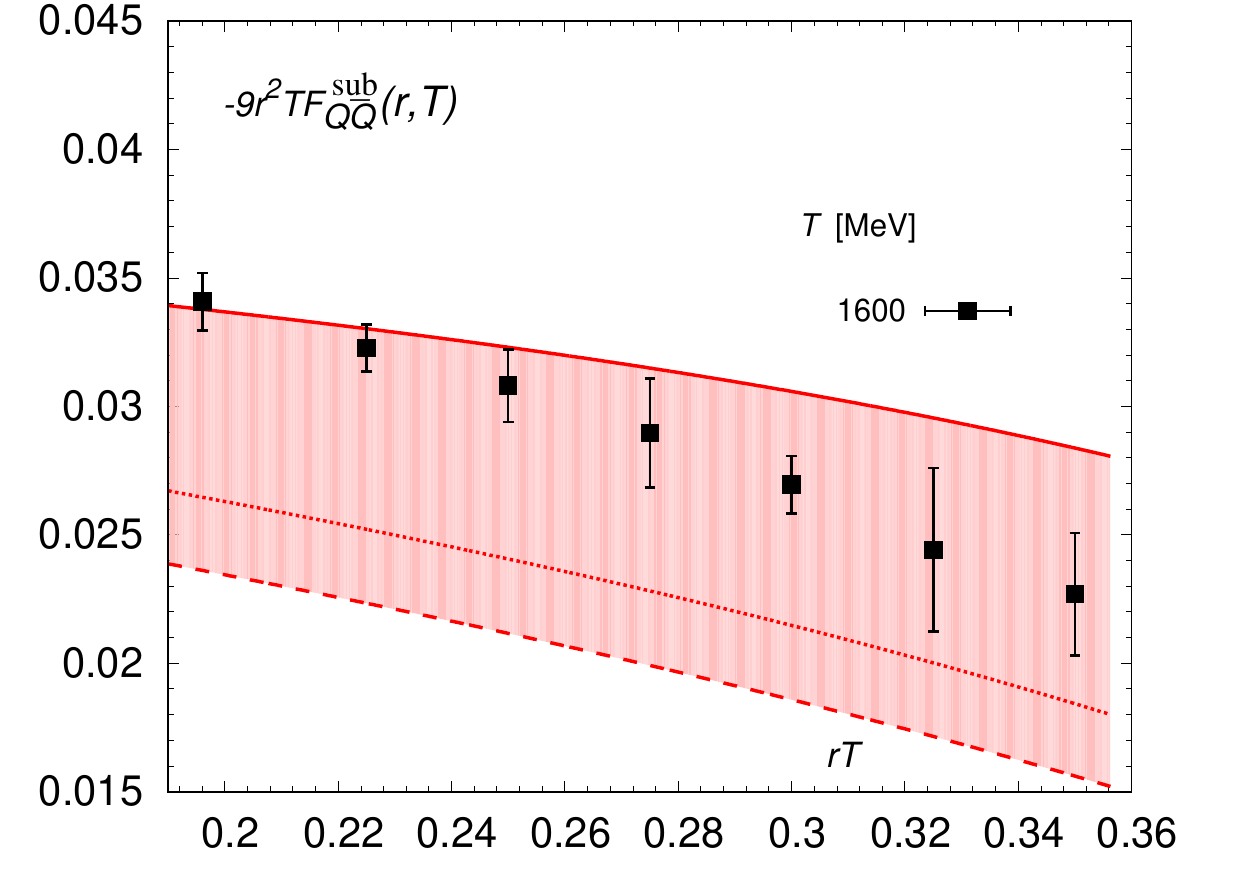}
\caption{
The $\qbq$ free energy multiplied by $-9r^2 T$ calculated for $T=1600$ MeV
in the continuum limit and compared with the NNNLO weak-coupling expression (lines).
The upper line corresponds to $\mu=\pi T$, the middle one to $\mu=2 \pi T$, 
and the lower line corresponds to $\mu=4 \pi T$.
}
\label{fig:g7}
\end{figure}

In order to compare the free energy $\Fqq^{\rm sub}$ with weak-coupling calculations, it is useful to distinguish two regimes.
In the regime $\as/r \ll T$, we can expand the exponentials in the right-hand side of Eq. \eqref{eq:pnrqcd fsfo} 
and compute them in perturbation theory, obtaining at leading order in $\as$

\ileq{
\Fqq^{\rm sub} = -\frac{\nc^2-1}{8\nc^2}\frac{\as^2}{r^2 T},
\label{eq:fqq_lo}
}

which is valid up to corrections of order \(g^5\).  
Weak-coupling calculations have been performed in this case up to next-to-next-to-next-to-leading order (NNNLO), 
which is accurate up to order $g^7$~\cite{Berwein:2017thy}. 
In \mbox{Fig.}~\ref{fig:g7}, we compare the NNNLO result for $\Fqq^{\rm sub}$ given by 
\mbox{Eqs.}~(54) and (55) of \mbox{Ref.}~\cite{Berwein:2017thy} with the continuum extrapolated lattice result at $T=1600$ MeV. 
We use three values of the renormalization scale: $\mu=\pi T$, $\mu=2 \pi T$, and $\mu=4 \pi T$. 
The running coupling has been calculated as before.
We see that the lattice result and the weak-coupling result agree for 
$r T>0.2$, although the latter has a large uncertainty. 
At very small distances $r$, the hierarchy $\as/r \ll T$ breaks down.

In the regime $\as/r \gg T$, the right-hand side of Eq. \eqref{eq:pnrqcd fsfo} is dominated by the color singlet contribution.
This happens at shorter and shorter distances as the temperature increases. 
We have seen in \mbox{Fig.}~\ref{fig:fqqs short} that our lattice data appear to be sensitive to this regime so that 
we could even define an effective coupling and compare it in \mbox{Fig.}~\ref{fig:aqqa} 
with the strong coupling from the zero temperature force between a static $\qbq$ pair.

\subsection{The free energy in the screening regime}
\label{sec:electric plc}

\begin{figure}
\includegraphics[width=8.7cm]{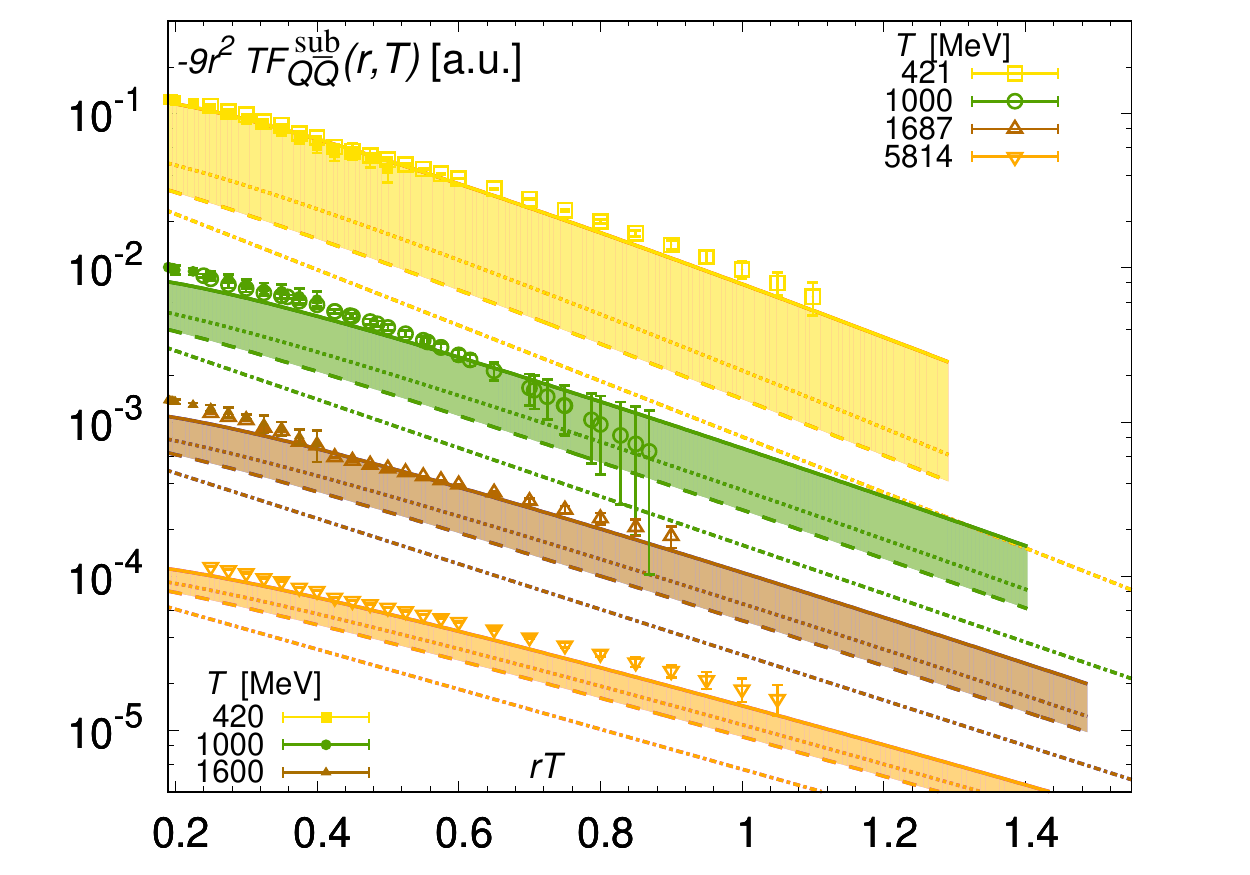}
\caption{\label{fig:sqq}
The free energy $\Fqq^{\rm sub}$ multiplied by $-9r^2 T$ calculated on the lattice and in weak-coupling EQCD. 
The open symbols show lattice data for $N_\tau=4$ with aspect ratio $N_\sigma/N_\tau=6$ corrected for cutoff effects. 
The result for \(T=1\,{\rm GeV}\) has \(N_\sigma/N_\tau=4\). 
The bands represent the NLO (solid) results, 
where the scale has been varied as $\mu=\pi T,\,2\pi T$, and $4\pi T$ (solid, dotted, and dashed lines). 
The dash-dotted lines correspond to the LO result evaluated at $\mu=4 \pi T$.
}
\end{figure}

In the following, we study to which extent the weak-coupling expression 
for the free energy is able to describe the lattice data in the electric screening regime. 
In \mbox{Sec.}~\ref{sec:electric}, we have shown that the singlet free 
energy is numerically consistent with the prediction of EQCD at NLO in the range $rT\sim 0.3-0.6$. 
However, studies of the screening regime of the Polyakov loop correlator on 
the lattice are a daunting task. 
First, the signal-to-noise ratio of the Polyakov loop correlator is 
very poor, so we need very large statistics. 
Second, we find that, due to the smallness of the signal, finite volume 
effects in the subtracted free energy become important already for rather small separations. 
Hence we require lattices with large volumes. 
We satisfy these constraints by using the new $N_\tau=4$ ensembles with aspect 
ratio $N_\sigma/N_\tau=6$ and total ensemble sizes of $2$-$6\times 10^4$ gauge configurations. 
The smallness of the signal can be understood from the leading-order 
expression of the Polyakov loop correlator. 
Because of cancellations between singlet and octet contributions, mentioned
in the previous subsection, the LO result for $\Fqq^{\rm sub}$ in the 
electric screening regime comes from the exchange of two electrostatic 
gluons~\cite{Gross:1980br,McLerran:1981pb,Nadkarni:1986cz},

\ileq{
\Fqq^{\rm sub}=-\frac{N_c^2-1}{8 N_c^2} \frac{\as^2}{r^2T}\exp(-2\md r).
}

Therefore, at large distances, $\Fqq^{\rm sub}$ is much smaller than $\Fs^{\rm sub}$. 

In \mbox{Fig.}~\ref{fig:sqq}, we show our results for the subtracted free 
energy as a function of $rT$ at different temperatures. 
The numerical results for different temperatures have been shifted vertically for better visibility.
The results at short distances, $r T<0.5$, and temperatures below 
\(2\,{\rm GeV}\) have been extrapolated to the continuum limit (see Appendix \ref{app:C}). 
The continuum extrapolated data are shown as filled symbols in \mbox{Fig.}~\ref{fig:sqq}.
At large distances, the errors are large for $N_{\tau}>4$ and thus continuum extrapolations are not possible.
To circumvent this problem, we correct the $N_{\tau}=4$ data for cutoff effects.
We define the correction factor \(K_4\) as the ratio between the continuum result for $\Fqq^{\rm sub}$ and the $N_{\tau}=4$ result: 
\(K_4=\Fqq^{\rm sub}({\rm cont})/\Fqq^{\rm sub}(N_\tau=4)\). 
Wherever the continuum result is available, the correction factor is known. 
Reliable continuum results are available only up to distances $rT \approx 0.45$. 
We assume that at distances $rT>0.45$ the correction factors are $r$ independent 
and equal to the correction factor at $rT=0.45$. 
Our analysis shows that cutoff effects are larger at larger distances (see Appendix \ref{app:C}).
As a consequence, the above procedure gives a continuum estimate that may be higher than the true continuum result
(the continuum limit is approached from above). 
For the highest temperature considered in our study, namely, $T=5814$ MeV, no continuum results are available. 
To make a continuum estimate at this temperature, we note that the correction factor decreases with increasing 
temperature because the cutoff effects become smaller at high temperatures (see Appendix \ref{app:C}).
The highest temperature where $K_4$ can be estimated is $T=1600$ MeV. 
We assume that the cutoff effects at $T=5814$ MeV are 2 times smaller than at $T=1600$ MeV, 
i.e., we use $K_4(T=5814~\text{MeV})=1-(1-K_4(T=1600~\text{MeV}))/2$.

Now let us compare the lattice results for $\Fqq^{\rm sub}$ with the
weak-coupling calculations at $r \sim 1/\md$, \mbox{i.e.} with the EQCD calculations. 
The free energy has been calculated at NLO in EQCD in \mbox{Ref.}~\cite{Nadkarni:1986cz}. 
Higher order corrections due to the running of the coupling are 
numerically important and have to be included in the analysis. 
These corrections have been computed in \mbox{Ref.}~\cite{Berwein:2017thy}.
Therefore, we use \mbox{Eq.}~(64) of \mbox{Ref.}~\cite{Berwein:2017thy} ,
together with the NLO result for the Debye mass, \mbox{Eq.}~\eqref{eq:md nlo}, to obtain the weak-coupling result. 
Keeping only the first term in \mbox{Eq.}~(64) of \mbox{Ref.}~\cite{Berwein:2017thy} 
corresponds to what we call the LO result, evaluated at the scale $\mu=4 \pi T$. 
As before, we use $\Lambda_{\rm \overline{MS}}=320\,{\rm MeV}$ and the two-loop running coupling. 
The weak-coupling results are shown in \mbox{Fig.}~\ref{fig:sqq}. 
The dash-dotted lines correspond to the LO result for $\mu=4 \pi T$. 
The NLO results at different temperatures are shown as bands.
We vary the scale $\mu$ from $\pi T$ to $4\pi T$ to account for the scale uncertainty. 
The widths of the bands correspond to the scale uncertainties.
We see again that the scale dependence reduces significantly at higher temperatures. 
The NLO result for $\Fqq^{\rm sub}$ is larger than the LO result and is significantly closer to the lattice data.
Deviations between the lattice data and the NLO result for the scale 
$\mu=\pi T$ are of the order of the statistical errors. 
Thus, for distances $r \sim 1/\md$ weak-coupling calculations in EQCD give a fair 
description of the lattice data, at least for the highest temperatures considered.
In the next section, we will discuss the very large distance behavior of the 
static correlators, $r \gg 1/\md$, where nonperturbative effects are important.

\section{Asymptotic screening regime}
\label{sec:screening}

In Secs.~\ref{sec:singlet} and~\ref{sec:plc}, we have shown that weak-coupling expansions 
describe lattice results fairly well both at short distances, 
$r\ll 1/T$, and at somewhat larger distances, $r\sim 1/\md$, in the electric screening regime, where resummation is necessary. 
At distances much larger than the Debye length, $1/\md$, the magnetic 
screening, which is inherently nonperturbative, becomes important. 
Perturbative calculations are not valid in this regime. 
In this section, we discuss the behavior of the static quark correlators 
at large distances, \mbox{i.e.} $r \gg 1/\md$. 
We expect that at these distances the correlators will be exponentially 
screened, \mbox{i.e.} behave like $\exp(-Mr)/r$, with $M$ being a nonperturbative screening mass. 
As we will see, the challenge in the determination of the screening masses
is the poor signal-to-noise ratio and possible contaminations from small distance physics. 

\begin{figure}
{
\centering
  \begin{minipage}{0.48\textwidth}
\centering
 \includegraphics[height=6cm,clip]{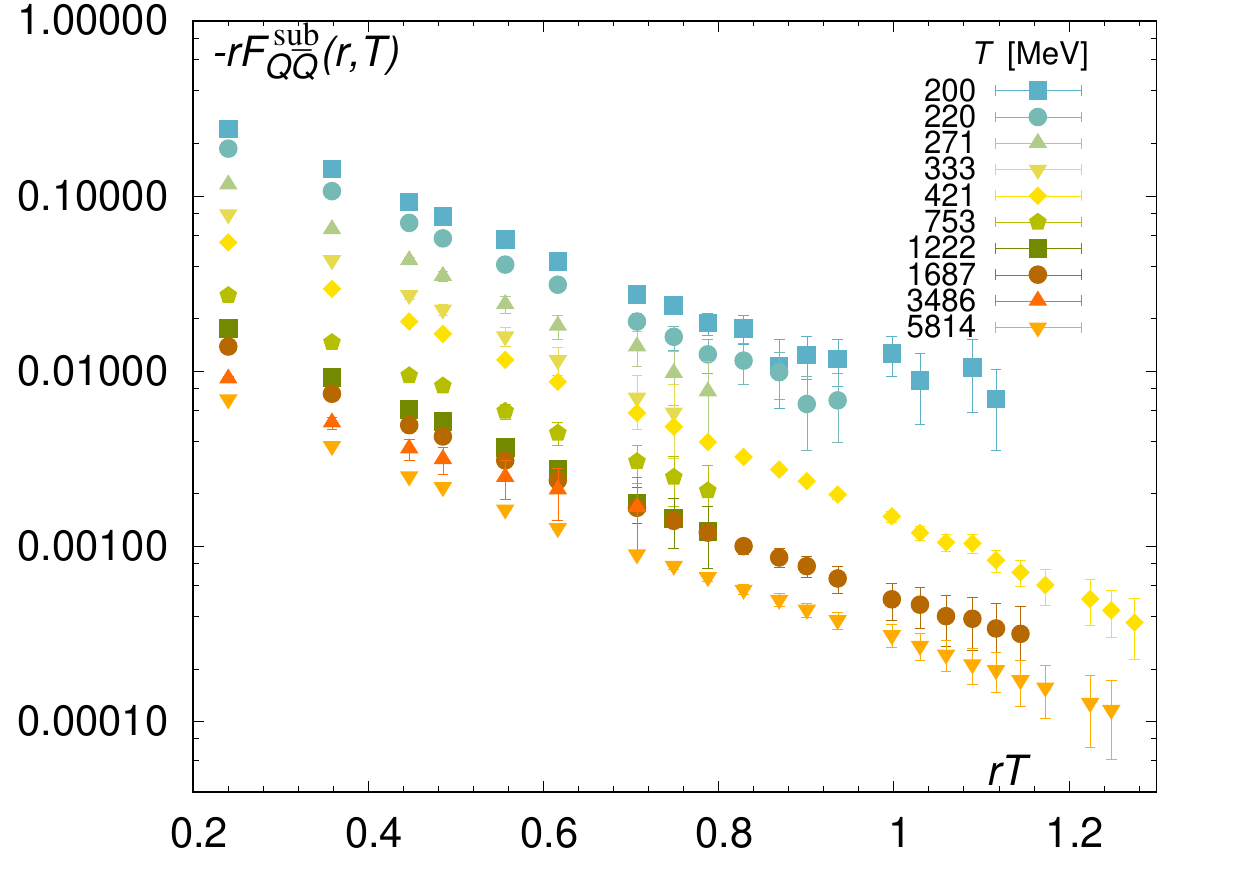}
  \vskip1ex
\end{minipage}
\hfill
\begin{minipage}{0.48\textwidth}
\centering
\includegraphics[height=6cm,clip]{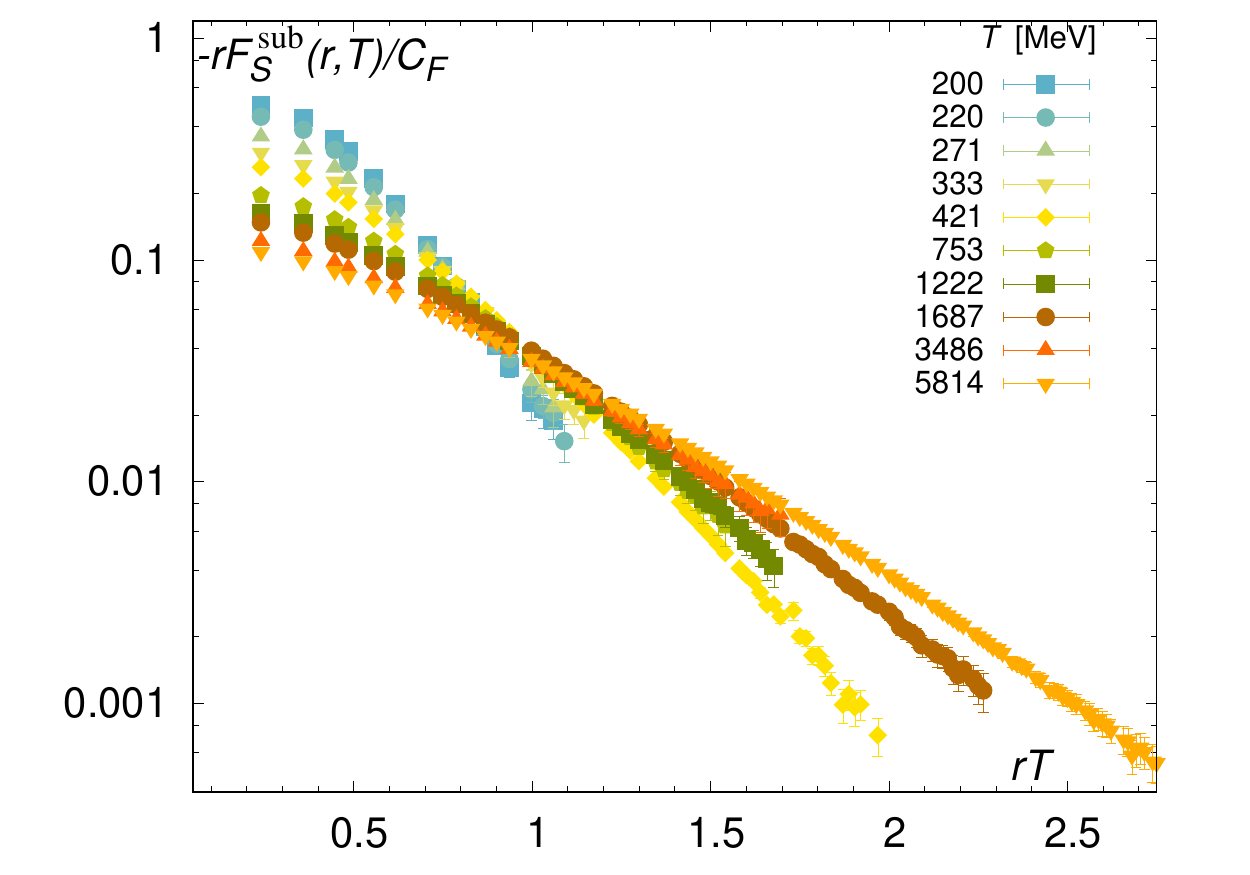}
\end{minipage}
}
\caption{\label{fig:screening functions}
The subtracted free energy (upper) and the subtracted singlet free 
energy (lower) multiplied by $-r$ calculated for different temperatures and $N_\tau=4$. 
}
\end{figure}

In \mbox{Fig.}~\ref{fig:screening functions} we show our numerical results for $\Fqq^{\rm sub}$ and $\Fs^{\rm sub}$ 
multiplied by $-r$ as function of $rT$ on a logarithmic scale. 
At large values of $rT$, the above quantities should appear as straight lines in the figure. 
This is indeed the case, \mbox{i.e.} at sufficiently large distances, both 
$\Fqq^{\rm sub}$ and $\Fs^{\rm sub}$ seem to decay exponentially. 
In the case of the singlet free energy, the data have some structure: 
the slope changes as we go from short to large distances. 
Furthermore, the slope increases with decreasing temperature, implying 
that the corresponding screening mass in temperature units increases at small temperatures. 
The fits of the correlators that we use for determining the screening 
masses clearly show this feature (see Appendix~\ref{app:D}). 
The lattice data for $-r\Fqq^{\rm sub}$ are relatively featureless.  
In particular, the slope seems independent of the temperature and the distance range considered. 
The data for $\Fqq^{\rm sub}$ also have much worse signal-to-noise ratio compared to the data for $\Fs^{\rm sub}$. 

\begin{figure}
{
\begin{minipage}{0.48\textwidth}
\centering
\includegraphics[height=6cm,clip]{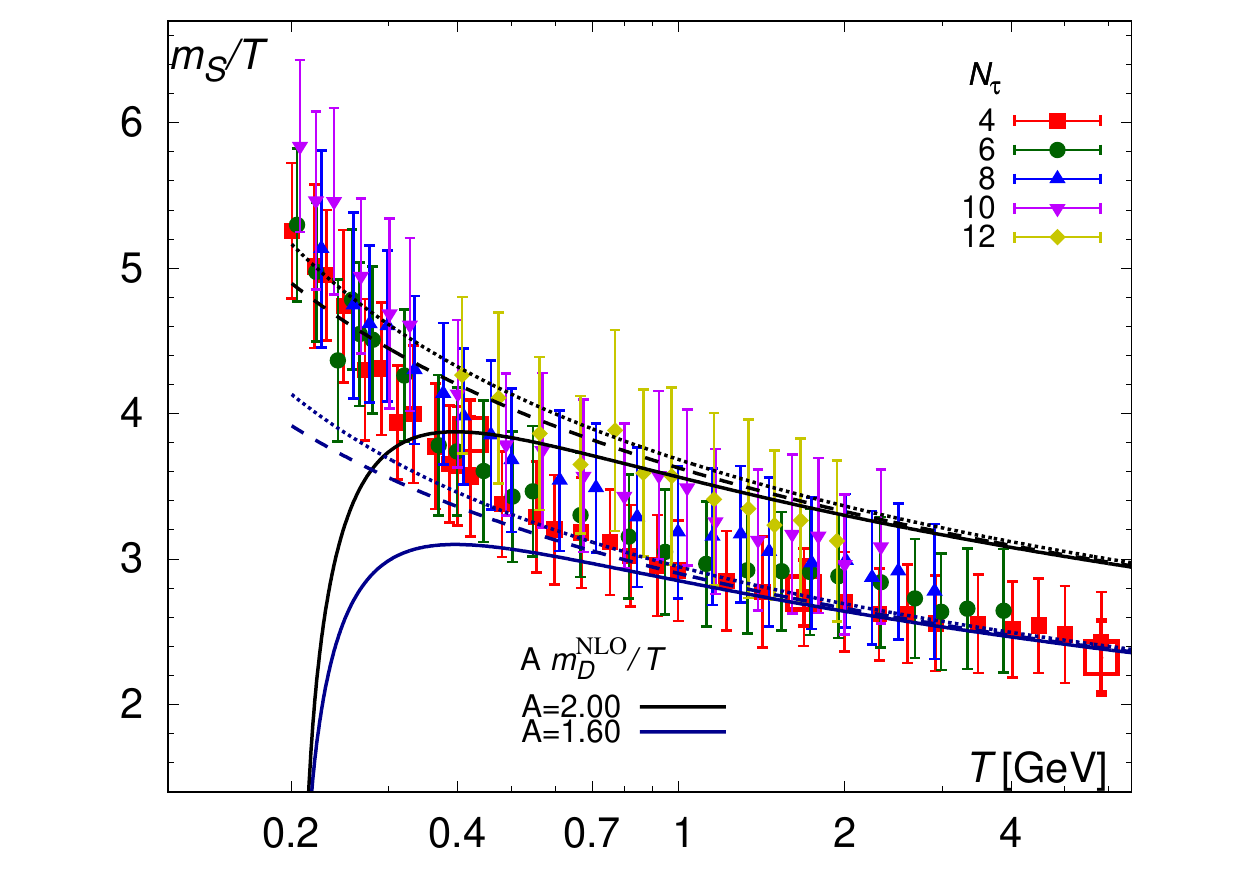}
\end{minipage}
}
\caption{\label{fig:ms T}
The asymptotic screening mass in temperature units $m_S/T$ from the exponential fits of $-r\Fs^{\rm sub}$. 
The lines correspond to the NLO Debye mass in temperature units calculated for $\mu=\pi T,~2\pi T$, 
and $4\pi T$ (solid, dotted and dashed) and multiplied by a constant to take into account nonperturbative effects. 
The open squares correspond to the screening masses determined using $N_{\tau}=4$ 
lattices with aspect ratio $6$.
}
\end{figure}

We discuss the details of the analysis of the asymptotic behavior and 
the extraction of the screening mass in Appendix~\ref{app:D}. 
We find that finite volume effects in the asymptotic behavior can be 
parametrized in terms of a constant shift between the correlators on 
lattices with different volumes (or aspect ratios $N_\sigma/N_\tau$).
After accounting for this constant shift, we find that screening masses 
are within errors independent of the volume for both $\Fqq^{\rm sub}$ and $\Fs^{\rm sub}$ (see Appendix \ref{app:D}).

We show the asymptotic screening mass from $\Fs^{\rm sub}$ in 
\mbox{Fig.}~\ref{fig:ms T} as a function of the temperature. 
The screening mass shows some cutoff dependence, which is only mildly temperature dependent.  
We do not see cutoff effects with a clear dependency on $N_{\tau}$ for $N_{\tau}\ge 8$, 
although screening masses with $N_{\tau}=4$ and $6$ appear to be systematically smaller.
The screening mass is considerably larger than the perturbative Debye mass 
at NLO, \mbox{Eq.}~\eqref{eq:md nlo}. 
After we rescale $\md$ by a constant $A=1.6$-$2.0$, we find that the 
temperature dependence of the screening mass and the rescaled Debye mass 
are very similar for high temperatures, $T> 700\,{\rm MeV}$. 
This rescaling factor is consistent with earlier observations in 
\mbox{Refs.}~\cite{Kaczmarek:2004gv,Kaczmarek:2007pb} using the LO Debye mass and much lower temperatures. 
A qualitatively similar enhancement was observed for the electric screening masses calculated from the Landau gauge gluon
propagators \cite{Heller:1997nqa,Karsch:1998tx}.
The two sets of lines in the figure represent the rescaled Debye mass with 
the scale $\mu$ varied from $\pi T$ to $4\pi T$. 

\begin{figure}
{
\begin{minipage}{0.48\textwidth}
\centering
\includegraphics[height=6cm,clip]{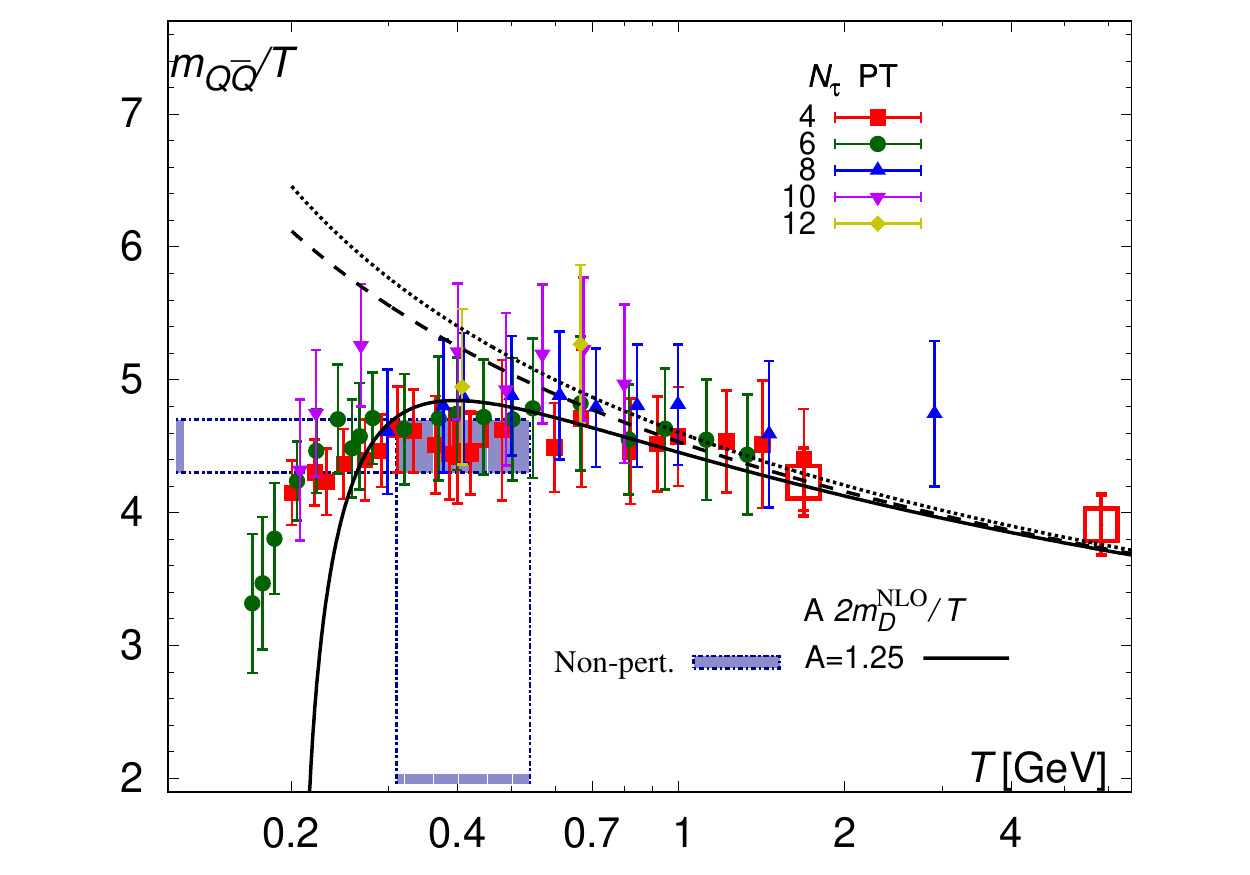}
\end{minipage}
}
\caption{\label{fig:mqq T}
The asymptotic screening mass in temperature units $m_{\qbq}/T$ from the exponential fits of $-r\Fqq^{\rm sub}$. 
The lines correspond to 2 times the NLO Debye mass in temperature units calculated for 
$\mu=\pi T,~2\pi T$, and $4\pi T$ (solid, dotted and dashed) and multiplied by a constant to take into account nonperturbative effects.
The open squares correspond to the screening masses determined using $N_{\tau}=4$ lattices with aspect ratio $6$. 
The horizontal band corresponds to the EQCD result for the screening mass obtained in 3D lattice calculations \cite{Hart:2000ha}.
}
\end{figure}

Because of the worse signal-to-noise ratio the fit of $-r\Fqq^{\rm sub}$ is more difficult.
We have to restrict the fits to smaller $r$ range. 
This implies that the contamination from small distance physics could be potentially larger here. 
On the other hand, because the slope is independent of $r T$, 
we do not see a large sensitivity to the choice of the lower fit range. 
Our estimate of the corresponding screening mass $m_{\qbq}$ is shown in \mbox{Fig.}~\ref{fig:mqq T}. 
At high temperatures, this screening mass is clearly larger than the one extracted from $\Fs^{\rm sub}$. 
This is expected and is in agreement with previous findings.
We compare our results with the screening mass determination on the lattice 
from EQCD~\cite{Hart:2000ha}, which is shown as the horizontal band in 
\mbox{Fig.}~\ref{fig:mqq T}.
Our results and the EQCD result are in good agreement.
At temperatures \(T>400\,{\rm MeV}\), the screening mass from the Polyakov loop 
correlator is only 25\% larger than the naive perturbative result $2\md$.
This is consistent with the EQCD-based analysis~\cite{Laine:2009dh}. 
At temperatures around $200$ MeV, we see a decrease in $m_{\qbq}$. 
Such a decrease in the screening mass has been observed also in the SU(3) gauge theory~\cite{Kaczmarek:1999mm}, 
where it is due to the (weakly) first-order nature of the deconfining transition 
and thus to the large correlation length of the order parameter (Polyakov loop). 

The poor signal-to-noise ratio is clearly a problem for the determination of the screening masses. 
To overcome this problem, the Wilson lines should be evaluated on smeared 
gauge configurations before calculating the correlators, as in \mbox{Ref.}~\cite{Borsanyi:2015yka}. 
We plan to perform such a calculation in the future.

\section{Conclusions}
\label{sec:conclusions}

In this paper we have studied the free energy of a static $\qbq$ pair and the static meson (singlet) correlator in (2+1)-flavor QCD 
with the aim to analyze at what distances the hot QCD medium modifies the interaction between the static quark and antiquark, 
and whether these modifications could be understood using perturbation theory. 
We have presented continuum extrapolated results for the free energy of a $\qbq$ 
pair as well as for the singlet free energy in the Coulomb gauge.

At asymptotically high temperatures, $g(T) \ll 1$, there are three distinct 
distance regimes: $r \ll 1/T$, $r \sim 1/m_D \sim 1/(g T)$, and $r\sim 1/(g^2 T)$.
The physics in these regimes is quite different. 
In the first case, we expect only small thermal corrections to the quark-antiquark interaction. 
For $r \sim 1/(g T)$, medium effects are significant and the interaction 
between the static quark and antiquark can be understood in terms of electric screening. 
At distances $r \sim 1/(g^2 T)$, nonperturbative effects due to static 
magnetic fields are important and the screening is nonperturbative. 

Weak-coupling calculations of the Polyakov loop correlator 
and of the singlet correlator can be organized using effective field theories. 
For $r\ll 1/T$, one can use pNRQCD, while for $r\sim 1/\md$, one can use EQCD. 
At larger distances, the screening masses have to be calculated 
nonperturbatively using lattice techniques~\cite{Kajantie:1997tt,Hart:1999dj,Hart:2000ha}.

For physically interesting temperatures the relevant coupling is $g(T) \sim 1$. 
Thus, the question arises if the above picture and weak-coupling calculations are relevant at all. 
A related question is how well the different energy scales are separated and at what distances screening sets in.
We have addressed these questions through a detailed comparison 
of lattice results with effective field-theory-based weak-coupling calculations. 
We have found that the weak-coupling picture provides a fair description for temperatures larger than $300$ MeV. 
For $rT\lesssim 0.3$, medium effects are small, as predicted by pNRQCD. 
For $0.3\lesssim rT \lesssim 0.6$, we see significant medium effects and the 
interaction between the static quark and antiquark is screened as predicted by perturbative EQCD. 
In this region, screening is controlled by the perturbative parameter $\md$. 
At larger separations, nonperturbative effects related to static 
chromomagnetic fields become important and the weak-coupling approach is no longer valid.  
In this region, we have determined the asymptotic screening masses
by fitting the correlators at large distances. 

The above results suggest that weak-coupling pNRQCD could be applicable for studying quarkonium properties at $T \gtrsim 300$ MeV. 
In a broader context, the success of the weak-coupling calculations for the 
Polyakov loop correlator and the singlet correlator is not completely surprising. 
Weak-coupling calculations have been found to agree well with lattice results 
for quark number susceptibilities~\cite{Bazavov:2013uja,Ding:2015fca,Bellwied:2015lba}, 
equation of state~\cite{Borsanyi:2016ksw,Bazavov:2017dsy},
and to some extent for the topological susceptibility~(\cite{Borsanyi:2016ksw,Petreczky:2016vrs}, see, however,~\cite{Bonati:2015vqz}). 
Furthermore, the observed Casimir scaling of the Polyakov loop expectation 
values in higher representations \cite{Petreczky:2015yta} is also naturally 
explained in the weak-coupling picture \cite{Berwein:2015ayt}.

As a byproduct of this analysis, we have improved the accuracy of the 
renormalization constants of the Polyakov loop and the entropy of the single static quark (see Appendix~\ref{app:renormalization}). 
We have also extended the study of cutoff effects for the $T=0$ static energy at 
short distances (see Appendix~\ref{app:B}). 
The latter may be of relevance for improved determinations of the strong coupling \(\as\).

\section*{ACKNOWLEDGMENTS}
This research was supported by the DFG cluster of excellence ``Origin and Structure of the Universe'' (www.universe-cluster.de).
N. B., A. V. and J. H. W. acknowledge the support by the Bundesministerium f\"ur Bildung und Forschung (BMBF) 
under the  Grant ``Verbundprojekt 05P2015 - ALICE at High Rate (BMBF-FSP 202) GEM-TPC Upgrade and Field theory based investigations of ALICE physics,''  
Project No. 05P15WOCA1.
This work has been supported in part by the U.S. Department of Energy through Award No. DE-SC0012704.
The simulations have been carried out 
on the computing facilities of the National Energy Research Scientific Computing Center (NERSC), a 
U.S. Department of Energy Office of Science User Facility operated under Contract No. DE-AC02-05CH11231, 
the Computational Center for Particle and Astrophysics (C2PAP) as well as on the SuperMUC at the 
Leibniz- Rechenzentrum (LRZ).
The lattice QCD calculations have been performed using the publicly available MILC code.
The data analysis was performed using the R statistical package~\cite{Rpackage}.

\appendix
\section{GAUGE ENSEMBLES}
\label{app:A}

In this Appendix, we give an account of the HISQ gauge ensembles 
underlying the presented lattice calculations. 
In \mbox{Sec.}~\ref{app:lattices}, we discuss the quark mass dependence in the new ensembles and finite volume effects. 
In \mbox{Sec.}~\ref{app:correlators}, we cover cuts in  molecular dynamics (MD) trajectories due to 
thermalization, the determination of statistical errors, and the normalization using the MD time histories of the Polyakov loop. 
We discuss our cuts for the correlators. 
In \mbox{Sec.}~\ref{app:renormalization}, we finally discuss the renormalization of 
$\Fqq$ and $\Fs$. 

\subsection{Lattices}
\label{app:lattices}

\begin{table}
\parbox{.98\linewidth}{
  \begin{tabular}{|c|c|c|c|c|c|}
    \hline
    $ \beta $ & $am_s$ & T (MeV) & $\max rT$: $\Fs,~\Fqq$ & \#TUs & Ref. \\
    \hline
    6.515 & 0.0604  &  122 & 0.21,\quad 0.15 &  32500 & \cite{Bazavov:2016uvm} \\
    6.608 & 0.0542  &  133 & 0.26,\quad 0.18 &  19990 & \cite{Bazavov:2016uvm} \\
    6.664 & 0.0514  &  141 & 0.26,\quad 0.24 &  45120 & \cite{ Bazavov:2016uvm} \\
    6.700 & 0.0496  &  146 & 0.21,\quad 0.21 &  15900 & \cite{ Bazavov:2016uvm} \\
    6.740 & 0.0476  &  151 & 0.31,\quad 0.24 &  29410 & \cite{Bazavov:2014pvz} \\
    6.770 & 0.0460  &  156 & 0.31,\quad 0.21 &  15530 & \cite{ Bazavov:2016uvm} \\
    6.800 & 0.0448  &  160 & 0.34,\quad 0.26 &  36060 & \cite{Bazavov:2014pvz} \\
    6.840 & 0.0430  &  166 & 0.35,\quad 0.24 &  17370 & \cite{ Bazavov:2016uvm} \\
    6.880 & 0.0412  &  173 & 0.37,\quad 0.28 &  46350 & \cite{Bazavov:2014pvz} \\
    6.950 & 0.0386  &  185 & 0.39,\quad 0.30 &  50550 & \cite{ Bazavov:2016uvm} \\
    7.030 & 0.0378  &  199 & 0.47,\quad 0.39 &  65940 & \cite{Bazavov:2014pvz} \\
    7.100 & 0.0332  &  213 & 0.46,\quad 0.30 &   9640 & \cite{Bazavov:2013uja} \\
    7.150 & 0.0320  &  223 & 0.45,\quad 0.35 &   9600 & \cite{Bazavov:2014pvz} \\
    7.200 & 0.0296  &  233 & 0.38,\quad 0.31 &   4010 & \cite{Bazavov:2013uja} \\
    7.280 & 0.0284  &  251 & 0.55,\quad 0.47 &  58210 & \cite{Bazavov:2014pvz} \\
    7.373 & 0.0250  &  273 & 0.57,\quad 0.55 &  85120 & \cite{Bazavov:2014pvz} \\
    7.596 & 0.0202  &  334 & 0.72,\quad 0.57 &  98010 & \cite{Bazavov:2014pvz} \\
    7.650 & 0.0202  &  350 & 0.54,\quad 0.39 &   3230 & \cite{Bazavov:2013uja} \\
    7.825 & 0.0164  &  408 & 0.94,\quad 0.60 & 134600 & \cite{Bazavov:2014pvz} \\
    8.000 & 0.0140  &  474 & 0.68,\quad 0.54 &   3110 & \cite{Ding:2015ona} \\
    8.200 & 0.01167 &  562 & 1.05,\quad 0.76 &  30090 & \cite{Ding:2015ona} \\
    8.400 & 0.00975 &  667 & 1.06,\quad 0.72 &  29190 & \cite{Ding:2015ona} \\
    \hline
  \end{tabular}
  \caption{\label{tab:nt 12}
  Parameters of $N_{\tau}=12$ ensembles with aspect ratio $4$ and $m_l=m_s/20$. Adjacent correlators are separated by $10$ TUs.
  }
}
\end{table}

\begin{table}
\parbox{.98\linewidth}{
  \begin{tabular}{|c|c|c|c|c|c|}
    \hline
    $ \beta $ & $am_s$ & T (MeV) & $\max rT$: $\Fs,~\Fqq$ & \#TUs & Ref. \\
    \hline
    6.285 & 0.0790  &  116 & 0.22,\quad 0.18 &   9260 & \cite{ Bazavov:2016uvm} \\
    6.341 & 0.0740  &  123 & 0.25,\quad 0.22 &  39190 & \cite{ Bazavov:2016uvm} \\
    6.423 & 0.0670  &  133 & 0.25,\quad 0.19 &  10360 & \cite{ Bazavov:2016uvm} \\
    6.488 & 0.0620  &  142 & 0.37,\quad 0.30 & 102690 & \cite{Bazavov:2014pvz} \\
    6.515 & 0.0604  &  146 & 0.37,\quad 0.33 & 107530 & \cite{Bazavov:2014pvz} \\
    6.575 & 0.0564  &  155 & 0.37,\quad 0.33 & 106020 & \cite{Bazavov:2014pvz} \\
    6.608 & 0.0542  &  160 & 0.46,\quad 0.33 & 112890 & \cite{Bazavov:2014pvz} \\
    6.664 & 0.0514  &  169 & 0.51,\quad 0.46 & 155440 & \cite{Bazavov:2014pvz} \\
    6.740 & 0.0476  &  181 & 0.57,\quad 0.55 & 200250 & \cite{Bazavov:2014pvz} \\
    6.800 & 0.0448  &  192 & 0.55,\quad 0.61 & 279830 & \cite{Bazavov:2014pvz} \\
    6.880 & 0.0412  &  208 & 0.66,\quad 0.68 & 341490 & \cite{Bazavov:2014pvz} \\
    6.950 & 0.0386  &  222 & 0.68,\quad 0.73 & 243480 & \cite{Bazavov:2014pvz} \\
    7.030 & 0.0378  &  239 & 0.68,\quad 0.68 & 137730 & \cite{Bazavov:2014pvz} \\
    7.150 & 0.0320  &  267 & 0.79,\quad 0.76 & 145440 & \cite{Bazavov:2014pvz} \\
    7.280 & 0.0284  &  301 & 0.93,\quad 0.71 & 105990 & \cite{Bazavov:2014pvz} \\
    7.373 & 0.0250  &  328 & 0.97,\quad 0.68 &  50840 & \cite{Bazavov:2014pvz} \\
    7.596 & 0.0202  &  400 & 1.09,\quad 0.81 &  51710 & \cite{Bazavov:2014pvz} \\
    7.825 & 0.0164  &  489 & 1.15,\quad 0.83 &  54000 & \cite{Bazavov:2014pvz} \\
    8.000 & 0.0140  &  569 & 1.09,\quad 1.00 &   6780 & \cite{Ding:2015ona} \\
    8.200 & 0.01167 &  675 & 1.12,\quad 0.57 &  27500 & \cite{Ding:2015ona} \\
    8.400 & 0.00975 &  800 & 1.07,\quad 0.58 &   7540 & \cite{Ding:2015ona} \\
    8.570 & 0.008376 & 924 & 1.31,\quad 0.93 &   3000 & \cite{Bazavov:2013uja} \\
    \hline
  \end{tabular}
  \caption{\label{tab:nt 10}
  Parameters of $N_{\tau}=10$ ensembles with aspect ratio $4$ and $m_l=m_s/20$. Adjacent correlators are separated by $10$ TUs.
  }
}
\end{table}

\begin{table} 
\parbox{.98\linewidth}{
  \begin{tabular}{|c|c|c|c|c|c|}
    \hline
    $ \beta $ & $am_s$ & T (MeV) & $\max rT$: $\Fs,~\Fqq$ & \#TUs & Ref. \\
    \hline
    6.050 & 0.1064   &  116 & 0.31,\quad 0.35 &  47130 & \cite{Bazavov:2014pvz} \\
    6.125 & 0.0966   &  125 & 0.31,\quad 0.31 &   8980 & \cite{Bazavov:2014pvz} \\
    6.175 & 0.0906   &  131 & 0.28,\quad 0.22 &   1520 & \cite{Bazavov:2014pvz} \\
    6.195 & 0.0880   &  133 & 0.39,\quad 0.35 &  16900 & \cite{Bazavov:2014pvz} \\
    6.245 & 0.0830   &  140 & 0.41,\quad 0.31 &  26100 & \cite{Bazavov:2014pvz} \\
    6.260 & 0.0810   &  142 & 0.39,\quad 0.35 &  10240 & \cite{Bazavov:2014pvz} \\
    6.285 & 0.0790   &  146 & 0.41,\quad 0.37 &  31440 & \cite{Bazavov:2014pvz} \\
    6.315 & 0.0760   &  150 & 0.45,\quad 0.35 &   7850 & \cite{Bazavov:2014pvz} \\
    6.341 & 0.0740   &  154 & 0.41,\quad 0.45 &  17130 & \cite{Bazavov:2014pvz} \\
    6.354 & 0.0728   &  156 & 0.41,\quad 0.47 &  12540 & \cite{Bazavov:2014pvz} \\
    6.390 & 0.0694   &  161 & 0.47,\quad 0.43 &  26950 & \cite{Bazavov:2014pvz} \\
    6.423 & 0.0670   &  167 & 0.52,\quad 0.53 &  20310 & \cite{Bazavov:2014pvz} \\
    6.445 & 0.0652   &  170 & 0.53,\quad 0.55 &  13630 & \cite{Bazavov:2014pvz} \\
    6.460 & 0.0640   &  173 & 0.53,\quad 0.43 &  16430 & \cite{Bazavov:2014pvz} \\
    6.488 & 0.0620   &  178 & 0.56,\quad 0.55 &  17900 & \cite{Bazavov:2014pvz} \\
    6.515 & 0.0604   &  182 & 0.56,\quad 0.56 &  24800 & \cite{Bazavov:2014pvz} \\
    6.550 & 0.0682   &  189 & 0.59,\quad 0.47 &  23560 & \cite{Bazavov:2014pvz} \\
    6.575 & 0.0564   &  193 & 0.64,\quad 0.65 &  26300 & \cite{Bazavov:2014pvz} \\
    6.608 & 0.0542   &  200 & 0.69,\quad 0.69 &  24190 & \cite{Bazavov:2014pvz} \\
    6.664 & 0.0514   &  211 & 0.69,\quad 0.72 &  19000 & \cite{Bazavov:2014pvz} \\
    6.740 & 0.0476   &  227 & 0.69,\quad 0.72 &  20160 & \cite{Bazavov:2014pvz} \\
    6.800 & 0.0448   &  240 & 0.80,\quad 0.80 &  16760 & \cite{Bazavov:2014pvz} \\
    6.880 & 0.0412   &  259 & 0.84,\quad 0.75 &  27400 & \cite{Bazavov:2014pvz} \\
    6.950 & 0.0386   &  277 & 0.85,\quad 0.71 &  24700 & \cite{Bazavov:2014pvz} \\
    7.030 & 0.0378   &  299 & 0.94,\quad 0.83 &  25570 & \cite{Bazavov:2014pvz} \\
    7.150 & 0.0320   &  334 & 0.85,\quad 0.65 &  19340 & \cite{Bazavov:2014pvz} \\
    7.280 & 0.0284   &  377 & 1.08,\quad 0.81 &  31990 & \cite{Bazavov:2014pvz} \\
    7.373 & 0.0250   &  410 & 1.19,\quad 0.96 & 119380 & \cite{Bazavov:2014pvz} \\
    7.500 & 0.0222   &  459 & 0.82,\quad 0.50 &   4990 & \cite{Bazavov:2013uja} \\
    7.596 & 0.0202   &  500 & 1.29,\quad 0.85 & 127990 & \cite{Bazavov:2014pvz} \\
    7.825 & 0.0164   &  611 & 1.31,\quad 0.83 & 122200 & \cite{Bazavov:2014pvz} \\
    8.000 & 0.0140   &  711 & 1.27,\quad 0.75 &  18370 & \cite{Bazavov:2013uja} \\
    8.200 & 0.01167  &  843 & 1.10,\quad 0.52 &   3080 & \cite{Bazavov:2013uja} \\
    8.400 & 0.00975  & 1000 & 1.14,\quad 0.69 &   3000 & \cite{Bazavov:2013uja} \\
    8.570 & 0.008376 & 1155 & 1.16,\quad 0.69 &  10250 & \cite{Bazavov:2016uvm} \\
    8.710 & 0.007394 & 1299 & 1.21,\quad 0.72 &  10030 & \cite{Bazavov:2016uvm} \\
    8.850 & 0.006528 & 1461 & 1.48,\quad 0.84 &  10000 & \cite{Bazavov:2016uvm} \\
    9.060 & 0.004834 & 1743 & 1.29,\quad 0.62 &  10810 & \cite{Bazavov:2016uvm} \\
    9.230 & 0.004148 & 2011 & 1.29,\quad 0.67 &  10250 & \cite{Bazavov:2016uvm} \\
    9.360 & 0.003691 & 2242 & 1.47,\quad 0.64 &   8120 & \cite{Bazavov:2016uvm} \\
    9.490 & 0.003285 & 2500 & 1.51,\quad 0.71 &   8010 & \cite{Bazavov:2016uvm} \\
    9.670 & 0.002798 & 2907 & 1.58,\quad 0.67 &   8050 & \cite{Bazavov:2016uvm} \\
    \hline
  \end{tabular}
  \caption{\label{tab:nt 8}
  Parameters of $N_{\tau}=8$ ensembles with aspect ratio $4$ and $m_l=m_s/20$. Adjacent correlators are separated by $10$ TUs.
  }
}
\end{table}

\begin{table} 
\parbox{.98\linewidth}{
  \begin{tabular}{|c|c|c|c|c|c|}
    \hline
    $ \beta $ & $am_s$ & T (MeV) & $\max rT$: $\Fs,~\Fqq$ & \#TUs & Ref. \\
    \hline
    5.900 & 0.1320   &  134 & 0.47,\quad 0.47 & 31070 & \cite{Bazavov:2014pvz} \\
    6.000 & 0.1138   &  147 & 0.55,\quad 0.58 & 31640 & \cite{Bazavov:2014pvz} \\
    6.050 & 0.1064   &  154 & 0.53,\quad 0.55 & 29990 & \cite{Bazavov:2014pvz} \\
    6.100 & 0.0998   &  162 & 0.62,\quad 0.69 & 29990 & \cite{Bazavov:2014pvz} \\
    6.150 & 0.0936   &  170 & 0.55,\quad 0.76 & 27990 & \cite{Bazavov:2014pvz} \\
    6.195 & 0.0880   &  178 & 0.69,\quad 0.78 & 29990 & \cite{Bazavov:2014pvz} \\
    6.215 & 0.0862   &  181 & 0.55,\quad 0.74 &  5390 & \cite{Bazavov:2014pvz} \\
    6.245 & 0.0830   &  187 & 0.76,\quad 0.83 & 30000 & \cite{Bazavov:2014pvz} \\
    6.285 & 0.0790   &  194 & 0.55,\quad 0.60 &  5750 & \cite{Bazavov:2014pvz} \\
    6.341 & 0.0740   &  205 & 0.76,\quad 0.90 & 30000 & \cite{Bazavov:2014pvz} \\
    6.423 & 0.0670   &  222 & 0.90,\quad 0.87 & 29990 & \cite{Bazavov:2014pvz} \\
    6.515 & 0.0604   &  243 & 0.91,\quad 0.94 & 29990 & \cite{Bazavov:2014pvz} \\
    6.575 & 0.0564   &  258 & 1.01,\quad 0.91 & 29990 & \cite{Bazavov:2014pvz} \\
    6.608 & 0.0542   &  266 & 1.03,\quad 0.91 & 29990 & \cite{Bazavov:2014pvz} \\
    6.664 & 0.0514   &  281 & 1.01,\quad 0.78 & 29990 & \cite{Bazavov:2014pvz} \\
    6.800 & 0.0448   &  320 & 1.07,\quad 0.91 & 29990 & \cite{Bazavov:2014pvz} \\
    6.950 & 0.0386   &  370 & 1.13,\quad 0.90 & 29990 & \cite{Bazavov:2014pvz} \\
    7.150 & 0.0320   &  446 & 1.30,\quad 0.97 & 25500 & \cite{Bazavov:2014pvz} \\
    7.280 & 0.0284   &  502 & 1.25,\quad 0.83 & 27140 & \cite{Bazavov:2014pvz} \\
    7.373 & 0.0250   &  547 & 1.25,\quad 0.78 & 24120 & \cite{Bazavov:2014pvz} \\
    7.500 & 0.0222   &  613 & 1.35,\quad 0.91 &  5670 & \cite{Bazavov:2013uja} \\
    7.596 & 0.0202   &  667 & 1.35,\quad 0.78 & 33520 & \cite{Bazavov:2014pvz} \\
    7.825 & 0.0164   &  815 & 1.67,\quad 1.00 & 30440 & \cite{Bazavov:2014pvz} \\
    8.000 & 0.0140   &  948 & 1.39,\quad 0.85 &  3200 & \cite{Bazavov:2013uja} \\
    8.200 & 0.01167  & 1125 & 1.30,\quad 0.76 & 10100 & \cite{Bazavov:2016uvm} \\
    8.400 & 0.00975  & 1334 & 1.31,\quad 0.82 & 10160 & \cite{Bazavov:2016uvm} \\
    8.570 & 0.008376 & 1540 & 1.39,\quad 0.71 & 10190 & \cite{Bazavov:2016uvm} \\
    8.710 & 0.007394 & 1732 & 1.39,\quad 0.96 & 10220 & \cite{Bazavov:2016uvm} \\
    8.850 & 0.006528 & 1949 & 1.36,\quad 0.74 & 10060 & \cite{Bazavov:2016uvm} \\
    9.060 & 0.004834 & 2324 & 1.52,\quad 0.96 & 10070 & \cite{Bazavov:2016uvm} \\
    9.230 & 0.004148 & 2681 & 1.43,\quad 0.85 & 10060 & \cite{Bazavov:2016uvm} \\
    9.360 & 0.003691 & 2989 & 1.64,\quad 0.90 &  8240 & \cite{Bazavov:2016uvm} \\
    9.490 & 0.003285 & 3333 & 1.55,\quad 0.62 &  8130 & \cite{Bazavov:2016uvm} \\
    9.670 & 0.002798 & 3876 & 1.41,\quad 0.62 & 10290 & \cite{Bazavov:2016uvm} \\
    \hline
  \end{tabular}
  \caption{\label{tab:nt 6}
  Parameters of $N_{\tau}=6$ ensembles with aspect ratio $4$ and $m_l=m_s/20$. Adjacent correlators are separated by $10$ TUs.
  }
}
\end{table}

\begin{table} 
\parbox{.98\linewidth}{
  \begin{tabular}{|c|c|c|c|c|c|}
    \hline
    $ \beta $ & $am_s$ & T (MeV) & $\max rT$: $\Fs,~\Fqq$ & \#TUs & Ref. \\
    \hline
    5.900 & 0.132000 &  201 & 0.90,\quad 1.17 & 65350 & \cite{Bazavov:2016uvm} \\
    6.000 & 0.113800 &  221 & 1.09,\quad 1.06 & 62610 & \cite{Bazavov:2016uvm} \\
    6.050 & 0.106400 &  232 & 1.12,\quad 1.12 & 62400 & \cite{Bazavov:2016uvm} \\
    6.125 & 0.096600 &  249 & 1.14,\quad 1.06 & 63510 & \cite{Bazavov:2016uvm} \\
    6.215 & 0.086200 &  272 & 1.06,\quad 0.94 & 26650 & \cite{Bazavov:2016uvm} \\
    6.285 & 0.079000 &  291 & 1.14,\quad 1.00 & 25380 & \cite{Bazavov:2016uvm} \\
    6.354 & 0.072800 &  311 & 1.12,\quad 0.94 & 19480 & \cite{Bazavov:2016uvm} \\
    6.423 & 0.067000 &  333 & 1.14,\quad 0.87 & 21930 & \cite{Bazavov:2016uvm} \\
    6.515 & 0.060300 &  364 & 1.14,\quad 0.90 & 31330 & \cite{Bazavov:2016uvm} \\
    6.575 & 0.056400 &  386 & 1.22,\quad 0.87 & 22770 & \cite{Bazavov:2016uvm} \\
    6.608 & 0.054200 &  399 & 1.30,\quad 1.09 & 39400 & \cite{Bazavov:2016uvm} \\
    6.664 & 0.051400 &  421 & 1.46,\quad 1.22 & 75770 & \cite{Bazavov:2016uvm} \\
    6.800 & 0.044800 &  480 & 1.14,\quad 0.71 & 37860 & \cite{Bazavov:2016uvm} \\
    6.950 & 0.038600 &  554 & 1.22,\quad 0.79 & 38090 & \cite{Bazavov:2016uvm} \\
    7.150 & 0.032000 &  669 & 1.25,\quad 0.62 & 31800 & \cite{Bazavov:2016uvm} \\
    7.280 & 0.028400 &  753 & 1.30,\quad 0.87 & 42810 & \cite{Bazavov:2016uvm} \\
    7.373 & 0.025000 &  819 & 1.62,\quad 1.00 & 65010 & \cite{Bazavov:2016uvm} \\
    7.500 & 0.022200 &  918 & 1.60,\quad 0.87 & 42950 & \cite{Bazavov:2016uvm} \\
    7.596 & 0.020200 & 1000 & 1.54,\quad 0.83 & 69920 & \cite{Bazavov:2016uvm} \\
    7.825 & 0.016400 & 1222 & 1.68,\quad 0.90 & 65380 & \cite{Bazavov:2016uvm} \\
    8.000 & 0.014000 & 1422 & 1.52,\quad 0.71 & 27510 & \cite{Bazavov:2016uvm} \\
    8.200 & 0.011670 & 1687 & 1.54,\quad 0.87 & 20790 & \cite{Bazavov:2016uvm} \\
    8.400 & 0.009750 & 1999 & 1.48,\quad 0.71 & 20950 & \cite{Bazavov:2016uvm} \\
    8.570 & 0.008376 & 2308 & 1.68,\quad 0.90 & 20280 & \cite{Bazavov:2016uvm} \\
    8.710 & 0.007394 & 2597 & 1.60,\quad 0.71 & 20200 & \cite{Bazavov:2016uvm} \\
    8.850 & 0.006528 & 2921 & 1.52,\quad 0.71 & 19210 & \cite{Bazavov:2016uvm} \\
    9.060 & 0.004834 & 3487 & 1.66,\quad 1.00 & 20950 & \cite{Bazavov:2016uvm} \\
    9.230 & 0.004148 & 4021 & 1.70,\quad 0.87 & 21240 & \cite{Bazavov:2016uvm} \\
    9.360 & 0.003691 & 4484 & 1.44,\quad 0.56 & 10620 & \cite{Bazavov:2016uvm} \\
    9.490 & 0.003285 & 5000 & 1.62,\quad 0.79 & 10320 & \cite{Bazavov:2016uvm} \\
    9.670 & 0.002798 & 5814 & 1.64,\quad 0.87 & 10340 & \cite{Bazavov:2016uvm} \\
    \hline
  \end{tabular}
  \caption{\label{tab:nt 4}
  The parameters of $N_{\tau}=4$ ensembles with aspect ratio $4$ and $m_l=m_s/20$. Adjacent correlators are separated by $10$ TUs.
  }
}
\end{table}

\begin{table*}[t] 
\parbox{1.0\linewidth}{
  \begin{tabular}{|c|c|c|c|c|c|c|}
    \hline
    \multicolumn{3}{|c|}{$N_\tau=12$:} &
    \multicolumn{4}{|c|}{$m_l=m_s/5$} 
    \\
    \hline
    $ \beta $  & $am_s$ & T (MeV) & 
    $N_x=N_y$,\quad $N_z$ & $\max rT$: $\Fs,~\Fqq$ & \#TUs & $ L^{\mathrm{bare}} $   \\
    \hline
    8.570 & 0.008376 &  770 & 48,\quad 48 & 1.03,\quad 0.49 & 6320 & 0.033869(75) \\
    8.710 & 0.007394 &  866 & 48,\quad 48 & 1.06,\quad 0.55 & 6490 & 0.037745(84) \\
    8.850 & 0.006528 &  974 & 48,\quad 48 & 1.09,\quad 0.58 & 6340 & 0.41749(101) \\
    9.060 & 0.004834 & 1162 & 48,\quad 96 & 1.15,\quad 0.60 & 7430 & 0.048343(66)  \\
    9.230 & 0.004148 & 1340 & 48,\quad 48 & 1.17,\quad 0.53 & 7280 & 0.053533(85) \\
    9.360 & 0.003691 & 1495 & 48,\quad 96 & 1.25,\quad 0.62 & 7910 & 0.057509(70)  \\
    9.490 & 0.003285 & 1667 & 48,\quad 48 & 1.15,\quad 0.53 & 9780 & 0.061575(63)  \\
    9.670 & 0.002798 & 1938 & 48,\quad 96 & 1.25,\quad 0.72 & 7650 & 0.067389(55)  \\
    \hline
    \hline
    \multicolumn{3}{|c|}{$N_\tau=10$:} &
    \multicolumn{4}{|c|}{$m_l=m_s/5$} 
    \\
    \hline
    $ \beta $  & $am_s$ & T (MeV) & 
    $N_x=N_y$,\quad $N_z$ & $\max rT$: $\Fs,~\Fqq$ & \#TUs & $ L^{\mathrm{bare}} $    \\
    \hline
    8.710 & 0.007394 & 1039 & 40,\quad 40 & 1.17,\quad 0.59 & 15320 & 0.068687(92) \\
    8.850 & 0.006528 & 1169 & 40,\quad 80 & 1.28,\quad 0.87 & 7690 & 0.074258(86) \\
    9.060 & 0.004834 & 1395 & 40,\quad 40 & 1.31,\quad 0.64 & 15490 & 0.083071(70)  \\
    9.230 & 0.004148 & 1608 & 40,\quad 80 & 1.39,\quad 0.60 & 7630 & 0.090309(95) \\
    9.360 & 0.003691 & 1794 & 40,\quad 40 & 1.30,\quad 0.61 & 15800 & 0.095748(78) \\
    9.490 & 0.003285 & 2000 & 40,\quad 80 & 1.26,\quad 0.62 & 7990 & 0.101253(107) \\
    9.670 & 0.002798 & 2326 & 40,\quad 40 & 1.38,\quad 0.74 & 15760 & 0.108795(110) \\
    \hline
  \end{tabular}
  \caption{\label{tab:fine}
  The parameters of the very fine $N_{\tau}=12$ and $10$ ensembles with $m_l=m_s/20$ and expectation values of bare Polyakov loops. 
  }
}
\end{table*}

\begin{table*}[t]
\parbox{1.0\linewidth}{
  \begin{tabular}{|c|c|c|c|c|c|c|}
    \hline
    \multicolumn{3}{|c|}{$N_\tau=16$:} &
    \multicolumn{4}{|c|}{$N_\sigma=4N_\tau$} 
    \\
    \hline
    $ \beta $  & $am_s$ & T (MeV) & 
    $m_l/m_s$ & $\max rT$: $\Fs,~\Fqq$ & \#TUs & $ L^{\mathrm{bare}} $    \\
    \hline
    7.825 & 0.00164  & 306  & 1/20 & 0.39,\quad 0.31 & 10780 & 0.002975(14) \\
    8.000 & 0.001299 & 356  & 1/5  & 0.42,\quad 0.29 & 11460 & 0.004185(17) \\
    8.200 & 0.001071 & 422  & 1/5  & 0.42,\quad 0.29 & 10660 & 0.005908(18) \\
    8.400 & 0.000887 & 500  & 1/5  & 0.60,\quad 0.34 & 11410 & 0.007813(26) \\
    8.570 & 0.008376 & 577  & 1/20 & 0.61,\quad 0.36 & 10400 & 0.009680(28) \\
    8.710 & 0.007394 & 650  & 1/20 & 0.65,\quad 0.40 & 10190 & 0.0113707(21) \\
    8.850 & 0.006528 & 731  & 1/20 & 0.67,\quad 0.42 &  4480 & 0.013246(32) \\
    9.060 & 0.004834 & 872  & 1/20 & 0.82,\quad 0.42 &  8350 & 0.015904(33)  \\
    9.230 & 0.004148 & 1005 & 1/20 & 0.76,\quad 0.99 &  3610 & 0.018594(62) \\
    9.360 & 0.003691 & 1121 & 1/20 & 0.82,\quad 0.34 &  3530 & 0.020562(52) \\
    9.490 & 0.003285 & 1250 & 1/20 & 0.86,\quad 0.36 &  6790 & 0.022477(46) \\
    9.670 & 0.002798 & 1454 & 1/20 & 0.86,\quad 0.76 &  8320 & 0.025508(52) \\
    \hline
  \end{tabular}
  \caption{\label{tab:nt 16}
  The parameters of the new $N_{\tau}=16$ ensembles with aspect ratio 
  \(4\) and expectation values of bare Polyakov loops. 
  }
}
\end{table*}

\begin{table*}[t] 
\parbox{1.0\linewidth}{
  \begin{tabular}{|c|c|c|c|c|c|c|c|c|c|}
    \hline
    \multicolumn{2}{|c|}{$N_\tau=12$:} &
    \multicolumn{4}{|c|}{$m_l=m_s/20$} &
    \multicolumn{4}{|c|}{$m_l=m_s/5$} 
    \\
    \hline
    $ \beta $  & T (MeV) & 
    $am_s$ & $\max rT$: $\Fs,~\Fqq$ & \#TUs & $ L^{\mathrm{bare}} $ & 
    $am_s$ & $\max rT$: $\Fs,~\Fqq$ & \#TUs & $ L^{\mathrm{bare}} $   \\
    \hline
    8.000 & 474 & 
    0.01400 & 0.68,\quad 0.54 &  3110 & 0.019196(90) & 
    0.01299 & 1.06,\quad 0.77 & 71670 & 0.019254(25) \\
    8.200 & 563 & 
    0.01167 & 1.05,\quad 0.76 & 30090 & 0.024060(43) & 
    0.01071 & 1.09,\quad 0.73 & 71390 & 0.024062(25) \\
    8.400 & 667 & 
    0.00975 & 1.06,\quad 0.72 & 29190 & 0.029300(41) & 
    0.00887 & 1.15,\quad 0.70 & 71170 & 0.029252(24) \\
    \hline
    \hline
    \multicolumn{2}{|c|}{$N_\tau=10$:} &
    \multicolumn{4}{|c|}{$m_l=m_s/20$} &
    \multicolumn{4}{|c|}{$m_l=m_s/5$} 
    \\
    \hline
    $ \beta $  & T (MeV) & 
    $am_s$ & $\max rT$: $\Fs,~\Fqq$ & \#TUs & $ L^{\mathrm{bare}} $ & 
    $am_s$ & $\max rT$: $\Fs,~\Fqq$ & \#TUs & $ L^{\mathrm{bare}} $   \\
    \hline
    8.000 & 569 & 
    0.01400 & 1.09,\quad 1.00 &  6780 & 0.040275(211) & 
    0.01299 & 1.21,\quad 0.80 & 82770 & 0.040215(36) \\
    8.200 & 675 & 
    0.01167 & 1.12,\quad 0.57 & 27500 & 0.047833(97) & 
    0.01071 & 1.31,\quad 0.83 & 72180 & 0.048010(33) \\
    8.400 & 800 & 
    0.00975 & 1.07,\quad 0.58 &  7540 & 0.055774(114) & 
    0.00887 & 1.27,\quad 0.83 & 72770 & 0.055918(38) \\
    \hline
    \hline
    \multicolumn{2}{|c|}{$N_\tau=8$:} &
    \multicolumn{4}{|c|}{$m_l=m_s/20$} &
    \multicolumn{4}{|c|}{$m_l=m_s/5$} 
    \\
    \hline
    $ \beta $  & T (MeV) & 
    $am_s$ & $\max rT$: $\Fs,~\Fqq$ & \#TUs & $ L^{\mathrm{bare}} $ & 
    $am_s$ & $\max rT$: $\Fs,~\Fqq$ & \#TUs & $ L^{\mathrm{bare}} $   \\
    \hline
    7.825 & 611 & 
    0.0164  & 1.31,\quad 0.83 & 122200 & 0.072566(29) & 
    0.01542 & 1.36,\quad 0.91 & 51510 & 0.072616(44) \\
    8.000 & 711 & 
    0.01400 & 1.27,\quad 0.75 &  18370 & 0.083107(317) & 
    0.01299 & 1.53,\quad 0.95 & 100770 & 0.082665(32) \\
    8.200 & 843 & 
    0.01167 & 1.10,\quad 0.52 &   3080 & 0.093920(224) & 
    0.01071 & 1.59,\quad 1.08 & 100410 & 0.094084(41) \\
    8.400 & 1000 & 
    0.00975 & 1.14,\quad 0.69 &  3000 & 0.105302(286) & 
    0.00887 & 1.49,\quad 0.91 & 100830 & 0.105537(48) \\
    \hline
    \hline
    \multicolumn{2}{|c|}{$N_\tau=6$:} &
    \multicolumn{4}{|c|}{$m_l=m_s/20$} &
    \multicolumn{4}{|c|}{$m_l=m_s/5$} 
    \\
    \hline
    $ \beta $  & T (MeV) & 
    $am_s$ & $\max rT$: $\Fs,~\Fqq$ & \#TUs & $ L^{\mathrm{bare}} $ & 
    $am_s$ & $\max rT$: $\Fs,~\Fqq$ & \#TUs & $ L^{\mathrm{bare}} $   \\
    \hline
    7.030 & 399 & 
    0.0356 & 1.25,\quad 0.96 & 30390 & 0.083622(82) & 
    0.0356 & 1.13,\quad 0.83 & 30830 & 0.083627(91) \\
    7.825 & 824 & 
    0.0164  & 1.67,\quad 1.00 & 54480 & 0.151915(64) & 
    0.01542 & 1.34,\quad 0.78 & 59450 & 0.151974(63) \\
    8.000 & 948 & 
    0.01400 & 1.39,\quad 0.85 &   3200 & 0.165828(296) & 
    0.01299 & 1.56,\quad 0.94 & 102210 & 0.165650(53) \\
    8.200 & 1125 & 
    0.01167 & 1.30,\quad 0.76 & 10100 & 0.180802(171) & 
    0.01071 & 1.74,\quad 1.09 & 102360 & 0.181024(57) \\
    8.400 & 1334 & 
    0.00975 & 1.31,\quad 0.82 & 10160 & 0.195736(153) & 
    0.00887 & 1.80,\quad 1.13 & 102700 & 0.195837(50) \\
    \hline
    \hline
    \multicolumn{2}{|c|}{$N_\tau=4$:} &
    \multicolumn{4}{|c|}{$m_l=m_s/20$} &
    \multicolumn{4}{|c|}{$m_l=m_s/5$} 
    \\
    \hline
    $ \beta $  & T (MeV) & 
    $am_s$ & $\max rT$: $\Fs,~\Fqq$ & \#TUs & $ L^{\mathrm{bare}} $ & 
    $am_s$ & $\max rT$: $\Fs,~\Fqq$ & \#TUs & $ L^{\mathrm{bare}} $   \\
    \hline
    7.030 & 598 & 
    0.0356 & 1.58,\quad 1.03 & 52640 & 0.229883(83) & 
    0.0356 & 1.46,\quad 0.90 & 50230 & 0.229996(76) \\
    7.825 & 1222 & 
    0.0164  & 1.68,\quad 0.90 & 65380 & 0.315334(66) & 
    0.01542 & 1.70,\quad 1.03 & 50390 & 0.315231(93) \\
    8.000 & 1422 & 
    0.01400 & 1.52,\quad 0.71 & 27510 & 0.331131(96)) & 
    0.01299 & 1.60,\quad 0.94 & 97570 &  0.331104(44) \\
    8.200 & 1687 & 
    0.01167 & 1.54,\quad 0.87 & 20790 & 0.348063(122) & 
    0.01071 & 1.87,\quad 1.03 & 99810 & 0.348141(59)  \\
    8.400 & 1999 & 
    0.00975 & 1.48,\quad 0.71 & 20950 & 0.363987(151) & 
    0.00887 & 1.62,\quad 0.90 & 100000 & 0.364152(46) \\
    \hline
  \end{tabular}
  \caption{\label{tab:hvy}
  The parameters of the $N_{\tau}=12,~10,~8,~6$ and $4$ ensembles with different light sea quark masses and expectation values of bare Polyakov loops. 
  }
}
\end{table*}

\begin{table*}[t] 
\parbox{.90\linewidth}{
  \begin{tabular}{|c|c|c|c|c|c|c|c|c|}
    \hline
    \multicolumn{3}{|c|}{$N_\tau=4$:} &
    \multicolumn{3}{|c|}{$N_\sigma/N_\tau=4$} &
    \multicolumn{3}{|c|}{$N_\sigma/N_\tau=6$} 
    \\
    \hline
    $ \beta $  & T (MeV) & $am_s$ & 
    $\max rT$: $\Fs,~\Fqq$ & \#TUs & $ L^{\mathrm{bare}} $ & 
    $\max rT$: $\Fs,~\Fqq$ & \#TUs & $ L^{\mathrm{bare}} $   \\
    \hline
    6.664 &  421 & 0.0514 &
    1.46,\quad 1.22 &  75770 & 0.180018(82) & 
    1.92,\quad 1.27 & 513260 & 0.180141(15) \\
    8.200 & 1687 & 0.01071 &
    1.87,\quad 1.03 & 99810 & 0.348141(59) & 
    2.26,\quad 1.27 & 203980 & 0.348081(16) \\
    9.670 & 5814 & 0.002798 &
    1.64,\quad 0.87 &  10340 & 0.447998(137) & 
    2.77,\quad 1.37 & 600760 & 0.448130(11) \\
    \hline
  \end{tabular}
  \caption{\label{tab:vol}
  The parameters of the $N_{\tau}=4$ ensembles with different 
  aspect ratios and expectation values of bare Polyakov loops. 
  }
}
\end{table*}

As mentioned in \mbox{Sec.}~\ref{sec:setup}, we make use of gauge ensembles 
from the HotQCD Collaboration and from an earlier publication of the 
TUMQCD Collaboration (\mbox{cf. Refs.}~\cite{Bazavov:2011nk,Bazavov:2014pvz, Ding:2015fca, Bazavov:2013uja, Bazavov:2013yv,Bazavov:2016uvm}). 
The parameters of these gauge configurations, the collected statistics that 
have been used for the correlators in this study, as well as references to the 
original publications are listed in Tables~\ref{tab:nt 12} --~\ref{tab:nt 4}. 
These lattices have light sea quark masses of $m_l=m_s/20$ and correspond to a 
pion mass of $m_\pi\sim 160\,{\rm MeV}$ in the continuum limit. 
Temperatures in the tables have been fixed using the $r_1$ scale. 
Most ensembles are available in the form of multiple streams of molecular 
dynamics (MD) time evolution with different starting lattices and different 
seeds of the pseudorandom number generator.
The different streams of each ensemble are indexed by single letters in 
alphabetical ordering, \mbox{i.e.} the first stream is the ``a'' stream. 
Although all streams have roughly similar length for most ensembles, there 
are cases where additional streams have been branched off the first streams much later and are therefore substantially shorter. 
The first 30 time units (TUs) of each stream of HotQCD lattices (40 TUs for 
TUMQCD lattices) use the rational hybrid molecular dynamics (RHMD) algorithm 
instead of the rational hybrid Monte Carlo (RHMC) algorithm~\cite{Clark:2004cp}. 
This approach avoids small acceptance rates due to large changes of the 
action during early thermalization and shortens the thermalization phase to typically 20-40 TUs. 
To be safe, we later omit at least 100 TUs of the initial lattices of each stream in the analysis.
For an unclear reason, we found a problem with thermalization of the \(N_\tau=6,\ \beta=7.825\) gauge ensemble. 
We generated a new ensemble with the same parameters from scratch, which we use in our analysis (\mbox{cf. Table}~\ref{tab:hvy}). 
In order to take the continuum limit for extremely high temperatures, 
\(T \sim 1-2\,{\rm GeV}\), we generated new very fine ensembles with \(N_\tau=12\) and~\(10\) (\mbox{cf. Table}~\ref{tab:fine}). 
In order to ensure control of cutoff effects at the smallest separations, we 
generated another set of gauge ensembles with \(N_\tau=16\) (\mbox{cf. Table}~\ref{tab:nt 16}). 

\begin{figure*}
\centering
 \includegraphics[height=4.1cm,clip]{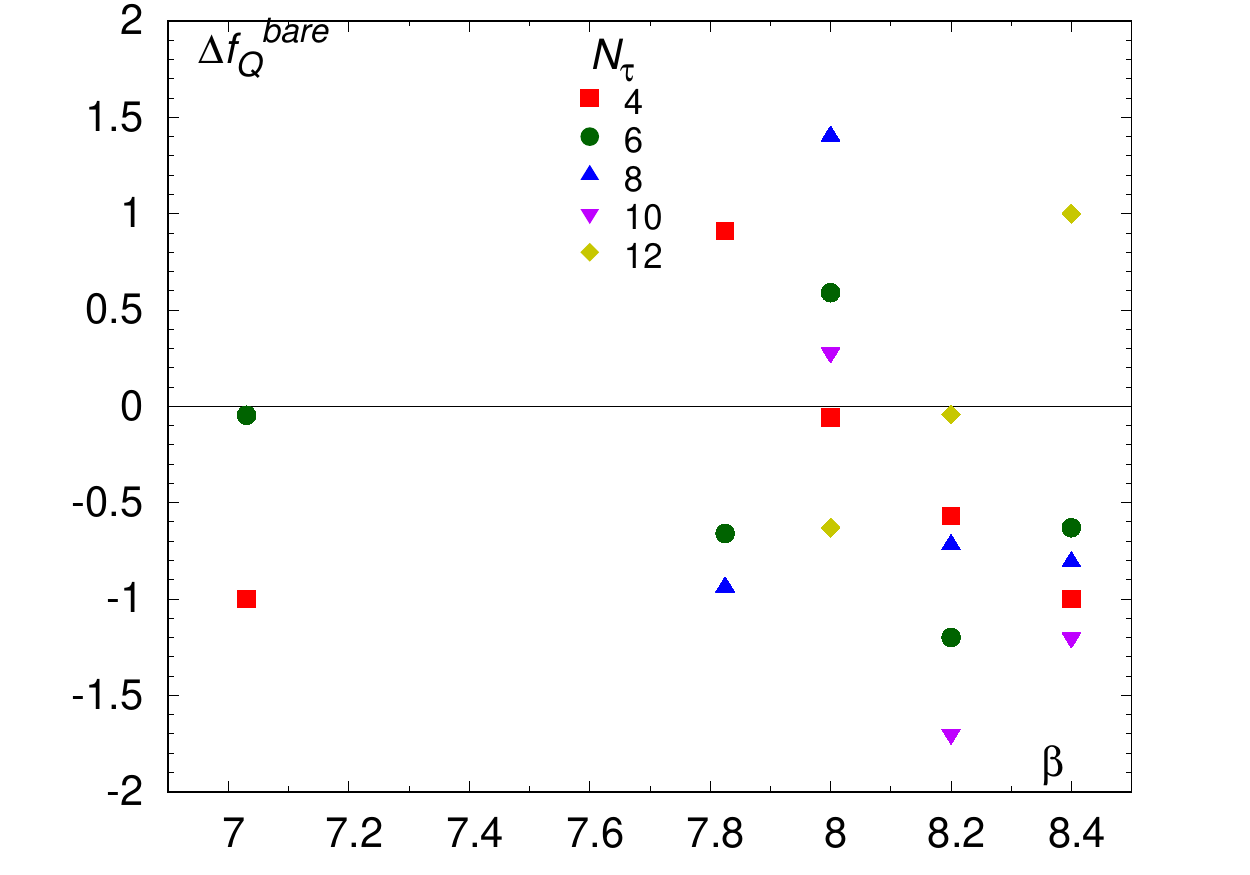}
\hfill
\centering
 \includegraphics[height=4.1cm,clip]{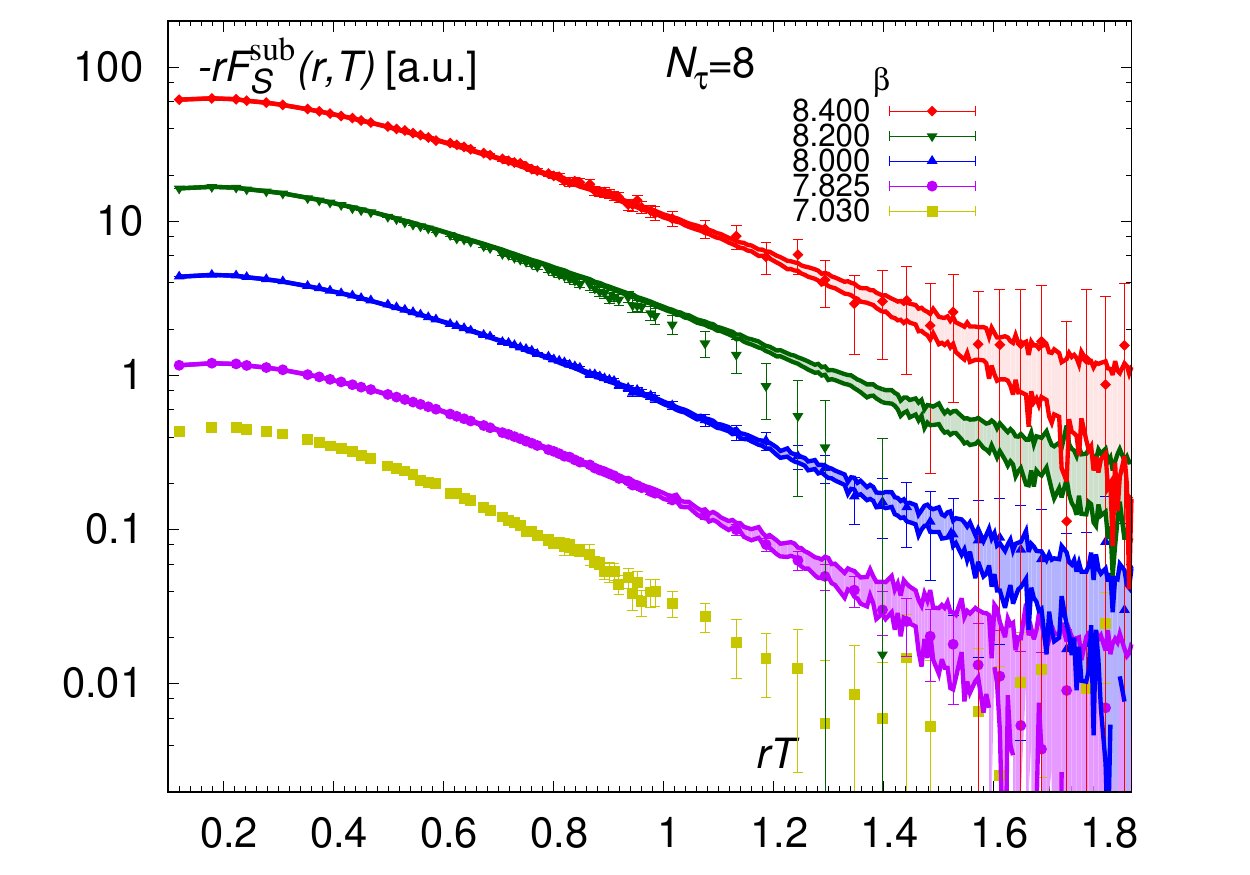}
\hfill
\centering
 \includegraphics[height=4.1cm,clip]{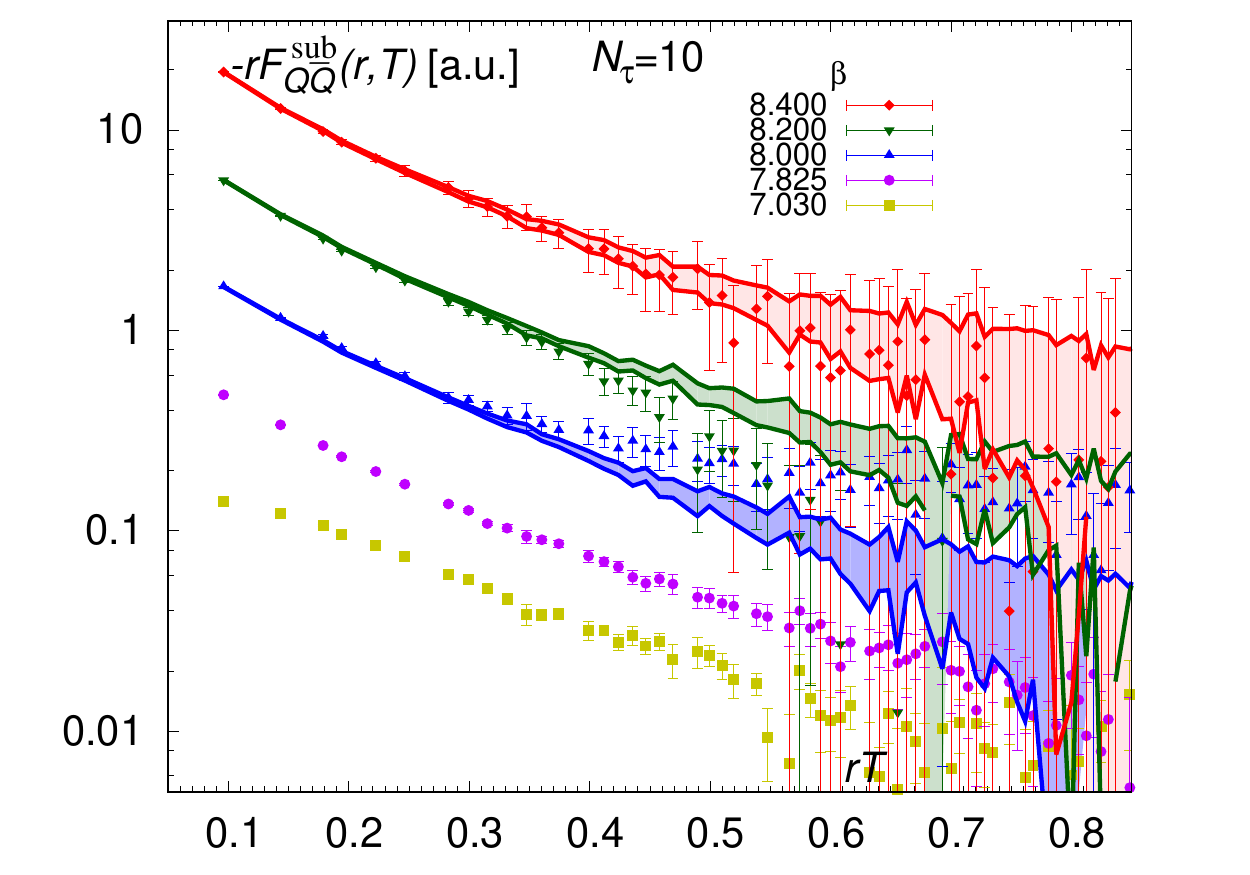}
\caption{\label{fig:hvy}
The ratio $\Delta f_Q^{\rm bare}$ (left), the screening 
function $S_1$ for the subtracted singlet free energy $\Fs^{\rm sub}$ (middle), 
and for the subtracted quark-antiquark free energy $F_{Q\bar Q}^{\rm sub}$ 
(right) are shown for different light sea quark masses. 
Results for different $\beta$ and $N_\tau$ have been shifted for reasons of visibility. 
The bands correspond to the ensembles with larger light sea quark masses.  
The observed deviation between ensembles with different masses can be understood in terms of larger statistical uncertainties of the lighter ensembles. 
For $N_\tau=4,~6$, and $12$, the ensemble sizes of most lighter ensembles are sufficiently large such that results are in good agreement. 
}
\end{figure*}

\begin{figure}
\centering
 \includegraphics[height=6.2cm,clip]{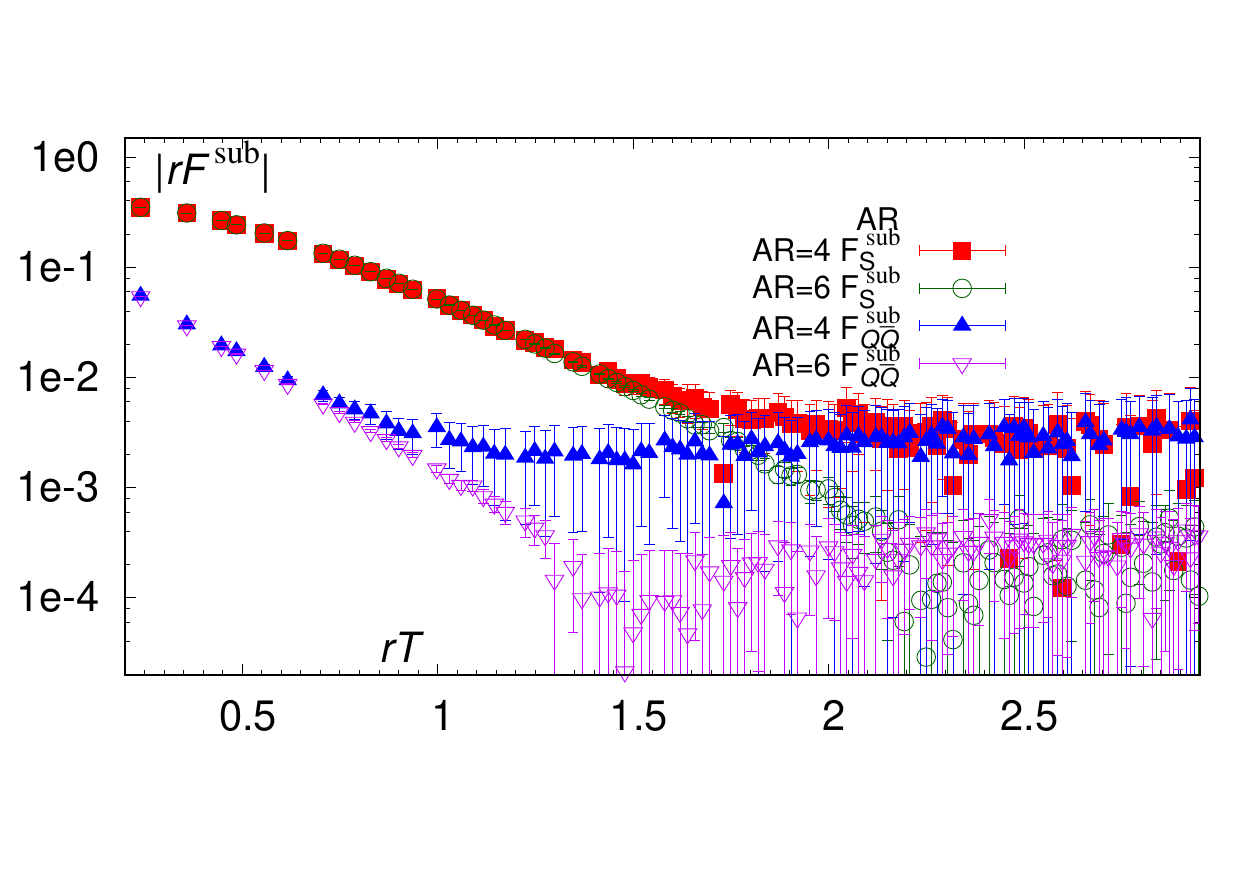}
\caption{\label{fig:vol}
The screening function $S_1$ for the subtracted singlet free energy
$\Fs^{\rm sub}$ and for the subtracted free energy $\Fqq^{\rm sub}$. 
The asymptotic behavior exhibits a strong volume dependence (\(N_\tau=4\) and \(\beta=6.664\)). 
Cancellation between quark-antiquark free energies and \(2\Fq\) is better in the larger volume.
}
\end{figure}

Furthermore, we generated new lattices with a larger light sea quark mass, 
corresponding to a pion mass of $m_\pi\sim 320\,{\rm MeV}$ in the continuum limit. 
These new lattices have $N_\tau=4,~6,~8,~10,~12$ and $16$ and 
$\beta=7.03,~7.825,~8.0,~8.2$, and $8.4$ and therefore span a temperature 
range from $T=306$ to $2000\,\rm MeV$. 
Since these lattices have high temperatures ($m_\pi \sim 2\,T_c\lesssim T$), 
numerical results are only mildly affected by the larger quark mass. 
We summarize these ensembles in \mbox{Table}~\ref{tab:hvy}. 
In the left panel of \mbox{Fig.}~\ref{fig:hvy}, we show a ratio of the 
difference between the bare single quark free energies with different light 
sea quark masses \(m_1\) and \(m_2\) over the combined uncertainty, obtained 
by adding their statistical errors in quadrature,

\ileq{
 \Delta \fq^{\rm bare} = \frac{\fq^{\rm bare}(m_{1})-\fq^{\rm bare}(m_{2})}
 {\sqrt{(\delta \fq^{\rm bare}(m_{1}))^2+(\delta \fq^{\rm bare}(m_{2}))^2}}.
}

\(\Delta \fq^{\rm bare}\) fluctuates in a \(\pm 1.5\sigma\) interval 
($\sigma$ being the standard deviation) and shows no signs of a systematic quark mass dependence. 
In the middle and right panels, we show screening functions for two 
representative correlators. 
The short distance behavior of the screening functions defined as

\ileq{
S_1=-r F^{\rm sub},~F=\Fs, \Fqq,
}

is independent of the quark mass within statistical uncertainties and we do 
not observe a unique trend towards stronger or weaker screening as a function of the quark masses. 
Hence, differences can be understood in terms of the much larger statistical 
errors of the HotQCD ensembles at $m_l=m_s/20$, leading to an incomplete 
cancellation with the squared bare Polyakov loop. 
From these observations, we have to conclude that the statistical 
errors on small ensembles may be unreliable.

Moreover, we study finite volume effects in the Polyakov loop and in the 
correlators on a set of ensembles with $N_\tau=4$ and $m_l=m_s/20$ or \(m_l=m_s/5\).  
We compare gauge ensembles with aspect ratio (AR) $N_\sigma/N_\tau=4$ to 
gauge ensembles with $N_\sigma/N_\tau=6$ for three different lattice spacings (\mbox{cf. Table}~\ref{tab:vol}).
We see small changes in the expectation value of Polyakov loop, which are, 
given the size of the errors on the $N_\sigma/N_\tau=4$ ensembles, 
statistically not significant (within one and one and a half statistical errors). 
We show a comparison of screening functions for different volumes in \mbox{Fig.}~\ref{fig:vol}. 
The singlet correlators for different volumes are numerically consistent up to \(rT\lesssim 1.25\). 
However, the screening behavior of Polyakov loop correlators for the two 
volumes differs quite dramatically already at \(rT\lesssim 0.5\). 
Since \(\Fqq^{\rm sub}\) and \(\Fs^{\rm sub}\) approach in the same volume 
the same asymptotic constant, the deviation due to different asymptotic 
values becomes relevant for \(\Fqq^{\rm sub}\) at smaller separations.

\subsection{Correlators}
\label{app:correlators}

As mentioned in \mbox{Sec.}~\ref{sec:setup}, we calculate the expectation 
values and the errors of the correlators with the jackknife method. 
We generally have one measurement for every ten TUs of molecular 
dynamics (MD). 
However, there are some gaps in the MD histories, since some ensembles were 
not available as coherent MD chains of gauge configurations. 
We simply ignore these gaps and attach the measurements in ascending order 
of MD time in each stream. 
We combine different streams of gauge ensembles into one ensemble
for the calculation of the correlators.
We omit the first 1000 TUs in the first stream, if the stream did not 
use a starting lattice with the same lattice spacing. 
Otherwise, we omit the first 100 TUs in each stream to ensure that 
correlations between different streams are sufficiently reduced and that 
the MD histories are not affected by the brief intermezzo of RHMD evolution 
at the start of a stream. 
We also combine the Polyakov loop time histories of all streams in 
the order as the correlators.
Since the Polyakov loop is measured after each single TU, there are roughly 
10 times as many Polyakov loop measurements as correlator measurements. 
We divide the Polyakov loop histories evenly over the jackknife bins of the 
correlators.

We sort data into a fixed number of jackknife bins (30) using bins of equal 
size and omit leftover measurements. 
We distribute the Polyakov loop history over an equal number of jackknife bins 
and divide the correlator on each jackknife bin by the square of the Polyakov loop on the same bin. 
Since the expectation value of the correlator approaches for asymptotically 
large $r$ the squared expectation value of the Polyakov loop up to effects 
due to the finite volume, this ratio, which we call the normalized
correlator (see main text) has reduced statistical fluctuations 
for large separations $r$ in correlators at high temperature. 
The logarithms of these correlators are the subtracted free energies, 
which--barring finite volume effects--decrease exponentially for 
asymptotically large separations $r\to\infty$. 
We list the maximal separations at which we could obtain a statistically 
significant signal for \(\Fs^{\rm sub}\) and \(\Fqq^{\rm sub}\) in 
Tables~\ref{tab:nt 12}--\ref{tab:vol} using \(rT\) units (usually in the fourth column). 
For \(\Fs^{\rm sub}\), we require \(\abs{\Fs^{\rm sub}/\delta \Fs^{\rm sub}}<5\) 
for any smaller separation; for \(\Fqq^{\rm sub}\), we require at most one 
instance of \(\abs{\Fqq^{\rm sub}/\delta \Fqq^{\rm sub}}<2\). 
In some ensembles with insufficient statistics, the cancellation may fail, 
and thus, we would see a fake signal at large separation. 
We avoid this by demanding that the modulus of the last accepted points is 
larger than the modulus of the omitted points and that \(-F^{\rm sub}\) as 
well as screening functions \(-rF^{\rm sub}\) and \(-r^2F^{\rm sub}\) 
have negative slope for sufficiently large \(rT\). 
We find that this estimator for the largest accessible separations of the 
\(\qbq\) pair varies only mildly within about 10\% between lattices with the 
same \(N_\tau\) and similar temperatures as well as similar ensemble sizes.
Since the cancellation between the correlator and the squared Polyakov loop 
is an infinite volume property, incomplete cancellation may as well be a hint 
of finite volume effects, which can be confirmed for the respective 
\(N_\tau=4\) lattices by inspecting \mbox{Fig.}~\ref{fig:vol}.

\vskip1ex
If we have data sets with different masses for the same pair of 
parameters~(\(\beta,~N_\tau\)), we use only the correlator that gives us the 
best integrated signal-to-noise ratio on the full range that we would use. 
Hence, our final set of ensembles in the analysis contains two different quark masses.

\subsection{Renormalization}
\label{app:renormalization}

\begin{table}[t]
\parbox{1.0\linewidth}{
  \begin{tabular}{|c|c||c|c||c|c|}
    \hline
    $ \beta $  & $c_Q$ &     $ \beta $  & $c_Q$ &     $ \beta $  & $c_Q$ \\
    \hline
    7.150 & -0.3826(13) & 7.825 & -0.3403(12) & 8.850 & -0.2804(22) \\
    7.200 & -0.3794(09) & 8.000 & -0.3296(09) & 9.060 & -0.2700(22) \\
    7.280 & -0.3745(16) & 8.200 & -0.3171(08) & 9.230 & -0.2616(21) \\
    7.373 & -0.3687(07) & 8.400 & -0.3044(15) & 9.360 & -0.2555(21) \\
    7.596 & -0.3545(11) & 8.570 & -0.2952(19) & 9.490 & -0.2496(21) \\
    7.650 & -0.3512(10) & 8.710 & -0.2879(22) & 9.670 & -0.2418(21) \\
    \hline
  \end{tabular}
  \caption{\label{tab:new cq}
  The renormalization constant \(c_Q=aC_Q\) from the direct renormalization 
  procedure. 
  }
}
\end{table}

\begin{figure}
\includegraphics[height=6.2cm,clip]{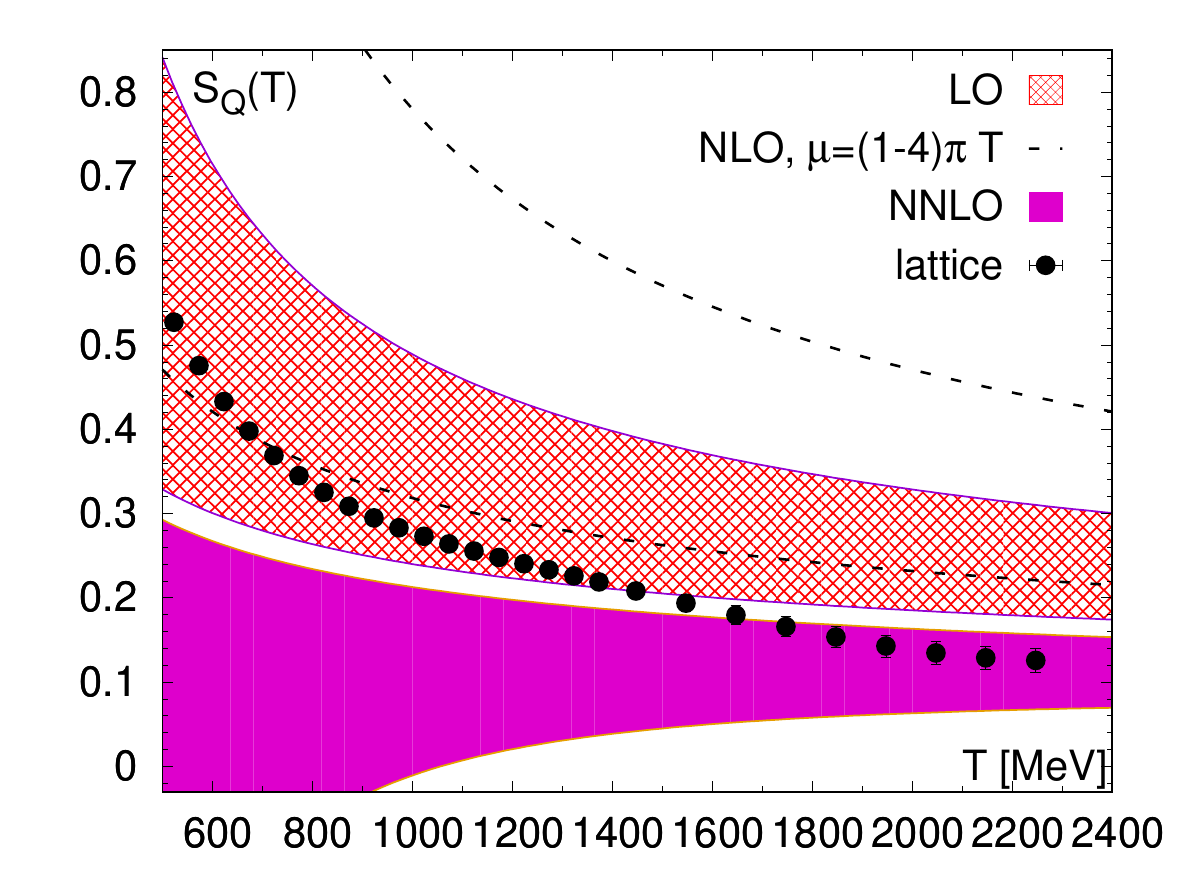}
\caption{\label{fig:Sqcont}
The continuum extrapolated entropy \(S_Q\) in the high temperature region. 
}
\end{figure}

We determine the renormalization constant $2C_Q$ for the $\qbq$ free 
energies by applying the direct renormalization scheme to the Polyakov loop. 
This procedure is covered in detail in \mbox{Ref.}~\cite{Bazavov:2016uvm}.
We obtain a set of initial values of $C_Q$ from the static energy and include 
the new \(T=0\) ensembles with \(m_l/m_s=1/5\)~\cite{Bazavov:2017dsy}. 
Here, the potential is fixed to the value $V(r_i)=C_i/r_i$ with $C_0=0.954$, $C_1=0.2065$, and $C_2=-0.2715$. 
The Sommer scale $r_0$ and the scales $r_1$ and $r_2$ are defined through 
$dV/dr(r_0)=1.65$, $dV/dr(r_1)=1.0$, and $dV/dr(r_2)=1/2$.
We repeat the direct renormalization procedure of 
\mbox{Ref.}~\cite{Bazavov:2016uvm} after including all new ensembles. 
Because of the availability of higher \(\beta\) at \(T=0\) and for larger 
\(N_\tau\), we do not require iterative direct renormalization anymore. 
We choose those with smaller uncertainties if multiple ensembles with the 
same \(\beta\) and \(N_\tau\) are available. 
In Table~\ref{tab:new cq}, we summarize the new results for \(c_Q=aC_Q\) with substantially reduced uncertainties. 
Results for omitted \(\beta\) are commensurate with $c_Q$ from Table V in \mbox{Ref.}~\cite{Bazavov:2016uvm}. 
As mentioned in \mbox{Sec.}~\ref{sec:setup}, we renormalize $\Fqq$ and $\Fs$
by adding twice the renormalized single quark free energy $2\Fq$ to $\Fqq^{\rm sub}$ and $\Fs^{\rm sub}$.
Including the new ensembles, we repeat the extrapolation to the continuum limit 
(using global fits including \(N_\tau \geq 6\) and \(N_\tau^{-2}\) scaling) (\mbox{cf. Ref.}~\cite{Bazavov:2016uvm}).
In the high temperature region \(T > 250\,{\rm MeV}\), we obtain a continuum 
limit that is reliable up to \(T \gtrsim 2.3\,{\rm GeV}\) with 
\(\chi^2=97.3\) on 95 degrees of freedom after including the full uncertainty of \(c_Q\) in the global fit. 
For higher temperatures, the continuum extrapolation is not reliable due to lack of data with \(N_\tau>8\). 
From $F_Q$, we can calculate the entropy of a static quark

\ileq{
S_Q=-\frac{\partial F_Q}{\partial T}.
}

In \mbox{Fig.}~\ref{fig:Sqcont}, we show the continuum limit of \(S_Q\), 
which agrees with the weak-coupling result at next-to-next-to-leading order (NNLO) for \(T\gtrsim 1.7\,{\rm GeV}\). 
The values for $\Fq$ in the renormalization of the \(\qbq\) free energies are taken from the same continuum extrapolation.

\section{CUTOFF EFFECTS AT SHORT DISTANCES}
\label{app:B}

In this Appendix, we discuss the details of the improvement procedure for 
reducing the cutoff effects at small separations of only a few lattice 
steps, \mbox{i.e.}, \(r/a < 5\), which are due to the explicit breaking 
of the rotational symmetry in the discretized theory.
We develop this procedure for the static energy at zero temperature and 
apply its result to the quark-antiquark (and singlet) free energies. 
Lattice results for the static energy are discussed in detail in \mbox{Refs.}~\cite{Bazavov:2011nk,Bazavov:2014pvz,Bazavov:2017dsy}.
In \mbox{Sec.}~\ref{sec:free theory}, we analyze cutoff effects in the static energy for the tree-level interaction and the tree-level improvement. 
In \mbox{Sec.}~\ref{sec:static energy}, we analyze cutoff effects in the static energy for the full interacting theory with the HISQ action and nonperturbative improvement. 
In \mbox{Sec.}~\ref{sec:finite T improvement}, we apply these corrections to the free energies under study in this paper 
and give a quantitative account of the nonperturbative improvement that we achieve.

\subsection{Cutoff effects for the tree level interaction} 
\label{sec:free theory}

\begin{figure}
\centering
 \includegraphics[width=7.1cm,clip]{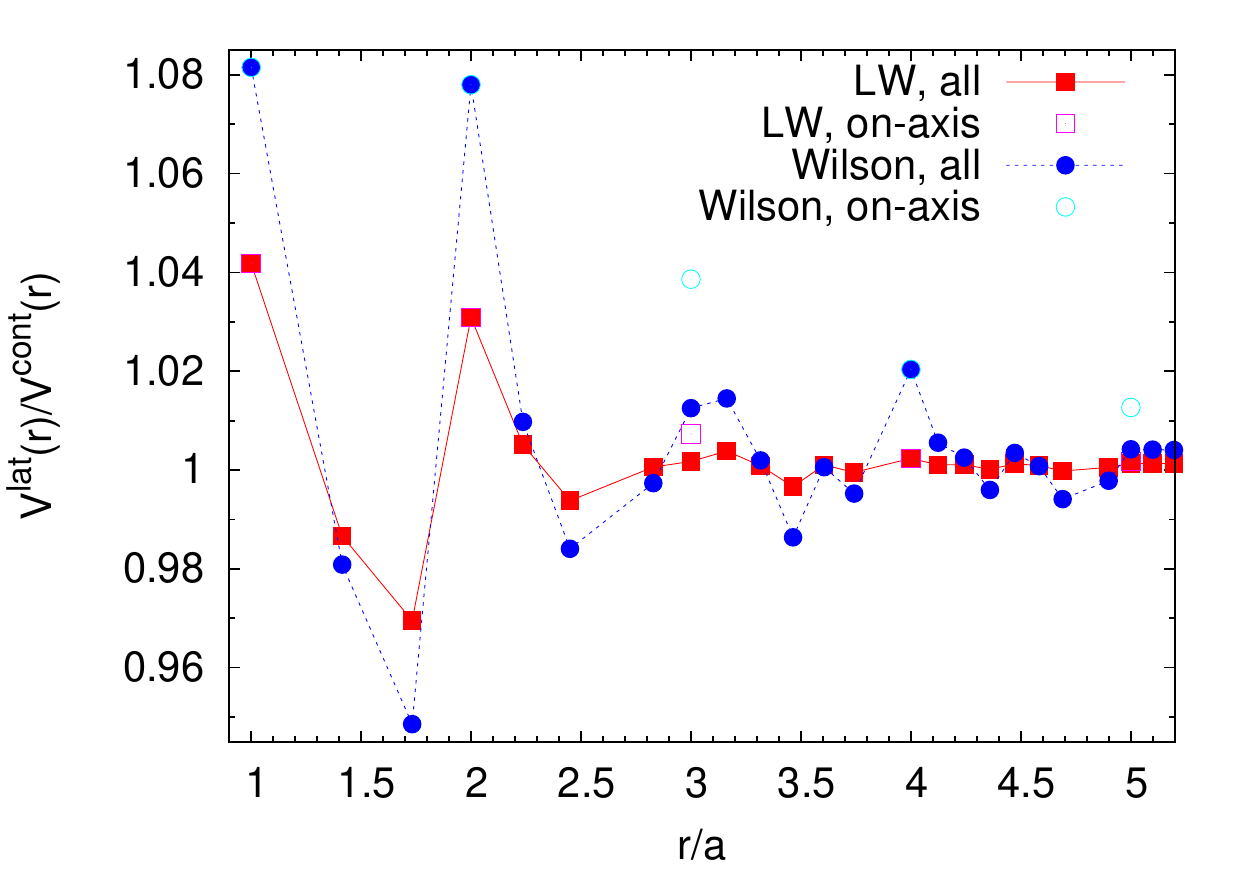}
\caption{\label{fig:tree-level}
Cutoff effects in the tree-level lattice static energy with L\"uscher-Weisz 
action are usually much smaller than for the Wilson gauge action. 
Both results are normalized by the corresponding continuum result.
}
\end{figure}

Our understanding of the cutoff effects in the static energy of a \(\qbq\) 
pair is largely based on the calculation at the tree level, \mbox{i.e.} for one-gluon exchange without a running coupling. 
We use this to estimate the cutoff effects on the static energy calculated with the HISQ action for \(\beta = 10/g^2 \geq 7.03\).

Using the lattice version of the temporal gluon propagator 
\(D_{00}(k_0,\bm k)\), it is straightforward to calculate the tree-level 
static energy on the lattice, which is given by

\ileq{
\Vs^{\rm lat,\,free}(r)=\int\frac{d^3 k}{(2 \pi)^3} 
D_{00}(k_0=0,\bm{k}) e^{i \bm{k}\cdot\bm{r}}
 \equiv \frac{1}{4 \pi r_I}.
\label{eq:tree}
}

The above equation defines the improved distance $r_I$. 
Sometimes, in the following, to remark the difference from the bare nonimproved distance \(r/a\), we will indicate the latter with \(r_b/a\). 
In \mbox{Fig.}~\ref{fig:tree-level}, we show the static energy at tree 
level calculated for the standard Wilson gauge action and for the improved 
L\"uscher-Weisz (LW) gauge action, normalized by the continuum result of \(1/(4\pi r)\). 
Because of the breaking of rotational symmetry on the lattice, we have to 
distinguish between on- and off-axis separations. 
Most of the separations shown in the figure are off-axis. 
For the values of \(r_b/a\), which correspond to both on- and off-axis 
separations (\mbox{e.g.}, \(r_b/a = 3\)), the average of the two is shown. 
As one can see from the figure, cutoff effects are largely reduced for the 
improved gauge action and they are very small for \(r_b/a > \sqrt{6}\). 
We use the tree-level result to reduce cutoff effects in the lattice 
calculations, namely, for each value of \(r\), we perform the replacement $r \rightarrow r_I$.
This is called the tree-level improvement.

\subsection{Cutoff effects in the static energy calculated with the HISQ action}
\label{sec:static energy}

\begin{table} 
\parbox{.98\linewidth}{
  \begin{tabular}{|c|c|c|c|c|c|c|c|}
    \hline
    \multicolumn{7}{|c|}{\(m_l=m_s/20\):} \\
    \hline
    $ \beta $ & a\,(\rm{fm}) & \(N_\sigma,N_\tau\) & $am_s$ & \#TUs & \#MEAS & Ref. \\
    \hline
    6.664 & 0.117 & \(32^4\) & 0.0514   & 4000 & 400 & \cite{Bazavov:2011nk}  \\
    6.740 & 0.109 & \(48^4\) & 0.0476   & 4000 & 800 & \cite{Bazavov:2011nk}  \\
    6.800 & 0.103 & \(32^4\) & 0.0448   & 4000 & 400 & \cite{Bazavov:2011nk}  \\
    6.880 & 0.095 & \(48^4\) & 0.0412   & 8200 & 820 & \cite{Bazavov:2011nk}  \\
    6.950 & 0.089 & \(32^4\) & 0.0386   & 19400 & 1940 & \cite{Bazavov:2011nk}  \\
    7.030 & 0.082 & \(48^4\) & 0.0356   & 7800 & 780 & \cite{Bazavov:2011nk}  \\
    7.150 & 0.070 & \(64^3\times 48\) & 0.0320  & 4047 & 667 & \cite{Bazavov:2011nk}  \\
    7.280 & 0.065 & \(64^3\times 48\) & 0.0284  & 3978 & 650 & \cite{Bazavov:2011nk}  \\
    7.373 & 0.060 & \(64^3\times 48\) & 0.0250  & 4623 & 1000 & \cite{Bazavov:2014pvz} \\
    7.596 & 0.049 & \(64^4\) & 0.0202   & 4757 & 1000& \cite{Bazavov:2014pvz} \\
    7.825 & 0.040 & \(64^4\) & 0.0164   & 4768 & 1000 & \cite{Bazavov:2014pvz} \\
  \hline
    \hline
    \multicolumn{7}{|c|}{\(m_l=m_s/5\):} \\
    \hline
    $ \beta $ & a\,[\rm{fm}] & \(N_\sigma,N_\tau\) & $am_s$ & \#TU & \#MEAS & Ref. \\
    \hline
    8.000 & 0.035 & \(64^4\) & 0.01299 & 4616 & 1000 & \cite{Bazavov:2017dsy} \\
    8.200 & 0.029 & \(64^4\) & 0.01071 & 4616 & 1000 & \cite{Bazavov:2017dsy} \\
    8.400 & 0.025 & \(64^4\) & 0.00887 & 4616 & 1000 & \cite{Bazavov:2017dsy} \\
  \hline
  \end{tabular}
  \caption{\label{tab:T=0}
  Parameters for the $T=0$ ensembles. Adjacent correlators are separated by $10$ TUs.
  In the sixth column we indicate the number of correlator measurements performed.
  }
}
\end{table}

\begin{figure*}
\centering
 \hskip-2em
 \includegraphics[height=6.6cm,clip]{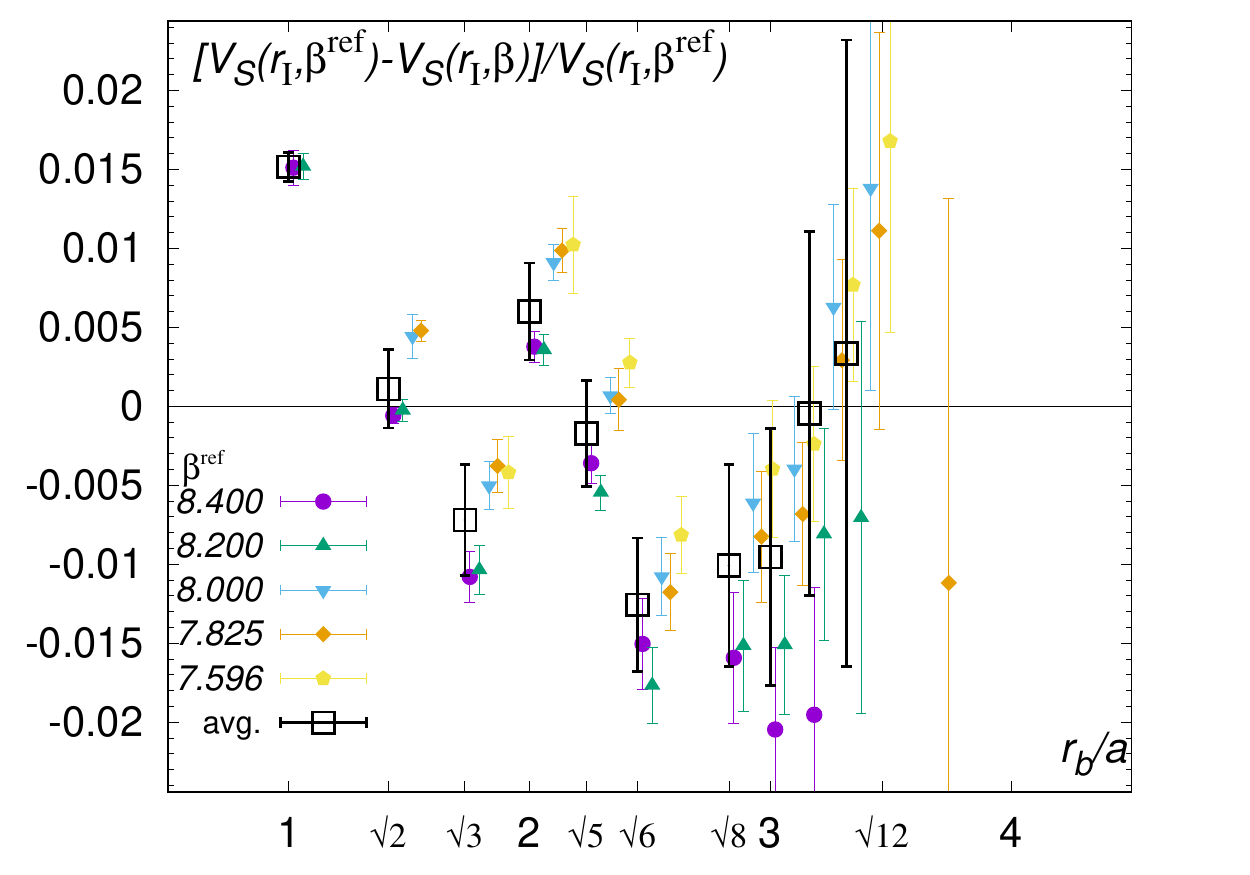}
 \hskip-2em
 \includegraphics[height=6.6cm,clip]{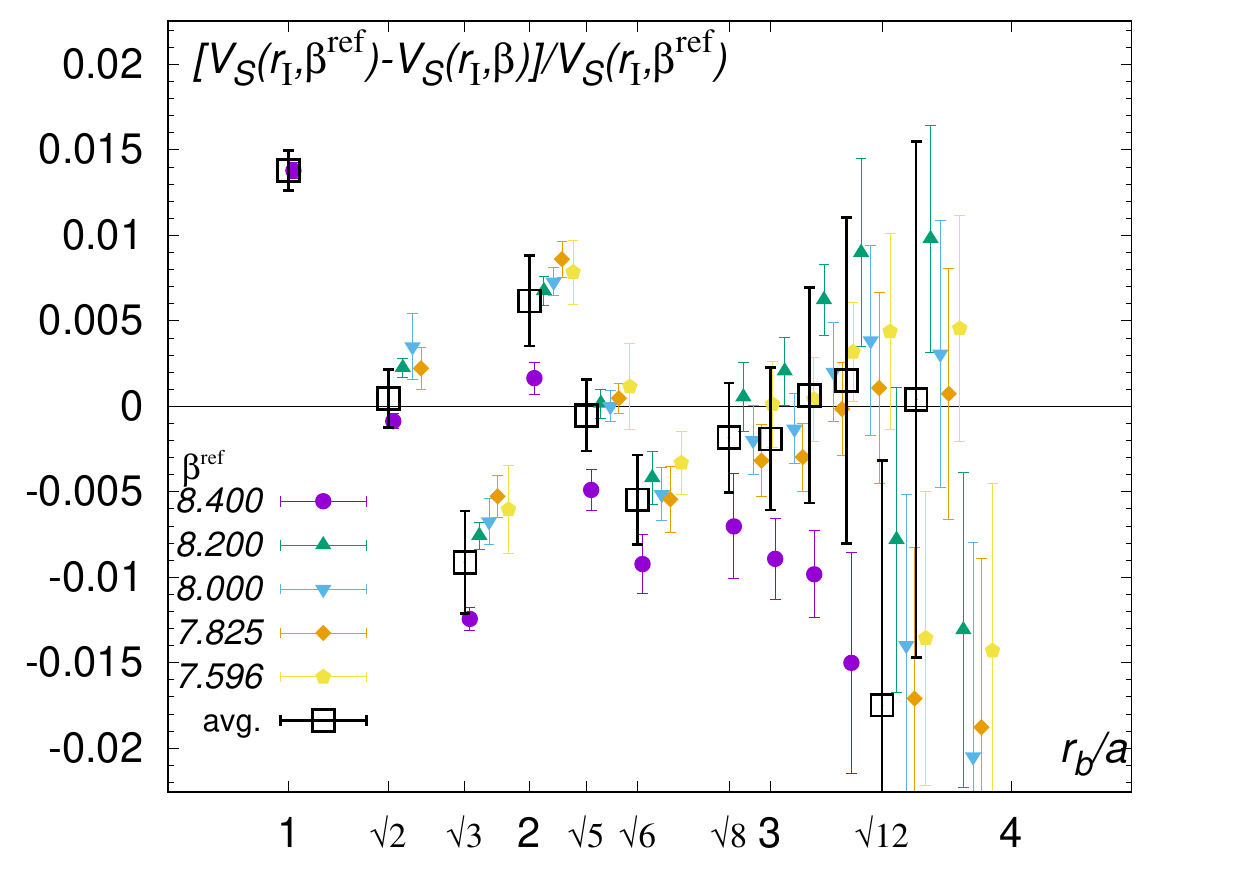}
 \hskip-2em

 \hskip-2em
 \includegraphics[height=6.6cm,clip]{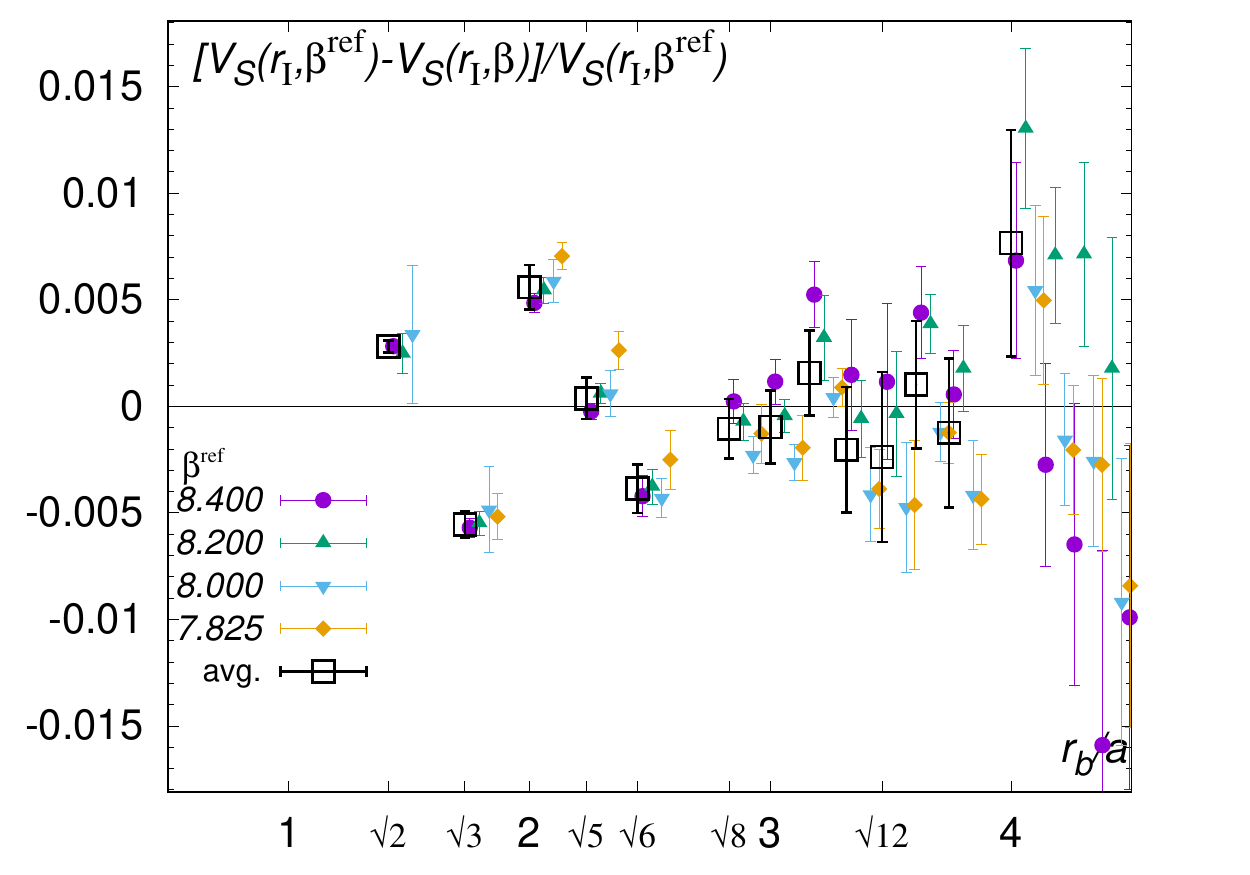}
 \hskip-2em
 \includegraphics[height=6.6cm,clip]{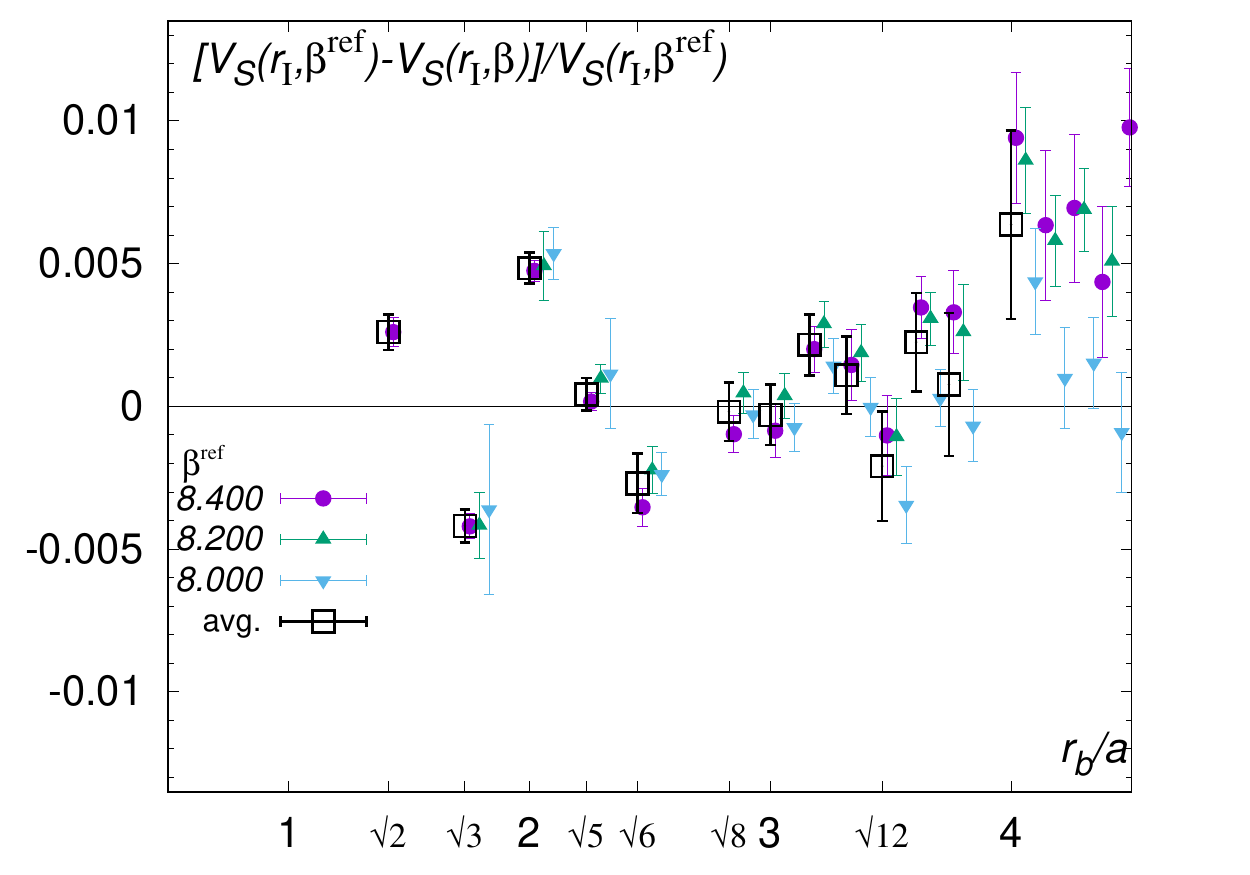}
 \hskip-2em

 \hskip-2em
 \includegraphics[height=6.6cm,clip]{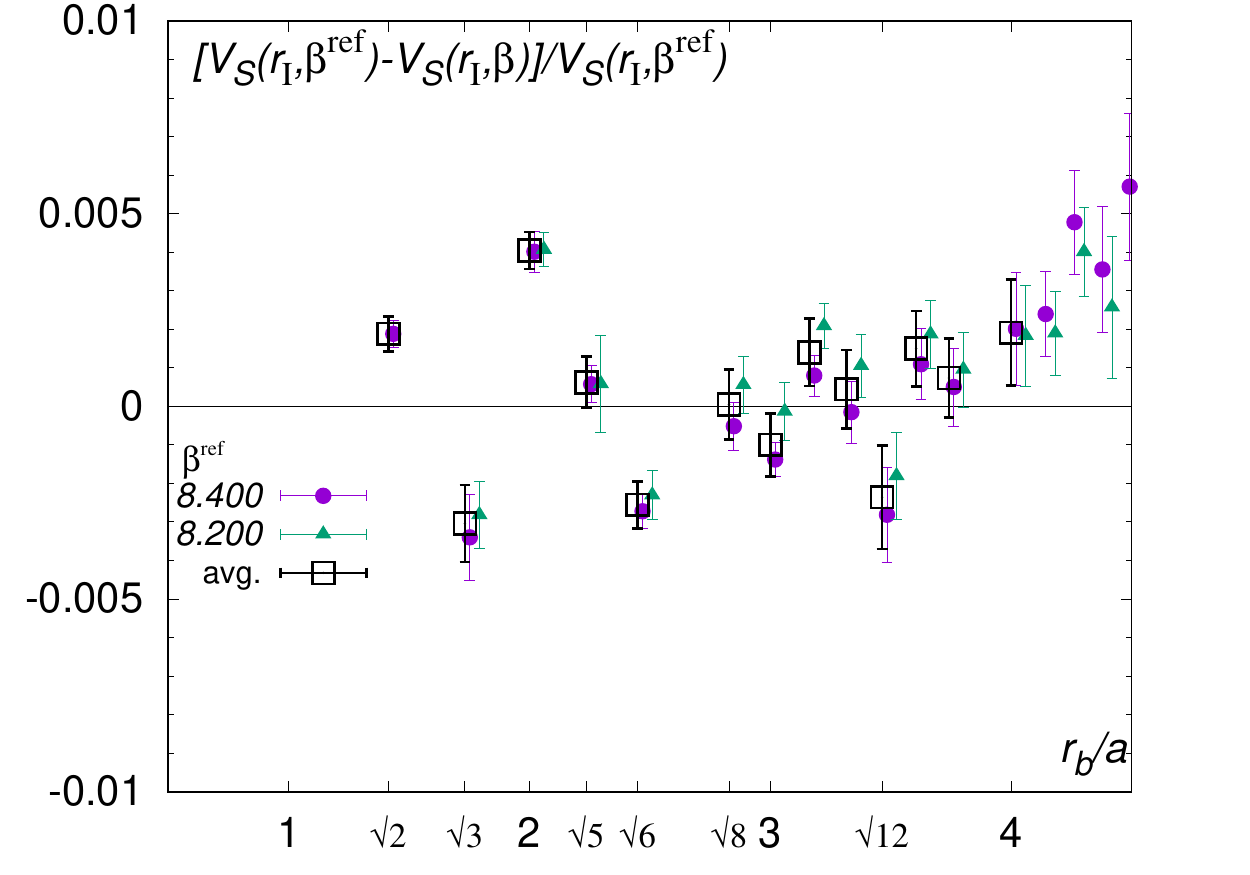}
 \hskip-2em
 \includegraphics[height=6.6cm,clip]{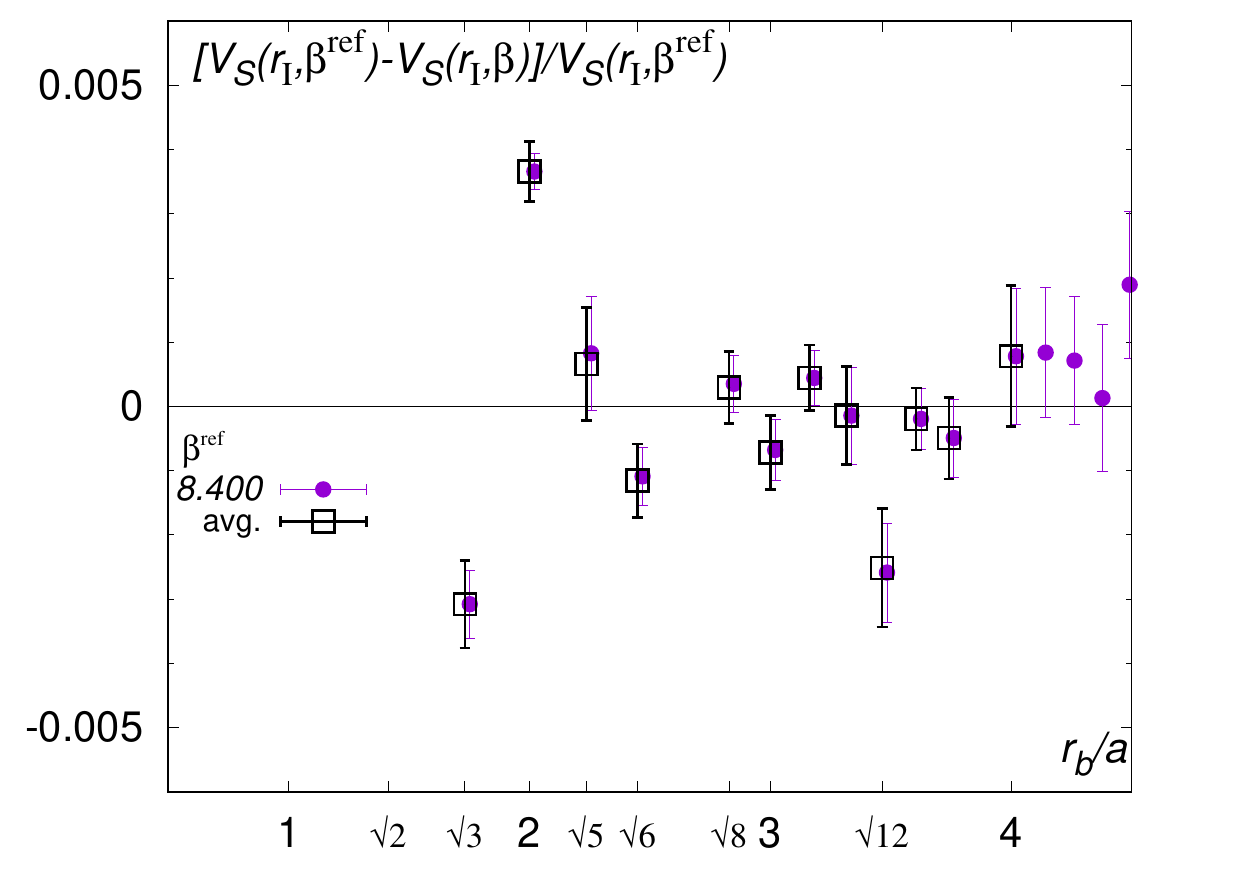}
 \hskip-2em
\caption{\label{fig:naive HISQ improvement}
The normalized deviation of the static energy from continuum estimates at different \(\beta\) values. 
(Top) The results for \(\beta=7.280\) (left)and \(\beta=7.373\) (right). 
(Center) The results for \(\beta=7.596\) (left), 
and \(\beta=7.825\)) (right). 
(Bottom) The results for \(\beta=8.000\) (left) 
and \(\beta=8.200\) (right).
Points for different \(\beta^{\rm ref}\) are displaced for better visibility. 
The averages (large open squares) are computed using the squared errors for each \(\beta^{\rm ref}\) as weights. 
}
\end{figure*}

The cutoff effects in the static energy with HISQ action have been 
studied in \mbox{Ref.}~\cite{Bazavov:2014soa}.
It has been shown that for the HISQ action the cutoff effects follow a 
similar pattern as in the free theory, namely, cutoff effects are about the 
same and are very small for \(r_b/a > \sqrt{6}\). 
Using the tree-level improvement, the cutoff effects can be reduced to 1\% 
or less (\mbox{cf. Fig.} 2 of \mbox{Ref.}~\cite{Bazavov:2014soa}). 
To reduce the cutoff effects even further, we follow the procedure outlined 
in \mbox{Ref.}~\cite{Bazavov:2014soa}, but implement new refinements, 
which are possible due to the fact that lattices for 
\(\beta=8.0,~8.2\), and \(8.4\) are available by now. 
We describe these refinements in the following.

In \mbox{Ref.}~\cite{Bazavov:2014pvz}, the static energy has been calculated 
for lattice spacings down to \(a = 0.04\,{\rm fm}\) for \(m_l = m_s/20\). 
In the recent study~\cite{Bazavov:2017dsy} of the QCD equation of state, the 
static energy was calculated for lattice spacings down to 
\(a = 0.025\,{\rm fm}\) with \(m_l = m_s/5\), and it was shown that the quark 
mass effects on the static energy are negligible at distances \(r<0.8\,r_1\). 
The lattice spacing was fixed using the scales \(r_1\) or \(r_2\) with 
\(r_1 = 0.3106\,{\rm fm}\) and \(r_2 = 0.1413\,{\rm fm}\). 
The lattice parameters for the static energy calculations are summarized in 
\mbox{Table}~\ref{tab:T=0}. 
The static energy calculated on a finer lattice with 
\(\beta^{\rm ref}>\beta\) can be used as a reference to estimate the cutoff 
effects in the static energy on any coarser lattice with smaller values of \(\beta\). 
By using the improved distance, we account for the tree-level cutoff effects 
in this comparison.
More precisely, the static energy for \(r_b/a>\sqrt{6}\) at \(\beta^{\rm ref}\) 
provides a continuum estimate of the static energy that lattice data at 
smaller \(\beta\) can be compared to. 
To compare the static energies calculated at different \(\beta\) values, 
one has to use the renormalization constants \(2C_Q\) given in \mbox{Table}~\ref{tab:new cq}. 
However, the results of the renormalization constants are not sufficiently 
accurate for comparison of the static energies at subpercent precision. 
Therefore, we have to shift the reference static energy by a constant within 
the errors of the renormalization constant for \(\beta^{\rm ref}\) such 
that the ratio of the potential is close to unity in the distance range 
\(r_b/a>\sqrt{6}\), where cutoff effects are small. 

Since we only have the static energy at discrete sets of data points for 
\(\beta^{\rm ref}\), we must interpolate between the data using the improved distance \(r_I\). 
The simplest interpolating function is the Cornell potential, \(V(r)=a/r+\sigma r+C\), 
and it was used in \mbox{Ref.}~\cite{Bazavov:2014soa}. 
However, now we find that it is not sufficiently flexible to provide accurate interpolations. 
So we modify the Cornell \textit{Ansatz} and supplement it by a polynomial in \(r\) up to sixth order. 
We also try a simpler form, namely, \(V(r) = a/(r \ln(kr))+\sigma r+C\), which works well in some cases. 
We also try a spline fit of \(r\Vs(r)\). 
We interpolate the data for \(r_b/a\geq\sqrt{5}\), assuming that cutoff 
effects are small enough to permit an interpolation with percent level accuracy. 
In order to extend the available range to smaller \(r\) and, hence, to 
determine estimates of the cutoff effects at smaller separations, we 
extrapolate towards 10\% smaller separations, \mbox{i.e.} down to \(r_b/a=2.00223\). 
We estimate the continuum limit with the modified Cornell \textit{Ansatz} and 
estimate the systematic uncertainty from the difference between fits with 
this modified Cornell \textit{Ansatz} and the spline \textit{Ansatz}. 
We add the statistical and systematic errors in quadrature.  

Some sample results demonstrating this analysis are shown in \mbox{Fig.}~\ref{fig:naive HISQ improvement}. 
We define the correction factors

\ileq{
K\left(\frac{r_b}{a},a,\beta^{\rm ref}\right) = 
\frac{\Vs(r_I,a)-\Vs(r_I,\beta^{\rm ref})}{\Vs(r_I,\beta^{\rm ref})}
\label{eq:correction factors},
}

and observe significant deviations of \(K\) from zero for \(r_b/a<\sqrt{6}\), 
although these become smaller for finer lattices. 
Up to an overall sign (due to normalization) \(1+K\) and 
\(\Vs^{\rm lat,\,free}/\Vs^{\rm cont,\,free}\) are quite similar, although 
deviations from zero in \mbox{Fig.}~\ref{fig:naive HISQ improvement} 
are smaller than deviations from one in \mbox{Fig.}~\ref{fig:tree-level}. 
In principle, we could divide the lattice data by \(1+K\) for any choice of 
\(\beta^{\rm ref}\) to remove the residual cutoff effects. 
The finest lattice, \mbox{i.e.} \(\beta^{\rm ref}=8.4\) in the present case, 
can be used as the reference lattice to estimate the cutoff effects in the 
static energy for any other value of \(\beta\). 
In the analysis of \mbox{Ref.}~\cite{Bazavov:2014pvz}, only the finest 
available lattice with \(\beta^{\rm ref}=7.825\) had been used. 
However, this approach is not optimal for a number of reasons. 
First of all, the same distance in physical units on a finer lattice 
corresponds to a larger number of lattice steps than on a coarser lattice.  
Thus, the signal-to-noise ratio for the same physical distances is worse for 
finer lattices than for coarser lattices. 
Second, for larger distances, \(r \gtrsim 0.8 r_1\), the lattices with 
smaller light quark mass should be used to avoid systematic errors due to using different quark masses. 
Finally, the continuum estimates for different \(\beta^{\rm ref}\) suffer 
from independent statistical fluctuations, which may be reduced by combining different \(\beta^{\rm ref}\). 
This leads to a procedure with strong similarities to step-scaling techniques. 
We vary the reference lattice from \(\beta=8.4\) to \(\beta=7.596\) and 
determine estimates for the cutoff effects for lattices with \(8.2 \geq \beta \geq 6.488\).

For small distances on the finer lattices, \(K(\beta^{\rm ref})\) is fairly 
independent of the choice of \(\beta^{\rm ref}\). 
However, \mbox{Fig.}~\ref{fig:naive HISQ improvement} shows that this 
observation does not hold similarly well on coarser lattices and for larger distances on the finer lattices. 
The spread between different determinations of \(K(\beta^{\rm ref})\) for 
different continuum estimates, \mbox{i.e.} different \(\beta^{\rm ref}\), can be understood as follows. 
The spread becomes larger for distances where the renormalized static 
energy is close to zero and, thus, the ratio in 
\mbox{Eq.}~\eqref{eq:correction factors} suffers from large fluctuations. 
Nevertheless, the overall pattern of the short distance cutoff effects does 
not change for different choices of \(\beta^{\rm ref}\). 
Hence, we determine a weighted mean \(\braket K\) from \(K(\beta^{\rm ref})\) 
with different \(\beta^{\rm ref}\) using squared errors as weights.
We obtain errors for \(\braket K\) from the (weighted) variance and from 
weighted averages of the errors of the \(K(\beta^{\rm ref})\). 
We add both types of errors in quadrature and show these in 
\mbox{Fig.}~\ref{fig:naive HISQ improvement}.

\begin{figure*}
\centering
 \hskip-2em
 \includegraphics[height=6.6cm,clip]{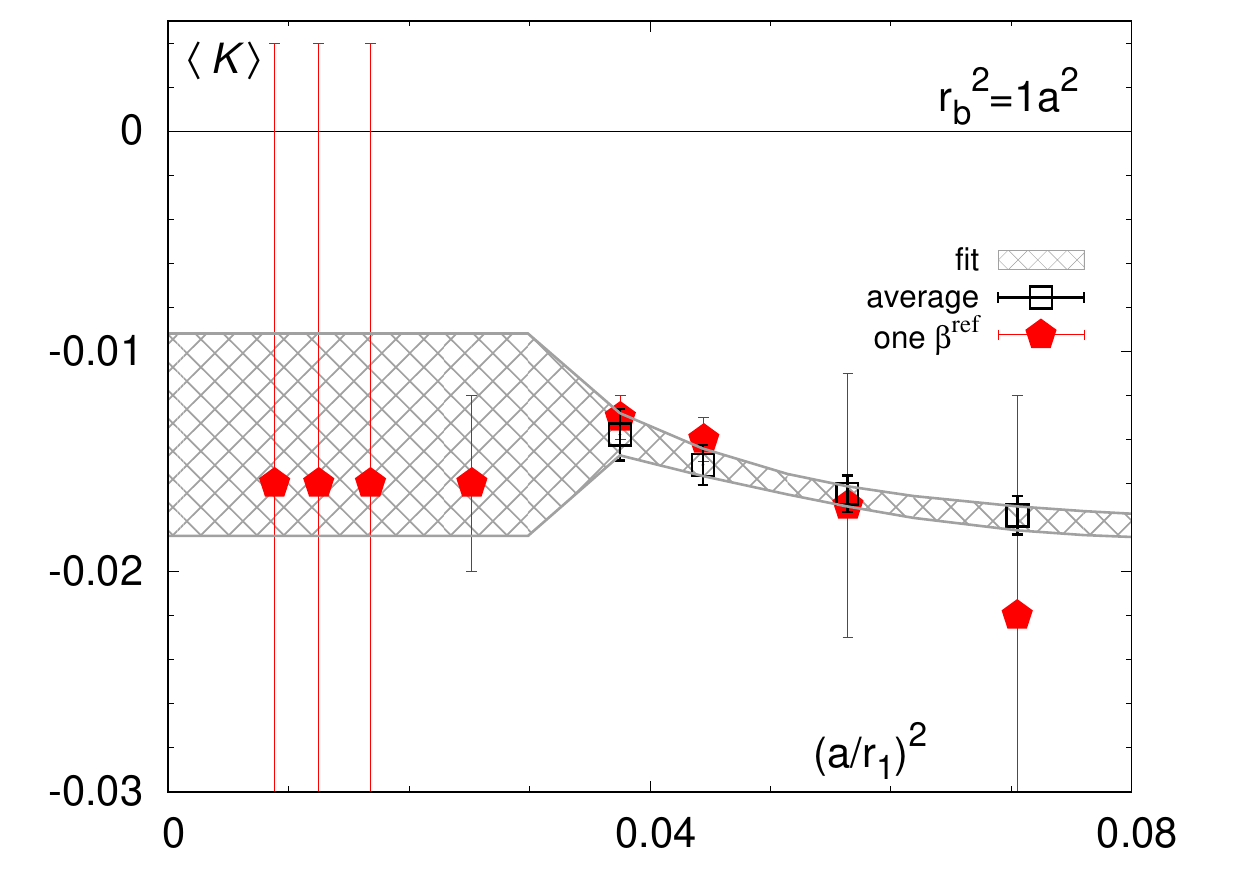}
 \hskip-2em
 \includegraphics[height=6.6cm,clip]{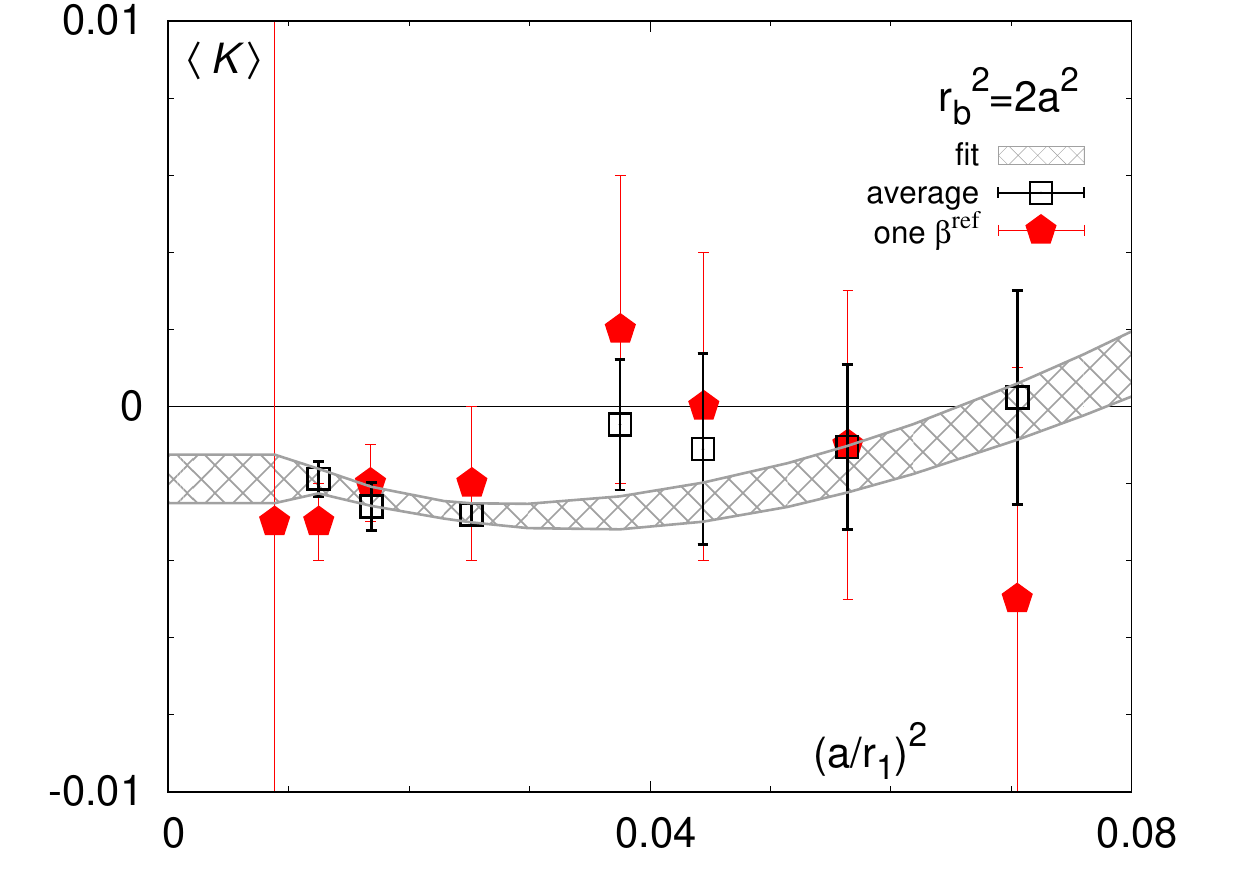}
 \hskip-2em
 
 \hskip-2em
 \includegraphics[height=6.6cm,clip]{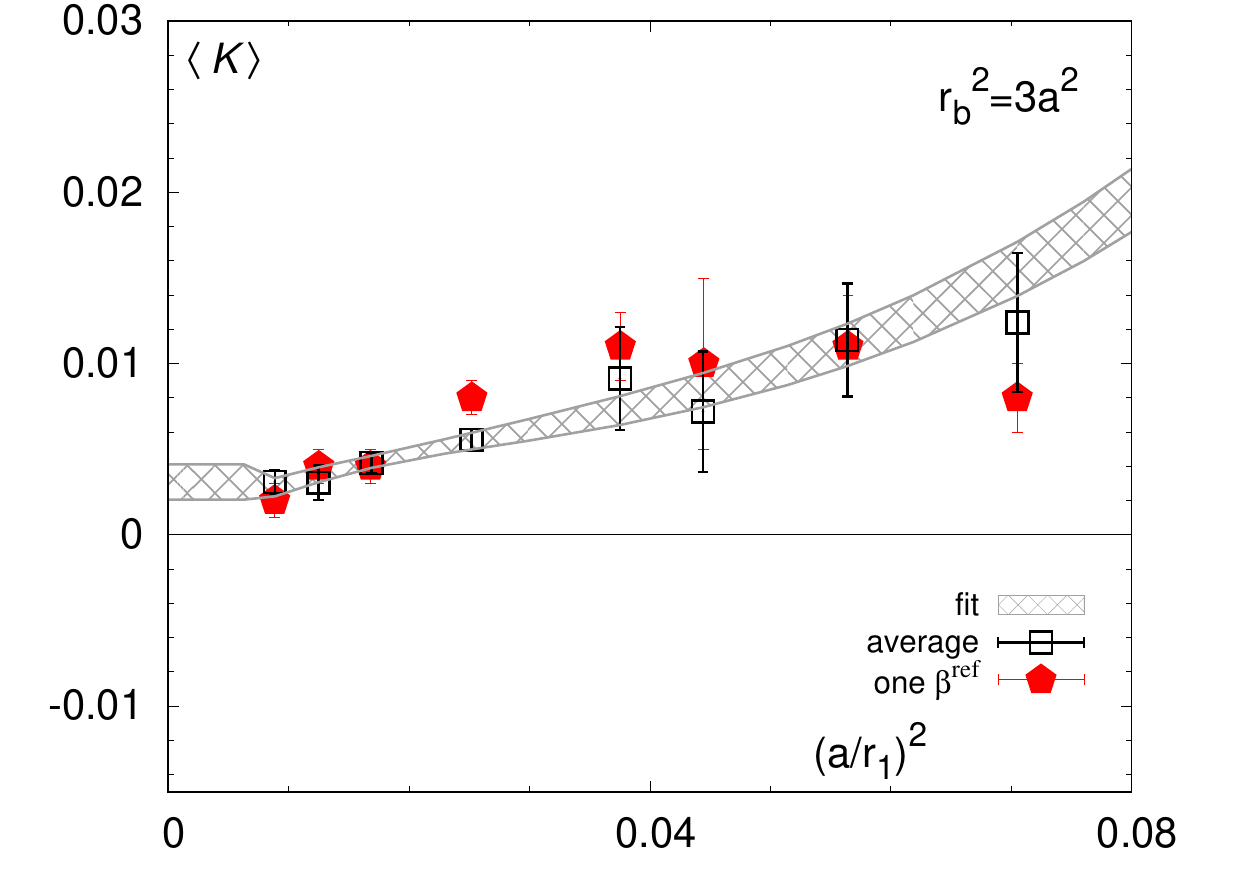}
 \hskip-2em
 \includegraphics[height=6.6cm,clip]{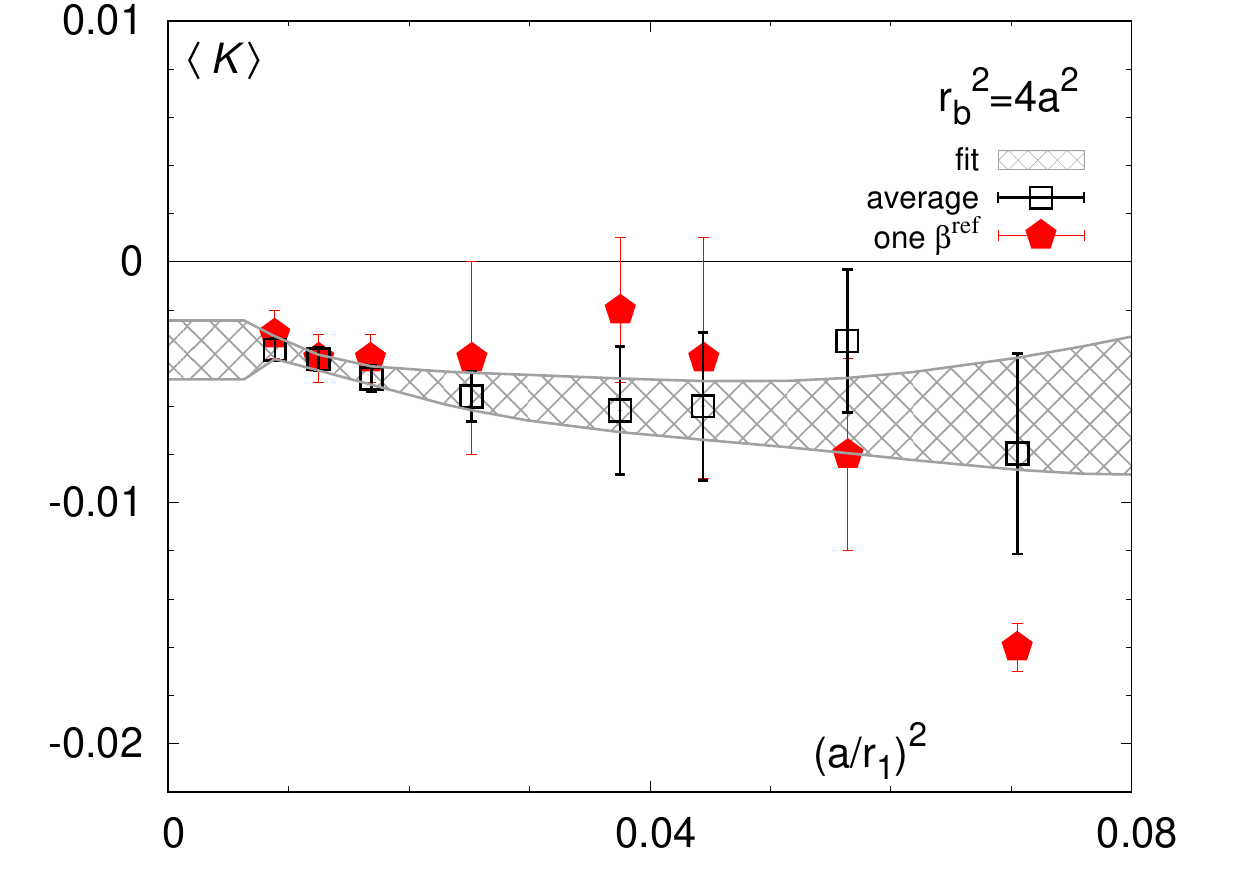}
 \hskip-2em
\caption{\label{fig:interpolated HISQ improvement}
Interpolation of the correction factors \(\braket K\). 
The points for ``one \(\beta^{\rm ref}\)'' have been obtained with 
\(\beta^{\rm ref}=8.4\) for \(\beta>7.373\) and with 
\(\beta^{\rm ref}\geq 7.825\) for \(\beta \leq 7.373\). 
}
\end{figure*}

After determining the \(\beta^{\rm ref}\)-averaged correction factors 
\(1+\braket K\), we put together sets of \(\braket{K}\) for different 
lattice spacings \(a\) and interpolate between these. 
For each \(r_b/a\), we use an \textit{Ansatz} of the form 

\ileq{
 \Braket{ K\left(\frac{r_b}{a},a\right)} = \sum\limits_{i,j=1}^{i+j<5} 
 b_{ij}\left(\frac{r_b}{a}\right) a^{2i} g_0^{2j}
\label{eq:interpolation of cutoff effects},
}

where terms \(b_{i0}a^{2i}\) can be omitted because the tree-level improvement 
discussed above removes lattice artifacts of order $g^0 a^2$. 
Moreover, as a term \(b_{0j}g_0^{2j}\) would persist in the continuum limit 
where rotational symmetry is not broken, such a term is forbidden. 
We perform a bootstrapped fit and vary the number of higher order terms kept 
in \mbox{Eq.}~\eqref{eq:interpolation of cutoff effects}. 
We accept only fits for which we obtain \(\cdf \leq 2.4\) and 
use their mean as our final interpolation result and the mean of their 
bootstrap errors as the final statistical error. 
If we do not obtain fits with reasonable \(\cdf\), we successively 
omit the last point at the small \(\beta\) end--the coarsest lattice--and 
repeat the same set of fits (only for \(r_b/a=1\) we had to omit the last 
point in order to obtain a smooth interpolation). 
We estimate the systematic error of the interpolation as the variance of the set of accepted fits. 
The final \(\cdf\) is then obtained from the final fit and the 
input data using an average of the degrees of freedom for the set of accepted fits. 
We show some sample results of this interpolation in 
\mbox{Fig.}~\ref{fig:interpolated HISQ improvement}, which describe  
\(\braket K\) quite well for all considered lattices.  

\begin{figure*}
\centering
 \hskip-2em
 \includegraphics[height=6.6cm,clip]{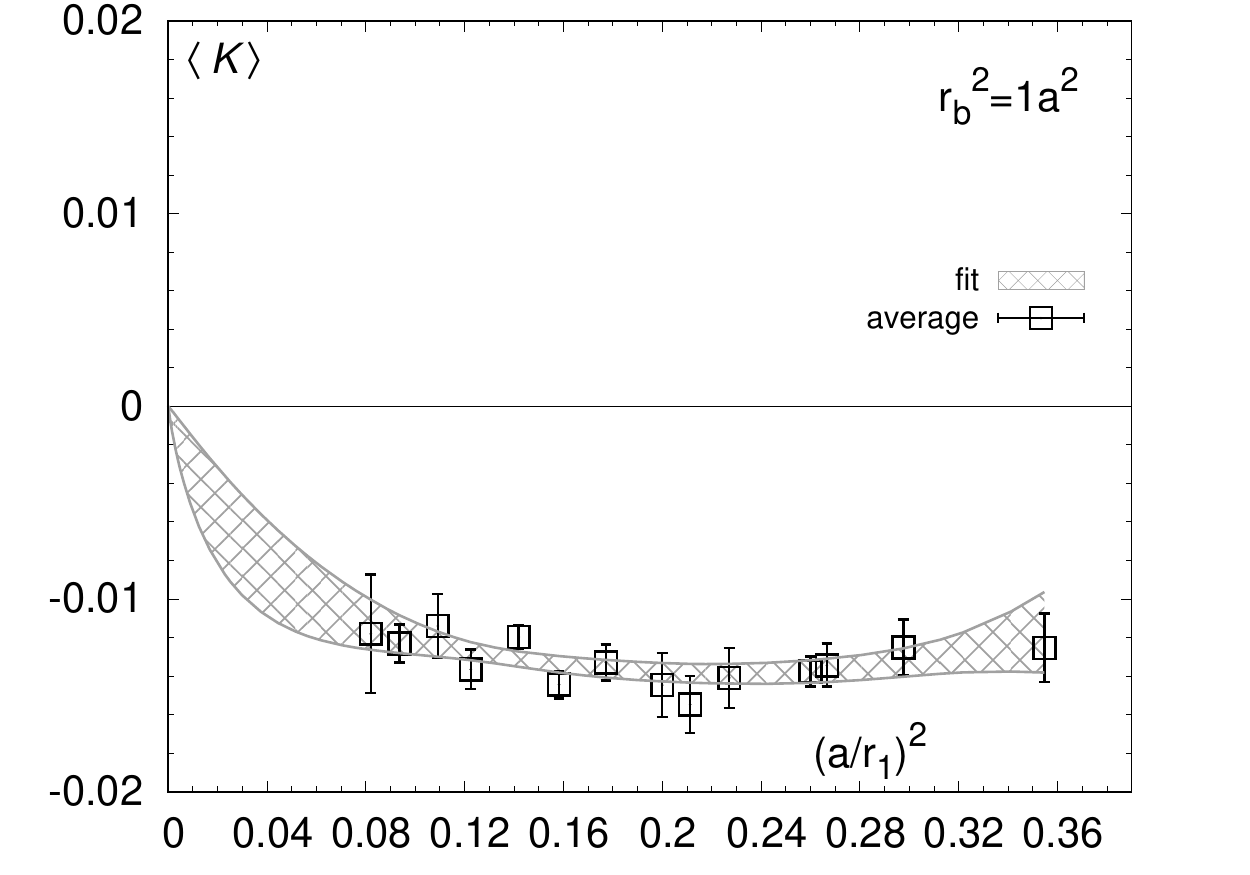}
 \hskip-2em
 \includegraphics[height=6.6cm,clip]{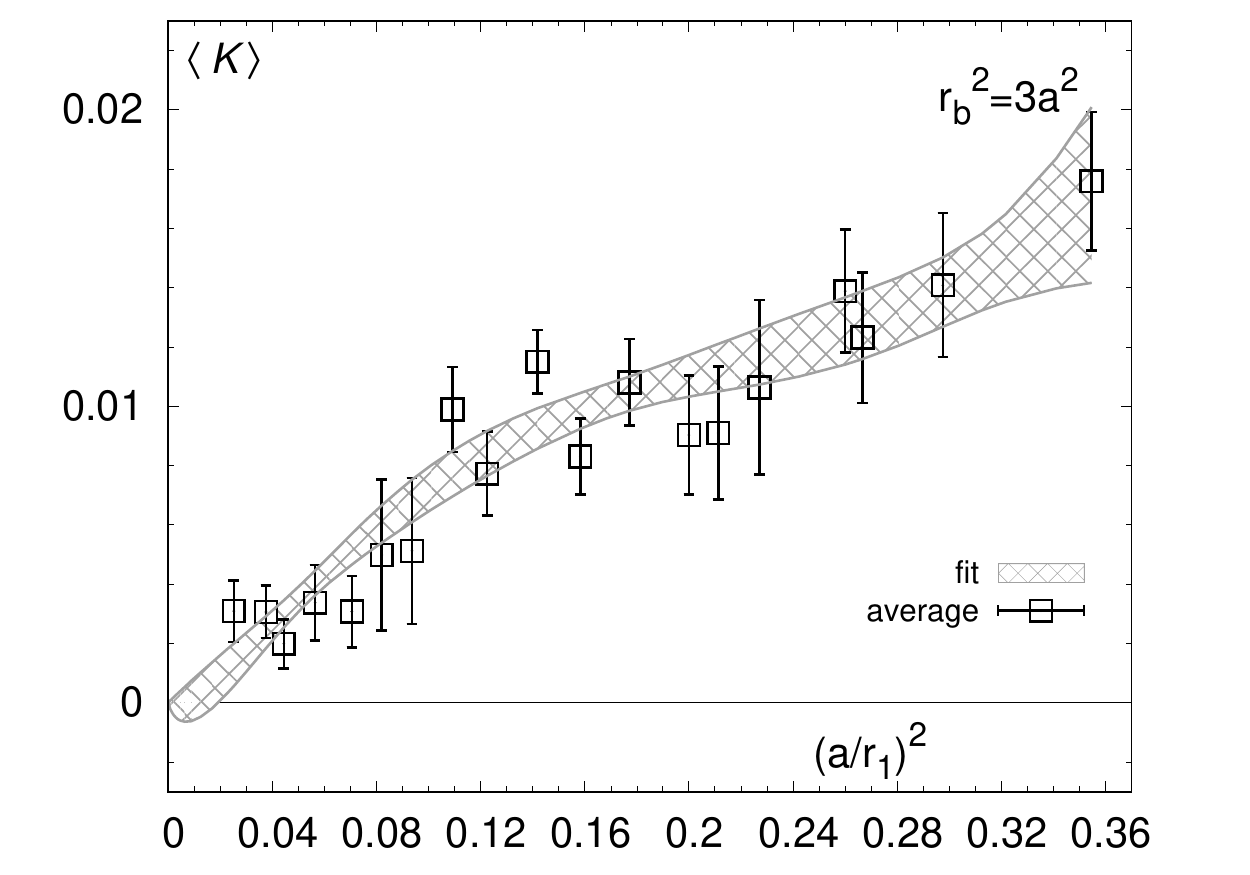}
 \hskip-2em
\caption{\label{fig:shifted HISQ improvement}
Continuum interpolation of the correction factors \(\braket K\) with 
a shifted renormalization scheme. 
The enlarged errors in the range \((a/r_1)^2 >0.1 \) are not shown in the 
figure. 
The approach to the continuum limit is less steep in the shifted scheme.  
}
\end{figure*}

\begin{figure*}
\centering
 \hskip-2em
 \includegraphics[height=6.6cm,clip]{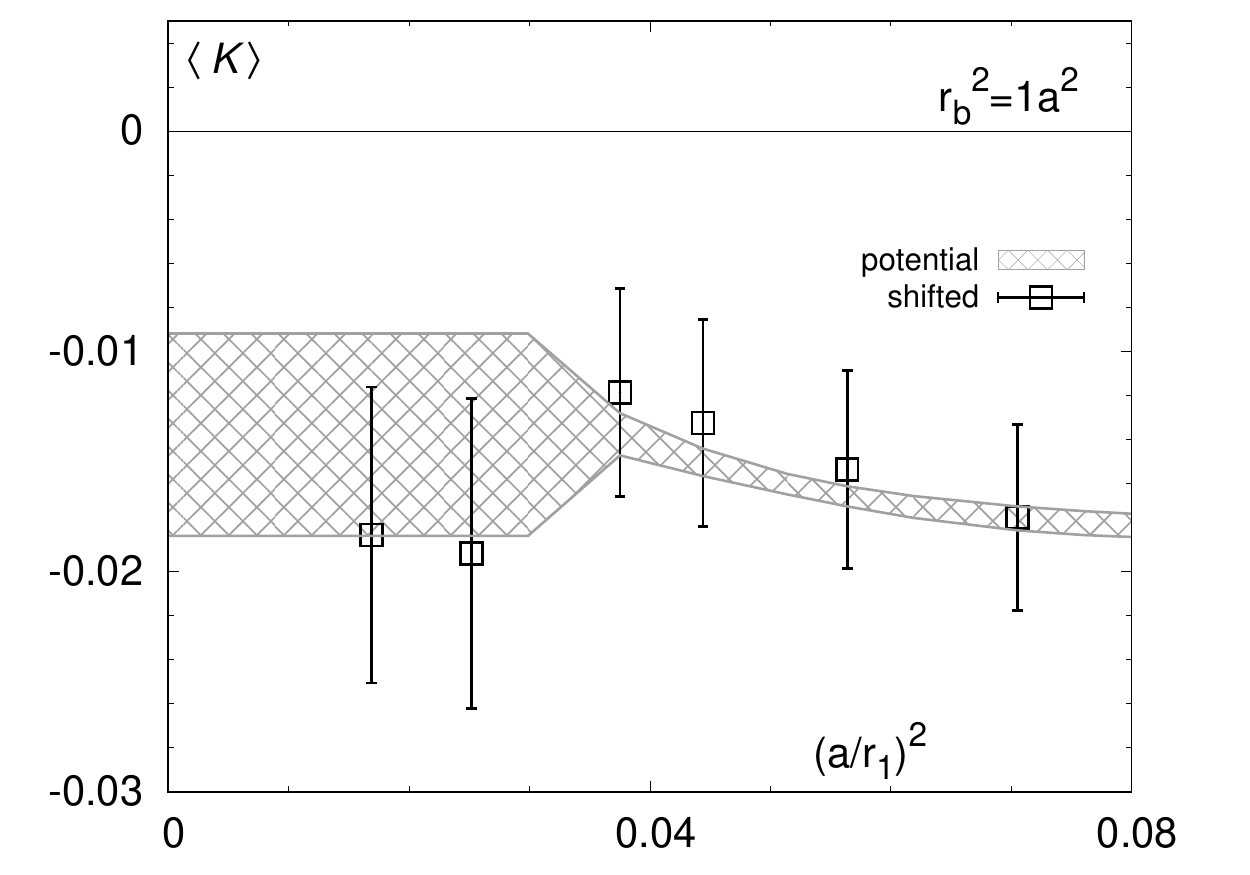}
 \hskip-2em
 \includegraphics[height=6.6cm,clip]{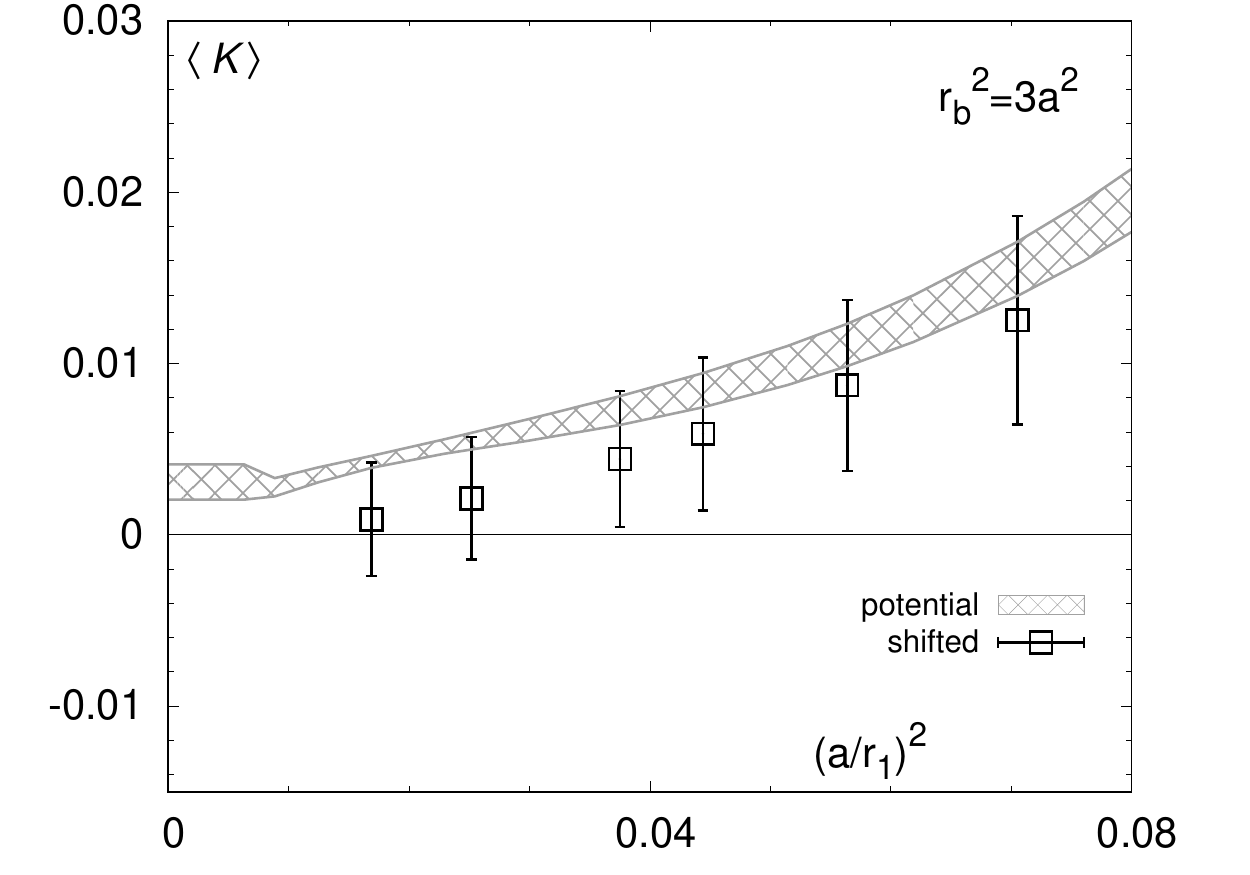}
 \hskip-2em
\caption{\label{fig:comparison HISQ improvement}
Comparison of the correction factors \(\braket K\) from two different renormalization schemes. 
The results from both schemes are compatible with each other.
}
\end{figure*}

However, we do not always see an unambiguous onset of scaling behavior for very fine lattices. 
This may indicate that the correction factors \(\braket K\) do not approach the continuum limit smoothly. 
For this reason, we conservatively estimate that the last \(\braket{K}(r_b/a,\beta)\) that we can obtain applies for any larger \(\beta\) 
and assume a 33\% relative systematic uncertainty. 
For all values of \(r_b/a\), we observe a strong increase in the fluctuations of \(\braket K\) for the coarsest lattices. 
This effect is due to the choice of the renormalization scheme, in which \(\Vs \sim 0\) for such distances.
We address this growth of fluctuations by using an alternative 
renormalization scheme, namely, we shift the static energy by 
\(-0.5\,{\rm GeV}\) with respect to the usual potential scheme. 
Thus, we defer \(\Vs \sim 0\) and the growth of fluctuations to substantially larger distances.
In this scheme, we can determine correction factors for the previously 
problematic range without a large growth of the errors.
We repeat the same type of analysis in the shifted renormalization scheme, 
but use \(\beta^{\rm ref} \in [7.030,7.825]\) as reference lattices. 
We observe here that errors seem to be underestimated in the range 
\((a/r_1)^2 \in(0.1,0.15)\), which corresponds to \(\beta \in (6.5,6.7)\). 
We require that the averaged correction factors \(\braket K\) have relative 
errors of at least 5\% for \(\beta<6.7\) in order to obtain stable fits. 
We show some sample results in \mbox{Fig.}~\ref{fig:shifted HISQ improvement}. 
\(\braket K\) is significantly smaller in the shifted scheme and the growth 
of \(\braket K\) for coarse lattices has been much diminished. 
In order to compare to the original potential scheme, we have to convert the 
correction factors back to the original scheme. 
We show the comparison between both schemes in \mbox{Fig.}~\ref{fig:comparison HISQ improvement}. 
The typical differences between both determinations of \(\braket K\) are 
about 0.2\%-0.4\%, but are mostly covered by the statistical errors that remain 
after converting the shifted scheme back to the standard renormalization.
The fair agreement between both schemes indicates that the cutoff 
effects have been identified appropriately and the systematic error due to scheme dependence is about 0.2\%.

If we can determine \(\braket K\) for given \(\beta\) and \(r_b/a<4\), 
we assign this systematic error of 0.2\% to any data (in the following, 
denoted as \(aE\)), which we add in quadrature to the statistical error. 
If we fail to obtain \(\braket K\) at the same \(\beta\) for larger  
\(r_b/a\) and at the same \(r_b/a\) for smaller \(\beta\), we assume that 
we have reached the limit of the validity of a controlled determination of \(\braket K\). 
In these cases, we refrain from using the correction factors and instead 
double the statistical error and add a 1\% relative error in quadrature. 
If the data are too close to zero, \mbox{i.e.} \(aE<0.01\), we add an 
additional systematic error estimate of \(0.001\) absolute to \(aE\).
In a case where we cannot determine a correction factor at all (for the 
coarsest lattices, \mbox{i.e.} \(\beta<6.195\)), we double the statistical 
error and add a 2\% relative error in quadrature. 
If the data are too close to zero in these cases, \mbox{i.e.} \(aE<0.05\), 
we add the same additional systematic error estimate of \(0.001\) absolute to \(aE\).
At last, we take a weighted average of the corrected results obtained in 
both schemes using the squared errors as weights. 
We use the same weights for averaging the errors and add the difference 
between both as systematical error in quadrature. 

Subsuming the aforementioned considerations in our definition of \(\braket K\), we define the improved static energy as 

\ileq{
 V_{S,I} = \frac{\Vs}{1+\braket K}.
}

\(\braket{K}\) for fine lattices, \mbox{i.e.}, \(a < 0.075\,{\rm fm}\), is usually within \(\pm 1\%\).  
The first point, \(r_b/a=1\), which also has the strongest tree-level 
correction, \((r_I-r_b)/r_b\approx -4\%\), is an exception with \(\braket{K} \gtrsim -2\%\).
Combined uncertainties due to statistical errors of the bare static 
energy and due to systematic errors of the correction factors are usually 
smaller than \(0.5\%\) relative error for fine lattice spacings. 
Although the latter are the dominant \(r\)-dependent source of uncertainty, 
errors due to renormalization with \(2C_Q\) are much larger than 
those due to correction for the cutoff effects on fine lattices.

\subsection{Improvement at finite temperature} 
\label{sec:finite T improvement}

\begin{figure*}
\centering
 \hskip-2em
 \includegraphics[height=6.6cm,clip]{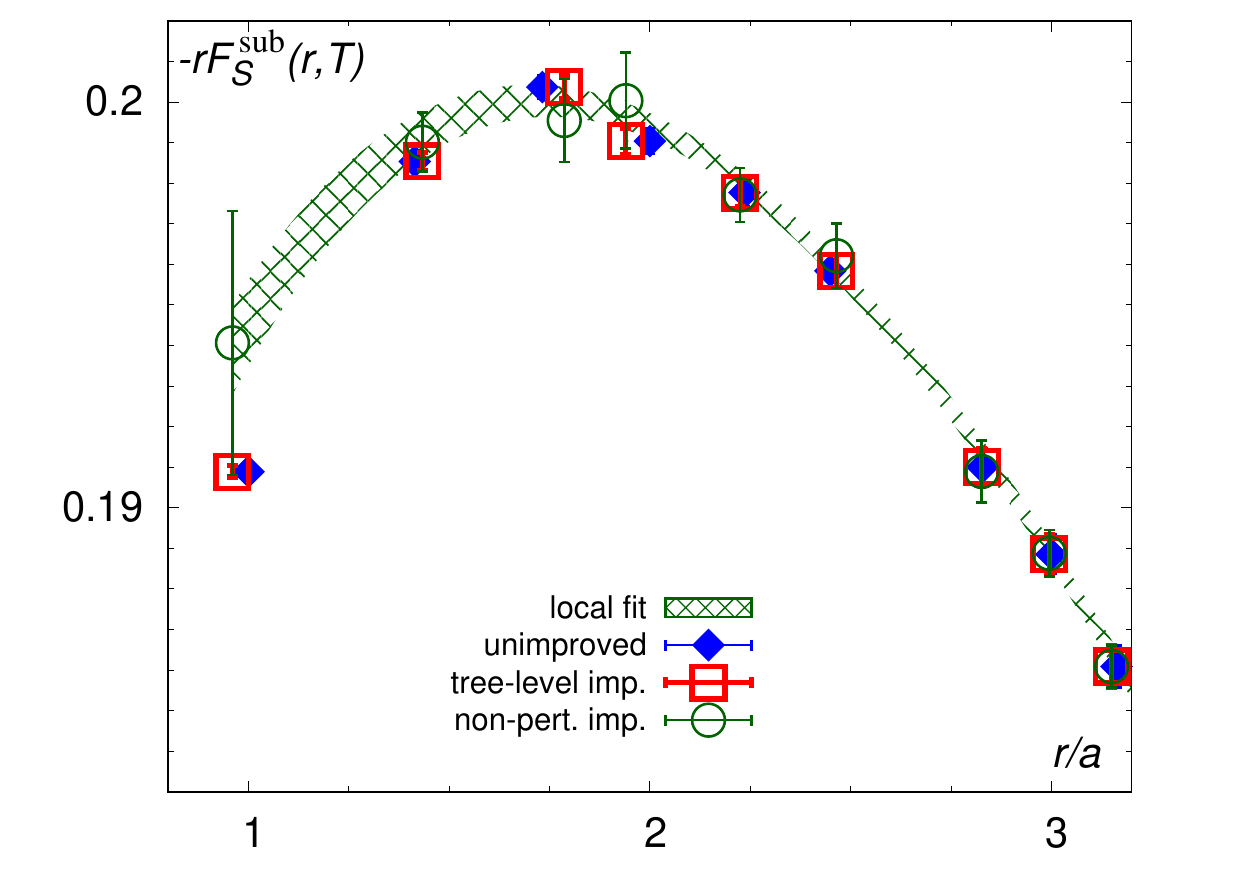}
 \hskip-2em
 \includegraphics[height=6.6cm,clip]{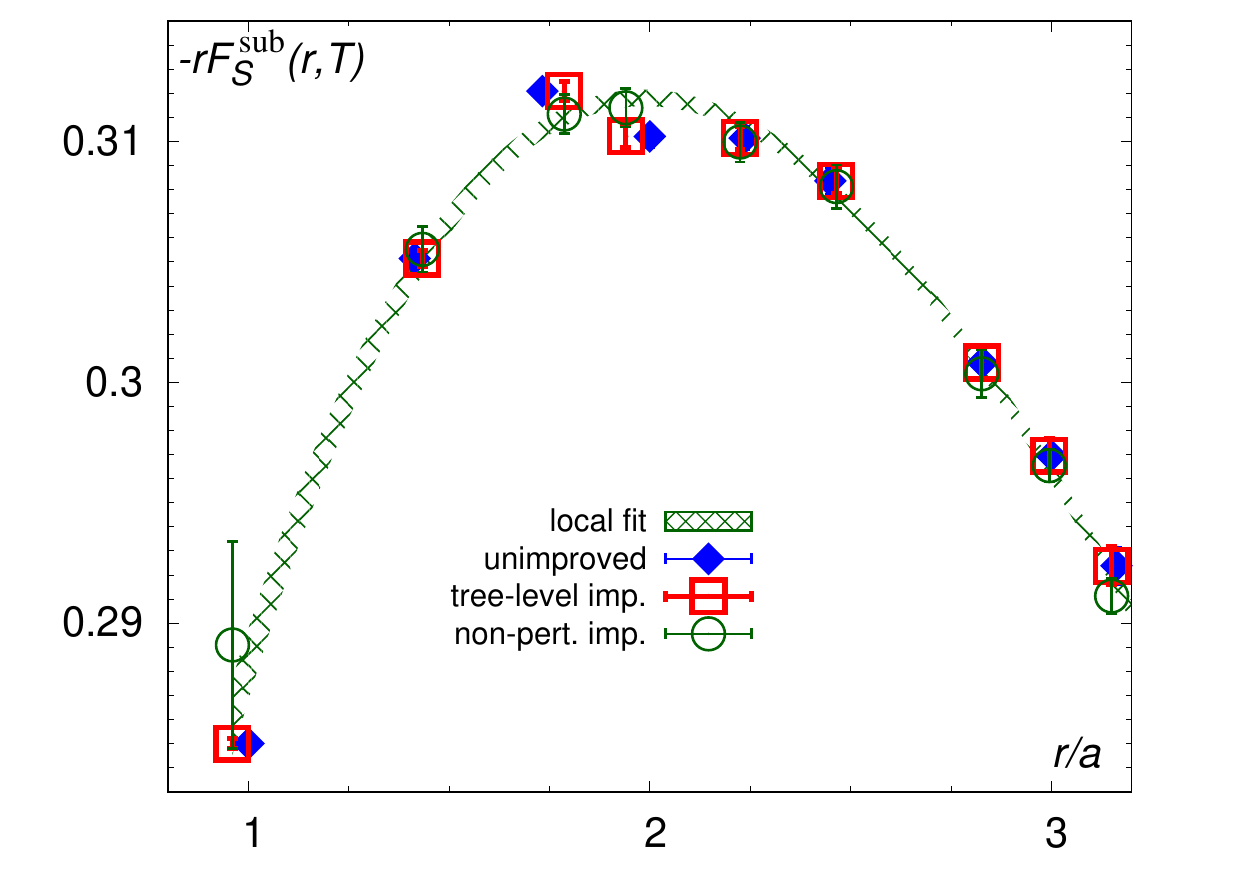}
 \hskip-2em

 \hskip-2em
 \includegraphics[height=6.6cm,clip]{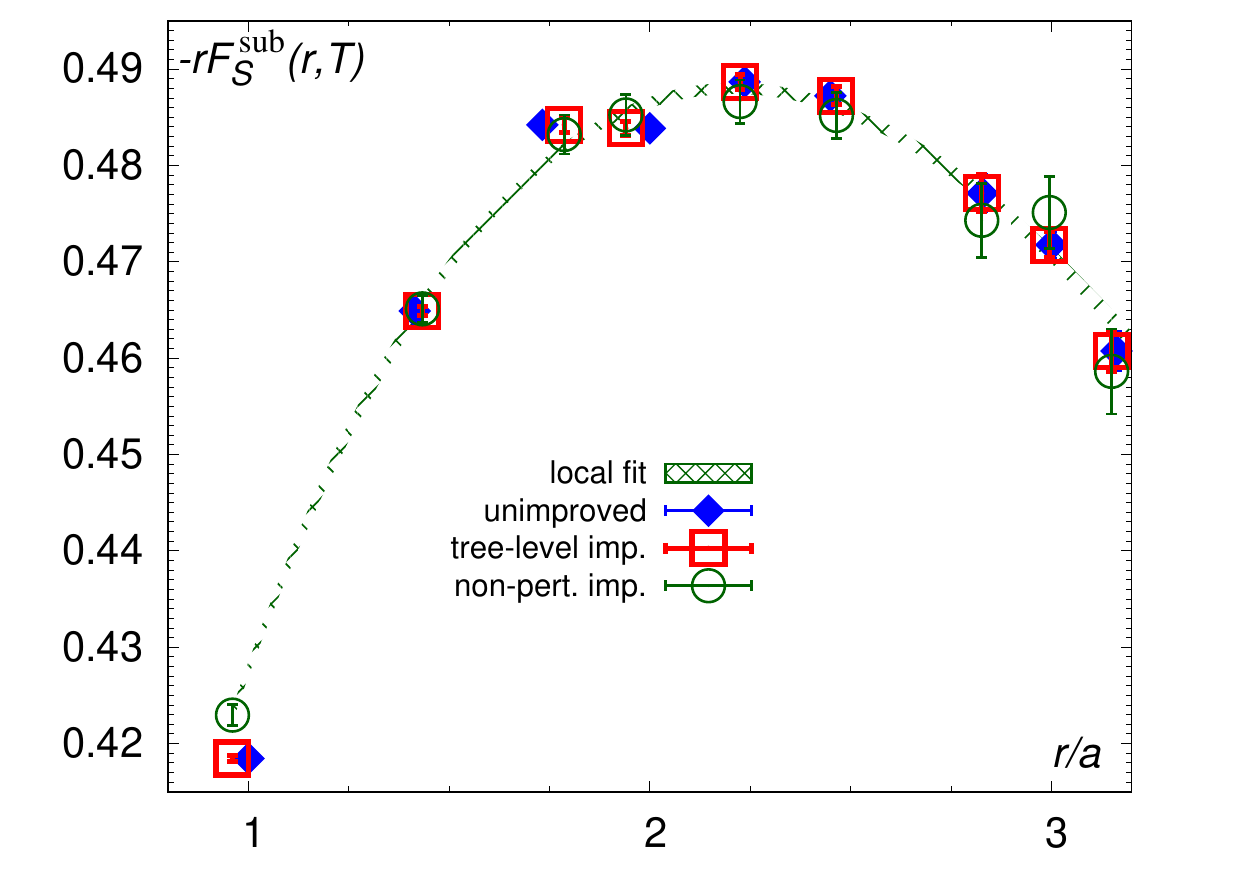}
 \hskip-2em
 \includegraphics[height=6.6cm,clip]{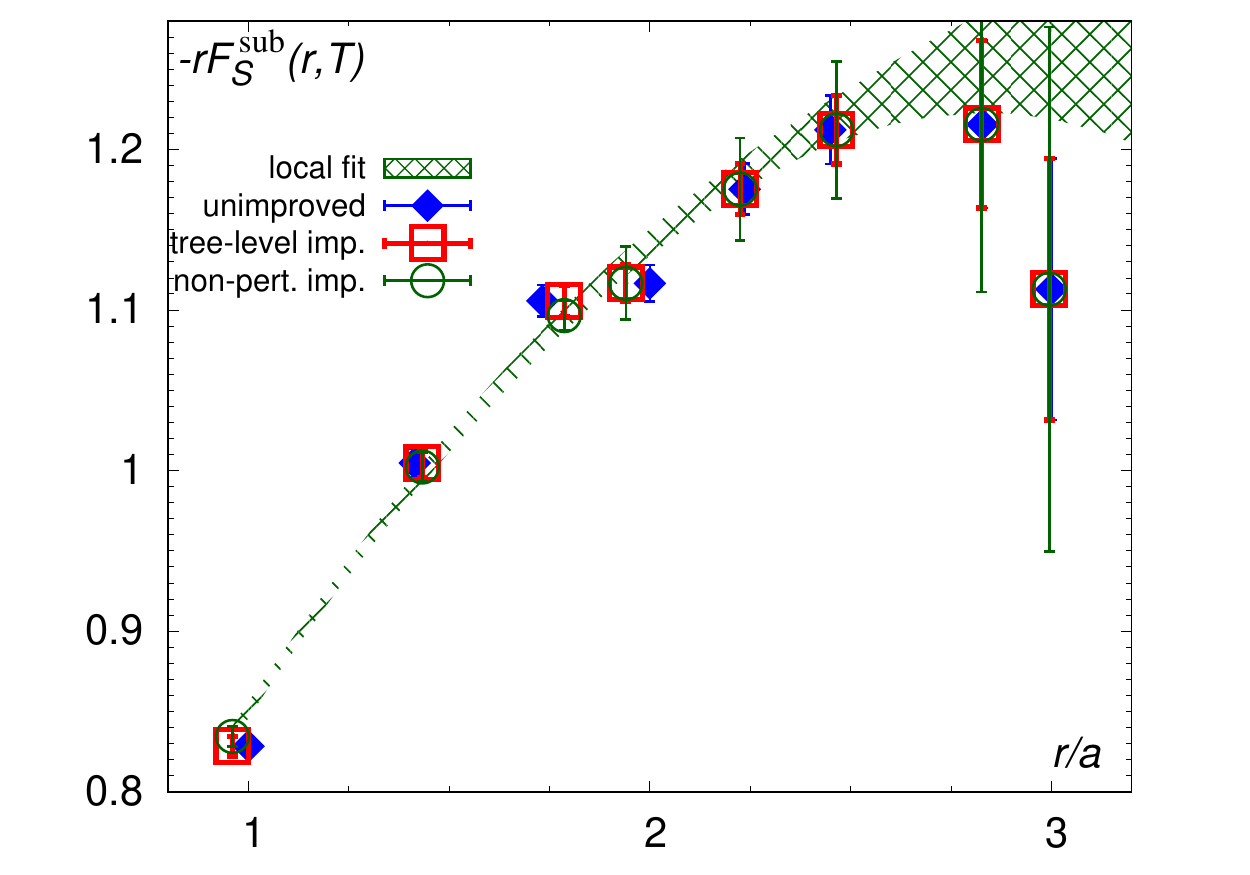}
 \hskip-2em

\caption{\label{fig:singlet free energy improvement}
Improvement of \(\Fs\). 
We show results for \(N_\tau=12\) for \(\beta=9.67\) (left) and~\(8.2\) (right) in the top row and for \(\beta=7.373\) (left) and~\(6.664\) (right) in the bottom row. 
}
\end{figure*}

\begin{figure*}
\centering
 \hskip-2em
 \includegraphics[height=6.6cm,clip]{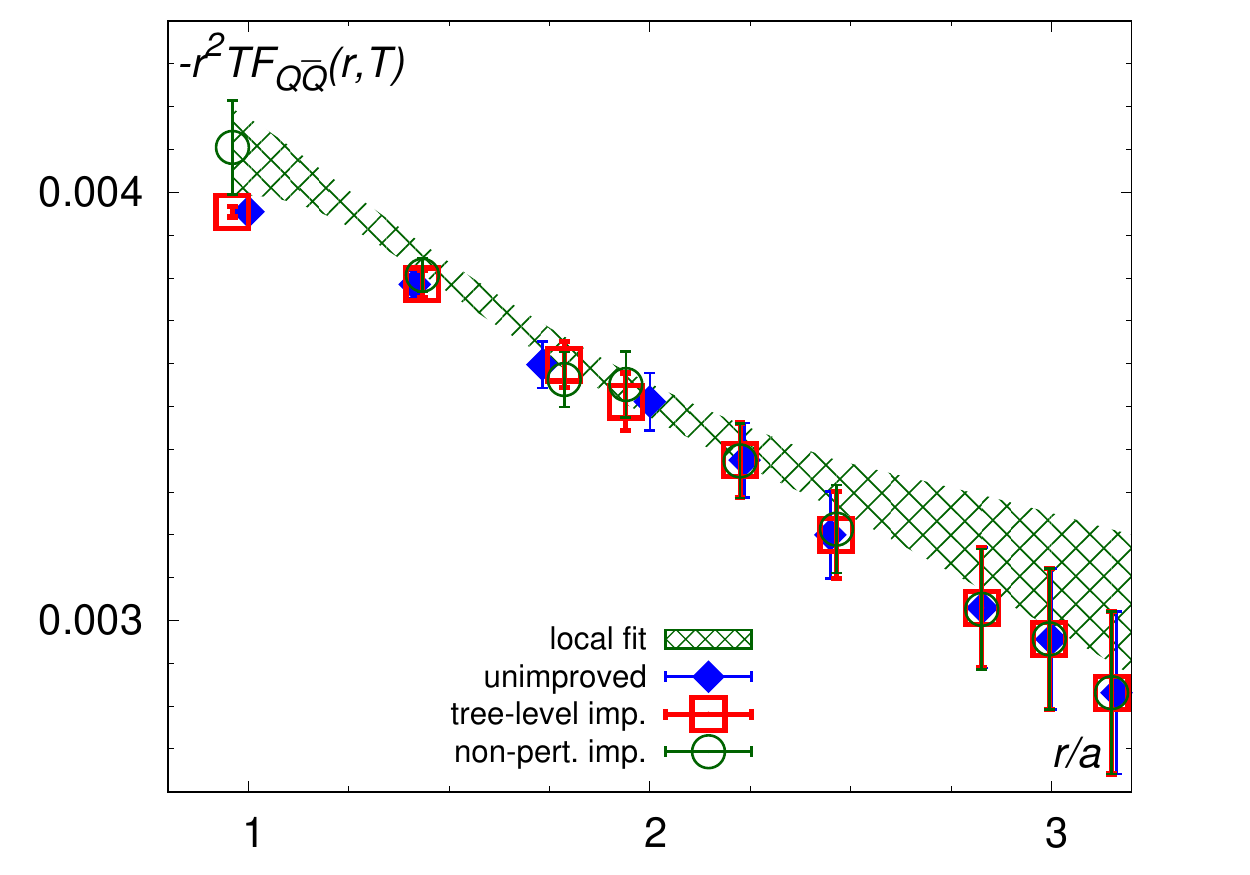}
 \hskip-2em
 \includegraphics[height=6.6cm,clip]{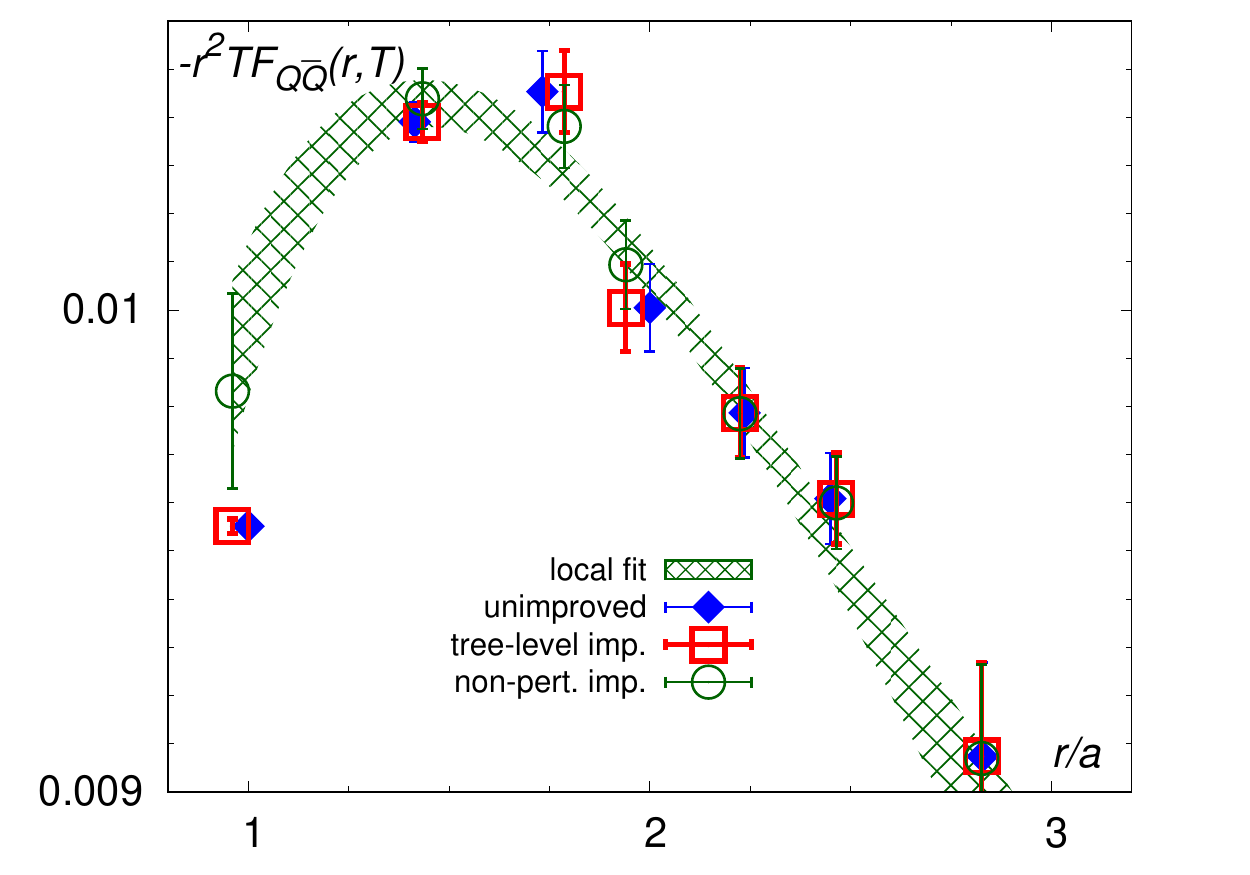}
 \hskip-2em

 \hskip-2em
 \includegraphics[height=6.6cm,clip]{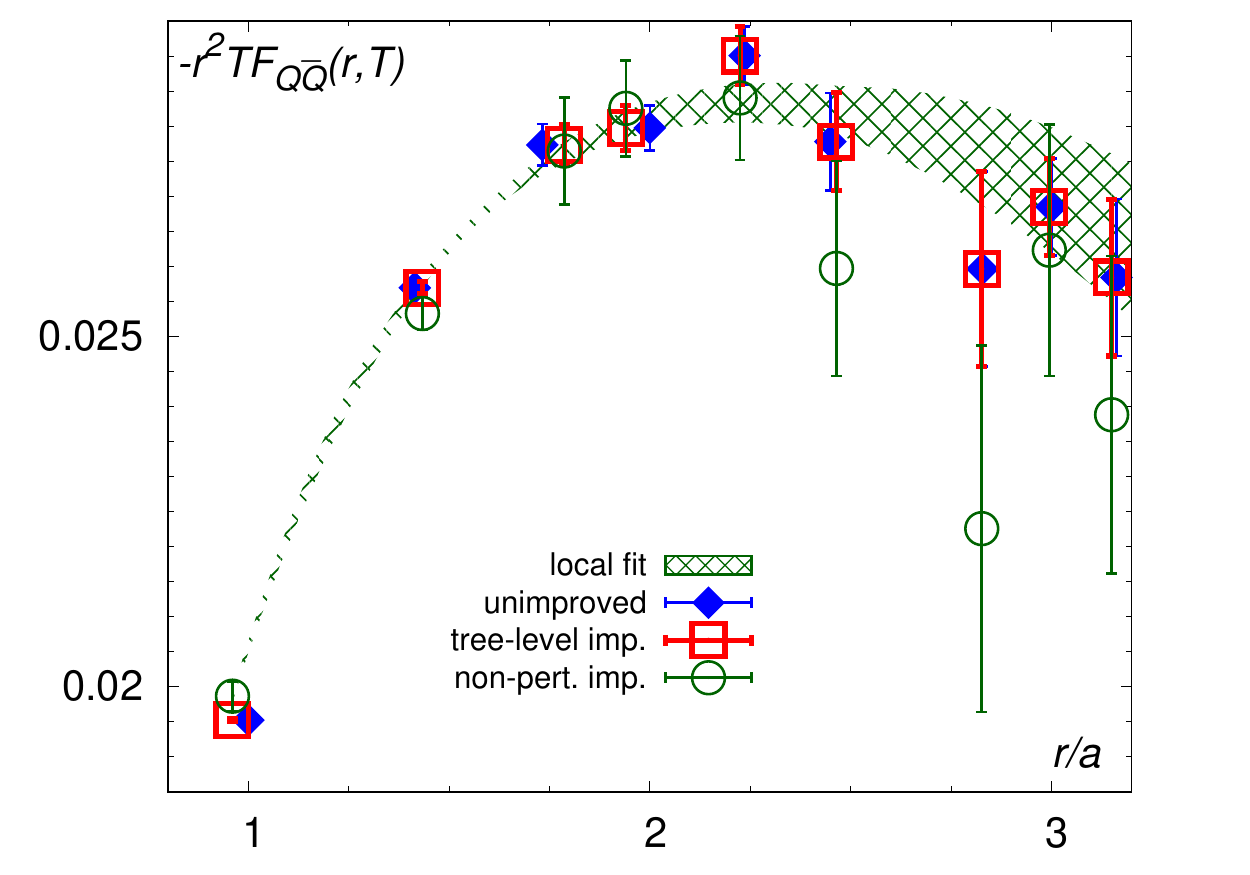}
 \hskip-2em
 \includegraphics[height=6.6cm,clip]{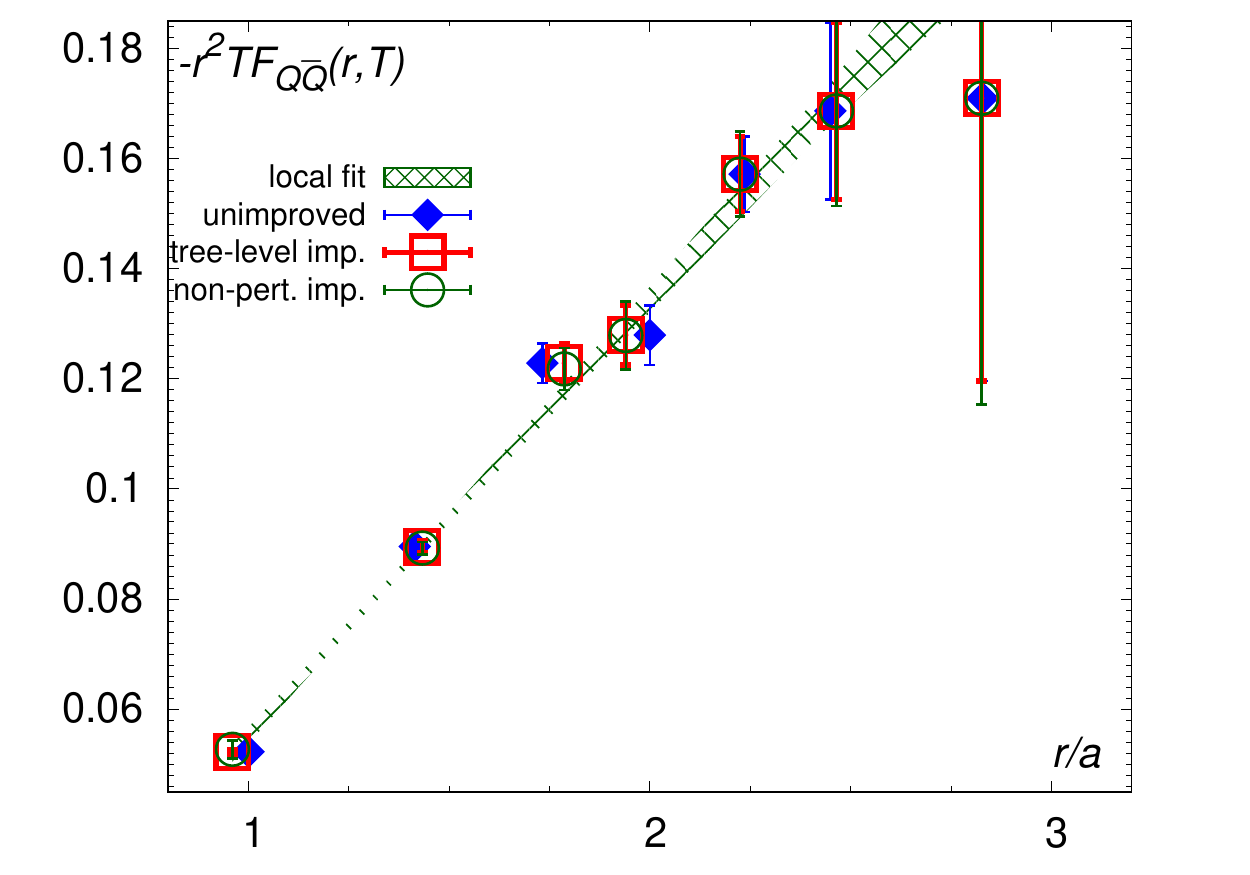}
 \hskip-2em

\caption{\label{fig:free energy improvement}
Improvement of \(\Fqq\). 
We show results for \(N_\tau=12\) for \(\beta=9.67\) (left) and~\(8.2\) (right) in the top row and for \(\beta=7.373\) (left) and~\(6.664\) (right) in the bottom row.  
}
\end{figure*}

The improvement procedure that we have developed so far is only relevant 
for cutoff effects at short distances, where short is meant in terms of 
only a few units of the lattice step, \mbox{i.e.} typically  \(r_b/a\ \lesssim 4\). 
When looking at short distances for finite temperature observables, we 
usually mean short in units of the inverse temperature, \mbox{i.e.} \(rT \ll 1\), \mbox{respectively}, \(r_b/a \ll N_\tau\). 
In the following, we argue why the same short distance improvement is appropriate for static correlators at finite temperature. 
At the shortest distances \(r_b/a\leq2\) cutoff effects usually exceed the 
small statistical errors of the free energies and cause nonsmooth behavior. 
This is a major problem for smooth interpolations including short as well as larger distances.

In \mbox{Sec.}~\ref{sec:singlet}, we have shown that the singlet free energy is 
vacuumlike up to small corrections for \(rT <0.3\). 
Thus, we expect to obtain a similar improvement with the correction factors that 
have been determined for the static energy at \(T=0\).
Namely, we define 

\ileq{
 F_{S,I} = \frac{\Fs}{1+\braket K}.
}

However, for \(N_\tau \leq 6\), it is not obvious if these improvement 
procedures are appropriate, since \(r_b/a \sim 2-3\) is already within the 
electric screening regime. 
We show a few sample singlet screening functions with tree-level or 
nonperturbative improvement and without improvement in 
\mbox{Fig.}~\ref{fig:singlet free energy improvement}. 
For \(N_\tau<8\), modifications due to improvement are small 
compared to medium effects and do not alter the result significantly. 
On the finer lattices, \mbox{i.e.}, \(N_\tau \geq 8\), improvement smoothens 
the short distance regime for \(r_b/a<\sqrt{6}\) and substantially helps 
with obtaining interpolations of the data. 
The systematical uncertainties of the improvement factors are usually at least 
as large as the statistical errors of the finite temperature results at short distances.
These larger errors are beneficial for smooth interpolations of the improved data. 
Since we use the same tree-level and nonperturbative corrections for 
\(\Vs\) and \(\Fs\), we expect that leading corrections are subject to 
strong cancellations in the difference \(\Vs-\Fs\).
We simply use unimproved distances \(r_b/a\) and omit nonperturbative 
correction factors. 

In \mbox{Sec.}~\ref{sec:plc}, we have shown that the Polyakov loop correlator 
can be understood through its decomposition into color singlet and octet in 
pNRQCD for \(rT < 0.3\) at all temperatures in the study.
Using \mbox{Eq.}~\eqref{eq:fqqrec}, we can obtain a correction 
prescription for \(\Fqq\) on the basis of the correction factors 
\(\braket K\) for the \(T=0\) static energy. 
We write the improved normalized Polyakov loop correlator 
\(C_{P,I}^{\rm sub}\) in terms of the improved static energies, 
\(V_{S,I}\) neglecting any Casimir scaling violations, 

\ileq{
C_{P,I}^{\rm sub} = \frac{1}{\nc^2} e^{-\tfrac{V_{S,I}-2\Fq}{T}}
 +  \frac{\nc^2-1}{\nc^2} e^{-\tfrac{V_{S,I}-2\Fq}{(\nc^2-1)T}}. 
}

We know \(V_{S,I}=\Vs/(1+\braket K)\) and we can use \(\braket K \ll 1\) 
to expand \(C_{P,I}^{\rm sub}\) in \(\braket K\), eventually obtaining 

\ileq{
C_{P,I} 
= C_P + \frac{\braket K}{\nc^2-1}\frac{\Vs}{T} [e^{-\frac{\Vs}{T}}-C_P]
+\mathcal O(\braket K^2). 
\label{eq:CPI}
}

The need for corrections and the validity of the pNRQCD formula are both 
restricted to short distances. 
In this range, we have seen that the reconstruction using \(\Fs\) works as 
well as using \(\Vs\), but provides access to the full \(\beta\) range. 
Hence, we replace \(\Vs\) by \(\Fs\) to obtain our final expression

\ileq{
 F_{\qbq,I}^{\rm sub}
= -T\ln\left( C_P^{\rm sub} + \frac{\braket K}{\nc^2-1}\frac{\Fs}{T} 
\left[C_S^{\rm sub}-C_P^{\rm sub}\right] \right), 
\label{eq:FQQI}
}

which is valid to leading order in \(\braket K\). 
We show a few sample screening functions with tree-level or 
nonperturbative improvement and without improvement in 
\mbox{Fig.}~\ref{fig:free energy improvement}. 
The effects of improvement are hardly relevant for \(\Fqq\) on most lattices. 
In the cases where these effects are visible, the corrections generally lead 
to a smoother free energy at short distances.

\section{INTERPOLATIONS AND EXTRAPOLATIONS}
\label{app:C}

In this Appendix, we discuss the details of the interpolations of the 
subtracted quark-antiquark (and singlet) free energies and how we 
extrapolate $\Fqq^{\rm sub}$ (and $\Fs^{\rm sub}$) to the continuum limit. 
We do not repeat here the discussion of the physical underpinning of 
the presented analysis, which we discuss in detail in 
Sec.~\ref{sec:analysis}.
Here we also discuss the interpolations and extrapolations of 
\(\Vs-\Fs=\Vs^{\rm bare}-\Fs^{\rm bare}\). 
In the following, we refer to either $\Fqq$ or $\Fs$ just as the free energy 
$F$ unless we apply different treatment to both and use these quantities in 
temperature units, namely, we fit \(f^{\rm sub}=F^{\rm sub}/T\) and 
\([\Vs-\Fs]/T\). 
This Appendix also covers the calculation of the $r$ derivative of $F$. 
We usually propagate statistical errors through the fits by generating 100 
bootstrap samples of mock data assuming Gaussian errors and perform the fits 
on the mock data. 
We perform fits using either linear regression, the built-in smoothing 
splines of the R statistical package~\cite{Rpackage}, or use for nonlinear 
fits the \textit{nlme} package~\cite{nlme}. 

We arrive at our continuum result through the following procedure.
\begin{enumerate}
\item 
We consider the subtracted free energies \(F^{\rm sub}\) defined in 
\mbox{Eqs.}~\eqref{eq:defFqqsub} and~\eqref{eq:Fssub} and \(\Vs-\Fs\). 
We omit data at large \(rT\) due to signal loss (\mbox{cf.} 
Appendix~\ref{app:A}) and perform the tree-level and 
nonperturbative improvement if needed (\mbox{cf.} Appendix~\ref{app:B}). 
\item 
We interpolate in \(rT\) and \(T\) using two different data-driven schemes. 
We use local fits in the first scheme and global fits in the second scheme. 
\item 
We calculate weighted averages of the results from the two schemes and 
estimate systematic errors due to model dependence. 
\item 
We extrapolate the combined results to the continuum limit for each 
individual \((rT,T)\) point using different subsets of the \(N_\tau\) and 
different \textit{Ans\"atze} for scaling behavior. 
\item 
After inspecting scaling behavior for sets of representative \((rT,T)\) 
points, we compile the final continuum result using different extrapolations 
in those regimes where they are appropriate. 
\item 
Finally, we obtain the renormalized continuum result by adding \(2\Fq\) (\mbox{cf.} Appendix~\ref{app:A}). 
\end{enumerate}

As stated, we follow a data-driven scheme in fitting the \(rT\) and \(T\) 
dependence. 
Namely, we iteratively try to find the most simple, smooth function of \(rT\) 
that incorporates the main features known from weak-coupling results (LO 
short distance behavior and exponential damping) and describes the largest 
subset of the statistically relevant data. 
We reduce the fit range at the large distance end in small steps if we fail to 
fit the data with any reasonable choice of smooth function. 
If the data set becomes too short, the statistical errors of the data at the 
shortest distances and of the nonperturbative corrections must have been 
underestimated. 
This implies that data are nonsmooth at the shortest distances.
In these cases, we have to enlarge the errors of all data by a small amount 
that permits a smoother curve and restart the procedure. 
We end the procedure when meeting a stopping criterion in terms of both the 
residues at the shortest distance and globally and in terms of a reasonable 
determination of the propagated errors of the interpolating function. 
We discuss the stopping criteria for each of the fits in the following sections.

\subsection{Local fits}
We perform local interpolations first in the quark-antiquark separation and 
later in the temperature. 
For the former, we explicitly account for correlations between data at 
different separations obtained within the same gauge ensemble by fitting 
on the individual jackknife bins. 
We have tested that a different ordering of the fits, \mbox{i.e.} first 
interpolating in the temperature for fixed distances \(rT\) and then 
interpolating in the distance \(rT\) for fixed temperature, does not change 
the results in a statistically significant way. 
However, the latter ordering does not permit one to account for the correlations 
within each ensemble.

\subsubsection{Distance interpolations}\label{sec:rT interpolations}
We independently interpolate \(F^{\rm sub}(r,T,a)\) or rather 
\(F^{\rm sub}(rT,\beta,N_\tau)\) as a function of \(rT\) for each \(N_\tau\) 
and each \(\beta\). 
We split the data into intervals of small and large separations, where 
we treat the first 16 points as data at small separations and the rest 
including some overlap region as data at large separations.  
Before interpolating data at large separations, we calculate residues at 
large separations by extrapolating with the fit to data at small separations. 
The residues are computed with data at large separations that did not enter 
into the fit at small separations. 
If these residues satisfy \(\cdf<1\) (\({\rm d.o.f.}\) being the data 
at large separation), we forgo the separate interpolation at large 
separations and just use the extrapolation. 
If we have separate interpolations for small and large separations, we match 
the two interpolations smoothly. 

For the local fits, we assume smoothness and screening of the subtracted free 
energies and use a modified, screened Cornell potential (MSCP) \textit{Ansatz}

\ileq{\label{eq:MSCP}
 F^{\rm sub}(rT) = C+\frac{e^{-M\,rT}}{(rT)^\alpha} 
 \sum\limits_{n=n_{\rm min}}^{n_{\rm max}} A_n (rT)^{n}.
}

We keep a parameter \(C\) to account for a possible offset due to incomplete 
cancellation in the subtraction of \(2\Fq\), but also try the same fits with 
\(C=0\), \mbox{i.e.} omitting the parameter altogether. 
We demand that the screening mass parameter \(M\) is positive, but also try 
fits with \(M=0\) for low temperatures, \mbox{i.e.}, \(T<235\,{\rm MeV}\). 
We generally use \(n_{\rm min}=0\) and limit the number of monomials by 
\(n_{\rm max} \leq 8\) if enough data points are available. 
If we fit for small separations with at least six data points and not very 
high temperatures, namely, with at most \(T \leq 500\,{\rm MeV}\), we limit 
the minimal number of monomials as \(n_{\rm max} \geq 1\). 
Otherwise, we use \(n_{\rm max} \geq 0\). 
In the denominator, we choose at short distances \(\alpha=2\) for 
\(\Fqq^{\rm sub}\)  due to the form of the LO result 
[\mbox{Eq.}~\eqref{eq:fqq_lo}], and \(\alpha=1\) for \(\Fs^{\rm sub}\).
At larger separations, we use \(\alpha=1\) both for \(\Fs^{\rm sub}\) and 
\(\Fqq^{\rm sub}\).
Lastly, we use \(\alpha=0\) for \(\Vs-\Fs\). 
In the latter case, we also limit the range to \(rT \leq 0.5\) for 
\(N_\tau>4\) and to \(rT\leq0.75\) for \(N_\tau=4\). 
Because of the small size of the gauge ensembles for \(N_\tau=16\), we enlarge 
the corresponding statistical errors by a \(1\%\) relative error estimate. 

We vary \(n_{\rm max}\) and thus perform nonlinear fits for different 
polynomial orders \(N_n=n_{\rm max}-n_{\rm min}+1\) in order to choose the 
most simple \textit{Ansatz} that matches the data for short or large separations. 
In a first cycle, we interpolate only the jackknife average to determine 
an appropriate fit function that permits meeting a stopping criterion. 
In a second cycle, we perform the interpolation on all jackknife bins, 
thus taking correlated fluctuations of data into account and propagating 
the statistical errors.  
If the need arises, we readjust the fit function as in the first cycle. 
We estimate the bias from the difference between the average of the fits on 
the individual bins of data and the fit on the average of the bins of data 
and add the bias to the statistical error in quadrature. 
As stopping criterion of the loop, we demand that the fit converges on at 
least 20\% of all bins with \(\cdf\leq 2.4\)\footnote{We use this 
purely heuristic limitation for acceptable \(\cdf\) to accelerate the 
process of identifying underestimated systematical errors and finding 
appropriate interpolating functions.} and that--after we exclude fits 
on bins with \(\cdf> 2.4\) from an average--the 
average of the fits satisfies \(\cdf\leq 1.2\). 
If we reach small enough \(\cdf\) for some \(n_{\rm max}\), we 
also attempt a fit with polynomial of order \(n_{\rm max}+1\) with the 
constant term \(C\) omitted. 
Between these, we choose the fit with smaller \(\cdf\). 

We perform a nested loop over different fits starting from the minimal 
permitted \(n_{\rm max}\) and increase towards the maximal permitted \(n_{\rm max}\) in an inner loop. 
If we fail to meet the stopping criterion, we omit points and restart the inner loop. 
For small (large) separations, we omit points at the large (small) \(r\) end of the fit range.  
If we fall below nine data points in this process, we restore the full fit 
range, but add a systematical error estimate on top of the previous errors to the data. 
We add 0.5\% times the minimal \(\cdf\) of the recent inner loop as relative error estimates and restart the inner loop. 
If we fail to fit for small (large) separations with at most 20\% (30\%) 
relative error added, we consider the fit as a failure. 
We restart the outer loop without a screening mass parameter \(M\) for small separations at low temperatures, \mbox{i.e.} \(T<235\,{\rm MeV}\), reasoning 
that we do not have enough data to resolve screening as well as subleading terms in the polynomial. 
For small separations, we aim at fits with sufficiently small residue on the first point. 
If we cannot achieve \(\chi^2(r/a=1)<1\), we restart the outer loop and permit \(\chi^2(r/a=1)<1.5\) or even \(2\). 
We can always get acceptable fits under the latter condition. 
We use the fits for large separations to extrapolate up to the first point, 
where the fit to \(F^{\rm sub}\) becomes compatible with zero within errors. 

We test for model independence of the fit procedure by performing smooth spline fits to \(rF^{\rm sub}(rT)\)
fits with 100 bins of bootstrap mock data treating all errors as independent.
For the latter, we use as few knots as possible and a smoothing parameter as 
large as permissible to still obtain \(\cdf \leq 1.2\).
We find fair agreement between all these fits within their respective 
uncertainties, albeit with underestimated errors for the bootstrapped 
nonlinear fit and with overestimated errors for the spline fit. 
We consider the MSCP fits on the jackknife bins as our final local \(rT\) interpolation of \(F^{\rm sub}\). 
We obtain the \(rT\) derivative analytically for every fit function and insert 
the fit parameters on the individual bins to compute the average and propagate the errors.  

For \(\Vs-\Fs\), we follow more or less the same procedure. 
We omit the screening mass altogether and do not use jackknife bins, since 
we obtain \(\Vs-\Fs\) by subtracting jackknife averages of statistically 
independent data sets whose sizes may differ substantially. 
In practice, the actual fluctuations appear to exceed the statistical errors, which have been added in quadrature. 
Since many data for \(\Vs-\Fs\) are very close to zero, we demand that 
absolute errors\footnote{Since \(\Vs-\Fs \sim 0.02\,T\) for very small 
distances and high temperatures, see \mbox{Sec.}~\ref{sec:vminusf}, this enlarged 
error cannot create fake plateaux at short distances.} 
are at least \(0.001\,T\). 
In the case of enlarging the errors, we add \(0.002\,T\) times the minimal 
\(\cdf\) of the recent inner loop as absolute error estimates and restart the inner loop. 
We do not extrapolate to larger separations here. 
Other than these, the local fits to \(\Vs-\Fs\) follow the same procedure as the fits to \(F^{\rm sub}\).

\subsubsection{Temperature interpolations}
\label{sec:T interpolations}

In the next step, we independently interpolate the MSCP-interpolated 
results for \(f^{\rm sub}=F^{\rm sub}/T\) or \([\Vs-\Fs]/T\) in \(T\) for each \(rT\) and fixed \(N_\tau\). 
We perform separate temperature interpolations for the \(rT\) derivatives. 
We use smoothing spline interpolations that penalize higher derivatives. 
We require for the internal smoothing parameter (\(\rm SM\)) of the built-in 
smooth splines of the R statistical package~\cite{Rpackage} \(\rm SM\geq 0\) 
for \(f^{\rm sub}\) and \(\rm SM\geq 0.2\) for the derivatives. 
We limit the maximal smoothing parameter \(\rm SM\) for the temperature 
interpolations to $0.45$ and include data for all available temperatures.
We perform separate interpolations in overlapping low 
($T<255\,{\rm MeV}$) and high ($T>180\,{\rm MeV}$) temperature intervals.
If the fits yield $\cdf>1.2$, we enlarge the errors of the input by 
a factor $\sqrt{\cdf}$ 
Here we have to stress that the input for these interpolations are 
intermediate mock data, whose errors may have been underestimated. 
If the fits in these intervals fail to produce $\cdf<3.6$\footnote{This 
restriction is heuristic. When enlarging the errors after obtaining 
\(\cdf>3.6\), the fits generally fail to reproduce the input values 
in the overlap region well.} without 
enlarging the errors, we remove the highest (or lowest) temperature in 
the overlap region from the fit and repeat the fits. 
We always keep at least one temperature common between both fits. 
This procedure is necessary since data at very low temperatures usually 
have much larger uncertainties than at higher temperatures. 
Thus, a global interpolation cannot resolve the increase of \(f^{\rm sub}\) 
towards low temperatures properly. 
In agreement with predictions of weak-coupling results (\mbox{cf.} 
Secs.~\ref{sec:singlet} and \ref{sec:plc}), the \(T\) dependence is much 
flatter at higher temperatures than at lower temperatures. 
In the overlap region, the second derivative is quite large, and thus, use of 
a smoothing parameter penalizing higher derivatives is hardly appropriate. 
In the full overlap region, we match the fits for low and high temperatures 
smoothly.

\subsection{Global fits}
We perform separate global fits to the free energies and to \(\Vs-\Fs\) for each \(N_\tau\). 
For the free energies, we perform the fits over the full temperature range 
as well as for low or high temperature intervals with 
\(T\leq 235\,{\rm MeV}\) or \(T\geq 175\,{\rm MeV}\), respectively. 
Because of the small size of the gauge ensembles for \(N_\tau=16\), we enlarge 
the corresponding statistical errors by a \(1\%\) relative error estimate. 
For \(\Vs-\Fs\), we limit the \(rT\) range to \(rT \leq 0.5\) for 
\(N_\tau>4\) and to \(rT\leq0.75\) for \(N_\tau=4\), but fit all available temperatures at once. 
In order to obtain stable fits, we have to enlarge the errors of \(\Vs-\Fs\) 
to 5\% relative error or \(0.001\, T\) absolute error, whichever is larger. 

For the global fits, we use a generalized temperature-dependent modified 
screened Cornell potential (GTDMSCP) \textit{Ansatz}. 
In the GTDMSCP \textit{Ansatz} we generalize \mbox{Eq.}~\eqref{eq:MSCP} as 

\ileq{\label{eq:GTDMSCP}
 F^{\rm sub}(rT) = \frac{e^{-M(T)\,rT}}{(rT)^\alpha} 
 \sum\limits_{m=m_{\rm min}}^{m_{\rm max}} \sum\limits_{n=n_{\rm min}}^{n_{\rm max}} A_{n,m} (rT)^n T^m.
}

We did not observe any systematic temperature dependence in the 
values of \(C\) from local fits using the MSCP \textit{Ansatz} 
[\mbox{Eq.}~\eqref{eq:MSCP}]. 
Hence, we must omit \(C\) in the global fits and consider effects 
leading to \(C\neq0\) in the local fits as temperature-dependent 
uncertainties that have to be averaged out in the global fit. 
In the GTDMSCP fits, we assume that the dominant temperature dependence is 
in the double polynomial, while we assume that the temperature dependence of 
the screening mass \(M(T)\) is rather mild.\footnote{In our conventions the screening mass \(M(T)\) is a screening mass in units of the temperature, 
namely \(m(T)=T\,M(T)\) is the screening mass in physical units. 
This trivial multiplicative temperature dependence is taken out in our \textit{Ansatz}.} 
We model the temperature dependence of the screening mass as 

\ileq{
M(T)=M_0\,\left[ 1+M_1 \ln\left(\frac{T}{T_1}\right) \right],
}

where we restrict the range of \(M_1\) to the interval \((-1,+1)\) 
and normalize the temperature logarithm by \(T_1=1\,{\rm GeV}\) to ensure 
a relatively mild temperature dependence of \(M(T)\). 
This choice of \(T_1\) restricts the logarithm to about \((-2,+2)\), which is 
sufficiently flexible to permit even negative screening masses at one end of the temperature range. 
As we typically obtain \(-0.5 \lesssim M_1 \lesssim 0.5\) in fits, our model is sufficiently flexible. 
Unless it comes out as zero, \(M_0\) is generally positive and larger than \(\abs{M_1}\) by an order of magnitude.

To model the dominant temperature dependence in the GTDMSCP \textit{Ansatz} in a 
very general form, we use sequences of consecutive powers of the temperature 
\(T^{m_i}\) and vary the lowest and highest powers. 
We use \({m_i}\) between \(m_{\min}\ge-4\) and \(m_{\rm max}\le+3\) and 
sequences with lengths \(N_m=m_{\rm max}-m_{\rm min}+1=1,\ 2,\ldots,\ 6\). 
After rescaling the temperatures as \(T/T_1\), we obtain mostly coefficients of natural size.
We construct all consecutive sequences including \(m_i=0\) and iteratively increase \(N_m\) when needed.   
We use \(n_{\rm max}=8\) as the highest power of \(rT\), 
but observe that the global fits generally favor lower powers of \(rT\). 
As leading inverse power of \(rT\), we use \(\alpha=1\) for \(\Fs^{\rm sub}\) 
and \(\alpha=2\) for \(\Fqq^{\rm sub}\). 
The underlying reasoning for this choice is the form of the LO result [\mbox{Eq.}~\eqref{eq:fqq_lo}]. 

\begin{table}
\parbox{.98\linewidth}{
  \begin{tabular}{|c|c|c|c|c|c|}
    \hline
    \(N_\tau\) & \(\beta\) range & \(rT\) range & \(N_m,N_n\) & \(\Delta\) & \(\cdf\) \\
    \hline
    4 & \([5.9,9.67]\) & \([0.240,1.920]\) & \(5,3\) & 0\% & 1032/930 \\
    6 & \([5.9,6.423]\) & \([0.160,0.898]\) & \(4,3\) & 0\% & 96/133 \\
    6 & \([6.195,9.67]\) & \([0.160,1.803]\) & \(4,3\) & 0\% & 1609/1463 \\
    8 & \([6.05,6.74]\) & \([0.120,0.685]\) & \(4,3\) & 0\% & 270/286 \\
    8 & \([6.488,9.67]\) & \([0.120,1.586\) & \(4,3\) & 0\% & 1558/1877 \\
    10 & \([6.285,6.95]\) & \([0.096,0.678]\) & \(3,3\) & 0.9\% & 221/217 \\
    10 & \([6.74,9.67]\) & \([0.096,1.389]\) & \(4,4\) & 0\% & 2397/2049 \\
    12 & \([6.515,7.2]\) & \([0.080,0.471]\) & \(4,3\) & 0.5\% & 170/202 \\
    12 & \([6.950,9.67]\) & \([0.080,1.253]\) & \(4,3\) & 0\% & 2398/2067 \\
    16 & \([7.825,9.67]\) & \([0.060,0.862]\) & \(2,3\) & 1.0\% & 1315/1185 \\
    \hline
  \end{tabular}
  \caption{\label{tab:global fs}
Summary of global fits of \(\Fs^{\rm sub}\). 
\(\Delta\) is a relative error added linearly to statistical errors before the fit. 
\(N_m=m_{\rm max}-m_{\rm min}+1\) and \(N_n=n_{\rm max}-n_{\rm min}+1\) are 
numbers of \(T\) and \(rT\) monomials in \mbox{Eq.}~\eqref{eq:GTDMSCP}. 
  }
}
\end{table}

\begin{table} 
\parbox{.98\linewidth}{
  \begin{tabular}{|c|c|c|c|c|c|}
    \hline
    \(N_\tau\) & \(\beta\) range & \(rT\) range & \(N_m,N_n\) & \(\Delta\) & \(\cdf\) \\
    \hline
    4 & \([5.9,9.67]\) & \([0.240,1.224]\) & \(4,3\) & 0\% & 350/387 \\
    6 & \([5.9,6.423]\) & \([0.160,0.898]\) & \(4,3\) & 0\% & 163/169 \\
    6 & \([6.195,9.67]\) & \([0.160,1.130]\) & \(4,3\) & 0\% & 723/695 \\
    8 & \([6.05,6.74]\) & \([0.120,0.718]\) & \(4,4\) & 1.2\% & 190/264 \\
    8 & \([6.488,9.67]\) & \([0.120,1.083\) & \(4,3\) & 0\% & 603/849 \\
    10 & \([6.285,6.95]\) & \([0.096,0.728]\) & \(3,4\) & 1.2\% & 227/199 \\
    10 & \([6.74,9.67]\) & \([0.096,0.917]\) & \(4,4\) & 0\% & 1025/923 \\
    12 & \([6.515,7.2]\) & \([0.080,0.391]\) & \(4,3\) & 0\% & 103/120 \\
    12 & \([6.950,9.67]\) & \([0.080,0.773]\) & \(4,3\) & 0\% & 944/821 \\
    16 & \([7.825,9.67]\) & \([0.060,0.492]\) & \(2,3\) & 1.0\% & 290/386 \\
    \hline
  \end{tabular}
  \caption{\label{tab:global fqq}
Summary of global fits of \(\Fqq^{\rm sub}\). 
\(\Delta\) is a relative error added linearly to statistical errors before the fit. 
\(N_m=m_{\rm max}-m_{\rm min}+1\) and \(N_n=n_{\rm max}-n_{\rm min}+1\) are 
numbers of \(T\) and \(rT\) monomials in \mbox{Eq.}~\eqref{eq:GTDMSCP}. 
  }
}
\end{table}

\begin{table} 
\parbox{.98\linewidth}{
  \begin{tabular}{|c|c|c|c|c|c|}
    \hline
    \(N_\tau\) & \(\beta\) range & \(rT\) range & \(N_m,N_n\) & \(\Delta\) & \(\cdf\) \\
    \hline
    4 & \([5.9,8.4]\) & \([0.250,0.75]\) & \(4,4\) & 0\% & 59/134 \\
    6 & \([5,9,8.4]\) & \([0.167,0.50]\) & \(5,6\) & 0.8\% & 188/136 \\
    8 & \([6.488,8.4]\) & \([0.125,0.50\) & \(5,5\) & 0.5\% & 353/297 \\
    10 & \([6.74,8.4]\) & \([0.100,0.50]\) & \(5,6\) & 0.5\% & 404/313 \\
    12 & \([6.950,8.4]\) & \([0.083,0.50]\) & \(4,5\) & 0\% & 340/330 \\
    \hline
  \end{tabular}
  \caption{\label{tab:global vmf}
Summary of global fits of \(\Vs-\Fs\). 
\(\Delta\) is a relative error added linearly to statistical errors before the fit. 
\(N_m=m_{\rm max}-m_{\rm min}+1\) and \(N_n=n_{\rm max}-n_{\rm min}+1\) are 
numbers of \(T\) and \(rT\) monomials in \mbox{Eq.}~\eqref{eq:GTDMSCP}. 
  }
}
\end{table}

We perform nonlinear fits for different choices of \(N_m\) and \(N_n\) to 
determine the most simple \textit{Ansatz} that matches the data. 
We use three nested loops for searching the optimal fit function. 
The two outer loops are the same as for the local fits, while the innermost 
loop is the loop over the sequences of consecutive powers of \(T\). 
In a first cycle, we perform the global fits on the jackknife average to 
identify an appropriate fit function. 
In a second cycle, we use bootstrap mock data to propagate the statistical errors. 
If the need arises, we readjust the fit function in the second cycle again. 
If we fail to reach convergence, we add 0.5\% times the minimal 
\(\cdf\) of the recent loop and restart the outermost loop. 
We observe that we almost universally need sequences from \(T^{-3}\) 
to \(T^0\) and sometimes in addition \(T^1\). 
For this reason, we eventually start from sequences of length \(m=4\) for our final fits. 
With regard to the \(rT\) dependence, we demand for the free energies at 
least two and at most three terms beyond the leading powers. 
We summarize the global fits of \(\Fs^{\rm sub}\) and \(\Fqq^{\rm sub}\) in Tables~\ref{tab:global fs} and~\ref{tab:global fqq}. 
Requirements for powers of the \(rT\) polynomials, enlarged errors, and 
achieved \(\cdf\) results are quite similar for the local fits. 

For \(\Vs-\Fs\), we omit the screening mass completely and demand at least 
three terms beyond the leading power, which we fix through \(\alpha=0\). 
As discussed in the subsection on local fits, \(\Vs-\Fs\) appears to be 
affected by fluctuations that are larger than the statistical errors.  
In order to obtain stable interpolations, we have to enlarge the errors 
globally to the larger of \(0.001\,T\) absolute error or \(5\%\) relative error before fitting. 
We summarize the global fits of \(\Vs-\Fs\) in Table~\ref{tab:global vmf}.

\begin{figure}
\centering
 \includegraphics[height=5.8cm,clip]{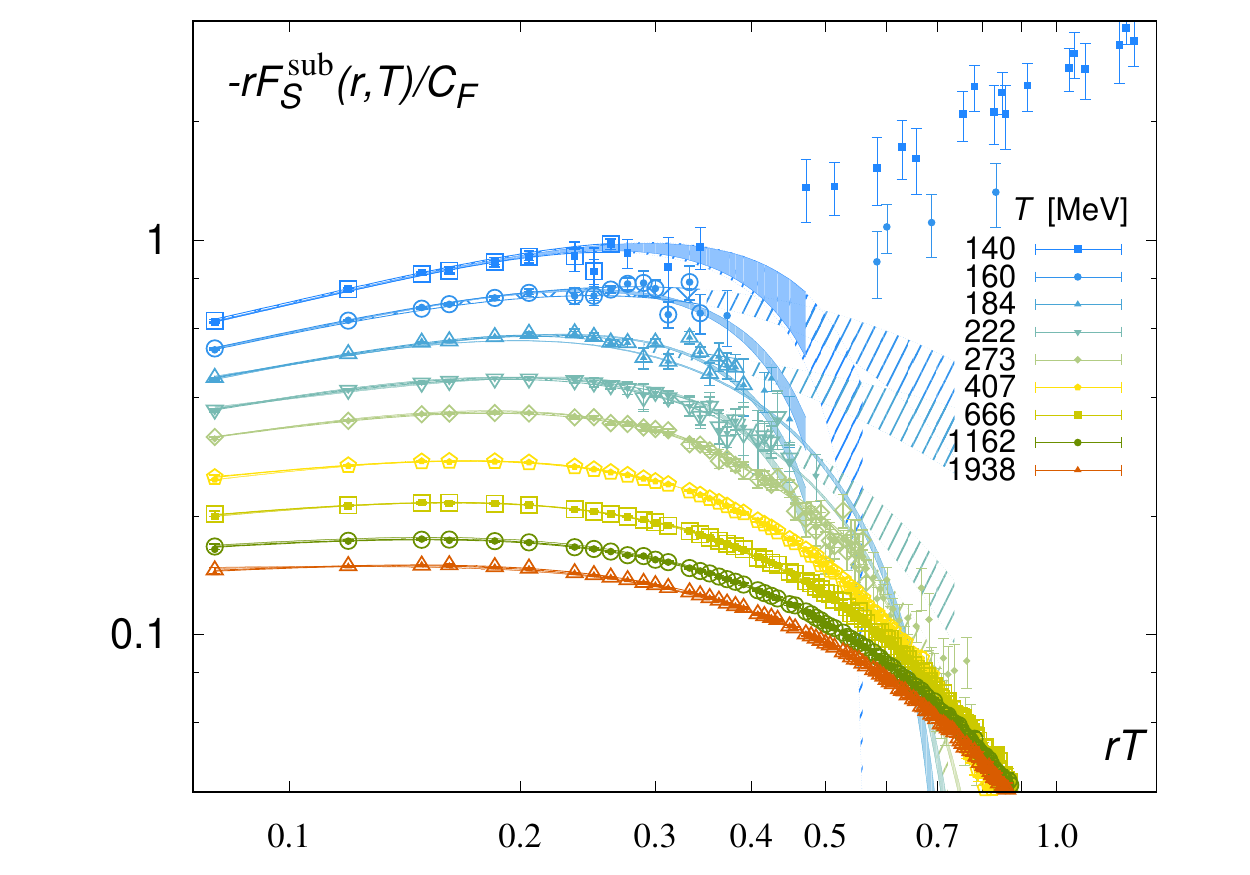}
\hfill
\centering
 \includegraphics[height=5.8cm,clip]{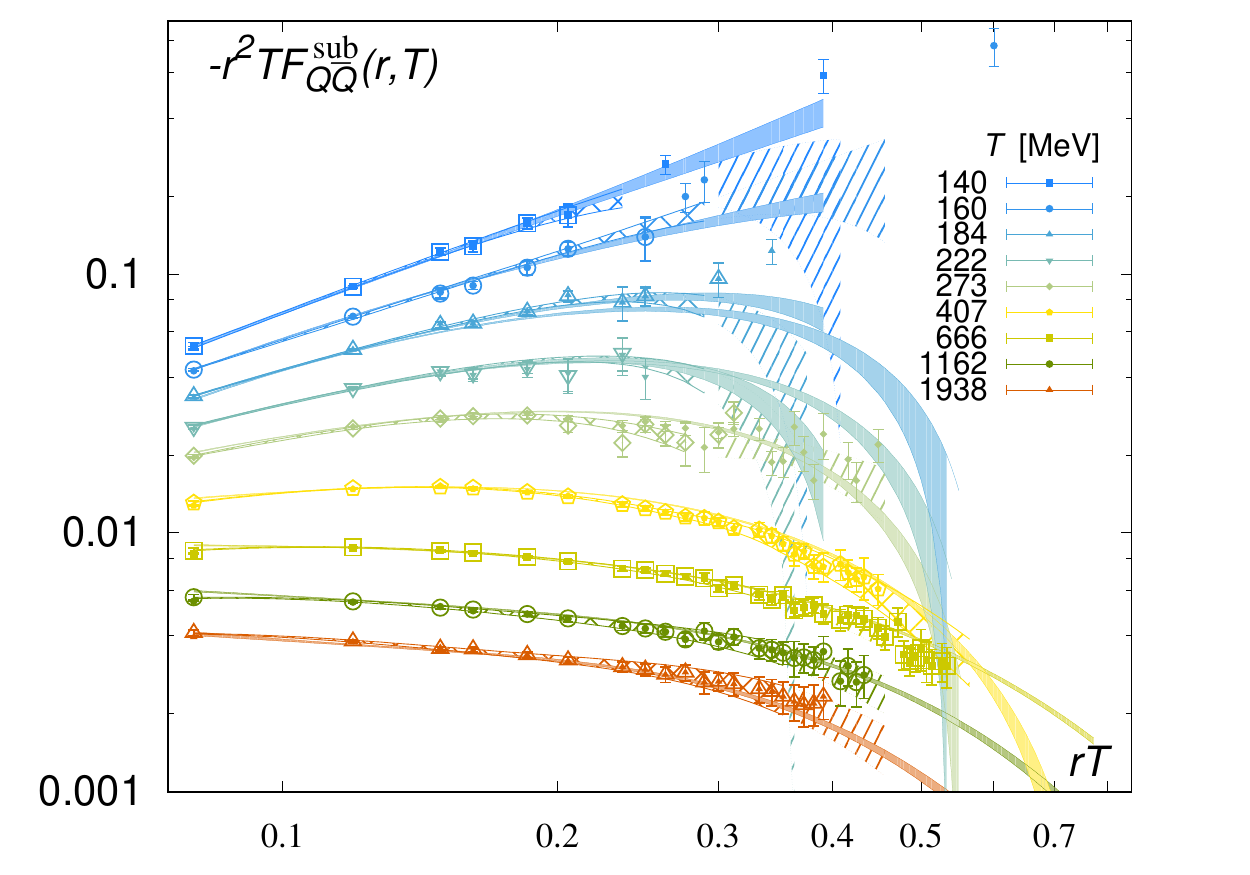}
\caption{\label{fig:local global}
Comparison of local and global fits for \(N_\tau=12\) as a log-log plot. 
We show the nonperturbatively corrected data entering the fits as 
large open symbols and the underlying data as small filled symbols. 
Both data sets are shown only for signal-to-noise ratio of at least five. 
The cross-hatched bands represent local interpolations, and the diagonal-hatched 
bands represent the associated extrapolation up to \(rT=0.75\). 
The solid bands represent global fits for low or high temperature ranges. 
In the overlap region, the shorter bands correspond to the low temperature global fit. 
}
\end{figure}

In \mbox{Fig.}~\ref{fig:local global}, we show a comparison of the low and 
high temperature global fits and the corresponding local fits before 
temperature interpolation with the underlying data for \(N_\tau=12\). 
Within the range of fitted data for \(T \leq 220\,{\rm MeV}\) we obtain fair 
agreement between local and low temperature global fits, but not with the other global fits. 
For \(T \geq 220\,{\rm MeV}\), we find fair agreement between local and 
global fits for high temperatures or on the full temperature range. 
Therefore, we switch for \(T > 220\,{\rm MeV}\) from the low temperature fits to the high temperature fits. 
We estimate systematical errors of the global fit from half of the 
difference in the range \(T\in(190,230)\,{\rm MeV}\) and add it to the statistical error. 
We obtain the \(rT\) and \(T\) derivatives analytically for every fit 
function and insert the fit parameters on the individual bins to compute the average and propagate the errors.

\subsection{Average of local and global fits}
The assumption of smoothness in the temperature leads to substantially 
smaller errors in the interpolating functions than in the actual data 
(see \mbox{Fig.}~\ref{fig:local global}). 
Hence, we assume that the errors of the interpolating function may be
underestimated and we have to account for that.
This is particularly relevant at the shortest and largest distances. 
At the shortest distances, there are only a few points with relatively large 
systematic errors, which have only little weight in the global fits.
At the largest distances, the local fits are oblivious of the smoothness 
condition between neighboring temperatures and are sensitive to incomplete 
subtraction.
These complementary systematic uncertainties can be directly observed in 
\mbox{Fig.}~\ref{fig:local global}. 

We eventually calculate the weighted average between the local and global 
fits before the continuum extrapolation. 
The former do not require that the subtracted free energies approach zero 
at large \(rT\), while the latter do not require that the screening mass 
parameter is positive. 
Hence, they have different systematic uncertainties for large \(rT\), which 
can be controlled better through an averaging procedure. 
We estimate the systematical error from the difference between local and 
global fits. ,
We restrict the growth of the error of the averaged result such that the 
\(1\sigma\) band of the averaged result is contained by the combined 
\(1\sigma\) bands of the local and global fits. 
Moreover, to account for any remaining uncertainties at short distances we 
demand that the relative error must not be smaller than the difference 
between relative errors of the corrected and uncorrected data for the nearest 
point in the \((rT,T)\) plane. 

This is still not sufficient for a reliable continuum extrapolation. 
We find that quite often interpolated data with \(N_\tau=10,\ 8\), and 
sometimes \(6\) are in the scaling window, while some interpolated data 
for \(N_\tau=12\) or \(16\) are not consistent with this scaling behavior 
(see the scaling plots in \mbox{Sec.}~\ref{sec:continuum}). 
This is direct evidence that interpolations with \(N_\tau>10\) still have underestimated uncertainties. 
For this reason, we enlarge the errors of interpolations by hand for \(N_\tau>10\). 
We add to interpolations with \(N_\tau=16\) a 1\% relative error for \(rT<0.2\) 
and a 2\% relative error for \(0.2\leq rT<0.3\). 
We add to interpolations with \(N_\tau=12\) a 0.5\% relative error for \(rT<0.2\) and a 1.5\% relative error for \(0.2\leq rT<0.3\).

\subsection{Continuum extrapolations and error estimates}
\label{sec:continuum}

\begin{figure*}
\centering
 \includegraphics[height=5.2cm,clip]{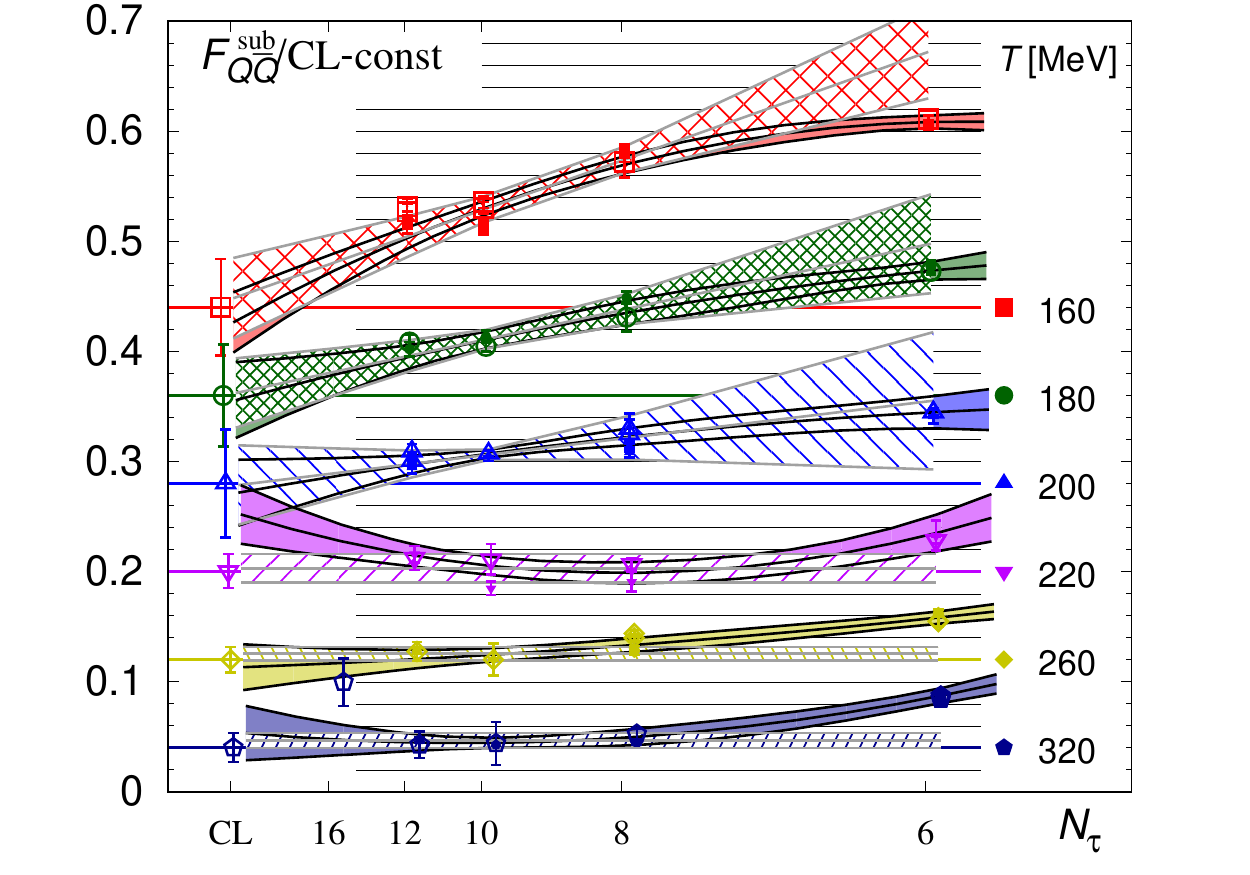}
 \includegraphics[height=5.2cm,clip]{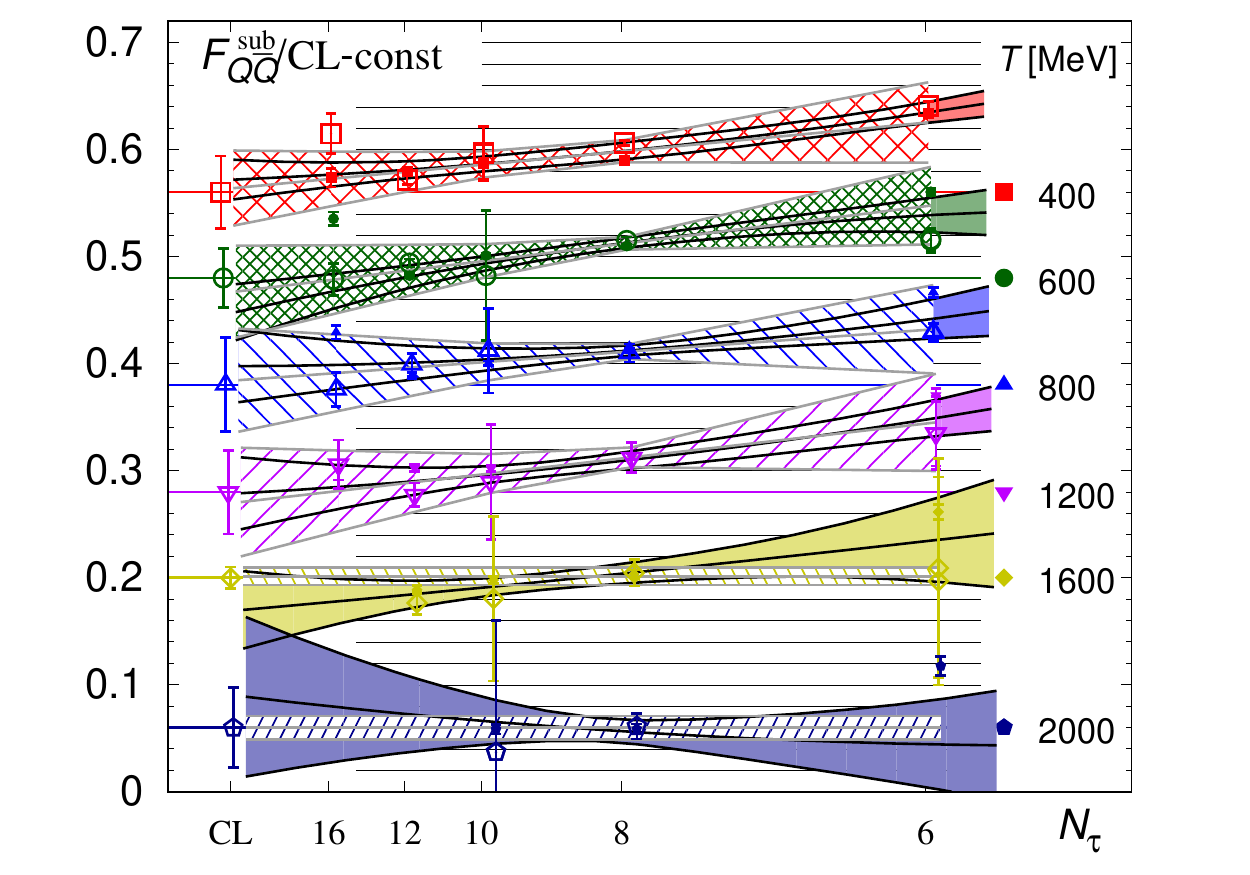}
 \vskip0ex
 \includegraphics[height=5.2cm,clip]{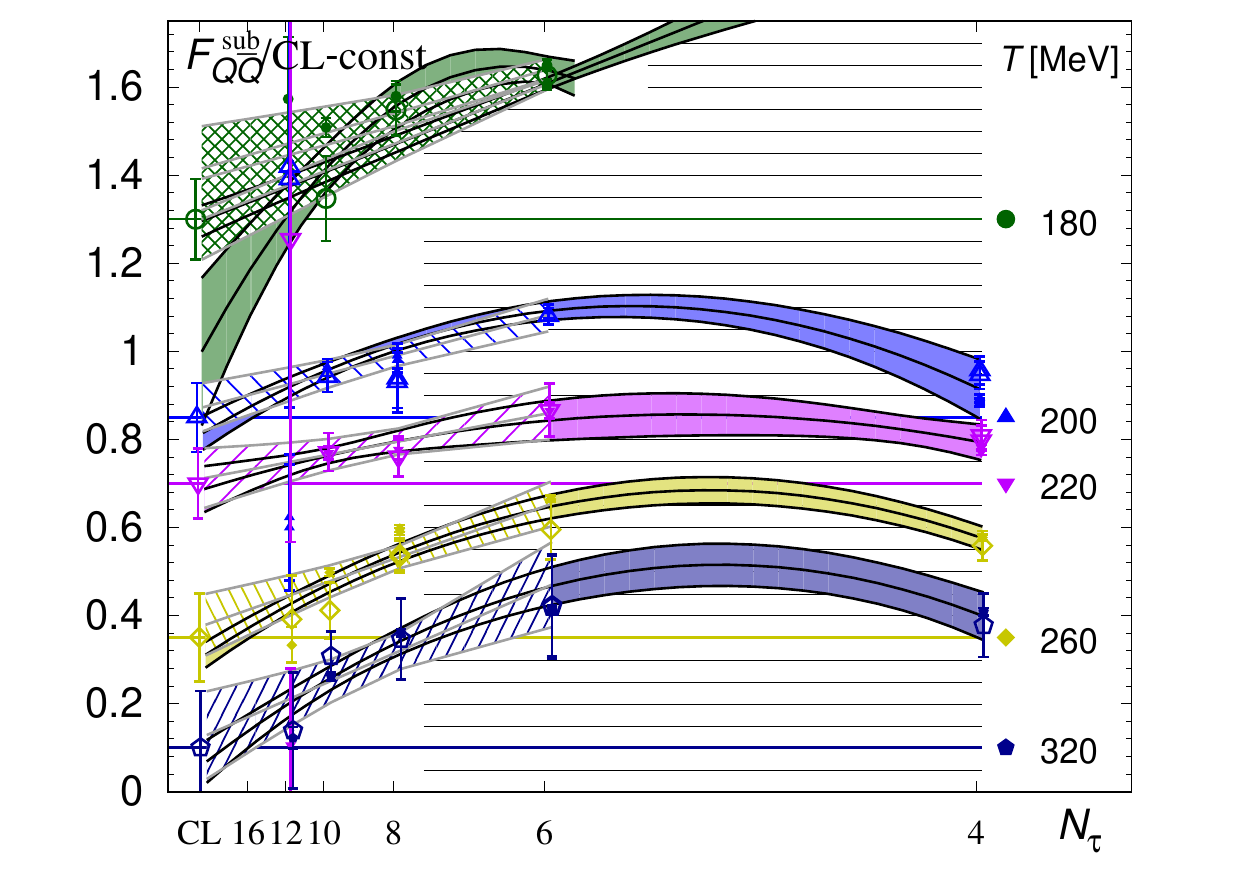}
 \includegraphics[height=5.2cm,clip]{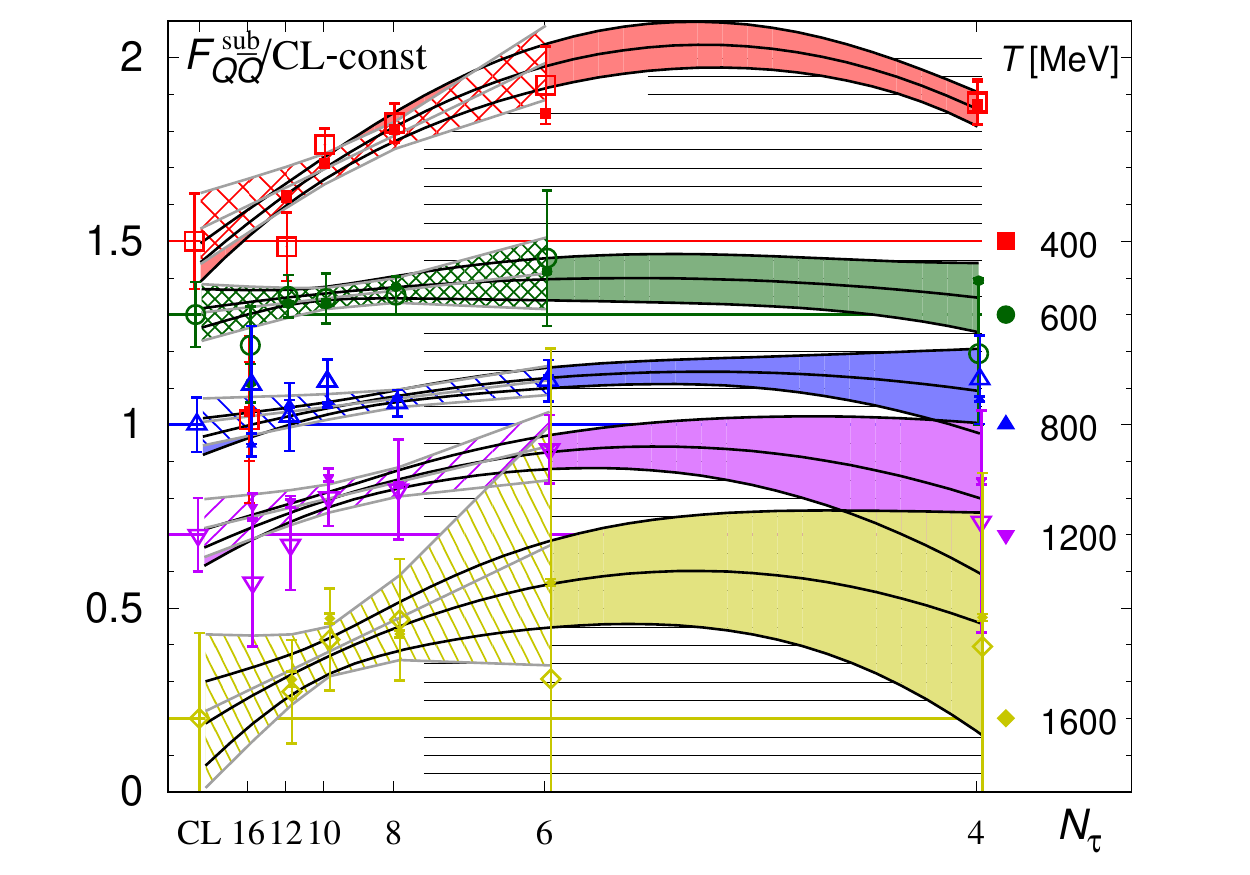}
\caption{\label{fig:fqqext}
Scaling behavior of the subtracted free energy \(\Fqq^{\rm sub}\) in units of 
the continuum result at various temperatures and at two representative 
separations (from top to bottom \(rT=0.16\) and~\(0.40\)) as function of 
\(N_\tau\). 
``CL'' marks the continuum limit \((N_\tau \to \infty)\). 
Ratios for each temperature are shifted by constants for better visibility. 
The bands with filled pattern indicate fits with \(b/N_\tau^2\), except 
for \(rT=0.16\) and \(T \in (200,400)\,{\rm MeV}\) or \(T>1200\,{\rm MeV}\). 
The solid bands indicate fits with  \(c/N_\tau^4\). 
The large open (small filled) symbols indicate results from local (global) fits. 
}
\end{figure*}

\begin{figure*}
\centering
 \includegraphics[height=5.2cm,clip]{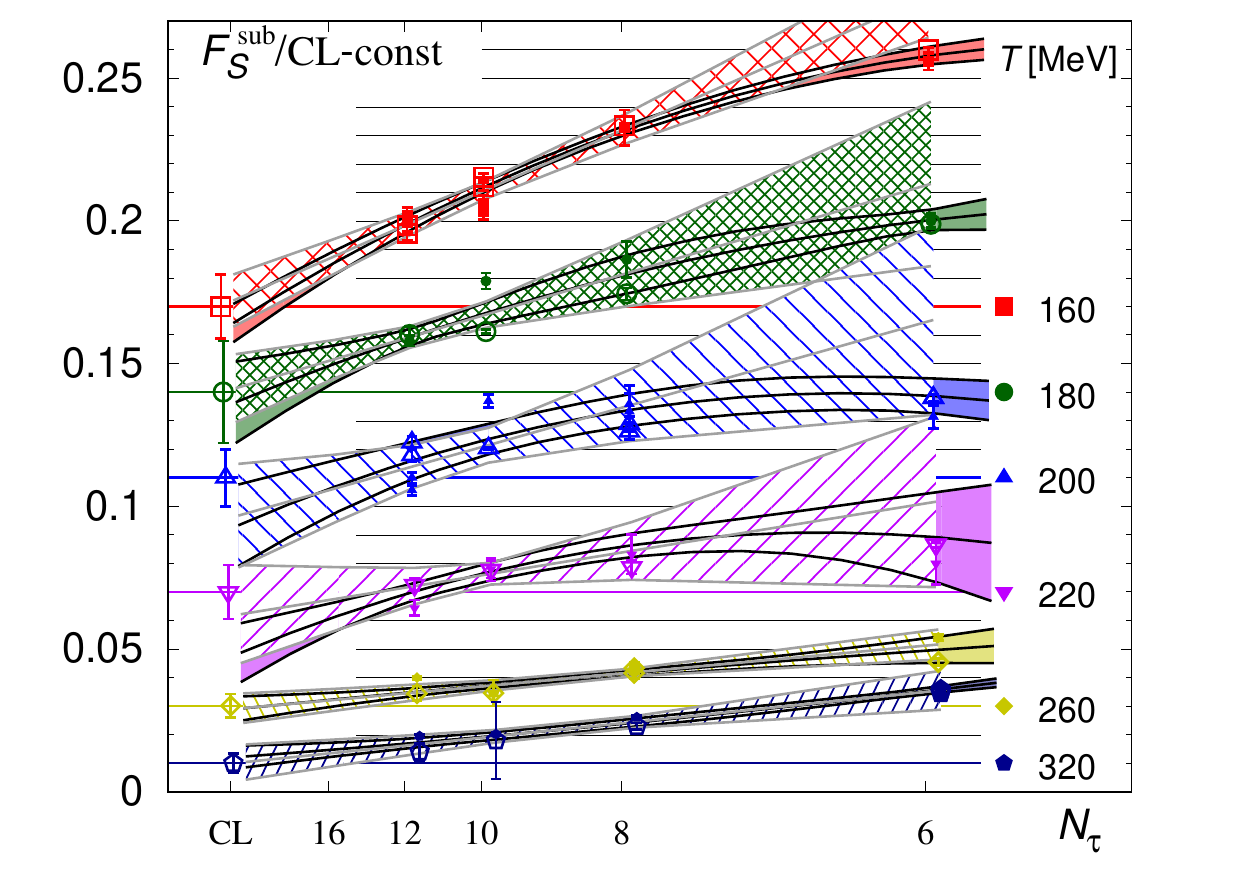}
 \includegraphics[height=5.2cm,clip]{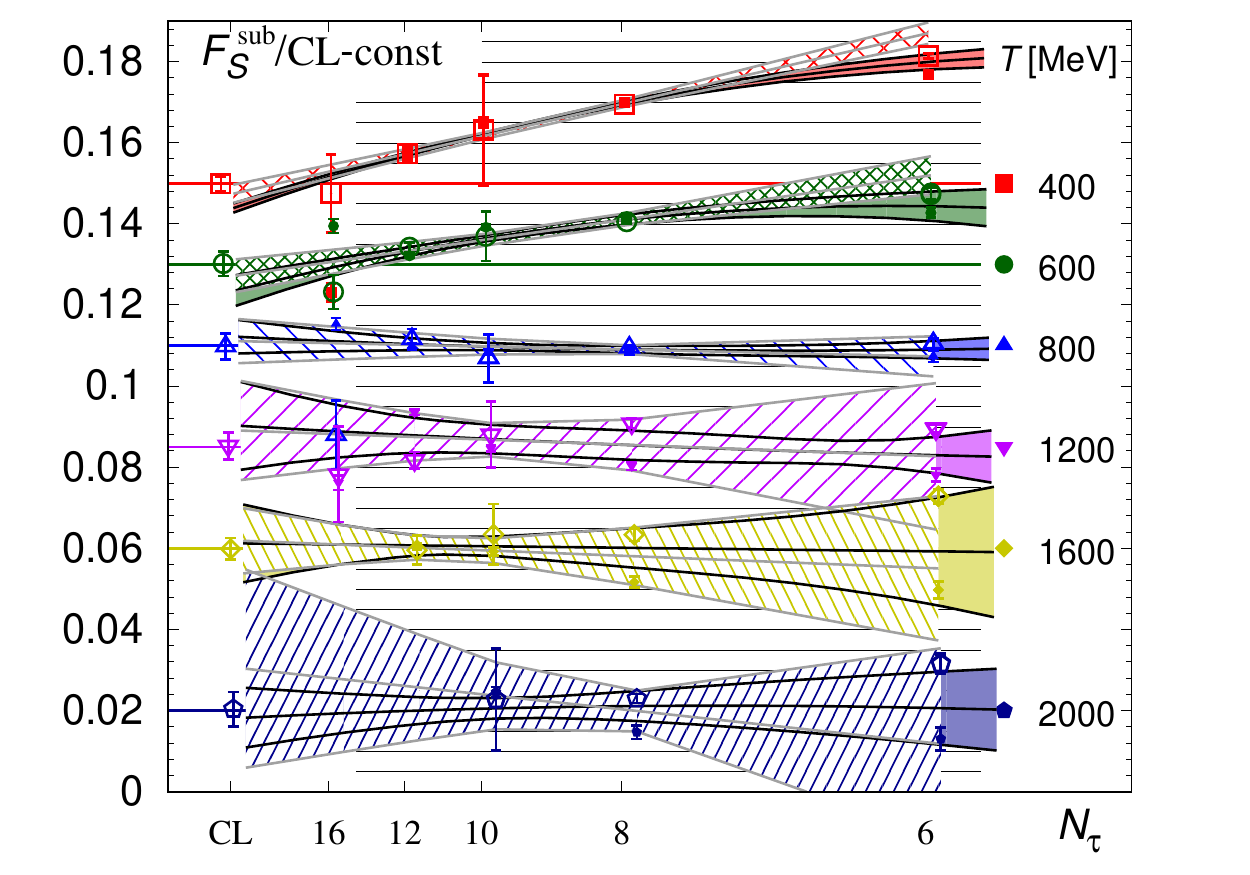}
 \vskip0ex
 \includegraphics[height=5.2cm,clip]{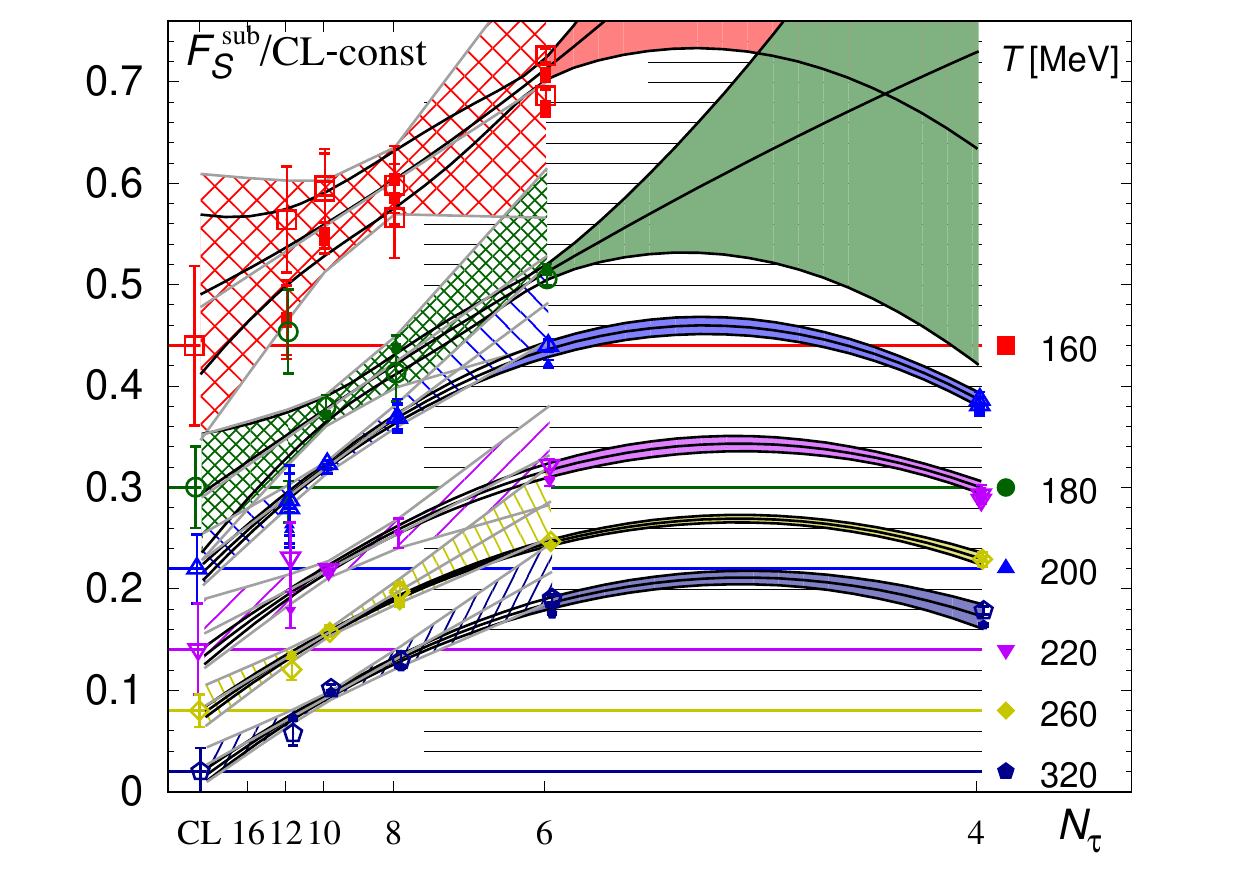}
 \includegraphics[height=5.2cm,clip]{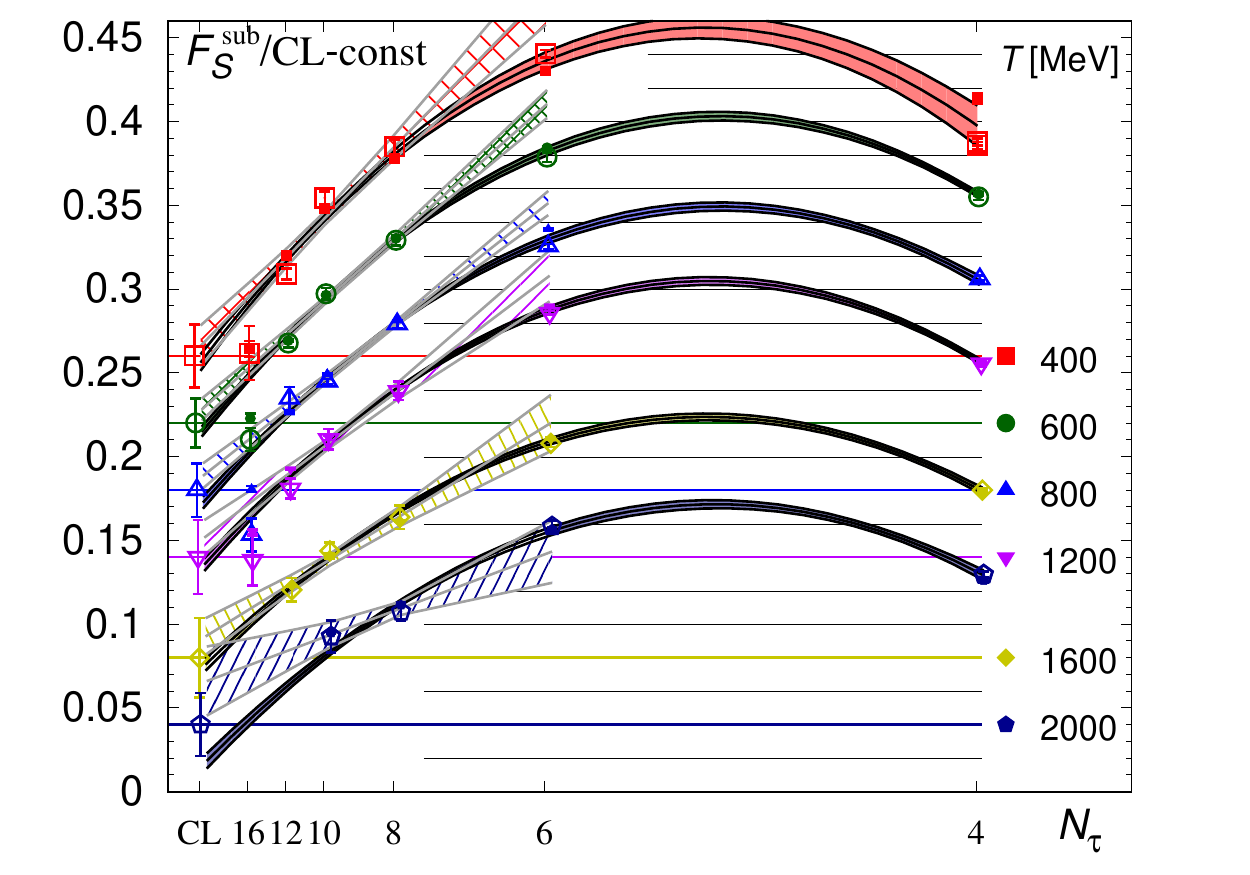}
\caption{\label{fig:fsext}
Scaling behavior of the subtracted singlet free energy \(\Fs^{\rm sub}\) in 
units of the continuum result at various temperatures and at two representative  
(from top to bottom \(rT=0.16\) and~\(0.40\)) as function of \(N_\tau\). 
``CL'' marks the continuum limit \((N_\tau \to \infty)\). 
Results for each temperature are shifted by constants for better visibility. 
The bands with filled pattern indicate fits with \(b/N_\tau^2\). 
The solid bands indicate fits with  \(c/N_\tau^4\). 
The large open (small filled) symbols indicate results from local (global) fits. 
}
\end{figure*}

\begin{figure*}
\centering
 \includegraphics[height=5.2cm,clip]{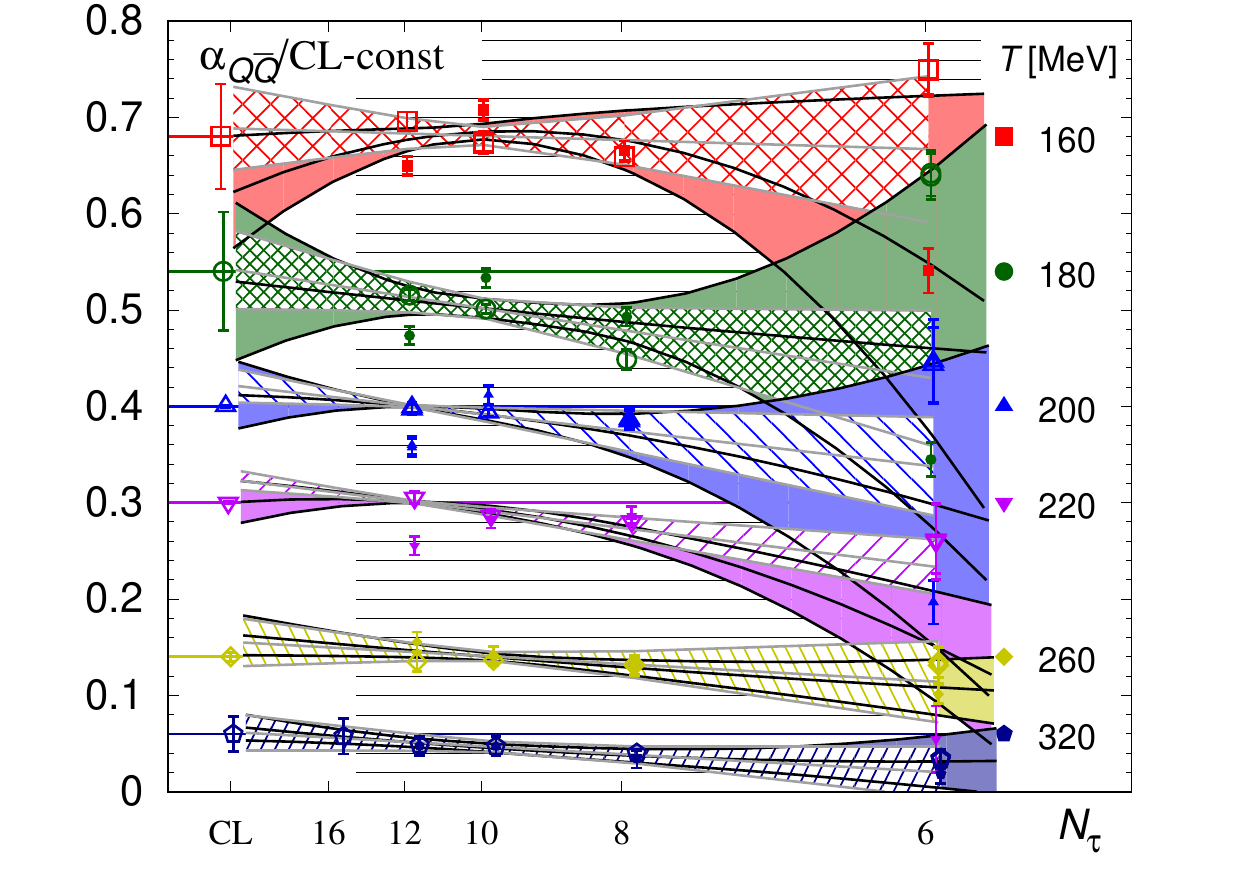}
 \includegraphics[height=5.2cm,clip]{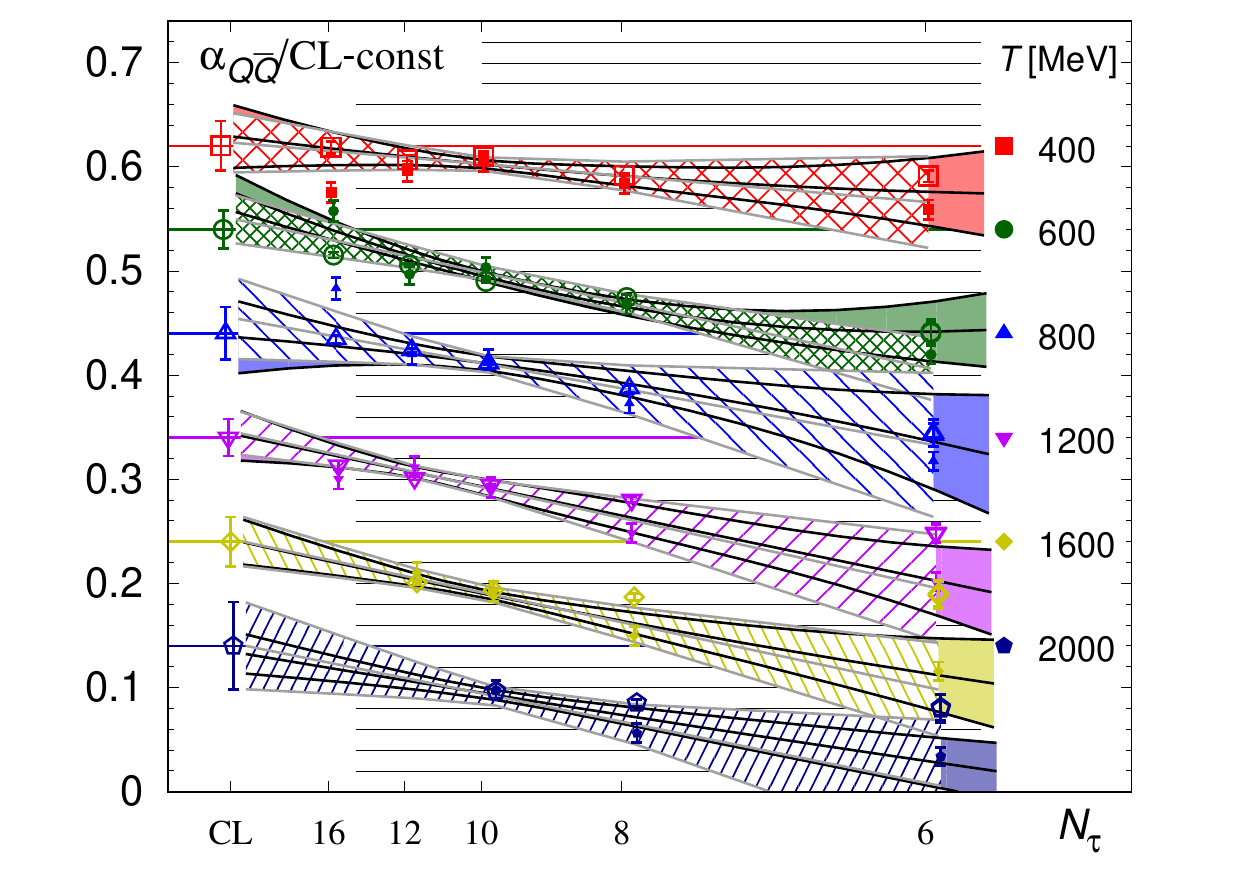}
 \vskip0ex
 \includegraphics[height=5.2cm,clip]{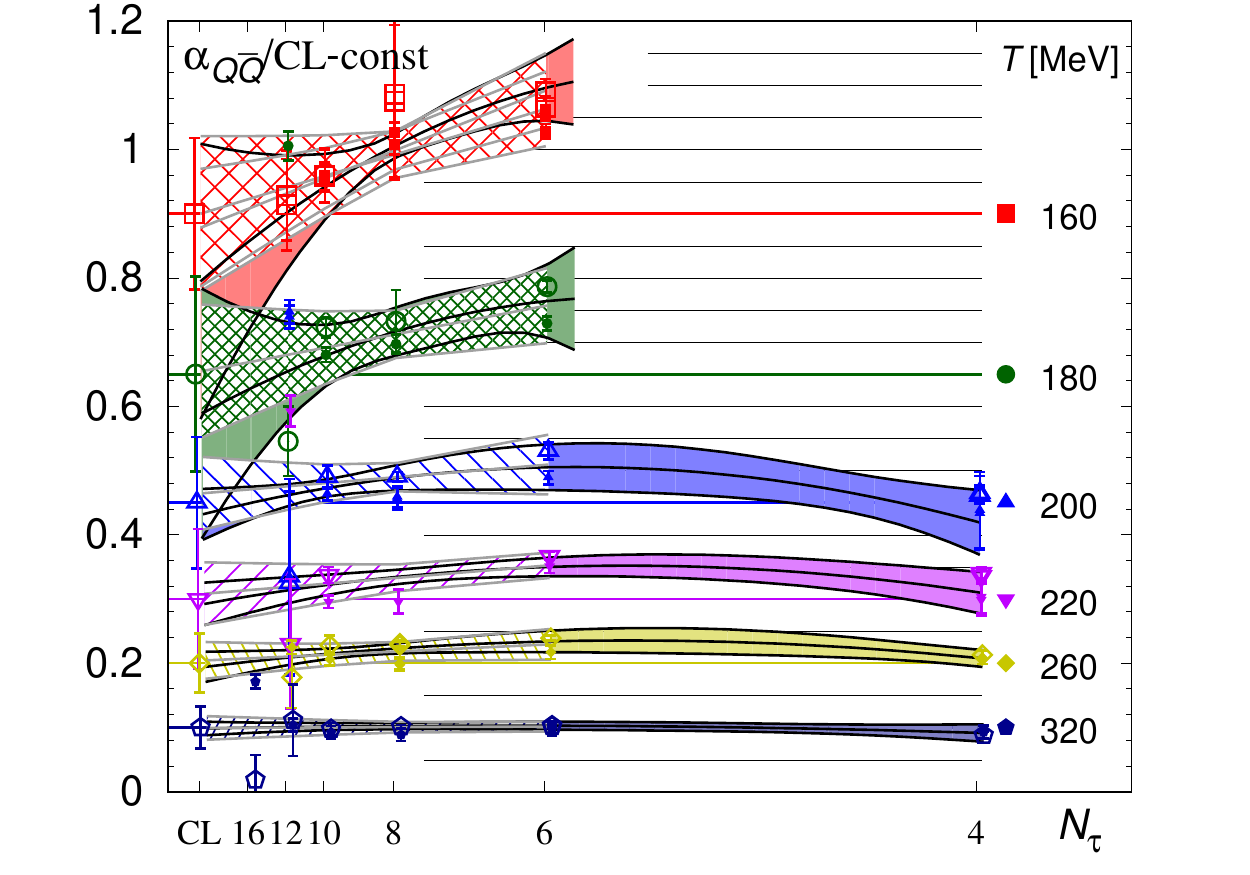}
 \includegraphics[height=5.2cm,clip]{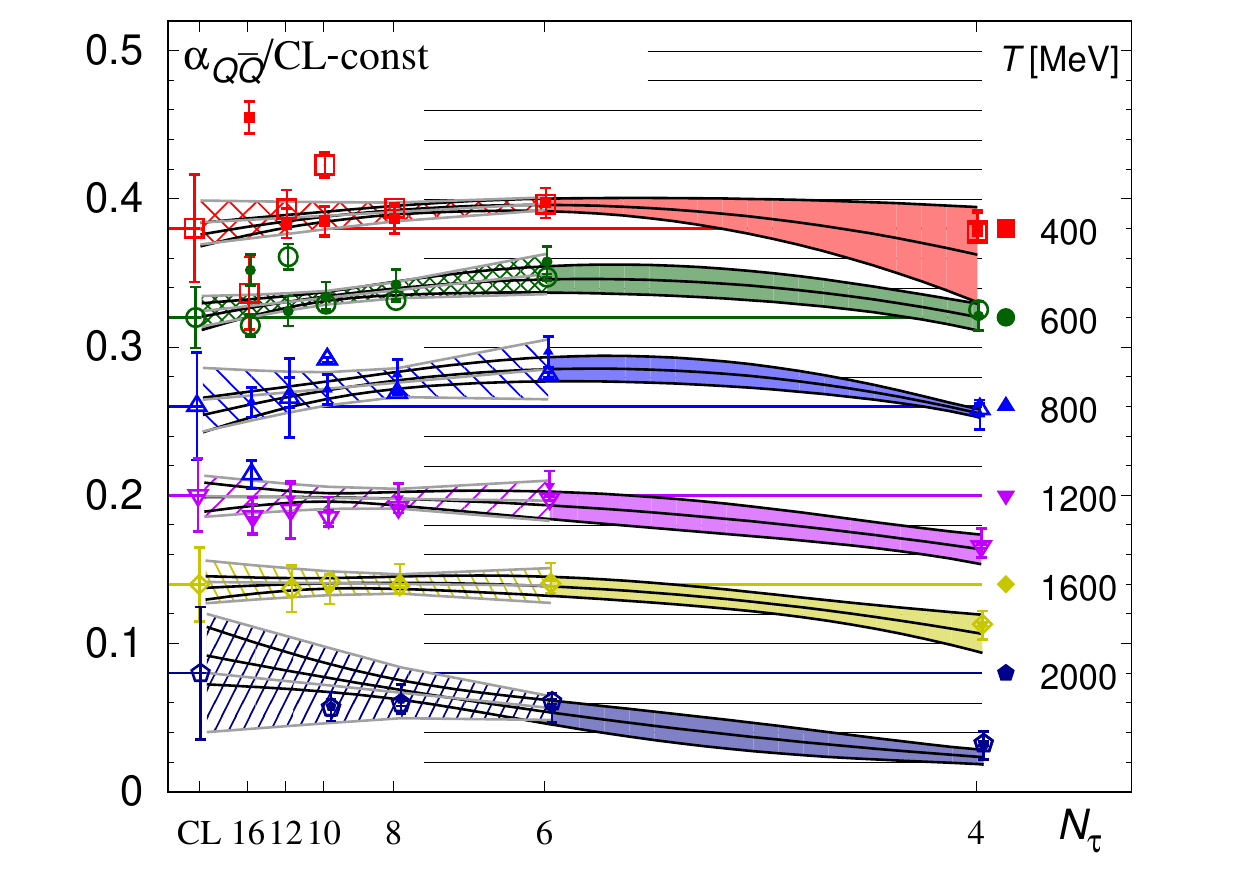}
\caption{\label{fig:aqqext}
Scaling behavior of the effective coupling \(\aqq[\Fs]\) in units of the 
continuum result at various temperatures and at two representative 
separations (from top to bottom \(rT=0.16\) and~\(0.40\)) as function of 
\(N_\tau\). 
``CL'' marks the continuum limit \((N_\tau \to \infty)\). 
Results for each temperature are shifted by constants for better visibility. 
The bands with filled pattern indicate fits with \(b/N_\tau^2\). 
The solid bands indicate fits with  \(c/N_\tau^4\). 
The large open (small filled) symbols indicate results from local (global) fits. 
}
\end{figure*}

\begin{figure*}
\centering
 \includegraphics[height=5.2cm,clip]{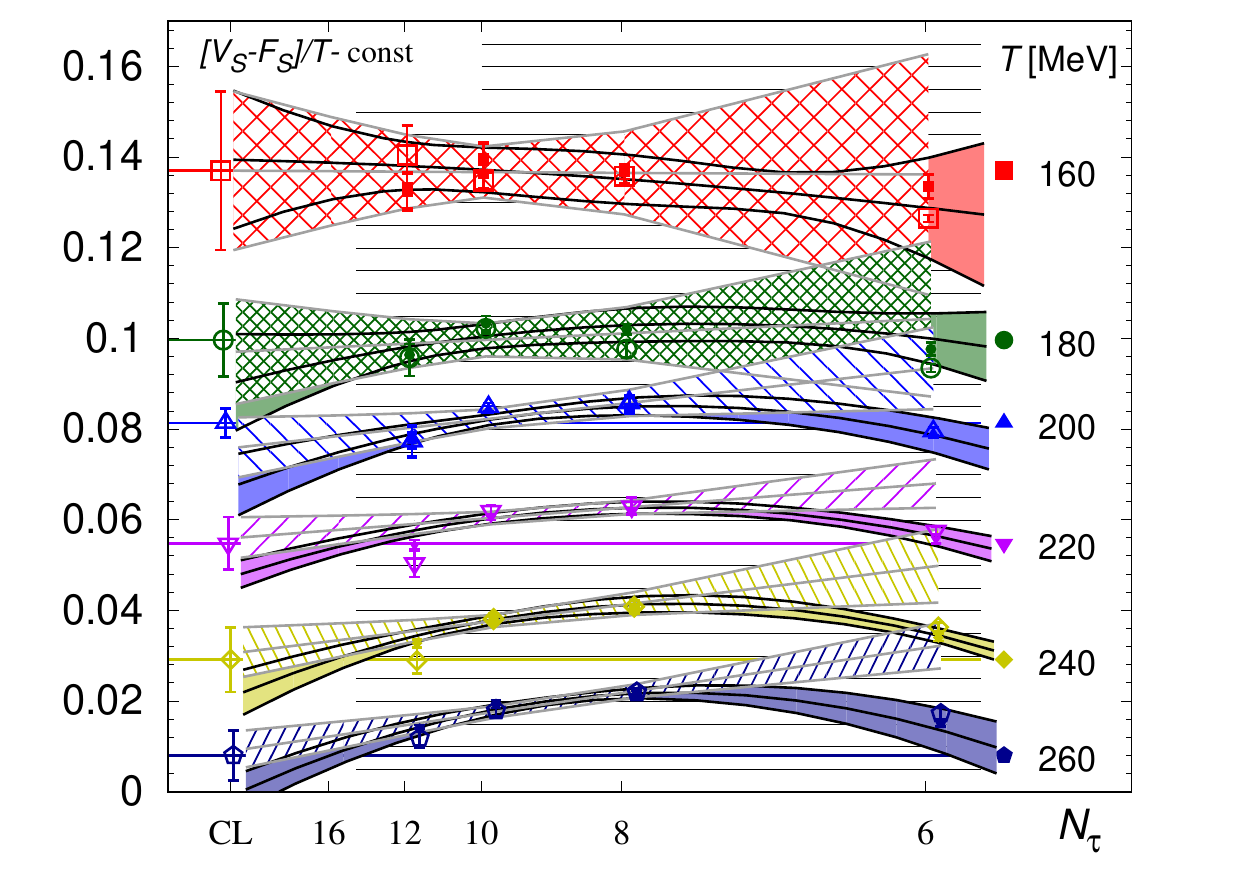}
 \includegraphics[height=5.2cm,clip]{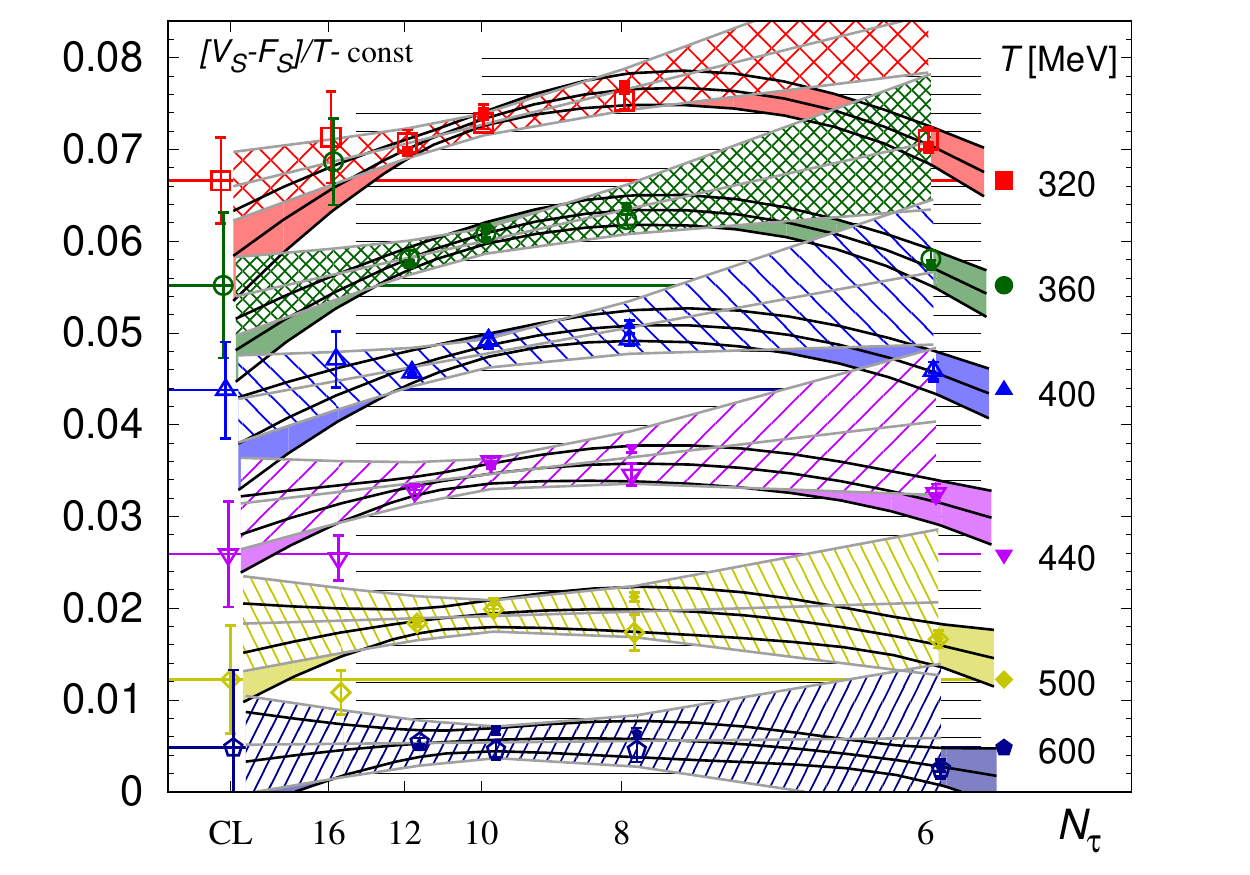}
 \vskip0ex
 \includegraphics[height=5.2cm,clip]{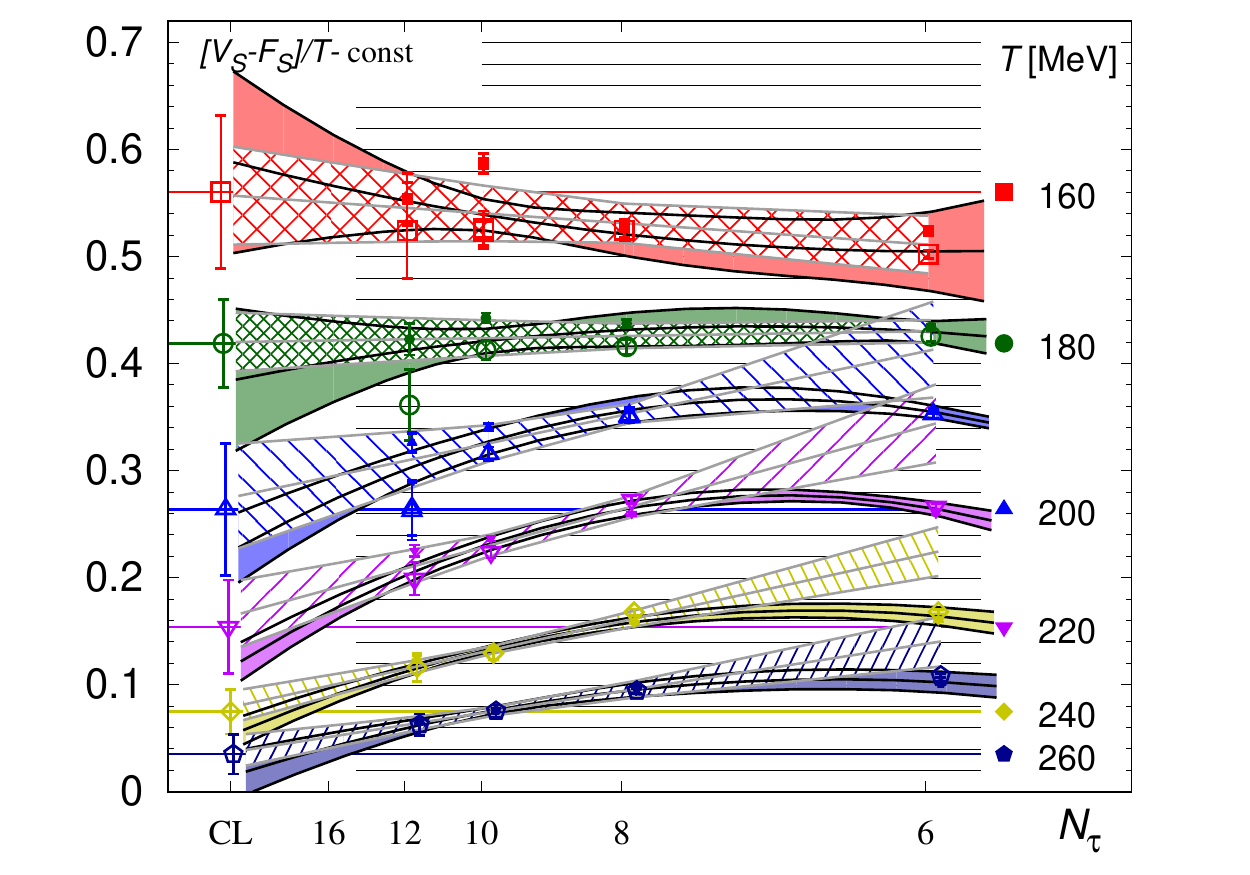}
 \includegraphics[height=5.2cm,clip]{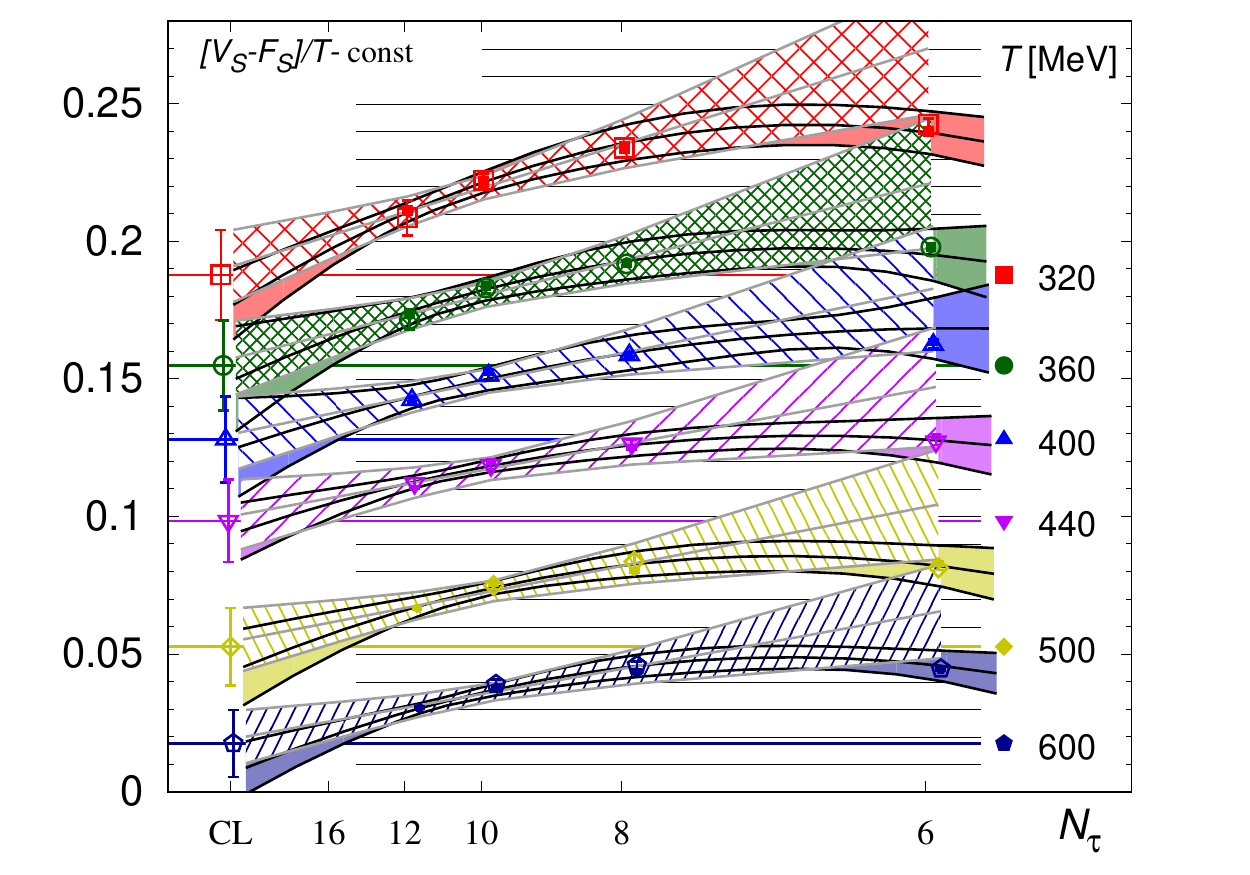}
\caption{\label{fig:vmfext}
Scaling behavior of \([\Vs-\Fs]/T\) in at various temperatures and at two 
representative separations (from top to bottom \(rT=0.17\) and~\(0.35\)) as 
function of \(N_\tau\). 
``CL'' marks the continuum limit \((N_\tau \to \infty)\). 
Results for each temperature are shifted by constants for better 
visibility. 
The bands with filled pattern indicate fits with \(b/N_\tau^2\). 
The solid bands indicate fits with  \(c/N_\tau^4\). 
The large open (small filled) symbols indicate results from local (global) 
fits. 
}
\end{figure*}

We extrapolate the subtracted free energies \(\Fqq^{\rm sub}\) and 
\(\Fs^{\rm sub}\), the derivative \(\partial \Fs/\partial (rT)\) (and 
thereby \(\aqq[\Fs]\)) and \(\Vs-\Fs\) to the continuum limit. 
With all continuum extrapolations, we attempt to cover as much of the 
\((rT,T)\) plane as we can control with reasonable uncertainties. 
In particular, since we are interested in the different regimes of the 
\(\qbq\) interactions, we have to anticipate a drastic variation of the 
cutoff effects of each quantity depending on the region of the \((rT,T)\) plane. 
Moreover, depending on the region, we have very different subsets of the data available.
Namely, for small separations, we can use only fine lattices with large 
\(N_\tau\), and for large separations, we have only data with smaller \(N_\tau\) 
due to the increasing noise problem with larger \(N_\tau\). 
For these reasons, we must assume that a global continuum extrapolation may 
fail to reflect the associated uncertainties of the data as well as the 
richness of the underlying physical structures appropriately. 
Instead, we perform independent continuum extrapolations using the averaged, interpolated results for each point of the \((rT,T)\) plane. 

\begin{table} 
\parbox{.98\linewidth}{
  \begin{tabular}{|c|c|c|}
    \hline
    \(\min(N_\tau)\) & \(\max(N_\tau)\) & \(p\)  \\
    \hline
    8,~10 & 12,~16 & 0 \\
    6,~8,~10 & 12,~16 & 2 \\
    4,~6 & 12,~16 & 4 \\
    \hline
  \end{tabular}
  \caption{\label{tab:cont sets}
Different types of continuum extrapolations. 
\(p\) is the highest inverse power of \(N_\tau\) considered, \mbox{i.e.}, 
\(p=0\) corresponds to only the coefficient \(a\), \(p=2\) corresponds to 
only coefficients \(a\) and~\(b\), and \(p=4\) corresponds to all 
coefficients \(a,~b\) and \(c\) in \mbox{Eq.}~\eqref{eq:extra}.
  }
}
\end{table}

Because of the structure of the HISQ/tree action, the leading cutoff effects 
are $\mathcal{O}(\as a^2,a^4)$. 
Since we take the continuum limit at fixed temperature and $T=1/(aN_\tau)$, 
this translates into a parametrization of leading cutoff effects through 
inverse powers of $N_\tau^2$. 
We use the \textit{Ansatz} 

\ileq{\label{eq:extra}
O(rT,T,N_\tau)=a(rT,T) +b(rT,T)/N_\tau^2 + c(rT,T)/N_\tau^4
}

and vary the set of $N_\tau$ values that are included in the fit. 
Furthermore, we either use just the \(a\) term (assuming data are already 
in the continuum limit within errors), or include the \(b/N_\tau^2\) term, 
assuming \(a^2\) scaling. 
In extrapolations that include data with \(N_\tau=6\) or~\(4\), we also 
include the \(c/N_\tau^4\) term. 
As we perform independent continuum extrapolations, we do not make any 
assumptions about the \((rT,T)\) dependence of \(a,~b\), and~\(c\). 
For this reason, our continuum extrapolations do not necessarily yield a 
smooth function in the \((rT,T)\) plane. 
We list the different types of fits used for continuum extrapolation in 
\mbox{Table}~\ref{tab:cont sets}.
If the number of data with different \(N_\tau\) exceeds the number 
of parameters of the interpolating function, we perform a linear regression 
fit and directly propagate the errors. 
After propagating the errors through the linear regression, we rescale them
by \(1/\sqrt{\cdf}\) in cases where we obtain \(\cdf<1\). 
Occasionally, this may lead to large spikes in the error of the continuum 
result if the underlying data are aligned particularly well due to 
fluctuations. 
Otherwise, we estimate the errors of the continuum limit with bootstrap 
mock data. 
We instead omit \(c/N_\tau^4\) if we have only three different \(N_\tau\) 
available. 

We have to consider the possibility that the sequential interpolations or the 
global fits together with the average and systematic error estimates may 
still underestimate the actual uncertainties of the correlators. 
If necessary, we enlarge the errors (of the averaged data underlying the 
continuum extrapolation) to reach $\cdf \leq 1.2$. 
For each datum, we determine in the \((rT,T)\) plane the closest point of the 
nonperturbatively corrected data set for the same \(N_\tau\). 
We permit enlarging the relative error of the averaged result up to the 
relative error of this closest corrected original datum. 
We simultaneously enlarge errors of the results for all \(N_\tau\) in four 
steps up to the permissible maximum for each and repeat the fit. 

In Figs.~\ref{fig:fqqext}--\ref{fig:vmfext}, we show the approach to 
the continuum limit for the various free energy observables, usually in terms 
of a ratio of the lattice result over the continuum limit. 
We show two representative distances: one in the vacuumlike regime 
and one in the electric screening regime.
We have shifted the ratios for different temperatures vertically for better 
visibility. 
The colored horizontal lines indicate the continuum value for each temperature 
to facilitate reading off the size of the cutoff effects for each \(N_\tau\).

\begin{table} 
\parbox{.98\linewidth}{
  \begin{tabular}{|c|c|c|}
    \hline
    \(rT;T {\rm [MeV]}\) & \([\min(N_\tau),\max(N_\tau)],p\) & \(\Delta_{\rm syst}\) \\
    \hline
    \((0.00-0.10]; [140-1400]\) & \([10,16],0\) or \([10,16],2\) & 0.5 \\
    \((0.10-0.12),[140-1800]\) & \([10,16],0\) or \([8,16],2\) & 1.0 \\
    \([0.12-0.15],[140-2200]\) & \([10,16],0\) or \([8,16],2\) & 2.0 \\
    \hline
    \((0.15-0.20],[140-200]\) & \([8,12],2\) or \([6,12],4\) & 0.5 \\
    \((0.15-0.20],(200-400)\) & \([8,16],0\) or \([6,12],2\) & 0.5 \\
    \((0.15-0.20],[400-1400)\) & \([8,16],2\) or \([6,16],2\) & 0.5 \\
    \((0.15-0.20],[1400-2200]\) & \([8,16],0\) or \([6,16],2\) & 0.5 \\
    \hline
    \((0.20-0.30],[140-200)\) & \([6,12],2\) or \([6,12],4\) & 0.5 \\
    \((0.20-0.30],[200-400)\) & \([6,12],2\) or \([4,12],2\) & 0.5 \\
    \((0.20-0.30],[400-1400)\) & \([8,16],2\) or \([6,16],2\) & 0.5 \\
    \((0.20-0.30],[1400-2200]\) & \([8,16],0\) or \([6,16],2\) & 0.5 \\    
    \hline
    \((0.30-0.40],(180-200)\) & \([6,12],2\) &  \\
    \((0.30-0.40],[200-1800]\) & \([6,12],2\) or \([4,12],2\) & 0.5 \\
    \hline
    \((0.40-0.50],[260-800]\) & \([6,12],2\) or \([4,16],4\) & 0.5 \\    
    \hline
  \end{tabular}
  \caption{\label{tab:fqqext}
  Final continuum extrapolation of \(\Fqq^{\rm sub}\). 
  We list \(rT\) and \(T\) ranges in the first column, suitable 
  extrapolations in the second column, and weights for the difference 
  (if applicable) as systematic error estimates in the third.
  }
}
\end{table}

We show the approach of $\Fqq^{\rm sub}$ to the continuum limit for a few 
representative pairs $(rT,T)$ in \mbox{Fig.}~\ref{fig:fqqext}. 
In general, we see that within an \(a^2\)-scaling window $\Fqq^{\rm sub}$ 
decreases towards the continuum limit and that \(a^4\)-scaling contributions 
come with  opposite sign. 
Within numerical uncertainties, \(\Fqq^{\rm sub}\) with \(N_\tau=6\) is always 
inside the \(a^2\)-scaling window for \(T > 180\,{\rm MeV}\), while 
\(\Fqq^{\rm sub}\) with \(N_\tau=4\) is never inside the \(a^2\)-scaling window.
In general, we see that the approach to the continuum limit becomes steeper 
for lower temperatures and larger separations and flatter for higher 
temperatures and shorter separations, \mbox{i.e.}, cutoff effects become 
larger from the upper left to the lower right corner of the \((rT,T)\) plane. 
For small distances up to \(rT\sim0.15\), cutoff effects are at most a few 
percent or not even unambiguously resolved at all. 
For intermediate distances up to \(rT\sim0.30\), cutoff effects grow up to 
about 20\% in the transition region and 5\% for temperatures beyond \(1\,{\rm GeV}\).
For large distances beyond \(rT\sim0.30\), cutoff effects are even larger, 
30\% and more for \(N_\tau=6\), and stay at such level even for the highest 
temperatures. 
We summarize in Table~\ref{tab:fqqext} which continuum extrapolations we 
combined in our final result for \(\Fqq^{\rm sub}\).

\begin{table}
\parbox{.98\linewidth}{
  \begin{tabular}{|c|c|c|}
    \hline
    \(rT;T {\rm [MeV]}\) & \([\min(N_\tau),\max(N_\tau)],p\) & \(\Delta_{\rm syst}\) \\
    \hline
    \((0.00-0.12),[140-1800]\) & \([10,12],0\) or \([8,16],2\) & 1.0 \\
    \hline
    \([0.12-0.15],[140-180)\) & \([8,12],2\) &  \\
    \([0.12-0.15],[180-200]\) & \([10,12],0\) or \([8,12],2\) & 0.5 \\
    \([0.12-0.15],(200-600]\) & \([10,16],0\) or \([816],2\) & 1.0 \\
    \([0.12-0.15],(600-2200]\) & \([10,16],0\) or \([8,16],2\) & 0.5 \\
    \hline
    \((0.15-0.20],[140-180]\) & \([8,12],2\) or \([6,12],4\) & 0.5 \\
    \((0.15-0.20],(180-600]\) & \([8,16],2\) or \([4,12],4\) & 0.5 \\
    \((0.15-0.16],(600-2200)\) & \([6,16],0\) &  \\
    \((0.16-0.20],[600-2200]\) & \([8,16],0\) or \([6,16],2\) & 0.5 \\
    \hline
    \((0.20-0.30],[140-200)\) & \([6,12],2\) or \([6,12],4\) & 0.5 \\
    \((0.20-0.30],[200-2200]\) & \([6,12],2\) or \([4,12],4\) & 0.5 \\
    \hline
    \((0.30-0.45],(160-200)\) & \([8,12],2\)  or \([6,12],2\) & 0.5 \\
    \((0.30-0.45],[200-1800]\) & \([8,12],2\) or \([4,12],4\) & 0.5 \\
    \((0.30-0.45],(1800-2200]\) & \([6,12],2\) or \([4,12],2\) & 0.5 \\
    \hline
    \((0.45-0.70],[200-2200]\) & \([6,12],2\) or \([4,12],4\) & 0.5 \\
    \((0.70-\ldots],[400-2200]\) & \([6,12],2\) or \([4,12],4\) & 0.5 \\
    \hline
  \end{tabular}
  \caption{\label{tab:fsext}
  Final continuum extrapolation of \(\Fs^{\rm sub}\).
  We list \(rT\) and \(T\) ranges in the first column, suitable 
  extrapolations in the second column, and weights for the difference 
  (if applicable) as systematic error estimates in the third.
  }
}
\end{table}

We show the approach of $\Fs^{\rm sub}$ to the continuum limit for the same 
representative pairs $(rT,T)$ in \mbox{Fig.}~\ref{fig:fsext}. 
As for \(\Fqq^{\rm sub}\) we see that within an \(a^2\)-scaling window 
$\Fs^{\rm sub}$ decreases towards the continuum limit and that 
\(a^4\)-scaling contributions come with opposite sign. 
Within numerical uncertainties, \(\Fs^{\rm sub}\) with \(N_\tau=6\) is always 
inside the \(a^2\)-scaling window for \(T > 180\,{\rm MeV}\), while 
\(\Fs^{\rm sub}\) with \(N_\tau=4\) is never inside the \(a^2\)-scaling window.
In general, we see as for \(\Fqq^{\rm sub}\) that the approach to the 
continuum limit becomes steeper for lower temperatures and larger 
separations and flatter for higher temperatures and shorter separations, 
\mbox{i.e.} cutoff effects become larger from the upper left to the 
lower right corner of the \((rT,T)\) plane.  
For small distances up to \(rT\sim0.15\), cutoff effects are at most a few 
percent or not even unambiguously resolved at all. 
In particular, the coefficient \(b\) seems to switch to the opposite sign above 
\(T\sim 600\,{\rm MeV}\) and \(rT \ll0.2\). 
We consider this as an indication that these data may be considered in the 
continuum limit within errors.
For intermediate distances up to \(rT\sim0.30\), cutoff effects grow up to 
about 15\% in the transition region and 4\% for temperatures beyond 
\(1\,{\rm GeV}\).
For larger distances beyond \(rT\sim0.30\), cutoff effects are even larger 
and  gradually increase up to 40\% for \(N_\tau=6\) at \(rT\sim 0.60\).
For the highest temperatures cutoff effects are about half this size. 
We summarize in Table~\ref{tab:fsext} which continuum extrapolations we 
combined in our final result for \(\Fs^{\rm sub}\).

\begin{table}
\parbox{.98\linewidth}{
  \begin{tabular}{|c|c|c|}
    \hline
    \(rT;T {\rm [MeV]}\) & \([\min(N_\tau),\max(N_\tau)],p\) & \(\Delta_{\rm syst}\) \\
    \hline
    \((0.00-0.12),[140-1800]\) & \([10,12],0\) or \([8,16],2\) & 1.0 \\
    \hline
    \([0.12-0.15],[140-180)\) & \([10,12],0\) &  \\
    \([0.12-0.15],[180-200]\) & \([10,12],0\) or \([8,12],2\) & 0.5 \\
    \([0.12-0.15],(200-600]\) & \([10,16],0\) or \([8,16],2\) & 1.0 \\
    \([0.12-0.15],(600-2200]\) & \([10,16],0\) or \([8,16],2\) & 0.5 \\
    \hline
    \((0.15-0.20],[140-180]\) & \([8,12],2\) or \([6,12],4\) & 0.5 \\
    \((0.15-0.20],(180-300]\) & \([8,12],0\) or \([6,12],2\) & 1.0 \\
    \((0.15-0.20],(300-1400]\) & \([8,16],2\) or \([4,12],4\) & 0.5 \\
    \((0.15-0.20],(1400-2200]\) & \([8,12],0\) or \([6,12],2\) & 0.5 \\
    \hline
    \((0.20-0.30],[140-200)\) & \([6,12],2\) or \([6,12],4\) & 0.5 \\
    \((0.20-0.30],[200-2200]\) & \([6,12],2\) or \([4,12],4\) & 0.5 \\
    \hline
    \((0.30-0.45],(160-200)\) & \([8,12],2\)  or \([6,12],2\) & 0.5 \\
    \((0.30-0.45],[200-1800]\) & \([8,12],2\) or \([4,12],4\) & 0.5 \\
    \((0.30-0.45],(1800-2200]\) & \([6,12],2\) or \([4,12],2\) & 0.5 \\
    \hline
    \((0.45-0.70],[200-2200]\) & \([6,12],2\) or \([4,12],4\) & 0.5 \\
    \((0.70-\ldots],[400-2200]\) & \([6,12],2\) or \([4,12],4\) & 0.5 \\
    \hline
  \end{tabular}
  \caption{\label{tab:aqqext}
  Final continuum extrapolation of \(\aqq[\Fs]\).
  We list \(rT\) and \(T\) ranges in the first column, suitable 
  extrapolations in the second column, and weights for the difference 
  (if applicable) as systematic error estimates in the third.
  }
}
\end{table}

We show the approach of $\aqq[\Fs]$ to the continuum limit for the same 
representative pairs $(rT,T)$ in \mbox{Fig.}~\ref{fig:aqqext}. 
Contrary to \(\fs\) we see that within an \(a^2\)-scaling window $\aqq[\Fs]$ 
increases towards the continuum limit at short distances, \(rT\ll0.4\), while 
it tends to decrease towards the continuum limits for larger distances. 
This change in behavior is clearly due to the physical regimes, namely, 
the vacuumlike regime and screening regime. 
The \(a^4\)-scaling contributions always lead to an increase towards the 
continuum limit.
Within numerical uncertainties, \(\aqq[\Fs]\) with \(N_\tau=6\) is always 
inside the \(a^2\)-scaling window. 
Even \(\aqq[\Fs]\) with \(N_\tau=4\) is close to the \(a^2\)-scaling window 
for the smallest separations. 
In general, we see for \(\aqq[\Fs]\) that the approach to the continuum limit 
is different in the vacuumlike regime and in the screening regime.
In the vacuumlike regime, it seems to become slightly steeper for higher 
temperatures and some intermediate separations, while in the screening regime, 
it behaves rather in the opposite way. 
For small and intermediate distances up to \(rT\sim0.30\), cutoff effects are 
at most about 5\% percent or not even unambiguously resolved at all. 
For larger distances beyond \(rT\sim0.30\), cutoff effects eventually change 
their sign and become even larger and similar to those of \(\fs\) in terms 
of sign and magnitude. 
We summarize in Table~\ref{tab:aqqext} which continuum extrapolations we 
combined in our final result for \(\aqq[\Fs]\).

\begin{table}
\parbox{.98\linewidth}{
  \begin{tabular}{|c|c|c|}
    \hline
    \(rT;T {\rm [MeV]}\) & \([\min(N_\tau),\max(N_\tau)],p\) & \(\Delta_{\rm syst}\) \\
    \hline
    \((0.00-0.125),[140-180)\) & \([10,12],0\) or \([10,12],2\) & 0.5 \\
    \((0.00-0.125),[180-600]\) & \([10,16],0\) &  \\
    \hline
    \([0.125-0.17],[140-180)\) & \([8,12],2\) &  \\
    \([0.125-0.17],[180-200]\) & \([10,12],0\) or \([8,16],2\) & 0.5 \\
    \([0.125-0.17],(200-600]\) & \([10,16],2\) or \([8,16],2\) & 1.0 \\
    \([0.125-0.17],(600-800]\) & \([10,12],0\) or \([8,12],2\) & 1.0 \\
    \hline
    \((0.17-0.20],[140-180]\) & \([10,12],2\) or \([8,12],2\) & 0.5 \\
    \((0.17-0.20],(180-600]\) & \([10,12],2\) or \([8,12],2\) & 0.5 \\
    \((0.17-0.20],(600-800]\) & \([10,12],2\) or \([6,12],4\) & 0.5 \\
    \((0.17-0.20],(800-1000]\) & \([8,12],2\) or \([6,12],4\) & 0.5 \\
    \hline    
    \((0.20-0.30],[140-200)\) & \([6,12],2\) or \([6,12],4\) & 0.5 \\
    \((0.20-0.30],[200-1000]\) & \([8,12],2\) or \([6,12],4\) & 0.5 \\
    \hline
    \((0.30-0.45],[140-200)\) & \([8,12],2\)  or \([6,12],2\) & 0.5 \\
    \((0.30-0.45],[200-1800]\) & \([8,12],2\) or \([6,12],4\) & 0.5 \\
    \hline
  \end{tabular}
  \caption{\label{tab:vmfext}
  Final continuum extrapolation of \(\Vs-\Fs\).
  We list \(rT\) and \(T\) ranges in the first column, suitable 
  extrapolations in the second column, and weights for the difference 
  (if applicable) as systematic error estimates in the third.
  }
}
\end{table}

Lastly, we show the approach of $\Vs-\Fs$ to the continuum limit for a 
similar set\footnote{
Here we use bare distances instead of improved distances. 
Hence, \(rT=0.16\) is not accessible with \(N_\tau=6\). 
The temperature range is limited by the \(\beta\) values of the available 
\(T=0\) lattices.} 
of representative pairs $(rT,T)$ in \mbox{Fig.}~\ref{fig:vmfext}. 
We cannot plot the ratio of results at finite lattice spacing and the 
continuum result, since data are actually crossing zero at different 
points in the \((rT,T)\) plane. 
In general, we see that within an \(a^2\)-scaling window $\Vs-\Fs$ decreases 
towards the continuum limit and that \(a^4\)-scaling contributions come with 
the opposite sign. 
Contrary to \(\Fs^{\rm sub}\), we see that the \(a^2\)-scaling window extends 
only to \(N_\tau=8\). 
Although \(\Vs-\Fs\) with \(N_\tau=6\) appears to be systematically outside 
the \(a^2\)-scaling window, the \(a^4\)-scaling contribution is rather mild. 
In general, we see for \(\Vs-\Fs\) that the approach to the continuum limit 
appears very flat for low temperatures, \(T<200\,{\rm MeV}\), and quite 
temperature independent for \(T>200\,{\rm MeV}\). 
Cutoff effects become systematically larger for larger separations. 
For small distances up to \(rT\sim0.15\), cutoff effects are not even 
unambiguously resolved and data at finite \(N_\tau\) fluctuate around the 
continuum result without clear scaling behavior. 
We consider this as an indication that these data may be considered in 
the continuum limit within errors.
For intermediate distances up to \(rT\sim0.20\) and temperatures above 
\(200~{\rm MeV}\), cutoff effects grow in size from \(0.01\,T\) to 
\(0.03\,T\) with increasing separation and show clear scaling behavior. 
For larger distances around \(rT\sim0.30\) and beyond, cutoff effects seem 
to stay at a similar level.
We summarize in Table~\ref{tab:vmfext} which continuum extrapolations we 
combined in our final result for~\(\Vs-\Fs\).

We universally find that cutoff effects at the shortest distances that we 
can access with multiple \(N_\tau\) are consistent with zero within errors.
Thus, we extend the continuum result to even shorter distances by taking the 
results for the (respective) largest \(N_\tau\) in the extremely short range 
as a continuum estimate. 
These continuum estimates are included in the continuum results that are 
plotted in the figures in the main body of the paper.

\subsection{Continuum estimates for \(T=0\) results}
For the continuum estimates of the zero temperature static energy we use 
the nonperturbatively improved results at finite lattice spacing. 
We use the lattices in Table~\ref{tab:T=0} and the improvement procedure 
discussed in Appendix~\ref{app:B}. 

After nonperturbative improvement, the results at different lattice spacings 
are consistent within their respective errors. 
Therefore, in the range where we have results available at multiple \(\beta\), 
each of these results is within its uncertainties in the continuum limit. 
For the shortest distances, \(r<0.029\,{\rm fm}\), we assume that the 
nonperturbatively improved results on the finest lattice are consistent 
with the continuum limit and treat these as our continuum estimate. 
Instead of following a similar approach as we did for \(T>0\) 
(\mbox{cf. Sec.}~\ref{sec:continuum}), we construct our entire continuum 
estimate for \(T=0\) from nonperturbatively improved results at finite 
lattice spacing.
The three reasons for this change of procedure are that we do not have to 
consider \(T\)- or \(rT\)-dependent cutoff effects, that the results on finer 
lattices anyways do not contribute strongly to the continuum limit at larger 
distances, and that the finest lattices have a larger light sea quark mass 
than the coarser lattices. 
The second reason can be understood from the fact that such distances 
correspond to rather large values of \(r/a\) on finer lattices, and thus, 
results on finer lattices have accumulated more noise at the same distance. 
We construct our continuum estimate by using only points with 
\(r/a \lesssim 6\) for each lattice spacing to keep the errors small.  
Quark mass dependence as the third reason has been discussed already in 
\mbox{Appendix}~\ref{app:B}. 
By restricting the lattice results with larger quark mass to distances 
much smaller than \(0.8\,r_1\), we avoid that the quark mass dependence is 
numerically important. 

\begin{table}
\parbox{.98\linewidth}{
  \begin{tabular}{|c|c|c|c|}
    \hline
    \(r_I\ {\rm (fm)}\) & \(r/a\) & \(a\ {\rm(fm)}\)& \(\beta\) \\
    \hline
    \((0.024  -0.100)\) & \([1,\sqrt{17}]\) & 0.025 & 8.400 \\
    \((0.070-0.150)\) & \([2,\sqrt{19}]\) & 0.035 & 8.000 \\
    \((0.100-0.220)\) & \([\sqrt{6},\sqrt{30}]\) & 0.040 & 7.825 \\
    \((0.200-0.360)\) & \([\sqrt{9},\sqrt{30}]\) & 0.065 & 7.280 \\
    \((0.320-0.650)\) & \([\sqrt{9},\sqrt{37}]\) & 0.109 & 6.740 \\
    \hline
  \end{tabular}
  \caption{\label{tab:tzero bands}
  Bands for the \(T=0\) results in the figures of the main body are 
  constructed from nonperturbatively improved \(\Vs\) at finite lattice 
  spacing in overlapping regions. 
  The two finest lattices have larger sea quark mass, namely, \(m_l=m_s/5\), 
  the other three lattices have \(m_l=m_s/20\).
  }
}
\end{table}

For the \(T=0\) results in the figures we obtain two bands as follows. 
We use results with five different lattice spacings, namely, with 
\(\beta=8.4,~8.0,~7.825,~7.28\), and \(6.74\). 
We plot the nonperturbatively corrected results for \(\Vs\) for each
lattice spacing in a different \(r\) range to ensure that we always use data 
at only a small number of lattice steps, \(r/a \lesssim 6\). 
For lattices with \(m_l=m_s/5\), we use even shorter separations, 
\(r/a \lesssim 4\). 
For the coarser lattices, we leave out the first few points, where cutoff 
effects are somewhat large, and thus, the nonperturbative improvement 
introduces a larger systematic error. 
The reason for these short distance cuts is that we aim at obtaining a 
smooth error band. 
We summarize this construction in Table~\ref{tab:tzero bands} and show the 
corresponding \(T=0\) result with black pentagons and a black band in 
\mbox{Fig.}~\ref{fig:main}. 
For the \(T=0\) result in Figs.~\ref{fig:aqq} and~\ref{fig:aqqa}, we use a 
derivative of this \(T=0\) result obtained numerically from a fit with an 
\textit{Ansatz} as a modified Cornell potential (see \mbox{App.}~\ref{app:B} for a 
discussion of these fits). 

However, this is still not enough for a comparison of \(\Fqq\) from high 
temperatures to the \(T=0\) limit. 
For this comparison, we make use of information obtained from the direct 
comparison between \(\Vs\) and \(\Fs\) in \mbox{Sec.}~\ref{sec:vminusf}.
Figure~\ref{fig:vs-fs} shows that \([\Vs-\Fs]/T\) is for sufficiently small 
\(rT\) within numerical uncertainties constant and independent of the 
temperature for \(T \gtrsim 300\,{\rm MeV}\). 
This constant is small compared to the numerical uncertainty introduced due 
to the renormalization constant \(2C_Q\) at high \(\beta\) values. 
Hence, within the numerical errors, the renormalized result for \(\Fs\) is a 
fair estimate of \(\Vs\) as long as \(T \gg 300\,{\rm MeV}\) and \(r \ll 0.2/T\). 
We employ the continuum result for \(\Fs\) in the range \(0.09\leq rT<0.17\) 
for \(T\geq 800\,{\rm MeV}\), namely, for the temperatures 
\(T=0.8,\ 1.0,\ 1.2,\ 1.4,\ 1.6,\ 1.8,\ 2.0\), and \(2.2\,{\rm GeV}\), as an 
estimate for the zero temperature result at even shorter distances, reaching 
down to \(r \sim 0.01\,{\rm fm}\). 
This estimate for \(V_S\) is shown as a light gray band 
in \mbox{Fig.}~\ref{fig:main}. 

\section{FITS OF THE LARGE DISTANCE BEHAVIOR OF THE CORRELATORS}
\label{app:D}

\begin{figure}
\centering
 \includegraphics[height=5.8cm,clip]{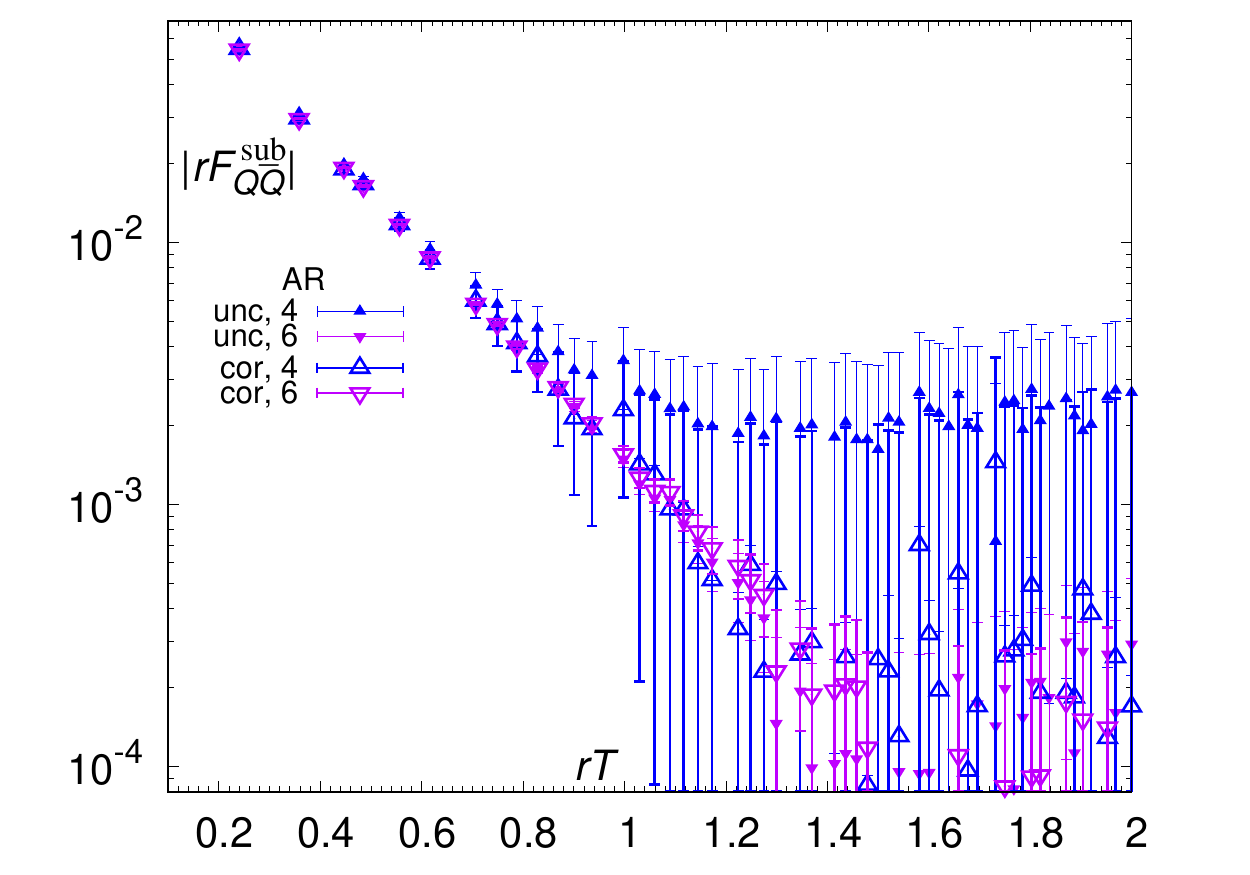}
\caption{\label{fig:vol sub S1}
Correction for finite volume [\(\beta=6.664\), \(N_\tau=4\), and aspect 
ratio (AR) $= N_\sigma/N_\tau$, equal to $4$ and $6$]. 
Uncorrected and finite-volume corrected data are shown with smaller filled and 
larger open symbols respectively. 
}
\end{figure}

In this Appendix, we discuss the details of the analysis of the asymptotic 
behavior of the subtracted free energies and the extraction of the asymptotic 
screening masses. 
For this analysis, we perform fits on all jackknife bins to account for 
correlated fluctuations of neighboring data points.  
The discussion of finite volume effects in Appendix~\ref{app:A} revealed 
that screening functions for static quark-antiquark free energies are quite 
susceptible to finite volume corrections. 
Thus, we fit $F^{\rm sub}$ for large separations as 

\ileq{\label{eq:oge ansatz}
 F^{\rm sub}(rT,\ldots) = C-\frac {AT}{rT} \exp[-M rT],
}

assuming that the one-particle exchange is the dominant process for large 
separations. 
The asymptotic cancellation between the squared Polyakov loop and the 
static quark-antiquark correlators strictly holds only in infinite volume. 
Hence, we account with the fit parameter \(C\) for possible offsets due to 
an incomplete cancellation with \(2\Fq\) at finite volume. 
We determine \(C\) by fitting a constant to the asymptotic tail of 
\(F^{\rm sub}\). 
We define this tail as made of all points with signal-to-noise ratio less 
than one and of all points at larger \(rT\).
By subtracting the parameter \(C\) from \(F^{\rm sub}\) we account for the 
asymptotic finite volume corrections and bring results with different volumes 
into much better numerical agreement. 

For \(\beta=6.664\), we show the effect of the correction on 
\(\Fqq^{\rm sub}\) in \mbox{Fig.}~\ref{fig:vol sub S1}.
There, we obtain a good signal in the larger volume up to \(rT\approx 1.3\). 
After the finite volume correction, the central value of the result in the 
smaller volume sits on the top of the finite volume corrected result in the 
larger volume (albeit with large errors that render analysis of the last few 
data points moot). 
For different \(\beta\), the picture is quite similar. 
With the singlet free energy, we can access much larger ranges, but the main 
conclusions stay the same (\mbox{cf. Table}~\ref{tab:vol} for the uncorrected 
results). 

Thus, we conclude that we may consider the other two fit parameters \(A\) and 
\(M\) as volume independent within errors. 
We determine \(\ln[rT(F^{\rm sub}-C)]= -\ln[AT]-M rT\) and fit in the range 
up to the first (uncorrected point with signal-to-noise ratio less than one) 
through linear regression.
By varying the minimal and maximal separations \((rT)_{\rm min}\) and 
\((rT)_{\rm max}\) that are included in the fit, we obtain a local 
definition of the two parameters \(A\) and \(M\) on each jackknife bin.  
We propagate the errors on each bin via regression and enlarge the errors 
by \(1/\sqrt{\cdf}\) if \(\cdf<1\). 
We neglect results where regression errors of either parameter are 100\%. 
We propagate the statistical errors with the jackknife method and add the 
estimate of the bias in quadrature. 
Finally, we calculate for each \((rT)_{\rm min}\) a weighted average of the 
parameters obtained with different \((rT)_{\rm max}\) and estimate the error 
by adding the average jackknife error and an estimate for the bias due to 
dependence on \((rT)_{\rm max}\) in quadrature. 
We interpret the fit parameter $M$ as the screening mass in units of the temperature. 

\begin{figure}
\centering
 \includegraphics[height=5.8cm,clip]{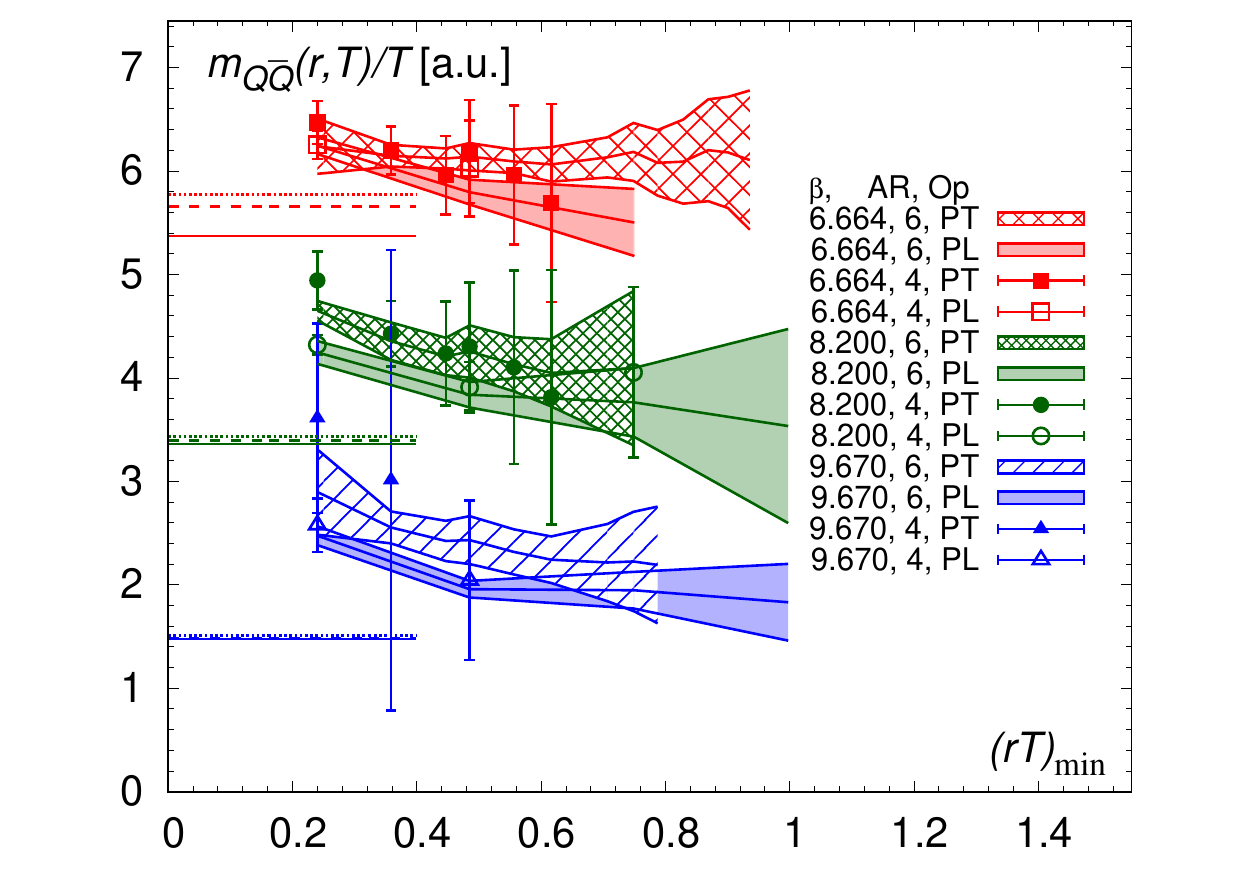}
\caption{\label{fig:vol mqq}
Volume dependence of the screening mass $m_{\qbq}$ from $\Fqq^{\rm sub}$. 
$\beta=6.664$ corresponds to a moderately high temperature 
($T=421\,{\rm MeV}$), $\beta=8.200$ corresponds to a high temperature 
($T=1687\,{\rm MeV}$), and $\beta=9.670$ corresponds to the highest temperature in the study ($T=5814\,{\rm MeV}$). 
We show the screening mass in units of the temperature, \(m_{\qbq}/T\). 
We shifted \(m_{\qbq}/T\) by a constant \(\pm1.5\) at the lowest or highest 
of the three temperatures for better visibility. 
The screening mass in the larger volume is shown with patterned bands for Polyakov loop point sources and with solid bands for Polyakov loop plane sources.
The screening mass in the smaller volume is shown with filled symbols for Polyakov loop point sources and with open symbols for Polyakov loop plane sources.
The triplets of horizontal lines indicate the naive weak-coupling 
expectation \(2\md/T\) for the scales \(\mu=\pi\,T,~2\pi\,T\), and 
\(4\pi\,T\) (solid, dotted, and dashed) using the NLO result [\mbox{Eq.}~\eqref{eq:md nlo}]. 
}
\end{figure}

We show the local screening mass $m_{\qbq}/T$ from $\Fqq^{\rm sub}$ in \mbox{Fig.}~\ref{fig:vol mqq}. 
There is fair numerical agreement between the screening masses for both aspect ratios.
For the smaller volume, we fail to determine $m_{\qbq}/T$ reliably 
for the highest temperature, which is caused by the much smaller signal 
and, thus, the much worse signal-to-noise ratio. 
At lower temperatures, the agreement between $m_{\qbq}/T$ for different 
volumes is quite remarkable, in particular, the screening mass assumes 
a plateau already at $(rT)_{\rm min}\sim$ $0.4-0.5$, which is approached from above. 
We observe that the difference between \(m_{\qbq}/T\) for \((rT)_{\rm min}=0.25\) and for \((rT)_{\rm min}=0.5\) is about 0.5. 
We use this difference to correct for screening masses \(m_{\qbq}/T\) 
determined at \((rT)_{\rm min}=0.25\) on lattices with aspect ratio 
\(N_\sigma/N_\tau=4\) and assign a systematic uncertainty of \(0.1\) to the corrected result. 
We show the corrected results in \mbox{Fig.}~\ref{fig:mqq T}, where we include the 
results for aspect ratio \(N_\sigma/N_\tau=6\) directly at \((rT)_{\rm min}=0.5\) as large open symbols. 
The screening mass is actually slightly above 2 times the perturbative 
Debye mass at NLO [\mbox{Eq.}~\eqref{eq:md nlo}], indicated by the horizontal 
lines in the lower left corner for the scales $\mu=\pi T,\ 2\pi T$, and $4\pi T$. 
We recall that in Fig.~\ref{fig:mqq T} we compare with 2 times the NLO Debye mass times a factor \(A=1.25\).
This indicates that for \(\Fqq^{\rm sub}\) we cannot clearly distinguish 
between the electrostatic and the asymptotic screening regimes numerically.

In order to test whether we have already reached the asymptotic behavior, we 
also include plane-averaged Polyakov loop correlators in the analysis, 

\ileq{
  C_{PL}(\beta,N_\tau,z) =
   \Braket{ P(N_\tau,\bm x)P^\dagger(N_\tau,\bm x+z\hat{e}_z)}
  \label{eq:defCPplane},
}

where the Polyakov loops are integrated over the \(xy\) plane. 
Since the operators are separated only in the \(z\) direction we have \(r=z\) and \(rT=zT\) in this case.
Since the integration is only two-dimensional the plane-averaged Polyakov loop correlator behaves as 

\ileq{
\Fqq^{PL}(rT,\ldots) = C^\prime- A^\prime T \exp[-M rT]
}

in the regime of asymptotic screening, where \(C^\prime\) is a constant that diverges in the continuum limit.  
We did not subtract the asymptotic constant along the MD time history, as we 
found that it is not very helpful given numerical fluctuations in the large distance tail. 
Using the same approach as discussed for the point-to-point Polyakov loop 
correlation functions, we determine the screening mass from the plane-averaged Polyakov loop correlation function. 
Results for both operators are fairly consistent, although the plane-averaged 
correlator yields for small separations at low temperatures a 
systematically higher and at high temperatures a systematically lower screening mass by a few percent.
These differences are of similar size as the errors in the smaller volume, 
as the difference between determinations in different volumes, and as well as our 
correction for the screening mass obtained at \((rT)_{\rm min}=0.25\) [\mbox{cf.} the discussion preceding \mbox{Eq.}~\eqref{eq:defCPplane}]. 
Therefore, we consider the screening masses as consistent and asymptotic for \((rT)_{\rm min}\geq0.5\).

\begin{figure}
\centering
 \includegraphics[height=5.8cm,clip]{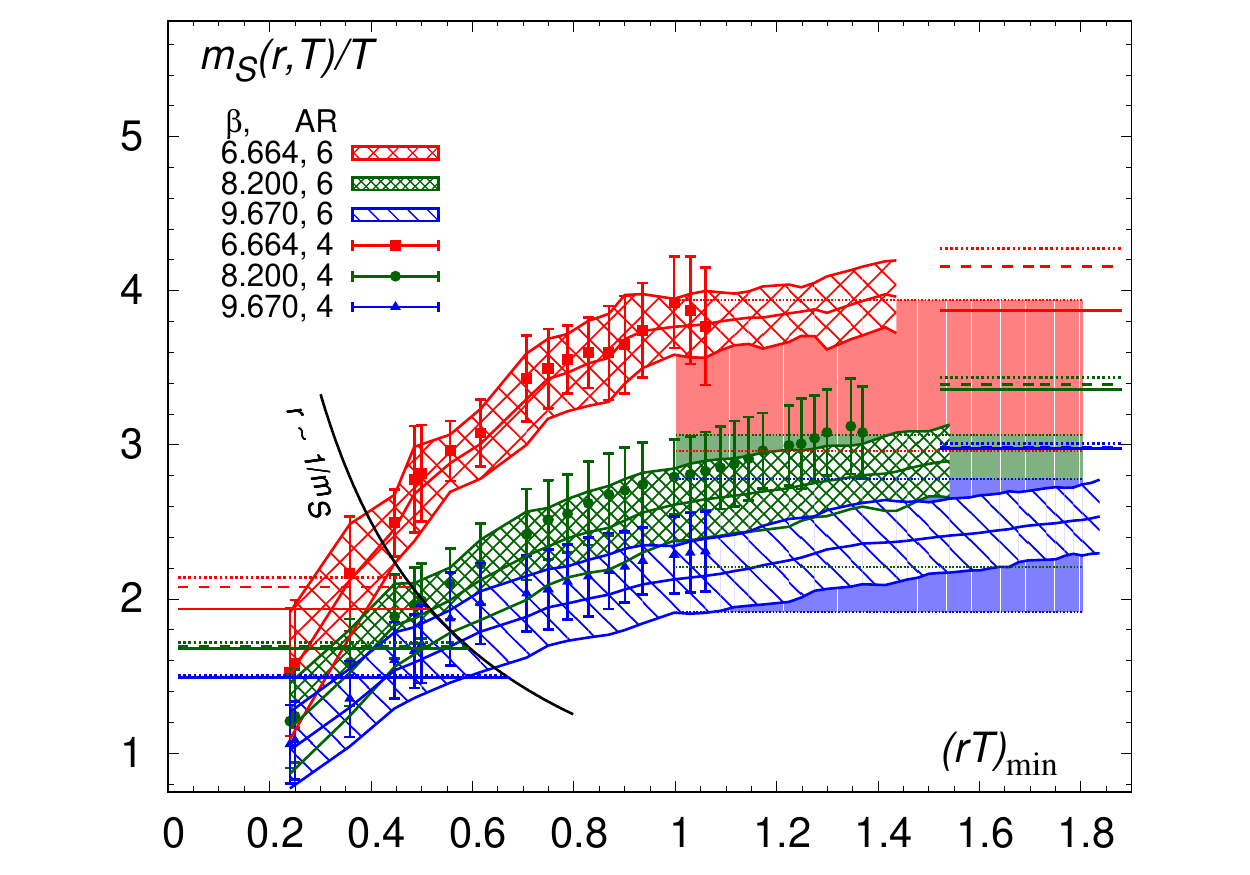}
\caption{\label{fig:vol msng}
Volume dependence of the screening mass $m_S$ from $\Fs^{\rm sub}$. 
$\beta=6.664$ corresponds to a moderately high temperature 
($T=421\,{\rm MeV}$), $\beta=8.200$ corresponds to a high temperature 
($T=1687\,{\rm MeV}$), and $\beta=9.670$ corresponds to the highest temperature in the study ($T=5814\,{\rm MeV}$). 
We show the screening mass in units of the temperature, \(m_{S}/T\). 
The screening mass in the larger volume is shown with patterned bands, while 
the screening mass in the smaller volume is shown with filled symbols.
The triplets of horizontal lines on the left side indicate the naive 
weak-coupling expectation \(\md/T\) for the scales \(\mu=\pi\,T,~2\pi\,T\), 
and \(4\pi\,T\) (solid, dotted, and dashed) using the NLO result [\mbox{Eq.}~\eqref{eq:md nlo}].
The triplets of horizontal lines on the right side indicate the naive 
weak-coupling expectation for exchange of a bound state \(2\md/T\) of two scalar modes, each with the mass \(\md/T\).  
The scales are the same. 
Lastly, the horizontal solid bands indicate the estimate of the asymptotic 
screening mass obtained by shifting the screening mass for \(rT_{\rm min}=0.5\) by a temperature-independent constant. 
}
\end{figure}

We show the local screening mass $m_{S}/T$ from $\Fs^{\rm sub}$ in \mbox{Fig.}~\ref{fig:vol msng}. 
Fits with \mbox{Eq.}~\eqref{eq:oge ansatz} as \textit{Ansatz} yield \(m_S < \md\) and \(\cdf > 1\) for \((rT)_{\rm min}<0.3\). 
Both outcomes are consistent with the observation in \mbox{Sec.}~\ref{sec:vminusf} 
that screening is partially compensated by other medium effects at short distances. 
For larger \((rT)_{\rm min}\) we always obtain \(\cdf<1\). 
$m_S/T$ for $(rT)_{\rm min}\sim$ 0.3--0.6 is within errors fairly consistent 
with the perturbative Debye mass at NLO as indicated by the three horizontal 
lines in the lower left corner of the figure representing $\md$ for the scales $\mu=\pi T,\ 2\pi T$ and $4\pi T$. 
Deviations can be attributed to the higher order terms discussed in \mbox{Sec.}~\ref{sec:electric}, which are missing in \mbox{Eq.}~\eqref{eq:oge ansatz}.
There is good numerical agreement between the screening masses for both volumes for $(rT)_{\rm min}\lesssim 1$. 
Even in the larger volume, we do not see an unambiguous plateau in \(m_S/T\) 
for any of the temperatures, although \(m_S/T\) is within errors consistent with a constant for \((rT)_{\rm min}\gtrsim1\). 
For this reason, we consider \((rT)_{\rm min}\gtrsim1\) as the asymptotic 
regime, although the onset of the asymptotic behavior seems to occur already somewhat earlier for lower temperatures.
For ensembles with poorer signal-to-noise ratio or low temperatures 
(\(T <2\tc\)), the fits may fail well before \((rT)_{\rm min}\sim1\) and we cannot determine $m_S/T$ in the asymptotic regime. 
As for the case of the \(\Fqq^{\rm sub}\), we estimate a correction to the screening mass determined at shorter distances. 
We observe that the difference between \(m_S/T\) determined for \((rT)_{\rm min}=0.5\) and \((rT)_{\rm min}=1.3\) is about 0.75.
We use this difference to correct for screening masses \(m_S/T\) 
determined at \((rT)_{\rm min}=0.5\) on lattices with aspect ratio 
\(N_\sigma/N_\tau=4\) and assign a systematic uncertainty of \(0.2\) to the corrected result. 
The corrected results are shown as the solid bands in \mbox{Fig.}~\ref{fig:vol msng} and appear to underestimate the actual 
asymptotic screening mass somewhat at low temperatures, \(T\lesssim3\tc\). 
The screening masses in \mbox{Fig.}~\ref{fig:ms T} are obtained with this procedure. 
We include the results for aspect ratio \(N_\sigma/N_\tau=6\) directly at \((rT)_{\rm min}=1.3\) as large open symbols. 
We consider this estimate of the asymptotic screening mass as sufficiently accurate for \(T>3\tc\).

\bibliography{ref}

\end{document}